# ATTRACTING COMMERCIAL ARTIFICIAL INTELLIGENCE FIRMS TO SUPPORT NATIONAL SECURITY THROUGH COLLABORATIVE CONTRACTS

ANDREW BOWNE

Submitted in fulfilment of the requirements for the degree of
Doctor of Philosophy

Adelaide Law School
The University of Adelaide
February 2023





**Disclaimer**

The views and conclusions contained in this thesis are those of the author and should not be interpreted as representing the official policies, expressed or implied, of the United States Government, the Department of Defense (DoD) or any federal agency. The appearance of external hyperlinks does not constitute endorsement by the DoD of the linked websites, or the information, products, or services contained therein. The DoD does not exercise any editorial, security, or other control over the information you may find at these locations.















# Table of Figures






ABSTRACT

The United States Department of Defense ('DoD') has determined it is not ready to compete in the Artificial Intelligence ('AI') era without significant changes to how it acquires AI. Unlike other military technologies driven by national security needs and developed with federal funding, this ubiquitous technology enabler is predominantly funded and advanced by commercial industry for civilian applications. However, there is a lack of understanding of the reasons commercial AI firms decide to work with the DoD or choose to abstain from the defence market.

Although there are several challenges to attracting commercial AI firms to support national security, this thesis argues that the DoD's contract law and procurement framework are among the most significant obstacles. This research indicates that the commercial AI industry actually views the DoD as an attractive customer. However, this attraction is despite the obstacles presented by traditional contract law and procurement practices used to solicit and award contracts.

Drawing on social exchange theory, this thesis introduces a theoretical framework – 'optimal buyer theory' – to understand the factors that influence a commercial AI firm's decision to engage with the DoD. It develops evidence-based best practices in contract law that reveal how the DoD can become a more attractive customer to commercial AI firms. This research builds upon research at the nexus of national security and defence contracts as it studies business decision-makers from AI firms through an explanatory sequential mixed methods design. In the study's first phase, participants are surveyed to discover the perceptions, opinions, and preferences at AI firms of all sizes, maturity, location, and experience within the DoD marketplace. In the second phase of the study, interviews from a sample of the participants


explain why the AI industry holds such perceptions, opinions, and preferences about contracts generally and the DoD, specifically, in its role as a customer.

This thesis concludes that commercial AI firms are attracted to contracts that are consistent with their business and technology considerations. These considerations align with contractual relationships that are collaborative, flexible, negotiated, iterative, and awarded promptly as opposed to those with fixed requirements and driven by regulations foreign to the commercial market. Additionally, it develops best practices for leveraging existing contract law, primarily other transaction authority, to align the DoD's contracting practices with commercial preferences and the machine learning development and deployment lifecycle. Armed with this understanding, the DoD can better attract commercial AI firms to support its national security objectives.





I certify that this work contains no material which has been accepted for the award of any other degree or diploma in my name, in any university or other tertiary institution and, to the best of my knowledge and belief, contains no material previously published or written by another person, except where due reference has been made in the text. In addition, I certify that no part of this work will, in the future, be used in a submission in my name, for any other degree or diploma in any university or other tertiary institution without the prior approval of the University of Adelaide and where applicable, any partner institution responsible for the joint award of this degree.

I give permission for the digital version of my thesis to be made available on the web, via the University's digital research repository, the Library Search and also through web search engines, unless permission has been granted by the University to restrict access for a period of time.

Andrew Bowne

Date: 11 February 2023



ACKNOWLEDGEMENTS

The work in this thesis represents the culmination of my research from 2019 to 2023. During this time, the world changed. Two events shaped this research and, indeed, human history, in profound ways. The first was the COVID-19 pandemic that impacted everyone and led to the loss of millions of lives. However, fear and suffering inspired incredible achievements in medical research that led to the rapid discovery, production, and distribution of life saving vaccines. The isolation and need for continued productivity were the catalyst for mass scale adoption of emergent video conferencing and collaborative software tools. The second event was Russia's invasion of Ukraine. Despite being completely overmatched by conventional weapons, Ukraine has been able to leverage modern technology, such as small drones, cyber prowess, satellites and even social media to defend itself. Both historic events brought into focus several themes: firstly, artificial intelligence is a powerful tool that can make possible the seemingly impossible; secondly, the speed and effectiveness of public procurement can make the difference between life and death; and finally, that collaboration across disciplines is increasingly important in the era of AI.

These themes are reflected in this research. Despite the losses experienced by so many due to these events, countless stories of perseverance, innovation, and unlikely partnerships inspired me throughout this time in my life and led to, I hope, a final thesis that is better than what I could produce prior to gaining perspective through bearing witness to this historic time.

Of course, I was not alone. There are so many people that I have the great fortune to have in my life that have supported, taught and motivated me. Unfortunately, I cannot name everyone, but I want to acknowledge that this research has benefitted from the kindness of multitudes. I am honoured and humbled to serve alongside the men and women of the armed



forces who have done so much to shape me into a better version of myself. I wish to thank them all. My service has given me incredible opportunities to meet people all over the world, so many that have done so much to make this world safer and better. I have learnt so much from so many, but my friends at the Department of the Air Force - Massachusetts Institute of Technology Artificial Intelligence Accelerator have made this assignment the highlight of my professional career and helped me become a better lawyer, teacher, thinker and leader.

I also wish to thank my supervisors, Melissa de Zwart, Colette Langos and Dale Stephens, who supported me, mentored me, pushed me and encouraged me to discover hidden truths. I am very grateful for their supervision and friendship. I also want to thank the University of Adelaide for making what I thought would be a painful experience as an international and remote student an incredibly fulfilling one. I thank Alex Wawryk for her support in navigating this process, and to Joanna Jarose for assisting me with copyediting my thesis, especially in helping 'translate' my American English and correcting my bad habits from years of using Bluebook citations. I sincerely thank all the participants in this research that shared their valuable time and perspectives with me to make this thesis possible.

Finally, I wish to thank my family for their unconditional love and support for me throughout all my endeavours. First, my parents, Laurie and Bruce, have done so much to motivate me towards pursuing my goals and, from a young age, cultivated a love of learning that has only grown over time. All my grandparents have supported me throughout my life. I dedicate this work to my grandfather, Skip Sergott, who passed away in 2022 at 93 years young. The prototypical lifelong learner, he read all he could find about artificial intelligence so he could understand and discuss my research with me. He set the example for me to serve in the military, pursue my academic goals, and most importantly, be a good husband, father, officer,



and citizen. I thank my brother, Jordan, and sister, Kathryn, who have always been my support and source of pride in all they have accomplished. I am extremely lucky to have married into an incredible family. Bonnie, Craig and Jeff welcomed me into their family and I am forever grateful for all they have done for me. Most importantly, I thank my wife, Stacey, and our children, Atticus and Alistair, with all my heart. Without their love, patience, understanding, encouragement, humour, and inspiration, I could not do what I do. They have been right beside me through everything and inspire me to be better and to make them proud. I have a very rich and happy life because of them and look forward to supporting their own pursuits, whatever they may be.

To all those who contributed to this dissertation or supported me throughout my life: thank you. It has been a challenging but truly rewarding journey. I am grateful to have made it with so many amazing people.





Artificial Intelligence ('AI') is a ubiquitous and disruptive technology enabler, augmenting human thought and action.[1]  Powered by new algorithms, plentiful data, and increasingly powerful and inexpensive computing, AI, particularly machine learning methods, can produce insights and innovations that have long eluded human thinkers.[2]  Until the past decade, AI was confined to academia and science fiction; however, AI is now omnipresent in virtually all industries.[3]  The rapidly improving ability of AI to solve problems and perform tasks that would otherwise require human intelligence is world altering, posing a challenge to the United States' technological advantage.[4]

The nature and diffusion of AI development and applications make it the quintessential 'dual-use' technology, creating advantages for commercial and military capabilities alike.[5] Accordingly, advancements in AI create advantages in the marketplace as well as on the battlefield.[6]  While there is no agreement on how AI will ultimately impact national security, the better question is when, not if, it will.[7]

The United States has consistently held that leveraging AI is critical to its national security and collective defence of its allies, specifically in its competition with China and Russia.[8]  However, the United States' military, led by the Department of Defense ('DoD'),[9] realises that it 'cannot maintain its competitive advantage without transforming itself into an AI-ready and data-centric' organisation.[10]  As concluded by the National Security Commission on Artificial Intelligence ('NSCAI'), the United States is at risk of falling behind its peer competitors without whole of nation efforts to address shortcomings in AI readiness.[11]  The DoD already lags far behind the commercial sector in integrating new technologies such as AI into its operations.[12]  While building organic talent pipelines, supporting fundamental research in federal labs and academia, and developing partnerships with allies are important components to implementing the national security strategy, leveraging the commercial sector is critically necessary for the DoD to achieve AI readiness at speed and scale.[13]

---

[1] Henry A Kissinger, Eric Schmidt and Daniel Huttenlocher, *The Age of AI and Our Human Future* (Little, Brown, 2021) 18.  Artificial intelligence ('AI') is defined below in section I.A.1.

[2] Ibid 14.

[3] See Kai-Fu Lee and Chen Qiufan, *AI 2041* (Currency, 2022) xi–xiii.

[4] National Security Commission on Artificial Intelligence, *Final Report* (March 2021) 7 ('*NSCAI Final Report*').

[5] See ibid 22.  'Dual use' refers to technology that is suitable for both civilian and military purposes. See ibid.

[6] See ibid 22–3.

[7] See generally ibid 7; Greg Allen and Taniel Chan, 'Artificial Intelligence and National Security' (Study. Belfer Center for Science and International Affairs, Harvard Kennedy School, July 2017) 7–8 <https://www.belfercenter.org/sites/default/files/files/publication/AI%20NatSec%20-%20final.pdf>; Robert H Latiff, *Future War: Preparing for the New Global Battlefield* (Alfred A Knopf, 2017); Paul Scharre, *Army of None: Autonomous Weapons and the Future of War* (Norton, 2018); Forrest E Morgan et al, *Military Applications of Artificial Intelligence* (RAND, 2020) 118.

This dissertation focuses on the legal, transactional and policy relationships between the DoD and the commercial firms that can advance AI applications for defence purposes.  The research is situated at the intersection of national security, applications in AI, and contract law as it applies to the DoD.  Through a better understanding of how commercial AI firms view the DoD as a customer, the DoD can better align its contract practice to attract industry partners that can help it meet its national security objectives. This dissertation draws on literature, surveys and interviews with AI business leaders, as well as analysis of the two primary legal frameworks governing DoD contracts to develop a theory of what contract attributes are best suited for the DoD to acquire AI-enabled capabilities from commercial AI firms. This dissertation presents an original theory called the *optimal buyer theory*. As the purpose of this dissertation is to understand how to optimise the DoD's engagement and buying practices for acquiring AI-enabled capabilities within the existing legal framework, the optimal buyer theory, like all optimisation problems, seeks to identify and select of the best options from the available alternatives. The available alternatives in this case is the choice of contract law framework available to the DoD to buy AI-enabled capabilities, namely the Federal Acquisition Regulation (FAR) and other transaction (OT) authority. The choice of legal framework, and the many choices the buyer has throughout the contract negotiation process, is optimal if it best aligns customer satisfaction (the DoD acquires what it needs at the right time and value) with the business preferences of the selling or servicing firm and the AI design, development, and deployment lifecycle.

---

[8] Department of Defense, *Summary of the 2018 Department of Defense Artificial Intelligence Strategy* (February 2019) 5 ('*2018 DoD AI Strategy Summary*'), claiming the '[f]ailure to adopt AI will result in legacy systems irrelevant to the defense of our people, eroding cohesion among allies and partners, reduced access to markets that will contribute to a decline in our prosperity and standard of living, and growing challenges to societies that have been built upon individual freedoms' and investments by nations such as China and Russia 'threaten to erode our technological and operational advantages and destabilise the free and open international order'; Joint Artificial Intelligence Center, Department of Defense, *2020 Department of Defense Artificial Intelligence Education Strategy* (September 2020) 2, explaining that the United States, together with its allies and industry partners, 'must continue to urgently implement its digital transformation strategy and adopt AI technologies at rapid speed and scale' to preserve and extend its competitive military advantage over potential adversaries.
[9] 10 USC § 111 (2018).
[10] DoD Responsible AI Working Council, Department of Defense, *United States Department of Defense Responsible Artificial Intelligence Strategy and Implementation Pathway* (June 2022) 8 ('*2022 DoD Responsible AI Strategy*').
[11] *NSCAI Final Report* (n 4) 1.
[12] Ibid 291.
[13] See ibid 65–72, 125, 233.

This research develops the theory that contract attributes, through law or practice, can affect the relative attractiveness of a buyer and thus, given sufficient flexibility to shape those attributes, can be optimised for attractiveness. Optimal buyer theory is contextual; it is advanced here to explain customer attractiveness from the perspective of commercial AI firms and help predict whether the process and execution of a contract will align with commercial preferences in supplying AI-enabled technologies. Optimal buyer theory is grounded in and builds upon social exchange theory. Social exchange theory accounts for both economic and sociological drivers in a relationship and explains why collaborative relationships are more attractive than purely



transactional contracts, particularly when competition is high and resources are scarce.[14]  The optimal buyer theory adds a contextual perspective to social exchange theory by considering the unique nature of AI development and capabilities as well as the preferences of commercial firms developing AI applications.  This theory provides a conceptual framework for assessing how the applicable contract law aligns with the technological and business concerns of commercial AI firms to improve the DoD's ability to acquire AI-enabled capabilities.  Of course, contracts are but a means to an end, not the end itself.  The end is to optimise contract law and practice to attract AI capabilities from the commercial market because by optimising its ability to attract AI talent and capabilities, the DoD can outpace its competitors to defend the United States and its allies.

As developed though this dissertation, the optimal buyer theory states that collaborative efforts, characterised by transparent communication, flexibility to experiment and iterate, negotiated terms that benefit all parties, and purpose-oriented efforts are most attractive to commercial AI firms and are best aligned with developing and deploying AI-enabled technologies.  This dissertation provides a decision tree as a conceptual framework to assist the DoD navigate the legal options at various decision nodes in the contracting process.  The DoD can use optimal buyer theory to align its contract practice with industry preferences and account for the unique contract challenges posed by AI development and deployment in a defence context.

This introductory chapter contextualises this research providing a background on AI: both the specific technology and the industry; its implications on national security; and the

---





challenges facing the DoD in contracting for AI-enabled capabilities.  This chapter also presents the research problem, research questions, scope and dissertation overview.

A *Artificial Intelligence and National Security Background*

1 *Artificial Intelligence is a Disruptive Technology Enabler*

AI is unique as a technology because it augments and enables technologies across various domains.  Although many have attempted to define AI, there is no widely accepted definition.  It has been described as a computer program that can accomplish tasks typically requiring human intelligence.[15]  However, there are limitations with that definition.  One, this definition raises the question of how to define 'intelligence'.[16]  Two, tasks that typically require human intelligence have temporal limits and lack objectivity; once technology exists to accomplish a task previously requiring human intelligence, is it no longer 'intelligent'?[17]  Russell and Norvig focus on rationality in reasoning and behaviour to explain the shifting nature and subjectivity of that definition.[18]  Other definitions focus less on what AI is — rather, the definition explains what AI can do.  The Organisation for Economic Co-operation and Development ('OECD') defines an 'AI system' as 'machine-based system that can, for a given set of human-defined objectives, make predictions, recommendations, or decisions influencing real or virtual environments.'[19]

---

[15] *2018 DoD AI Strategy Summary* (n 8) 5, defining AI as the 'ability of machines to perform tasks that normally require human intelligence – for example recognizing patterns, learning from experience, drawing conclusions, making predictions, or taking action – whether digitally or as the smart software behind autonomous physical systems.'  See also Kissinger, Schmidt and Huttenlocher (n 1) 14; Christopher Manning, Stanford Institute for Human-Centered Artificial Intelligence, 'Artificial Intelligence Definitions' (Stanford University, September 2020) <https://hai.stanford.edu/sites/default/files/2020-09/AI-Definitions-HAI.pdf>.

[16] Mark Coeckelbergh, *AI Ethics* (MIT Press, 2020) 64.

[17] See Morgan et al (n 7) 9.  Because AI is a rapidly evolving and its impact on the world is still largely unknown, attempts at defining AI lack consistency and often differ significantly from the technical understandings.  See Lyria Bennett Moses, 'The Legal Eye on Technology' (ALTI Forum, Amsterdam Law & Technology Institute, 31 January 2022) <https://alti.amsterdam/moses-legal-eye/>.

[18] Stuart Russell and Peter Norvig, *Artificial Intelligence: A Modern Approach* (Pearson, 4th ed, 2021) 1–2.

Additional confusion arises from the question of how AI reasons and acts. There are various subsets of systems that meet the definitions of AI. One related field is autonomy. Despite depictions of AI in science fiction, AI and autonomy are not synonymous: an AI system may perform autonomous decision-making, such as email spam filtering, though most developments to AI have human-machine interfaces.[20] Accordingly, autonomous systems sometimes use AI and AI sometimes results in automation, but not all AI is autonomous and not all automation uses AI.

One subset of AI is machine learning, which differs from the hand-crafted knowledge and expert systems of older AI applications as machine learning occurs when a 'computer observes some data, builds a model based on the data, and uses the model as both a hypothesis about the world and a piece of software that can solve problems.'[21] Thus, machine learning permits computers to learn by example without being explicitly programmed.[22] Examples of techniques of machine learning include supervised learning,[23] unsupervised learning,[24] deep learning,[25] and reinforcement learning.[26]  Each of those techniques, which essentially learn the relationship between inputs and outputs, are algorithms, or processes to formalise general mathematical reasoning as logical deduction.[27]  However, algorithms are but a component in a machine learning system — it is the data that is the primary driver of model behaviour.[28]  A way to visualise the relation of various components of the AI taxonomy is provided in Figure 1 below.[29]

---

[19] Organisation for Economic Co-operation and Development (OECD), *Recommendation of the Council on Artificial Intelligence*, OED/LEGAL/0449 (Adopted 22 May 2019) 7 [I] ('*Council on AI Recommendation*').
[20] Vijay Gadepally et al, 'AI Enabling Technologies: A Survey' (Paper, Massachusetts Institute of Technology, 2019) 2 <https://arxiv.org/pdf/1905.03592>. DoD policy defines autonomous weapon systems as a 'weapon system that, once activated, can select and engage targets without further intervention by a human operator. This includes, but is not limited to, operator-supervised autonomous weapon systems that are designed to allow operators to override operation of the weapon system, but can select and engage targets without further operator input after activation': Department of Defense, *Autonomy in Weapon Systems* (Directive No 3000.09, 25 January 2023) 21.
[21] Russell and Norvig (n 18) 651. All machine learning systems are AI, but not all AI systems use machine learning.
[22] Defense Innovation Board, *AI Principles: Recommendations on the Ethical Use of Artificial Intelligence by the Department of Defense (Supporting Document)* (Report, October 2019) 46 ('*AI Principles: Recommendations*').
[23] 'Supervised learning' is 'machine learning that maps inputs to outputs based on known input-output pairs from labelled data in a training sample': ibid 47.
[24] 'Unsupervised learning' is 'machine learning that learns the underlying structure or distribution of unlabelled input data': ibid.
[25] 'Deep learning' is 'the use of large multi-layer (artificial) neural networks that compute with continuous (real number) representations, a little like the hierarchically organised neurons in human brains. It is currently the most successful ML approach, usable for all types of ML, with better generalisation from small data and better scaling to big data and compute budgets': Manning (n 15).

*Figure 1: Taxonomy of Artificial Intelligence*

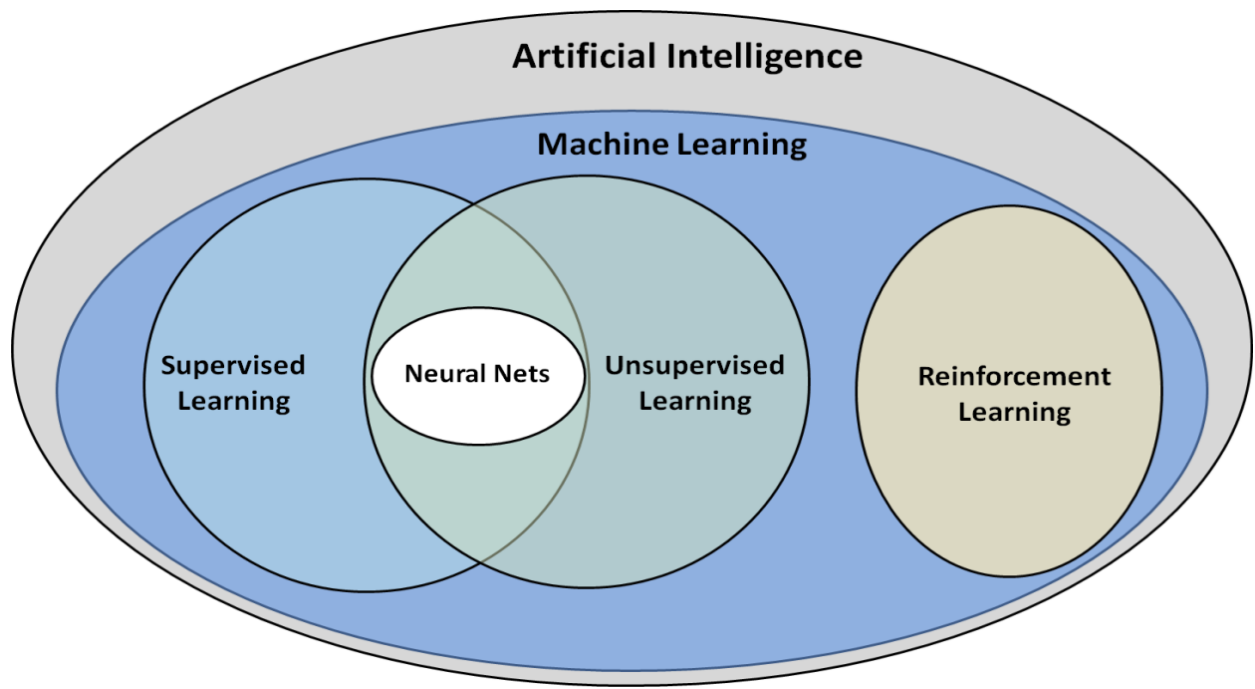

Machine learning represents the state-of-the-art in AI and is responsible for the current revolution in AI applications, attributed to the confluence of copious amounts of data, fast and efficient computing, advanced algorithms, and significant commercial investment.[30]  This research focuses on machine learning because it is the most relevant subset of AI in the national security context. While there is no universally accepted definition of AI, for the purposes of this dissertation a definition derived from the critical sources that reflects the technology relevant to the DoD is used in this dissertation. Hence, for the purposes of this dissertation, the following definition of AI will be relied upon:

> Artificial intelligence is the field of computer science whereby computers, whether in physical or digital space, are programmed to independently perceive or process data rationally, in a manner typically requiring human intelligence, to complete a task or goal.[31]

---

[26] 'Reinforcement learning' is a machine learning system where software agents learn to take actions in an environment through the requirement to maximise some notion of cumulative reward (often discounted for future rewards) through episodic training.  *AI Principles: Recommendations* (n 22) 47.

[27] Russell and Norvig (n 18) 9.  Algorithms are 'used to extract patterns, predict new events, fill in missing data, or look for similarities across datasets,' essentially converting the input information to actionable knowledge. Gadepally et al (n 20) 1–2.

[28] Manning (n 15).

[29] Figure developed by Kyle McAlpin from Gadepally et al (n 20) 13.

[30] See ibid; Allen and Chan (n 7) 7.

## 2 *AI is Different than Other Technologies*

Russell and Norvig claim 'AI is relevant to any intellectual task; it is truly a universal field.'[32] The ubiquity of AI distinguishes it from other technologies; it is a technology enabler, so other fields of technologies, such as robotics,[33] cyber capabilities,[34] hypersonic flight,[35] and space domain awareness,[36] are advanced by AI. The universal applicability of AI to various use cases makes it the quintessential dual-use technology – technology used in both civilian and military domains.[37] While there have been many dual-use technologies throughout history, AI is distinct as a dual-use capability in that it has great diffusion and potential for substantial destruction.[38]

---

[31] This definition was used during the survey and interviews of commercial AI firms. See below Chapter III.
[32] Russell and Norvig (n 18) 1.
[33] Ibid 938.
[34] Martin Libicki, 'A Hacker Way of Warfare' in Nicholas D Wright (ed), *AI, China, Russia, and the Global World Order: Technological, Political, Global, and Creative Perspectives* (Strategic Multilayer Assessment Periodic Publication, December 2018) 128.
[35] Lora Saalman, 'China's Integration of Neural Networks into Hypersonic Glide Vehicles' in Nicholas D Wright (ed), *AI, China, Russia, and the Global World Order: Technological, Political, Global, and Creative Perspectives* (Strategic Multilayer Assessment Periodic Publication, December 2018) 153.
[36] Daniel Jang et al, 'Space Situational Awareness Tasking for Narrow Field of View Sensors: A Deep Reinforcement Learning Approach' (Conference Paper, International Astronautical Congress, 12 October 2020) 2.
[37] Kissinger, Schmidt and Huttenlocher (n 1) 166.
[38] Ibid 166-7, explaining railroads, though dual-use and widely spread, had no destructive potential, and nuclear technologies have great destructive capacity but are relatively secure under government control.



The impact on civilian and military domains and potential for rapid and grand proliferation of a potentially destructive technology is not all that makes AI unique. It is technically different from legacy programming on a fundamental level. While traditional computer programming requires a human to (often painstakingly) code very detailed instructions to a computer to perform precisely in any possible situation, machine learning provides a way for computers to learn from experience with data.[39] This capability is useful in contexts where it is impossible to anticipate all future situations and sometimes programmers have no idea how to program the solution themselves.[40]

Because machine learning is accomplished through pattern recognition in data and not explicit programming rules, the speed of insight or action by computers is unprecedented.[41] Decisions in the stock market are made in microseconds — approaching the speed of light.[42] Along with the ability to generate, access and analyse vast amounts of data, there has been a Cambrian explosion of computational power as we approach the end of Moore's Law and algorithms are becoming increasingly efficient.[43] These three pillars of machine learning have made possible the seemingly impossible, stunning AI sceptics and igniting imagination.[44] Though far from the most impressive advancement in AI, one of the most consequential demonstrations of AI capability is AlphaGo, a computer that convincingly defeated Lee Sedol,

---

[39] Russell and Norvig (n 18) 651.
[40] Ibid. An example is facial recognition; humans perform this task subconsciously so it is challenging to program a way to accomplish this task, but through machine learning, a computer can perform this abstraction through experience; see ibid.
[41] See Darrell M West and John R Allen, *Turning Point* (Brookings, 2020) 110–1.
[42] Phil Mackintosh, 'Time is Relative: Where Trade Speed Matters, and Where It Doesn't' *Nasdaq* (online, 30 May 2019) <https://www.nasdaq.com/articles/time-relative%3A-where-trade-speed-matters-and-where-it-doesnt-2019-05-30>.
[43] See Gadepally et al (n 20) 2.
[44] See Ben Buchanan and Andrew Imbrie, *The New Fire: War, Peace, and Democracy in the Age of AI* (MIT Press, 2022) 2.



the world's top human player of Go, a complex strategy boardgame.[45]  Not only did AlphaGo

beat the reigning champion, it did so in a way that disrupted centuries of accepted wisdom about

how to excel at Go.[46]  The next iteration, AlphaGo Zero, was not trained on any human-played

games like its predecessor, but rather given the rules of Go and left to discover the best ways to

win on its own with no human influence.[47]  It beat the prior version–the one that beat Lee Sedol–

100 games to none.[48]  AlphaGo beating Lee Sedol became a 'Sputnik moment' for China,[49] as it

demonstrated the potential advantages AI could bring to the warfighting domain.[50]  Since

AlphaGo, China has demonstrated its intent to pursue AI dominance.[51]

The promise of AI has resulted in great competition for resources such as talent and data

that affect both private and public sectors.[52]  Because AI enables an organisation to gain insight

into its data and make predictions to aid in decision making, and make sense of various

information in quantities or formats that would be challenging for humans to understand, there

are infinite use cases that can be applied to sales, finance, law, healthcare, education,

transportation, as well as military applications.[53]  AI is a ubiquitous technology enabler,

enlivening domestic objects, tracking the spread of diseases, automating routine job functions,

---

[45] Michael Kanaan, *T-Minus AI: Humanity's Countdown to Artificial Intelligence and the New Pursuit of Global Power* (BenBella Books, 2020) 94.  Go is a board game with exponentially greater mathematical complexities than chess: at 92.
[46] Ibid 96.
[47] Ibid.
[48] Ibid 96–7.
[49] Karim Jebari, Irina Vartanova, and Pontus Strimling, 'Was AlphaGo Asia's "Sputnik Moment"?', *AI Futures* (Blog Post, 24 September 2019) <https://www.aifutures.org/20190924>.
[50] See Elsa B Kania, 'Artificial Intelligence in China's Revolution in Military Affairs' (2021) 44(4) *Journal of Strategic Studies* 515, 521–2.
[51] See generally Kania (n 50).
[52] See *NSCAI Final Report* (n 4) 2.
[53] See generally West and Allen (n 41) 27–157.



and optimising traffic flow.[54]  Governments around the world have recognised that many of these commercial uses have military analogues.[55]

AI is a disruptive technology that will be transformative to our daily lives.[56]  While there are other disruptive technologies, such as quantum computing, AI has the greatest future impact, and most other emerging technologies rely on AI-enabled technologies.[57]  The immense potential of AI applied across every domain makes it a technology unlike any other.  While AI will inevitably result in broad societal changes, this thesis focuses specifically on the role of AI within the context of national security.

## 3 *Applications to National Security*

The unique nature of AI, enabling and augmenting capabilities of humans and other technologies, has driven national strategies in power competition.[58]  The dual-use nature of AI means advances in the field will impact both economic and military power.[59]  Several likely military applications for AI include autonomy, increasing decision speed, cyber operations, targeting, and command and control.[60]  However, potential applications of AI appear limited only by imagination and resources. Quantum encryption,[61] synthetic weather radar,[62] and alternatives to satellite navigation[63] are areas of AI research with civilian and military interest. Broad-based research into computer vision,[64] natural language processing[65] and generative adversarial networks[66] advance dual-use applications.

---

[54] *NSCAI Final Report* (n 4) 20.
[55] See West and Allen (n 41) 107; Kissinger, Schmidt and Huttenlocher (n 1) 140.
[56] Rosario Girasa, *Artificial Intelligence as a Disruptive Technology* (Palgrave Macmillan Cham, 2020) 5.
[57] Ibid 5–6, explaining that other disruptive technologies, such as the Internet of Things, advanced robotics, autonomous vehicles, and renewable energy are made possible by AI technology.
[58] See *2018 DoD AI Strategy Summary* (n 8) 11.  Australia, the United Kingdom, and China are examples of advanced militaries that have specific policies published on the importance of AI to their respective national security strategies: see, eg, Department of Defence, *More, Together: Defence Science and Technology Strategy 2030* (2020) 2; Office for Artificial Intelligence, *National AI Strategy* (Command Paper 525, Version 1.2, September 2021) 10; Graham Webster et al, 'Full Translation: China's 'New Generation Artificial Intelligence Development Plan' (2017)', *Digichina* (Blog Post, 1 August 2017) <https://www.newamerica.org/cybersecurity-initiative/digichina/blog/full-translation-chinas-new-generation-artificial-intelligence-development-plan-2017/> [trans of: 国务院关于印发, 新一代人工智能发展规划的通知 (20 July 2017) <http://www.gov.cn/zhengce/content/2017-07/20/content_5211996.htm>].
[59] Allen and Chan (n 7) 2.
[60] See West and Allen (n 41) 108–21.

The United States recognises it is not the only great power intent on leveraging AI for military applications. The NSCAI issued a call to action, warning 'China possesses the might, talent, and ambition to surpass the United States as the world's leader in AI' and observes that China leads in some areas of AI already.[67] Meanwhile, Russia has successfully deployed AI in disinformation campaigns and to interfere with democracies.[68] Both countries view AI as the path to offset the United States' military power and neither autocracy is constrained by democratic norms.[69] As AlphaGo's victories show, 'AI can identify patterns of conduct that an adversary did not plan or notice' and 'then recommend methods to counteract them.'[70] Though the United States may still lead in AI innovation, once created, sophisticated AI capabilities can easily proliferate to other countries.[71]

---

[61] Ibid 123.
[62] Brian P Reen, Huaqing Cai and John W Raby, Computational Information Sciences Directorate, *Preliminary Investigation of Assimilating Global Synthetic Weather Radar* (Technical Report, Combat Capabilities Development Command Army Research Laboratory, September 2020) 2.
[63] See Albert Gnadt, 'Machine Learning-Enhanced Magnetic Calibration for Airborne Magnetic Anomaly Detection' (Conference Paper, AIAA SCITECH Forum, January 2022) 1.
[64] Lucas Liebenwein et al, 'Compressing Neural Networks: Towards Determining the Optimal Layer-wise Decomposition', (Conference Paper, Conference on Neural Information Processing Systems, December 2021) 1.
[65] See West and Allen (n 41) 125–6.
[66] Kissinger, Schmidt and Huttenlocher (n 1) 73.
[67] See *NSCAI Final Report* (n 4) 2, 7.
[68] Ibid 7; see P W Singer and Emerson T Brooking, *LikeWar: the Weaponization of Social Media* (Houghton Mifflin Harcourt, 2018) 253–5.
[69] See *NSCAI Final Report* (n 4) 22–3.
[70] See Kissinger, Schmidt and Huttenlocher (n 1) 140.
[71] Ibid.



China and Russia are authoritarian States that employ civil-military ('civ-mil') fusion.[72] Under this approach, there is no distinction between private and public industry – if the government wants access to technology, there is no way for a company to shield its intellectual property or data from the government.[73] As a result, there are no concerns about paying for intellectual property and the acquisition timeline for weapon systems is much faster than the DoD is able to field a new platform.[74] Aside from the rare use of the Defense Production Act, the United States cannot compel private companies to do work on its behalf.[75] Rather, it must attract the private sector to supply and support the public sector voluntarily. However, as discussed below, the private sector is fractured into companies that historically contract with the DoD and companies with limited experience working with the government.

## 4 *Comparing the Traditional Defence Industrial Base with Commercial AI Firms*

Despite the importance of AI applications to national security, and unlike any other defence technology, AI advancements are driven by the private sector.[76] Not only is private investment dwarfing public investment in AI overall, but 'there are also multiple Silicon Valley and Chinese companies who each spend more annually on AI R&D [research and development] than the entire United States government does on R&D for all of mathematics and computer science combined.'[77]  The Defense Innovation Board explained that for the first time in history, neither DoD nor the traditional defence industrial base ('DIB') control or even maintain favourable access to the advances of computing and AI, continuing a trend since the early 1980s.[78]  The leading commercial firms that are developing and deploying AI-enabled technology are not the defence firms that typically seek and perform contracts with the DoD, and the drivers and investments in the advancements of AI are not for military application, but for business and consumer applications.[79]  Commercial companies such as Tesla and Snapchat are arguably more advanced in the fields of autonomy and computer vision, respectively, than the DoD.[80]

---

[72] See *NSCAI Final Report* (n 4) 25.
[73] Ibid.
[74] The top contracting officer in the United States Air Force explained that China is acquiring weapons and equipment five to six times faster than the United States can, spending just one dollar for every twenty dollars spent by the United States for the same capability: Joe Saballa, 'China Weapons Acquisition Five Times Faster Than US: Defense Official', *The Defense Post* (online, 8 July 2022) <https://www.thedefensepost.com/2022/07/08/china-weapons-faster-us/>.
[75] The Defense Production Act permits the President to direct private companies to prioritise orders for the federal government. See 50 USC § 4511 (2018). The United States Constitution offers numerous protections from the federal government taking property or limiting freedom of speech or commerce: see, eg, *United States Constitution* amends I, V, XIV.
[76] *AI Principles: Recommendations* (n 22) 15.

The commercial AI industry differs from the DIB in talent, business model, and its relationship with the DoD.[81]  Top commercial AI companies compete intensely to attract top talent and can offer high salaries, cutting-edge advanced technologies, and less bureaucracy than a government agency or DIB contractor.[82]  While traditional defence contractors are often criticised for failing to modernise,[83] devoting only a small fraction of their research funding to AI,[84] in 2020, Amazon Web Services (AWS) spent over $42.7 billion in research and

development, with the majority going to advance AI and machine learning.[85]  Overall, the top

five commercial technology firms by R&D spend amounted to $130 billion in 2020.[86]  While

Amazon, Google, Meta, and Microsoft are global companies, there are hundreds of technology

companies that provide AI-enabled capabilities potentially suitable for the DoD.[87]

　　　Unlike during the early age of computing when the government — specifically the DoD

— was the primary attractor of talent, it is academia and private sector labs that are now

advancing AI and machine learning.[88]  This fact, coupled with the market control of AI

innovation residing in the commercial sector, demonstrates the necessity of the DoD's

collaboration with this emerging technology sector.  However, unlike China and Russia — which

can compel private industry to work on national security applications — the United States and its

allies are liberal democracies.[89]  Thus, the DoD must win the trust of commercial AI firms.[90]

　　　However, although it is critical for the DoD to attract commercial AI firms, there are

several obstacles.  The NSCAI concluded the DoD is too slow, too bureaucratic, lacks the

foundation to use AI even if it had it.[91]  This, in the Commission's estimation, makes it

economically irrational for many start-ups to even try to work with the DoD.[92]  Another potential

rationale for commercial AI firms to avoid working for the DoD is an ethical argument that the

cutting-edge research on AI should never be used on any military or surveillance applications.[93]

Some AI experts have committed to never working for the DoD.  Google made headlines when it

---

announced it was walking away from work it was doing with the DoD on its first large AI project, known as Project Maven.[94]

While the rationale of any commercial AI firm to work with the DoD appears multi-dimensional, this research examines the reasons AI companies may avoid working on national security applications for the DoD. Based on a review of the literature[95] and surveys and interviews of business leaders at commercial AI firms,[96] a significant consideration is how the DoD contracts for AI. The next section provides a background on the two legal frameworks available to the DoD that govern contracts.

B *Department of Defense Contract Law Background*

There are two distinct legal paradigms available to the DoD to award contracts for commercial AI-enabled capabilities. One option is contracting within the traditional procurement framework governed by the Federal Acquisition Regulations (FAR).[97] As discussed in Chapter 2, such contracts are the most commonly used option in the DoD, although FAR contracts have been the subject of years of acquisition reform efforts and criticism by Congress and industry.[98] The other option is known as other transaction (OT) agreements governed by a separate legal authority.[99]  This authority has a long history, but DoD contracting officials have only recently started to build practice in this law.[100]

---

[94] Ibid 116–22. Although Google completed the contract and opted to not compete for a larger contract bid on by its competitors, including Microsoft and Amazon, the fallout from Google ending its role in Project Maven resonated with many AI companies and the public: see ibid 122–6. In a survey of AI engineers that asked what event or issue influenced their understanding of the DoD's use of AI, Project Maven received the second most responses, behind drones and tied with Edward Snowden: see James Ryseff et al (n 81) 36.

[95] See below Chapter II.

[96] See below Chapter IV.

[97] Merve Hickok, 'Public Procurement of Artificial Intelligence Systems: New Risks and Future Proofing' (2 October 2022) *AI & Society* https://doi.org/10.1007/s00146-022-01572-2: 1–15, 4 [4].

[98] See J Ronald Fox, *Defense Acquisition Reform, 1960-2009: An Elusive Goal* (Center of Military History, 2011) 189, explaining that despite five decades of reform efforts by Congress to the defence acquisition system, it has remained largely resistant to change; Heide M Peters, Congressional Research Service, 'Defense Acquisition Reform: Background, Analysis, and Issues for Congress' (Report No R43566, 4th rev ed, May 2014) 5–6; Frank Kendall, *Getting Defense Acquisition Right* (Defense Acquisition University Press, 2017) 32–5.

The FAR is the baseline regulatory framework for all federal procurement, while agency-specific regulations, such as the Defense Federal Acquisition Regulations Supplement ('DFARS'), provide additional oversight on procurement contracts.[101]  However, the FAR distinguishes procurement — the acquisition of supplies or services — with research and development for 'which the work or methods cannot be precisely described in advance' and 'must provide an environment in which the work can be pursued with reasonable flexibility and minimum administrative burden.'[102]  OT authority is intended for research and development activities, including prototyping and experimentation.[103]  OT authority is characterised by flexibility and streamlined administrative requirements.[104]

While both FAR contracts and OT agreements can be used by the DoD to acquire AI-enabled capabilities, the differences in the legal frameworks are significant.  These differences in how a contract is competed, negotiated, performed, and closed out may impact the DoD's choice of contract authority.  Additionally, the choice of law may affect how private AI firms perceive the DoD as a customer and their willingness to compete for a contract opportunity.  As this thesis argues, factors that impact commercial AI firms' willingness to work for the DoD should be considered in the DoD buyer's decision on which legal framework to use.  Chapter II provides a

---

[99] See William E Novak, 'Artificial Intelligence (AI) and Machine Learning (ML) Acquisition and Policy Implications' (White Paper, Carnegie Mellon University Software Engineering Institute, February 2021) 20; 10 USCA §§ 4021–3 (2022) (formally codified at 10 USC §§ 2371, 2371b and 2373 (2018)).
[100] See Douglas Steinberg, 'Leveraging the Department of Defense's Other Transaction Authority to Foster a Twenty-First Century Acquisition Ecosystem' (2020) 49(3) *Public Contract Law Journal* 537, 557.
[101] See *Defense Federal Acquisition Regulations Supplement* ('*DFARS*') part 201.
[102] *Federal Acquisition Regulation* ('*FAR*') 48 CFR § 35.002.
[103] 10 USCA §§ 4021–3 (2022).
[104] Office of the Under Secretary of Defense for Acquisition and Sustainment, *Other Transactions Guide* (Department of Defense, November 2018) 2 ('*Other Transactions Guide*').



thorough explanation of the differences between the two legal frameworks.  Figure 2 below

presents a way to compare the two frameworks.



*Figure 2: Comparison of FAR Contracts and OT Agreements*

| Feature | FAR | OT |
|---|---|---|
| **Solicitation** | Starts with the requirement; detailed instructions on how the contractor is expected to perform.[105] | Starts with a problem statement that notifies what the contractor should solve.[106] |
| **Competition** | With limited exceptions, the DoD is required to promote full and open competition under the Competition in Contracting Act ('CICA') and FAR Part 6.[107] | The OT statutes require competition 'to the maximum extent practicable'.[108] |
| **Terms & Conditions** | Prescribed by the FAR and supplements; they are often unique to the government (such as termination for convenience, accounting standards, auditing requirements, etc.).[109] | Negotiated by the parties, often resulted in more commercial-like terms in areas like intellectual property and disputes, with several exceptions where federal law requires clauses reflecting public policy or other law (such as procurement integrity, domestic preference, fiscal restrictions imposed by Congress).[110] |
| **Flexible** | The DoD's ability to negotiate terms, payment timeline, and contract structure as well as changing the approach of tasks of the contractor during performance is limited by regulations and procurement law, including the changes clause and CICA. | Performance milestones can be iterative and experimental; modifications can be frequent and are not bounded by FAR changes clause or CICA; the DoD can accept payments from the contractor or other third parties.[111] |
| **Communication** | Limited to contracting officer during competition and modifications.[112] | Encourages engagement between industry and end-user throughout competition and performance.[113] |
| **Intended Use** | Procuring supplies and services. FAR contracts are not intended to be used for research and development as, unlike most contracts for supplies and services, most research and development contracts are directed towards objectives that cannot be precisely described in advance.[114] | Research and development, prototyping, experimentation, support dual-use projects, adapt novel business practices, and broaden the industrial base available to the government.[115] |

---

[105] 48 CFR § 2.101.

[106] *Other Transactions Guide* (n 104) 9, 13.

[107] 48 CFR Part 6.

[108] 10 USCA § 4022 (2022).

[109] See 48 CFR Part 52.

[110] See *Other Transactions Guide* (n 104) 49–51.

[111] Crane Lopes, 'Historical Institutionalism and Defense Public Procurement: The Case of Other Transactions Agreements' (Dissertation, Virginia Polytechnic Institute and State University, September 2018) 504.

[112] See 48 CFR §15.306 (limiting discussions between potential contractors and the government to the contracting officer during competition

[113] See Lopes (n 111) 513–4.

[114] 48 CFR § 35.002, § 35.003.

[115] 10 USCA §§ 4021–3 (2022); *Other Transactions Guide* (n 104) 3.



There are advantages and disadvantages of both legal frameworks. Despite massive

increases in the dollars obligated by the DoD on OT agreements since 2016 when the authority

became permanent, the FAR is by far the more commonly used framework in the DoD.[116] As

such, practitioners are more comfortable working on a FAR contract than an OT and much of the

training contracting officers and lawyers receive is focused on the FAR.[117] Due to the regulatory

nature of the FAR system, there is arguably more predictability in the performance of a FAR

contract as most clauses come directly and unchanged from the FAR clauses section.[118] There is

significant oversight and auditability of contracts executed under the FAR by multiple

agencies.[119] Because the competition requirements for FAR contracts establish layers of

transparency, there is potentially more assurances of fairness for contractors and clear avenues to

challenge the government's source selection.[120] However, such predictability, fairness, and

transparency, though laudable policy objectives for a public contract law framework, may

sacrifice flexibility to help attract commercial firms that previously have not business with the

DoD.[121]

Alternatively, OT agreements are inherently flexible; the required rules and clauses are

amount to dozens of pages compared to the thousands of pages of the FAR and thousands more

on agency supplements. With this flexibility, the DoD can adjust its approach to each individual contract to support its goals.[122]  As such, the DoD can negotiate more freely with industry, and reduce the cost of compliance for both parties.[123]  This flexibility applies not only to the terms and conditions of the agreement but also to business arrangements with multiple parties and funding from outside the federal government — all difficult or prohibited in FAR contracts.[124] However, OTs arguably lack the transparency and oversight of a FAR contract,[125] and there is little training and education specific to OT agreements provided to contracting professionals in the DoD.[126]  Finally, OT authority is limited to contracts for R&D, prototypes, production, and procurement for experimental purposes.[127]  Although this is broad authority, it is not available for all contract actions that can be accomplished under the FAR.

---

[116] See Rhys McCormick and Gregory Sanders, *Trends in Department of Defense Other Transaction Authority Usage* (Report, Center for Strategic and International Studies, May 2022) 7, noting OT dollars obligated increased 2030 percent between 2015 and 2020 but are still just $16 billion compared to nearly $400 billion in contracts.
[117] See Lopes (n 111) 407.
[118] See 48 CFR Part 52.
[119] See 48 CFR § 52.215-2.
[120] See Scott Amey, 'Other Transactions: Do the Rewards Outweigh the Risks?', *Project on Government Oversight* (Online Report, 15 March 2019) <https://www.pogo.org/report/2019/03/other-transactions-do-the-rewards-outweigh-the-risks>.
[121] See Government Accountability Office, *Other Transaction Agreements: DOD Can Improve Planning for Consortia Awards* (Report to Congressional Committees No GAO-22-105357, September 2022) 1.

While there is extensive literature devoted to praising or critiquing the FAR or OT frameworks, there is limited research analysing how these different legal frameworks support the DoD's goal of acquiring AI capabilities.  This research fills that gap by identifying how the attributes of each legal framework align with the AI development and deployment lifecycle and whether one framework is more attractive to the AI innovation ecosystem.

In this dissertation, the relative merits of the two contract law frameworks are synthesised with the opinions and preferences of commercial AI firms.  The goal of this research is to better understand the relationship between the DoD's use of its contract law and the business calculus of commercial AI firms.  This understanding can help the DoD assess how to leverage its legal authorities can attract commercial partners in developing and deploying AI capabilities.  By understanding the respective attributes and respective barriers and advantages posed by the two contract law frameworks in the DoD as perceived by commercial AI firms, the DoD can make decisions and policies that align with buying AI-enabled capabilities from the private sector. This dissertation recommends that the DoD utilise the contract law framework and tools that best align with the development and deployment of AI and maximises the attractiveness of the DoD as a customer to promote robust competition and innovation.

---

[122] See ibid 5: explaining that OT authority provides the DoD with a 'blank slate' to negotiate with industry.
[123] See Lopes (n 111) 34–5.
[124] Ibid 35–6.
[125] Amey (n 120).
[126] See Stan Soloway, Jason Knudson and Vincent Wroble, *Other Transactions Authorities: After 60 Years, Hitting Their Stride or Hitting the Wall?* (Report, IBM Center for The Business of Government, 2021) 44–5.
[127] 10 USCA §§ 4021–3 (2022).

This research finds that of the two legal paradigms that govern contracts in the DoD, OT agreements are best aligned with the design, development, integration, and deployment of AI-enabled technologies and the preferences of many commercial AI firms, making them comparatively attractive to FAR-based contracts. Moreover, the ability to attract commercial AI firms and the inherent flexibility of OT agreements allow the DoD to access and acquire technology that meets its mission requirements and supports its national security goals. However, the mere choice of law does not ensure the DoD will attract industry partners and achieve its national security objectives: the practice and likely culture of contract professionals will need to change. Leveraging a conceptual framework that explains relational dynamics, such as social exchange theory, can assist the DoD in identifying and promoting contract law practices that commercial AI firms are likely to find attractive.

## C *Social Exchange Theory*

Social exchange theory is principally an economic theory that helps explains relational dynamics between individuals, businesses, or organisations that interact in an exchange of resources.[128] It is used to explain relational exchanges assessed across several variables, including trust, cooperation, communication, reputation, dependence, and satisfaction.[129]

---

[128] Lambe, Wittman and Spekman (n 14) 4.
[129] Ibid 16–9.



Although social exchange theory is broad and applies in a variety of contexts, the foundational premise of the theory is that exchange interactions involve economic — the balance of rewards and costs — and social outcomes.[130]  These outcomes are compared by the parties with the predicted outcomes of available exchange alternations; thus, the more positive an exchange interaction is, the stronger the relationship and dependency becomes.[131]  In any exchange, both economic and social outcomes are judged together to assess the value of the relationship and compare it to alternatives.[132]  The satisfaction level of the interaction will be determined based on the comparison of the expected benefits derived from the exchange.[133]  If the expected benefits of a potential interaction are greater than the best possible alternative, the party is more likely to enter and maintain that relationship; conversely, if the expected or realised benefits of an interaction are less than expected or the best possible alternative, the party is more likely to avoid or end the relationship.[134]

Like any relationship, a business relationship will encounter situations where the parties have different preferences or objectives.  If each party only seeks their most favoured outcome (such as a couple deciding on which movie to see when each person desires a different movie, or a business-to-business deal where each is looking to unreasonably maximise economic advantage) the results are often poor, and the process is costly.[135]  However, when the parties view the relationship through lens of past and future interchanges and assess the benefits of collaboration, the perception of the value of the relationship increases.[136]  From a social exchange perspective, successful collaborations, where both parties experience greater outcomes by working together to achieve their mutual goals than their individual goals, occur with trust developed over multiple interactions.[137]  Thus, the initial transaction is crucial to determining whether the relationship will expand, diminish, remain the same or dissolve.[138]  Attraction is the catalyst to initiating a relationship.[139]  Attraction in the context of a buyer and supplier in an initial exchange is the supplier's subjective assessment of a buyer based on the anticipated outcome of the interaction.[140]  Determinants of attractiveness of a buyer vary based on a number of factors depending on the type of exchange, but include economic value, efficiency, exchange relations and communication, ethical behaviour, risk, market linkages, and image and reputation.[141]  The customer's barriers to entry, technological skills, compatibility with the supplier's business model, and market influence factor into a supplier's assessment of numerous determinants.[142]  Additionally, the ability to negotiate and flexibility to evolve during the relationship are critical to initial attraction, especially when the seller has numerous attractive alternatives.[143]  When alternatives present greater economic drivers, a customer must compensate through increasing their attractiveness in other determinants, such as cost efficiency, organisational and cultural fit, and reputation opportunities.[144]

---

[130] Ibid 5.
[131] See ibid 6.
[132] Ibid 8.
[133] Ibid 9.
[134] See ibid.
[135] Ibid 10.
[136] See ibid 9–10.

Because the United States seeks to leverage the commercial AI industry to support its AI strategy, and commercial AI firms have multiple alternative customers, the DoD must understand what attracts commercial AI firms and align its buying practices with those attributes.  Hence, social exchange theory is useful as a conceptual framework in developing questions, hypotheses, methodological approaches, and theoretical lens to analyse the research data. Additionally, the existing social exchange theory literature serves as the foundation for building a context-specific theory to help explain what attributes of the DoD as a customer attract – or repel – commercial AI firms. Understanding what attributes attract commercial AI firms to a customer can enable contract professionals in the DoD to adapt their choice of law and practice to optimise attractiveness.

---

[137] See ibid 12.
[138] Ibid 13.
[139] See Aino Halinen, *Relationship Marketing in Professional Services: A Study of Agency-Client Dynamics in the Advertising Sector* (Routledge, 1997) 59.
[140] Antonello La Rocca, Albert Caruana and Ivan Snehota, 'Measuring Customer Attractiveness' (2012) 41(8) *Industrial Marketing Management* 1241, 1244.
[141] Suvi Lehto, 'Buyer Organisation Attractiveness in Luxury Retail' (Master's Thesis, University of Turku, 25 May 2020) 24.
[142] See ibid 24–5.
[143] See ibid 47.
[144] Ibid 54.

D *Research Problem*

Despite AI's critical role in meeting the United States' national security objectives,[145] the NSCAI reached a grim conclusion: 'America is not prepared to defend or compete in the AI era.'[146] The Commission explained one of the underlying reasons the United States is losing its competitive edge against China in AI is the DoD's contracting practice is not aligned with business and technical considerations of commercial AI.[147] Because AI fundamentally differs from other technologies, contracting for AI systems is unlike buying other services or supplies. There is no consensus on how well existing contract law aligns with the DoD's objectives to contract for AI-enabled capabilities, nor whether the legal practice impacts commercial firms' decision to assist the DoD in becoming ready to compete in the AI era. This research seeks to understand how to optimise the DoD's engagement and buying practices for acquiring AI-enabled capabilities.

---

[145] Multiple administrations underscored AI as a priority technology to be developed and as an integral enabler in the national defence. See, eg, Executive Office of the President, *Maintaining American Leadership in Artificial Intelligence*, 89 Fed Reg 3967 (Executive Order No 13859, 11 February 2019) ('Executive Order 13859'), in which the Trump administration stated its intent to maintain American leadership in AI; Subcommittee on Networking and Information Technology Research and Development et al, *The Networking & Information Technology R&D Program and the National Artificial Intelligence Initiative Office Supplement to the President's FY2022 Budget* (Report, December 2021), in which the Biden administration stated AI innovations 'strengthen our national security and protect our economy' <https://www.nitrd.gov/pubs/FY2022-NITRD-NAIIO-Supplement.pdf>.
[146] *NSCAI Final Report* (n 4) 1.
[147] See ibid 306–8.

## E *Scope*

This dissertation focuses on the relational dynamics between commercial AI firms and the DoD to understand what barriers exist that may prevent commercial AI firms from providing AI-enabled capabilities to the DoD.  As the relationship between commercial AI firms and the DoD is manifested through contracts, the research predominantly examines existing legal frameworks available to the DoD to acquire AI systems.  However, other potential barriers to attracting firms are considered, specifically the perception of the DoD's ethical and responsible use of AI.

There are several issues that impact the DoD's ability to become AI-ready and able to compete with China.  These issues include receiving appropriated funds on time from Congress,[148] changes to rules on planning, programming, and budgeting,[149] modernised export control regulations,[150] and improved data and talent management practices.[151]  For the United States to achieve its strategic goals, concentrated efforts are required in each of these areas.[152]  However, those issues are not central to this dissertation.  The focus here is on one significant component of the challenge — attracting commercial AI firms to develop and deliver AI for military applications to the DoD. This dissertation examines the formation of the legally binding relationship between a commercial AI firm and the DoD that enables access to AI-enabled capabilities to advance the United States' national security objectives.

---

[148] See National Defense Industrial Association, 'Risks to National Security: A Full-Year Continuing Resolution for 2022' (White Paper, January 2022) 4 <http://www.ndia.org/CR>.

[149] See William Thornberry, 'How Congress Must Reform its Budget Process to Compete Against China in AI', *Hill* (online, 25 June 2021) <https://thehill.com/blogs/congress-blog/economy-budget/560345-how-congress-must-reform-its-budget-process-to-compete/>.

[150] See Dave Aitel, 'We Need a Drastic Rethink on Export Controls for AI', *Council on Foreign Relations* (Blog Post, 21 January 2020) <https://www.cfr.org/blog/we-need-drastic-rethink-export-controls-ai>; *NSCAI Final Report* (n 4) 226, 496-500, recommending action to incorporate AI in existing export control, International Traffic in Arms Regulations ('ITAR'), and Committee on Foreign Investment in the United States ('CFIUS') measures to prevent technology theft by China and Russia.

[151] *NSCAI Final Report* (n 4) 63.

[152] Ibid 272.

## F *Research Questions*

The research problem raises a question fundamental to this dissertation:

- *Why do commercial AI firms decide to contract with the DoD?*

Asking this question can lead to a better understanding of what factors commercial AI firms consider when deciding to devote its resources to solving problems for the DoD. Understanding why an innovative AI company would forgo potential opportunities in the private sector to compete for and perform a contract to support national security will help the DoD align, and adapt, if necessary, its contract practice to best attract commercial AI firms and leverage AI-enabled capabilities.

The research question posed in the alternative is also critical to this research:

- *Why do commercial AI firms decide to not contract with the DoD?*

Understanding factors that may deter commercial AI firms from forming a relationship with the DoD is important to the DoD as it can identify and seek to remove these barriers.

The above questions are fundamental, first principles questions about the research problem. These questions intentionally avoid assumptions. Although answering those questions is a necessary first step to addressing the problem, it will not solve the problem. Thus, secondary questions that propel the theory-based understanding to a practical impact follow the fundamental questions. As discussed above, the DoD has a choice of law to acquire AI-enabled



technologies. Therefore, once it is understood why commercial firms choose to contract with the DoD (and why they choose to not contract with the DoD), the questions turn to what the DoD can do to improve its attractiveness to commercial AI firms.

- *Does existing contract law applicable to the DoD align with acquiring AI-enabled technologies from commercial firms?*

Because there are two distinct legal frameworks for the DoD to contract for AI-enabled technologies, there is a series of questions that follow.

- *Which legal framework best aligns with the development and deployment of AI systems in the DoD?*

- *What contract attributes do commercial AI firms prefer?*

- *Why do commercial AI firms prefer certain contract attributes over others?*

- *What unique characteristics of AI development and deployment affect the formation and performance of a contract for the DoD?*

- *How does the choice of law affect the DoD's ability to contract for AI-enabled capabilities?*

Although the thrust of these questions is aimed at understanding whether FAR contracts or OT agreements are better aligned with the technological and business concerns of commercial AI firms, it is assumed that neither framework is truly optimal without changes to the law or practice. Thus, the research seeks to understand what contract attributes affect commercial AI firms' perception of the DoD and then assesses how each contract law framework can enhance or detract from those attributes.



### G *Dissertation Overview*

This dissertation provides insight into the research question based on analysis, interpretation, and synthesis of the surveys and interviews of business decision-makers at commercial AI firms. There are indications that there are challenges in attracting commercial AI firms to work on DoD contracts, but little research outside several anecdotes offers explanations from industry as to those challenges. This dissertation presents novel research into industry perspectives that, when viewed through the lens of social exchange theory, can help inform procurement practitioners and policy makers in the DoD on best practices in contracting to optimise the customer attractiveness of the DoD.

This research offers a better understanding of the factors that impact the commercial AI industry's perception of the attractiveness of the DoD as a customer. Through this understanding, the DoD can leverage best practices in engaging and contracting with commercial AI firms to procure AI-enabled capabilities to meet its strategic objectives. The purpose of this dissertation is to develop a better understanding of the firms developing and selling AI so contract professionals in the DoD can best assess and draft contract terms and contract vehicles that align with their preferences. It assesses the impact that the DoD's contract law has on its ability to attract commercial AI firms from the lens of both the problem (how the law impedes relationships between industry and government) and the solution (how contract lawyers for the DoD can align legal practice to better attract commercial AI firms). The DoD must leverage skilled understanding of its contract law to remain competitive in the race for AI supremacy.

This dissertation has six chapters, including this introduction. Chapter II reviews existing literature on three topics relevant to the research question. The first topic examines literature related to the nexus of AI and national security. The literature on this topic indicates that the



development of AI-enabled technologies, and the competition such development has sparked with global powers, pose national security challenges to the United States. The second topic views literature related to the DoD's contract law and procurement framework. There are two primary contract law paradigms available to the DoD to acquire commercial AI technology: contracts governed by the FAR[153] and contracts awarded using other transaction authority.[154] Relevant literature analysing the relative merits of each contract type and the applicable statutes and regulations are compared. While the FAR dominates contract actions in the DoD, the permissive and flexible law governing OT agreements is more closely aligned with commercial contracts and appears to be more compatible with the development of AI-enabled systems.[155] The third topic covers literature related to social exchange theory. Literature on this theory fits within this dissertation's purpose, as the core explanatory mechanism of social exchange theory is the relational contract, making it useful in understanding the dynamics between contracting parties, such as the DoD and commercial AI firms.[156] This dissertation uses social exchange theory and the related concept of customer attraction as a conceptual framework and theoretical lens to build the research design, form hypotheses, and interpret the research findings.

Chapter III presents the research design and methodology. This study uses a two-phase explanatory sequential design.[157] The first phase consists of collecting and analysing quantitative data derived from a survey of participants from the commercial AI industry. The second phase consists of qualitative interviews of a purposeful cross-section of the survey

---

participants. The qualitative interviews provide a greater understanding of how to interpret and explain the quantitative data.

Chapter IV summarises and interprets the results of the survey and interviews. This large sample provides primary source data regarding the commercial AI industry's perceptions, opinions, and preferences pertaining to contracting and the DoD as a customer. After the surveys were completed and analysed, a purposefully selected cross-section of the survey sample was interviewed. The survey and interview questions were aimed at collecting data to discover findings and answer the research questions. This chapter reports the findings from the survey and interviews and integrates the data with the law and literature to interpret and synthesise the major findings.

Chapter V presents an integrated analysis and synthesis of the major thematic findings from the data discussed in Chapter IV, describing the alignment of the findings with concepts from social exchange theory and existing contract law available to the DoD when buying AI systems. This chapter provides an answer to the research questions and assesses the hypotheses presented in Chapter II in light of the research.

Chapter VI offers recommendations for establishing best practices in contracting with commercial AI firms. This chapter provides a conceptual framework for assessing the attractiveness of a contractual relationship between the DoD and a commercial AI firm. This chapter also introduces an original theory, called 'optimal buyer theory,' that can help the DoD optimise its ability to attract innovative AI solutions from the private sector to better align with the preferences of commercial AI firms and the idiosyncrasies of developing AI-enabled technologies. This chapter presents optimal buyer theory in the context of the DoD's choice of



contract law framework and recommends actions at specific decision nodes in the contracting process that align theory with practice.

This dissertation concludes with reflections on the research findings and discusses the impact DoD contract lawyers can have in strengthening national security.





A *Overview*

The purpose of this study is to understand how to optimise the DoD's engagement and buying practices for acquiring AI-enabled capabilities within the existing legal framework available to the DoD. The research design used to address the research purpose is to explore the perceptions and opinions of commercial AI firms to better understand how well existing law and policy aligns with commercial AI business preferences and technology development. This understanding can lead to better alignment of contract law and practice to optimise the DoD's attractiveness as an AI customer. This study seeks to understand what factors make the DoD an attractive customer to commercial AI firms, and what barriers limit a commercial firm's engagement with the DoD. To develop a conceptual framework for answering the research question, building hypotheses, and crafting the survey instrument and interview questions, a critical review was conducted on the existing literature of three topics of study – the nature of AI and its relationship to defence; DoD contract law; and social exchange theory.

As the relationship between the national defence sector and the commercial AI sector is a current topic that has evolved since beginning of this dissertation, and will undoubtedly continue to evolve, literature is recent, and the field of study is growing. In contrast, much has been written on the topic of public procurement, and specifically on military contracts. Thus, a more historic perspective is used to review the state of the DoD's contract law frameworks. Finally, social exchange theory literature is examined to provide a theoretical lens that assists in understanding the relationship between commercial AI firms and the DoD and the impact the



contract — the legal mechanism codifying the nature of the relationship — can have on the perceived attractiveness of the potential buyer.

The first topic covers the nexus of national defence and AI, specifically the unique nature of the development and deployment of AI and how this rapidly growing field of technology differs from other defence technologies. The literature explaining how AI is different as a technology generally, and in the defence context specifically, underscores why traditional contracting practice may not adequately enable the DoD to attract commercial AI firms or field emerging technology at the speed and scale required to keep pace with its competition.

The second topic covers the legal and regulatory framework of defence contracts and procurement policy within the DoD. As the main topic of this dissertation, this section presents a critical review of primary sources, such as the statutes, regulations, policy, and official guidance that govern how the DoD solicits, evaluates, awards, and administers contracts, as well as commentary by leading academics and practitioners in the field. This literature review examines how the DoD meets its requirements via contracts. The goals and purposes of a DoD procurement often focus on competition as the fundamental assumption driving the DoD's procurement system is that competition leads to improved efficiency, innovation, quality, and performance.[158] While this study does not dispute this assumption, it challenges the existing legal framework that creates the process for the DoD to achieve effective competition as it relates to attracting commercial AI firms. Additionally, other legal and regulatory constructs, such as intellectual property rights and requirements development, are confronted for their potential negative impact on competition. Existing research in this field highlights the practical limitations posed by existing law and identifies gaps in evidence-based reform and the lack of a cohesive theoretical model that can help explain the effect the law and efforts to reform the procurement process has on commercial AI firms.

---

[158] Jacques S Gansler, William Lucyshyn and Michael Arendt, *Competition in Defense Acquisitions* (Report, University of Maryland School of Public Policy, February 2009) 3.

The third topic presents concepts from social exchange theory as a theoretical lens through which to evaluate the DoD's procurement system and its effect on commercial AI firms. Despite spending approximately half a trillion dollars annually on contracts and extensive and ongoing efforts to reform the procurement system, there have been a lack of theory-based models used to assess existing laws or practice or its reform.[159]  A number of social science theories were examined for their potential explanatory power on the phenomenon of commercial AI firms making the business calculus to engage with or abstain from the DoD's procurement process. This review led to the identification and application of social exchange theory to better interrogate the relationship between the DoD and commercial AI firms, with particular focus on the costs and rewards that affect decisions in initiating, maintaining, and terminating such relationships.

This chapter concludes with a synthesis of the three topics that explains the interconnectedness of these fields of study.  Through this synthesis, a conceptual framework emerged that was used to develop the research methodology covered in Chapter III and analyse the research findings covered in Chapter IV.  Ultimately, this conceptual framework resulted in the development of the optimal buyer theory and extended to help predict *a priori* the attractiveness of a contract law framework or a specific contract.

---

[159] Christopher R Yukins, 'A Versatile Prism: Assessing Procurement Law Through the Principal-Agent Model' (Fall 2010) 40(1) *Public Contract Law Journal* 63, 64.



## B *Review*

To conduct this literature review, many information sources were used. Primary sources of law and policy were identified and statutes, regulations, policy, strategy, caselaw, and executive orders as well as expert commentary were examined from the United States Code, Code of Federal Regulations, Congressional Record, Congressional Research Service, and the Government Accountability Office. Secondary sources included scholarly writing, textbooks, and news media.

The delimiting timeframes used varied by topic. Because of the recent advancements in the field of AI and its relationship with national security, searches for sources in the first topic were conducted within the last ten years with a preference and priority given for more contemporary sources. Research into contract law and the procurement system focused on existing law and recent commentaries, as well as legislative history, policy intent, and historic commentaries to assist in understanding how the current system operates and why previous efforts at reforming the system led to the current system. Thus, no delimiting timeframe was used for research into the second topic. Research into theory began as exploratory. Dissertations and articles in the fields of law, economics, science, psychology, sociology, innovation, and industrial organisation were reviewed to gain an understanding of potentially relevant theory.[160]

---

[160] Several other theories were considered as frameworks for the dissertation, including contract theory, principal-agent model, and public procurement for innovation. Contract theory posits optimal contracts are complete as they specify all actions the parties can take to maximise mutual benefit: see Klaus M Schmidt, 'Contributions of Oliver Hart and Bengt Homström to Contract Theory' (2017) 119(3) *Scandinavian Journal of Economics* 489, 490. In the principal-agent model, contract optimisation occurs when there is no asymmetry of information, but as agents have knowledge the principal does not, incentives are created to minimise the conflict: see Yukins (n 159) 63–5. Public procurement for innovation theory examines the interactions between public organisations and private actors in demand side innovation, identifying numerous barriers limiting industry participation in public procurement: see Hans Knutsson and Anna Thomasson, 'Innovation in the Public Procurement Process: A Study of the Creation of Innovation-Friendly Public Procurement' (2014) 16(2) *Public Management Review* 242, 250; Charles Edquist and Jon Mikel Zabala-Iturriagagoitia, 'Public Procurement for Innovation as Mission-oriented Innovation Policy' (2012) 41 *Research Policy* 1757, 1758. While each of these theoretical models provide useful ways to understand the research question, social exchange theory was chosen for its interdisciplinary utility and ability to help explain 1) the



The literature reviewed from these topics was integrated to create a theory of analysis used to assess the alignment of contract law with the business and technological considerations of AI development. This assessment led to findings and recommendations for the DoD to optimise its customer attractiveness through contracts. Through attracting engagement and competition from commercial AI firms, the DoD can access cutting edge AI advancements to field in its strategy to deter and defend against adversaries such as China and Russia.

### C *Structure: An Overview of the Literature Review Topics*

This study draws from literature in three distinct yet converging topics. The first topic is the state of artificial intelligence technology and its application and impact on defence and security. Literature comes from technical writings on AI, government reports, strategies, legislation, and commentary from academics, industry analysts, and defence leaders. The second topic is contract law and practice in the DoD. As contracts are the legal mechanisms by which the DoD leverages the commercial sector, a well-developed understanding of the law, policy, and practice of the procurement system is fundamental to examining the relationship between the DoD and commercial AI firms. The third topic is the overarching theoretical framework of social exchange theory literature, with a focus on the concept of customer attractiveness. The theory is used as a framework to explain how contract law and practice can affect the ability of the DoD to attract commercial AI firms to support national security.

The conceptual framework examines the relation and nexus of each of the topics. Below is a visual depiction of the literature at the nexus of national defence, AI, and the DoD's contract law and procurement practice. Using social exchange theory as the lens to view the literature,

relational dynamics between the buyer and seller, especially in a resource constrained environment, and 2) the factors that can affect the attractiveness of a buyer.



this research connects and synthesises existing literature to better understand the research question.

*Figure 3: Research Topic Framework*

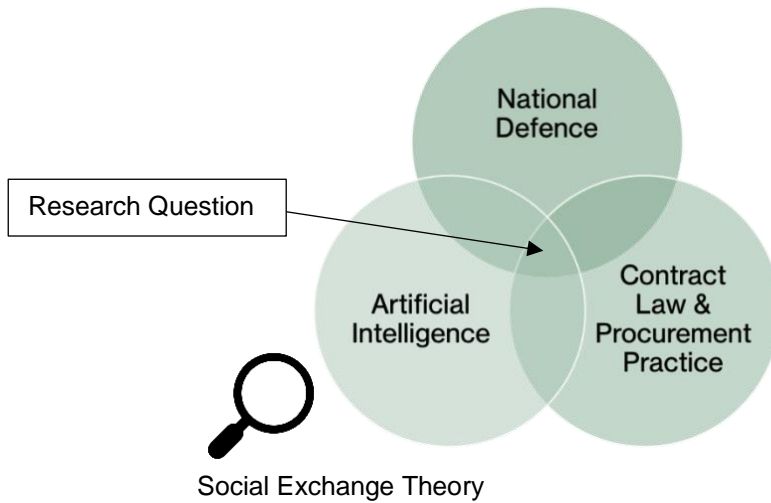

D *Topic I: Artificial Intelligence and National Defence Strategy*

1 *Introduction*

This first topic reviews literature that provides a background on the evolution of AI and how this technology differs from other technologies.  With a foundational understanding of how AI systems are developed, it becomes clear that AI capabilities will impact defence strategy and operations.  This topic examines the relationship between the DoD and the commercial firms that are developing AI-enabled capabilities, including the diminished role of DoD funding in private section research and development and the shift from defence-specific contractors to commercial firms as the source of military innovation.  The literature reviewed in this topic presents potential relational challenges between the DoD and commercial AI firms, leading to the review on the contract law frameworks in Topic II of this chapter.





Subject of science fiction since the nineteenth century, AI is a field that inspires awe, fear, and confusion, yet is now ubiquitous and mundane — just as it becomes increasing more powerful by the day.  There is robust debate on whether AI is over-hyped or under-valued.[161] What is not subject to debate, however, is the general consensus that AI is a revolutionary field that will greatly impact society, and likely disrupt the current competition and conflict paradigm.[162]  The impact AI will have on global society in the near future is alluded to in the economic projections: 'In 2018, the McKinsey Global Institute estimated that AI could add around 16 percent, or $13 trillion, to global output by 2030.  Since then, COVID-19 has further accelerated the use of AI'.[163]

Stuart Russell and Peter Norvig introduce the study of AI by tracing back to its historic roots to show how the principles of AI operation are grounded in a wide array of disciplines.[164] The fundamental concepts of AI are rooted in Aristotelian logic and rationality.[165]  During the Enlightenment, philosophers, including Hobbes, Descartes, and Bacon, further developed the concept of a rational machine that could calculate, use logic, and think outside of the physical realm.[166]  Mathematicians were able to build upon the logical and philosophical foundations to compute and predict outcomes with algorithms and processes, leading to the foundation of the field of computer science.[167]  Economists developed theories that provide application for AI that contributed to the notion of rational agents.[168]  While the previous disciplines investigate thought, reason, and decision-making in an abstract space, the field of neuroscience provides a physical and biological element to the study of intelligence.[169]  Psychology, the study of how one thinks,[170] combined with computer modelling to create the field of cognitive science in 1956, and demonstrated how computer models could be used to address memory, language, and logical thinking.[171]

---

[161] See, eg, Luciano Floridi, 'AI and Its New Winter: From Myths to Realities' (2020) 33 *Philosophy & Technology* 1–3; Sam Shead, 'Researchers: Are we on the Cusp of an "AI Winter"?', *BBC* (online, 12 January 2020) <https://www.bbc.com/news/technology-51064369>; Stephen C Slota et al, 'Good Systems, Bad Data?: Interpretations of AI Hype and Failures' (2020) 57(1) *Proceedings of the Association for Information Science and Technology* e275:1–11, 6; Melanie Mitchell, *Artificial Intelligence: A Guide for Thinking Humans* (Picador, 2019) 45.
[162] Kanaan (n 45) 232.
[163] Joshua P Meltzer and Cameron F Kerry, 'Strengthening International Cooperation on Artificial Intelligence', *Brookings* (Online Report, 17 February 2021) <https://www.brookings.edu/research/strengthening-international-cooperation-on-artificial-intelligence/>.
[164] See Russell and Norvig (n 18) 5–25.
[165] Ibid 5–6.
[166] Ibid 6.

These fields of study contributed to the knowledge that led to the first work recognised as AI by Warren McCulloch and Walter Pitts in 1943 that proposed a model of artificial neurons which demonstrated that any computable function could be computed by some network of connected neurons and net structures could implement all logical connectives.[172]  From the beginning of the study of AI, the field embraced the idea of an intelligent artificial agent behaving in a manner similar to humans that could operate autonomously in complex and changing environments.[173]

Despite decades of research into AI and the many applications that can be augmented by AI, it was not until the past decade that the algorithms that enable machines to learn have had the key ingredients to power meaningful AI advancement.[174]  Digitised data is available in an

---

unprecedented amounts,[175] computing power is faster and more cost-effective than ever,[176] performance of machine learning algorithms have improved to use on large-scale, complex problems, and many tools and libraries are free — these enablers have created the conditions for a Cambrian explosion of AI research and use cases.[177] This 'AI Revolution' has resulted in the advancement of AI at exponential speed, enabling AI methods such as deep learning to make complex predications by learning from data.[178] However, because of this rapid advancement, the state-of-the-art in AI is constantly evolving.[179]

Building on the discussion in Chapter I, there are several unique attributes of AI that require understanding to appreciate the impact it can have on society and military operations. As a technology, AI has similarities with modern software, though there are important differences that make development and contracting for AI-enabled capabilities challenging. First, AI is neither a single piece of software or hardware, but a 'constellation of technologies.'[180] Fundamental components of AI are interrelated and include data, hardware that provides the computing power, algorithms that learn through experience with new data, integration to other systems, and human talent that can build and test before deployment of the AI system.[181] A chart

---

[175] The top performing AI systems relied on the use of extra training sets, a trend that implicitly favours private sector actors with access to vast datasets such as Google, Amazon, and Netflix. Daniel Zhang et al, *The AI Index 2022 Annual Report* (AI Index Steering Committee, Stanford Institute for Human-Centered AI, March 2022) 51.

[176] Ibid (explaining the trend of lower training cost but faster training time supports the widespread commercial adoption of AI technologies).

[177] Melanie Reuter-Oppermann and Peter Buxmann, 'Introduction into Artificial Intelligence and Machine Learning' in Thomas Reinhold and Niklas Schörnig (eds), *Armament, Arms Control and Artificial Intelligence* (Springer, 2022) 11, 12.

[178] Brose (n 174) 61-2; Russell and Norvig (n 164) 27.

[179] See Edd Gent, 'Artificial Intelligence is Evolving All by Itself' (13 April 2020) *Science* doi: 10.1126/science.abc2274.

[180] National Security Commission on Artificial Intelligence, *Interim Report* (November 2019) 8 ('*NSCAI Interim Report*').

[181] See Gadepally et al (n 20) 1.



depicting the modern AI end-to-end pipeline and its components is included in Figure 4 below.[182]

*Figure 4: Modern AI canonical architecture*

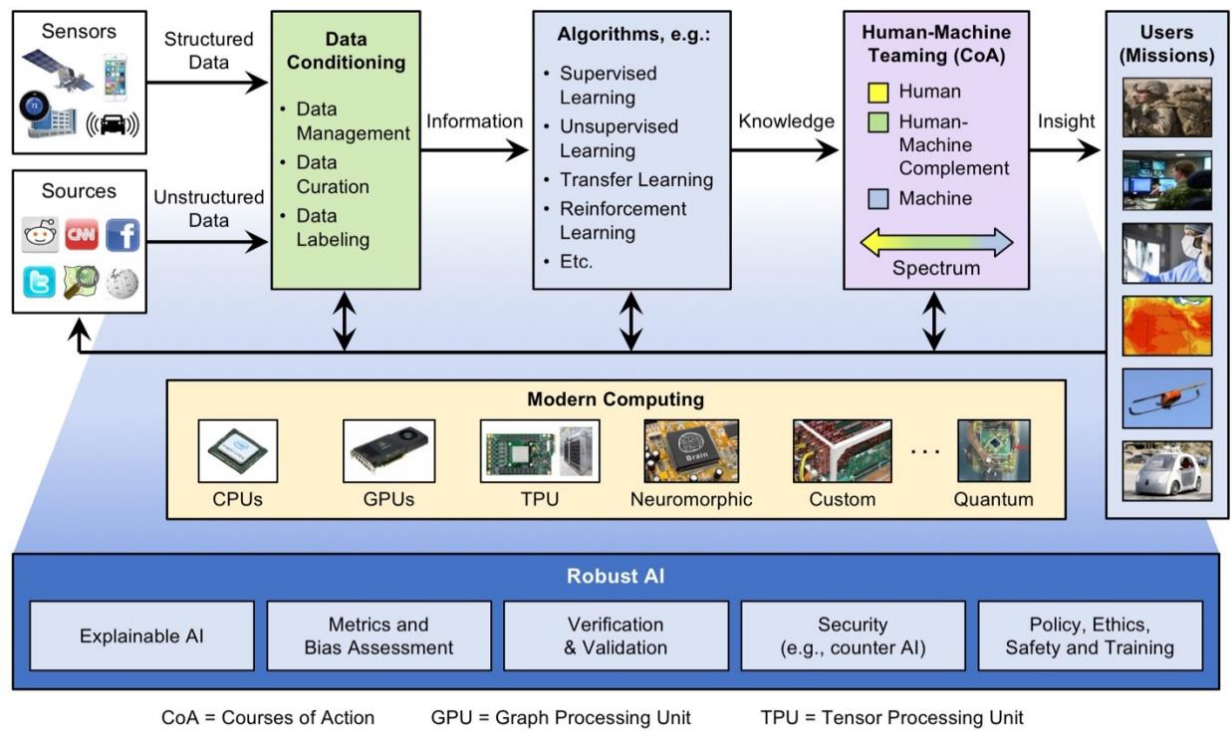

CoA = Courses of Action        GPU = Graph Processing Unit        TPU = Tensor Processing Unit

The development and lifecycle for conventional software starts with specifying the program's objective, implemented by a computer programmer writing code — the rules that the program follows for any possible input, or every different kind of problem the program is expected to encounter.[183]  Developing these detailed instructions is painstaking and limited; some tasks, like writing a program so that a computer can recognise faces, are impossible.[184] Machine learning is a powerful alternative for use cases involving describing what happened,

---

predicting what will happen, or making recommendations about what action to take.[185] Additionally, machine learning can recognise patterns that humans may have overlooked or which can only be detected by looking at more data than humans could ever process.[186] These functions are achieved through three subcategories of machine learning: supervised, unsupervised, and reinforcement learning.[187] Each of these techniques involves a process that starts with a body of data (the input) with the machine iteratively deriving rules to explain the data or predict future data, resulting in a model that serves as an algorithm for future computations of new data (the output).[188] The quality of the model is dependent on the quality of the data used to train with: representative examples of the intended use case in the training data is likely to yield refined models, just as biased data is likely to lead to biased models.[189] These techniques are used to execute a variety of complex tasks, including natural language processing, computer vision, deep learning, robotics, and autonomous decision making.[190]

In contrast to traditional computing programming, the lifecycle for a data-driven machine learning system does not implement the program's objective by writing code; rather, it is an iterative process of acquiring relevant data and training on the data before testing the machine learning system.[191] Machine learning builds statistical models based on data it observes and uses the model as both a hypothesis and as software that can solve problems.[192] This model continuously and iteratively trains itself using the available data to refine the algorithm that will ultimately produce a concise and environment-specific model.[193]  If successful, machine learning can accurately predict an event at the same or greater accuracy than humans.[194]  These models can make these predictions at high speeds — far outpacing any human.[195]  However, without properly formatted and conditioned data, the model will fail to achieve the intended objectives.[196]

---

[185] Thomas W Malone, Daniela Rus and Robert Laubacher, 'Artificial Intelligence and the Future of Work' (Research Brief No 17, MIT Work of the Future, December 2020) 6.
[186] Ibid 17.
[187] Ibid 7. Supervised learning refers to the use of manually labeled data to train the system. Unsupervised learning refers to systems that can identify patterns or anomalies using naturally occurring data without any specially added labels. Reinforcement learning is a technique where the objective is to train a system to take the best action by establishing a reward system, whereby the machine is programmed to maximise points and then rewarded points for optimising a task: see ibid.
[188] Ibid 6.
[189] Ibid.
[190] See ibid 6–10.
[191] *NSCAI Interim Report* (n 180) 56.
[192] Russel and Norvig (n 18) 651.

To mitigate the concerns about model failure due to data problems, a machine learning model still requires human interfacing throughout its lifecycle.[197]  Unlike traditional software, which is typically developed by a vendor and independently deployed by the user, there is much more interaction between the AI system's developer and the user as the user has knowledge and control of the training data.[198]  This difference requires frequent communication during the design and development of the machine learning system, and feedback during the deployment phase, where performance evaluation is complex; thus, machine learning is more of a sustainment and service than a product procurement like traditional software.[199]  Unlike traditional software programs, machine learning model behaviour is not easily specified in advance as it is dependent on the dynamic qualities of the data and various model parameters.[200]  Additionally, because a machine learning model is making predictions on previously unseen

data, and typically incorporates new data over time into training, it is important to continuously monitor performance and adjust the model features throughout its employment.[201] Ironically, humans play a greater role in the machine learning lifecycle than in traditional software programming. Humans are still responsible for defining the project, requiring a robust understanding of the task intended for modelling and the data available for training and deploying the model.[202] Humans are often better at labelling data appropriately in ambiguous circumstances.[203] Humans are also in the decision-making process where the knowledge from the model can be turned into actionable insight.[204]

Based on the literature reviewed discussing the technical development and deployment of AI systems, Figure 5 below synthesises the attributes of an AI and machine learning system compared to traditional computer programming.

*Figure 5: AI Attributes*

| Attribute | Traditional Computer Programming | Artificial Intelligence/Machine Learning |
|---|---|---|
| Speed | Incremental | Disruptive |
| Programming | Human generated code | Machine learning through data |
| Tasks | Software can only complete tasks it was explicitly programmed for, solving problems using simple logic | Can perform complex tasks such as identifying patterns, and making predictions, explanations, and recommendations based on data |
| Adaptability | No, every action is pre-programmed | Yes, model changes with new data |
| Lifecycle | Once deployed, program is sustained and updated, but complete | Iterative development and deployment cycles throughout lifecycle |
| Human-interfacing | Responsible for developing the program and sustaining deployed software | Critical in scoping the problem, providing feedback, labelling data, monitoring performance |
| Relationship between developer and user | Indirect; typically, little if any interaction | Interactive throughout lifecycle |
| Result of program | Known from outset of development and solved with simple rules | Unknown as program is data dependent and intended to make the most accurate predications possible on individualised problems |

---

[201] Ibid 6.
[202] See Gadepally et al (n 20) 2; Martin Erwig, *Once Upon an Algorithm* (MIT, 2017) 17.
[203] Gadepally et al (n 20) 12.
[204] See ibid 2.



The capabilities unlocked by AI will change how the DoD, its allies and its competitors operate.[205] The speed with which AI programs can process vast amounts of data can allow militaries to rapidly identify threats, targets, and anomalies.[206] AI-enabled systems can connect various platforms and domains that provide understanding of increasingly complex battlespaces and allow commanders to make decisions with more relevant information at greater speed.[207] AI applications for the DoD run the breadth of military functions, including intelligence, surveillance and reconnaissance; logistics; cyber operations; information operations; command and control; autonomous vehicles; weapon systems; and business processes.[208] However, because much of the underlying technology that enables this enhanced battlespace is developed by the commercial sector and available in the public domain, the speed at which the United States and its allies adopt, integrate, and field AI-enabled systems may determine the outcome of the next conflict, or deter it altogether.[209]

3 *The Commercial Sector's Role in Defence*

The current explosion in AI applications and capabilities happens to coincide with a geopolitical era of uncertainty.[210] After the end of the Cold War, the United States and its allies enjoyed decades of hegemonic power.[211] The United States demonstrated its military power in the Persian Gulf in the early 1990s; China noticed and began to transform its military strategy.[212]

---

[205] See ibid 9.
[206] Ibid 9–10.
[207] See ibid 10.
[208] See Congressional Research Service, *Artificial Intelligence and National Security* (CRS Report No R45178, rev ed, November 10 2020) 7–17 <https://fas.org/sgp/crs/natsec/R45178.pdf>.
[209] See ibid 10–11.
[210] See ibid 6.
[211] See Department of Defense, 'Summary of the 2018 National Defense Strategy of the United States of America: Sharpening the American Military's Competitive Edge' (Synopsis, 2018) 2–3 <https://dod.defense.gov/Portals/1/Documents/pubs/2018-National-Defense-Strategy-Summary.pdf> ('2018 NDS').
[212] See Kurt Campbell and Ely Ratner, 'The China Reckoning: How Beijing Defied American Expectations' (online, March/April 2018) *Foreign Affairs*, <https://www.foreignaffairs.com/articles/china/2018-02-13/china-reckoning>.



Because of the terrorist attacks conducted on 11 September 2001, and the ensuing wars in Afghanistan and Iraq, the United States shifted its focus to asymmetric warfare and counterterrorism, prioritising 20th century legacy platforms over innovative new systems to address 21st century threats.[213]  During that time, China and Russia took advantage and reconfigured their militaries and developed a new strategy to neutralise the United States' power projection, known as anti-access and area denial (A2/AD).[214]

Former Deputy Secretary of Defence Robert Work saw Russia and then China becoming increasingly aggressive with the use of sophisticated technologies and warned that the United States' technological advantage was being eroded at a relatively fast pace.[215]  Work developed a new 'offset strategy' that called for the United States to leverage AI and other advanced technologies to maintain competitive overmatch with China and Russia.[216]  This concept evolved into the 2018 National Defense Strategy ('NDS'), which sets out the strategic objectives for the DoD in the context of the return of great power competition.[217]  The underlying theme of the NDS is the role AI will play: whichever military force is quickest to adapt and field innovations in AI will hold an advantage.[218]  This warning proved prophetic when Russia, with a massive conventional weapon advantage, suffered catastrophic losses against a more nimble, technologically advanced Ukrainian force.[219]  General Mark Milley, the chairman of the Joint

---

[213] See Brose (n 174) 11–5.

[214] See ibid 32–4.  China's anti-access and area denial weapons focus on neutralising the United States' power projection by concentrating stand-off capabilities on forward operating bases and aircraft carriers, making defence most costly and offensive strikes more challenging: see ibid.

[215] Ibid 39.

[216] Ibid.

[217] See *2018 NDS* (n 211) 4.

[218] See ibid 3.

[219] See Valeriya Ionan, 'Ukraine's tech excellence is playing a vital role in the war against Russia', *Atlantic Council* (Blog Post, 27 July 2022) <https://www.atlanticcouncil.org/blogs/ukrainealert/ukraines-tech-excellence-is-playing-a-vital-role-in-the-war-against-russia/>; Mauro Gilli, 'Beware of Wrong Lessons From Unsophisticated Russia', *Foreign Policy* (online, 5 January 2023) <https://foreignpolicy.com/2023/01/05/russia-ukraine-next-war-lessons-china-taiwan-strategy-technology-deterrence/#mauro-gilli>.



Chiefs of Staff, explained that 'tenacity, will and harnessing the latest technology give the Ukrainians a decisive advantage' against Russia and that Ukraine's use of advanced AI applications in the battlespace is demonstrating 'the ways wars will be fought, and won, for years to come.'[220] The NSCAI noted that AI is the quintessential dual use technology — AI systems that 'can perceive, decide, and act more quickly, in a more complex environment, with more accuracy than a human — represents a competitive advantage in any field.'[221] From a military perspective, the ability to operate in complex and changing environments is necessary to achieve its objectives to deter conflict and outmatch opposing forces in conflict. Because many of these technological advancements were developed not by traditional defence companies, but commercial firms unaccustomed to, and perhaps uninterested in, working with the DoD,[222] access to this innovation is open to all competitors.[223]

This point — that technology capable of changing the character of war is developed by commercial firms — is significant from a defence technology perspective but continuing a long run of commercial success in the field of AI. Since the 1980s, advancements in AI came not from the traditional defence industrial base, but by private sector firms.[224] The source of innovation and funding are critical factors that makes AI unique in terms of military technology.[225] The private sector outspends the government to develop AI for private sector customers, leading to significant breakthroughs.[226] These AI firms offer lucrative salaries, advanced technical toolsets, and a workplace culture that is less bureaucratic than the public

sector and defence industrial base.[227]  Accordingly, the traditional defence companies are outcompeted for AI talent; moreover, they are not accustomed to pursuing risky investments in unproven technologies.[228]

The US Defense Innovation Board, an advisory organisation reporting to the DoD made up of technology experts in the commercial sector and academia, observed, '[t]his is the first time in recent history that neither the DoD nor the traditional defence companies it works with controls or maintains favourable access to the advances of computing and AI for both for civilian and for military relevant technologies.'[229]  Because the development of and investment in AI applications is focused on commercial business rather than the military, the DoD has to compete against private businesses that drive the market; typically, the DoD is the singular or primary customer of a military technology, like stealth, nuclear power, or global positioning systems in previous decades.[230]  The DoD recognises that '[n]ew commercial technology will change society and, ultimately, the character of war'[231]; however, it will 'need to collaborate with companies that do not think of themselves as defence contractors' if it wants access to the cutting edge of AI research.[232]

4 *Potential Barriers to Commercial Collaboration*

   a.  *Diminished Influence in Technology Advancements*

The DoD's challenge in attracting a different commercial market outside the defence industrial base is compounded by the diminished influence of DoD research and development funding.  This issue is also a relatively new phenomenon.  The Soviet's successful launch of

---

[227] Ibid 2.
[228] Ibid.
[229] *AI Principles: Recommendations* (n 22) 15.
[230] See ibid.
[231] *2018 NDS* (n 211) 3.
[232] Ryseff et al (n 81) 2.



*Sputnik* in 1957 led to a concentrated strategy of the U.S. government awarding lucrative contracts to technologists to spur innovation, creating Silicon Valley as the DoD's own start-up.[233]  Because there was a real belief that the United States could lose the Cold War to the Soviet Union, its procurement priorities were clear: pick and fund the people who could quickly build technology that would help ensure military superiority over any adversary.[234]  This hyper-focused strategy made trade-offs, with the procurement goals of speed and effectiveness dominating concerns of fairness and efficiency.[235]  This system rewarded innovators in Silicon Valley that resulted in advanced technologies and a culture that believed working with the DoD could make them rich and contribute to national security.[236]

However, over the decades, defence contracting evolved and began to focus on efficiency, oversight, and control with Congress programming costs years in advance.[237]  The DoD now struggles to keep pace with the private sector.[238]  Since the early 1990s, after the fall of the Soviet Union, Congress and the DoD increased its technological R&D spending by 10 percent; meanwhile, commercial industry increased R&D investments by 200 percent.[239]  Worryingly for the DoD, '[t]here are multiple Silicon Valley and Chinese companies who each spend more annually on AI R&D than the entire United States government does on R&D for all of mathematics and computer science combined.'[240]  The DoD has lost its past market dominance and no longer drives 'the focus or direction of technology development.'[241]  Now,

---

few companies receiving private funding from venture capital or investors innovate military-specific technology.[242] Rather, private-sector investment predominantly goes to AI companies that are focused on commercially viable applications with low risk recurring revenue, not on challenging defence applications.[243] The prevalent business model is to innovate general purpose AI technologies that can later be adapted by the DoD.[244]

b. *Potential Ethical Concerns in Developing AI for Military Use*

One of the prevailing narratives of why commercial AI firms may be reluctant to work with the DoD focuses on the ethical implications of developing AI for military applications. Thus, as discussed in Chapter I, the fallout after Google ended its work on Project Maven serves as an apt case study to view the relationship between technology firms in Silicon Valley and the

---

[242] Zachary Arnold, Ilya Rahkovsky and Tina Huang, *Tracking AI Investment: Initial Findings from the Private Markets* (Report, Center for Security and Emerging Technology, 2020) 24. The origin of the defence start-up 'Shield AI' is a case study exemplifying the challenge new firms face in securing private funding to develop AI systems for defence solutions. Shield AI is a start-up founded by former Navy SEAL Brandon Tseng who, after a military career spent in close quarters combat where one of the most dangerous encounters was entering a building without knowing what is waiting across the threshold. Desperate to solve this problem that claimed so many service members' lives, Shield AI developed AI-powered autonomous quadcopters that help provide situation awareness to teams before breaching an enemy compound. Even though this solution would solve a problem that plagued U.S. troops in Iraq and Afghanistan, 29 out of the 30 potential investors passed on the idea. The one that took a pitch from Shield AI wanted to convince Tseng to develop the drone for taking selfie photographs. Without outside investors, Shield AI was backed by family and friends to develop a prototype. It was able to secure a small contract with Defense Innovation Unit (DIU), which helped reduce the standard defence contracting obstacles by leveraging OT authority. See Elliot Ackerman, 'A Navy SEAL, a Quadcopter, and a Quest to Save Lives in Combat', *Wired* (online, 30 October 2020) <https://www.wired.com/story/shield-ai-quadcopter-military-drone/?utm_source=onsite-share&utm_medium=email&utm_campaign=onsite-share&utm_brand=wired>; Mitchell Weiss and A J Steinlage, *Shield AI* (Harvard Business Review Case Study 9-819-062, 3 October 2018) 8.
The reason investors initially balked at backing Shield AI appear to be some parts an erroneous belief that the military already had a solution for this problem, and some parts a belief by the commercial sector that defence applications are too risky of an investment: see Weiss and Steinlage (n 242) 6, 8, explaining that the complex defence procurement system ends up costing more, takes longer, and results in lower profits than the commercial system; this problem is exacerbated when the DoD is buying AI which can refresh multiple times during the lengthy contracting cycle.
[243] Arnold, Rahkovsky and Huang (n 242) 24. Only one percent of all disclosed private investment into commercial AI companies went to companies focused on military, public safety, and government applications, while the estimated total private investment into military applications amounted to no more than a rounding error: at 25.
[244] Ibid 25. Highlighting this dichotomy between technology firms and traditional defence contractors is the fact that Google is worth over twice the value of the entire defence industrial base. Linell A Letendre, 'Google…It Ain't Ford: Why the United States Needs a Better Approach to Leveraging the Robotics Industry' (2017) 77 *Air Force Law Review* 51, 56–7.



DoD.  Prior to the controversy at Google, Dr. Fei-Fei Li, the chief scientist for AI at Google

Cloud, recognised working on DoD contracts was a lightning rod issue.[245]  She urged her

colleagues to use caution when speaking about military applications not because of actual ethical

concerns, but because '[w]eaponized AI is probably one of the most sensitized topics of AI – if

not THE most.  This is red meat to the media to find all ways to damage Google.'[246]  Once

Google's employees protested the company's involvement in a DoD contract to provide real-

time analysis capabilities to the DoD, Google determined not to renew its contract.[247]

The decision was derided by national security experts, including General Joseph

Dunford, the Chairman of the Joint Chiefs of Staff, claiming Google's objection to providing AI-

enabled capabilities to the United States military was 'indirectly benefiting the Chinese

military.'[248]  Others pointed out that Project Maven was not offensive in nature, and could make

it easier for the military to distinguish combatants from civilians and save civilian lives,[249]

enabling the DoD to better meet its legal obligations.[250]

Despite the widespread media coverage on Project Maven, it appears that most

commercial AI firms are willing to work with the DoD, and the 'ethical concerns' of developing

AI for military purposes are rarely clearly distinguished from the broader debate about developing ethical AI.[251]  Since Google's decision to abstain from working with the DoD, many other tech giants have competed for and performed DoD contracts.[252]  Many technologists, including former Google CEO and NSCAI chairman Eric Schmidt, have called for more AI in defence.[253]  Google subsequently re-entered the defence market with a new section devoted to the public sector, including work with the DoD, joining other commercial technology firms such as Microsoft, Amazon, Oracle, and Palantir,[254] along with many smaller commercial AI firms.[255]

---

[245] Gianpiero Petriglieri, 'Google and Project Maven (B): An Eventful Week in June' (Case Study No 6408, Insead, 2018) 1.
[246] Ibid.
[247] Ibid 2.
[248] Zak Doffman, 'Google Accused by Top U.S. General and Senator of Supporting Chinese Instead of U.S. Military', *Forbes* (online, 16 March 2019) <https://www.forbes.com/sites/zakdoffman/2019/03/16/google-accused-by-u-s-general-and-senator-of-benefiting-chinese-instead-of-u-s-military/?sh=29d99cdb1899>.
[249] Petriglieri (n 245) 2.  Project Maven sought to detect military objects of interest in the battlespace using computer vision:  Cheryl Pellerin, 'Project Maven to Deploy Computer Algorithms to War Zone by Year's End', *DOD News* (online, 21 July 2017) <https://www.defense.gov/News/News-Stories/Article/Article/1254719/project-maven-to-deploy-computer-algorithms-to-war-zone-by-years-end/>.
[250] See *Protocol Additional to the Geneva Conventions of 12 August 1949, and Relating to the Protection of Victims of International Armed Conflicts*, opened for signature 8 June 1977, 1125 UNTS 3 (entered into force 7 December 1978) art 48 ('*Protocol I*'), requiring that parties to a 'conflict shall at all times distinguish between the civilian population and combatants and between civilian objects and military objectives and accordingly shall direct their operations only against military objectives'.

As this incident demonstrated, there are at least some commercial AI firms that will cite their employees' or financial backers' concerns about the ethics of developing technology on behalf of the military as a reason to avoid competing for contracts with the DoD.  While much has been written on the quandary of developing ethical AI, particularly in the context of AI used in military applications,[256] there is limited research beyond the anecdotal case study that examines the impact this concern has in attracting commercial AI firms to work with the DoD.  Only one contemporary study[257] addresses the question of how much ethics affects the willingness of commercial AI firms to work with the DoD.  The study, conducted at Georgetown University's Center for Security and Emerging Technology (CSET), found that AI professionals are largely undeterred from working with the DoD, at least due to ethical concerns.[258]  However, the study found that AI professionals at commercial firms cited discomfort with how the DoD will use the technology and concerns about causing harm as the most common reasons for not working on DoD-funded projects.[259]

---

[251] See Baker (n 195) 254–6, noting that the DoD must adhere to the law of armed conflict at all times, and must review an AI application for legal and ethical policy compliance before authorising use of such application — these binding laws, though reflective of the generally accepted ethical considerations for the use of AI in decision making, data management, and bias, do not apply to companies building and selling AI applications.

[252] See Melissa Heikkila, 'Why Business is Booming for Military AI Startups', *MIT Technology Review* (online, 7 July 2022) <https://www.technologyreview.com/2022/07/07/1055526/why-business-is-booming-for-military-ai-startups/>.

[253] Ibid.

[254] See Frank Konkel, 'Pentagon Awards $9B Cloud Contract to Amazon, Google, Microsoft, Oracle', *Nextgov* (online, 7 December 2022) <https://www.nextgov.com/cxo-briefing/2022/12/amazon-google-microsoft-oracle-awarded-9b-pentagon-cloud-contract/380596/>.

[255] See Heikkila (n 252); see Steven Levy, 'Inside Palmer Luckey's Bid to Build a Border Wall, *Wired* (online, 11 June 2018) <https://www.wired.com/story/palmer-luckey-anduril-border-wall/>.

[256] See Lindsey R Sheppard et al, *Artificial Intelligence and National Security: The Importance of the AI Ecosystem* (Report, Center for Strategic and International Studies, 2018) 56–7, noting that ethics are often reflected in law which can guide development of technology, but because the development of AI outpaces the legislative and regulatory processes, the law is reactionary); Paul Scharre, 'Autonomous Weapons and Stability' (PhD Thesis, King's College London, March 2020) 23.

[257] Published shortly after completion of this study's survey period.

This study, while providing insight to challenge the prevailing narrative that ethics is the primary concern of commercial AI firms in preventing them from working with the DoD, does not address the impact the DoD's unique contract law framework has on the willingness of commercial AI firms to provide AI capabilities to the DoD.  This study shows that if there is a reason that is limiting the engagement of commercial AI firms with the DoD, it is likely not ethical concerns.  Ethics is, and will remain, a critical component of the conversation regarding the advancement of AI, both within and outside the military context, but it is important for all parties to the debate to distinguish ethics — 'the set of values, principles, and techniques that employ widely accepted standards of right and wrong to guide moral conduct in the development and use of AI technologies'[260] — with image, public relations, and business concerns.[261]  Ethical concerns regarding the responsible use of AI applications are broadly applicable.[262]  This dissertation's original research findings, as discussed in Chapter IV, are consistent with the CSET study's conclusion that ethics is not a significant barrier for commercial AI firms working with the DoD, allowing the focus to shift to other factors that influence the decision of commercial AI firms on whether to work with the DoD.[263]

---

[258] See Catherine Aiken, Rebecca Kagan and Michael Page, '"Cool Projects" or "Expanding the Efficiency of the Murderous American War Machine?": AI Professionals' Views on Working with the Department of Defense' (Study, Center for Security and Emerging Technology, November 2020) 2–3.
[259] Ibid 2.
[260] David Leslie, *A Guide for the Responsible Design and Implementation of AI Systems in the Public Sector* (The Alan Turing Institute, 2019) 3.
[261] Vincent C Müller, 'Ethics of Artificial Intelligence and Robotics' [2021] (Summer) *The Stanford Encyclopedia of Philosophy* <https://plato.stanford.edu/entries/ethics-ai/>.
[262] See *2022 DoD Responsible AI Strategy* (n 10) 5–8, explaining the DoD's approach to developing AI ensures the safety of systems and their ethical employment through ethical guidelines, testing standards, accountability checks, employment guidance, human systems integration, safety considerations, and legal obligations; *Council on AI Recommendation* (n 19) 4, recommending establishing ethical principles for responsible stewardship of trustworthy AI).  These ethical standards, the first applicable to the DoD's development and deployment of AI applications and the second applicable to the international community on government and private AI applications, are consistent.

c. *Commercial Business Practices are Unaligned with Defence Contracting*

Given the impact AI systems are expected to have on national security, there are many government reports and strategies that call for the DoD to prioritise and focus efforts to build AI competency, understanding, development, and application.[264] Few, however, discuss the challenges the DoD faces in acquiring AI innovation at the speed and scale necessary to remain competitive with China, which has invested in AI research and developing a market conducive to AI adoption and improvement,[265] and Russia, which has demonstrated it can leverage AI to efficiently disrupt political narratives with misinformation.[266] The DoD's AI Strategy boldly proclaims it will partner with the 'best academics' and 'industry to align civilian AI leadership with defence challenges,' though it does not explain how it will do so.[267] Unlike China's 'military-civil fusion,' the DoD cannot force academia and industry to advance AI development for military purposes[268] — the mechanism for leveraging non-state actors towards state goals in the United States is to entice them via contracts.[269]

---

[263] While ethics does not appear to be as significant of an obstacle for the DoD to overcome in leveraging commercial AI innovation as contracts, the two concepts are related. See Bowne and McMartin, (n 155) 9, explaining implementation of responsible AI principles requires a new data licensing framework to account for the unique role data plays in the development and deployment of an AI system.

[264] See, eg, *2018 DoD AI Strategy Summary* (n 8); Executive Order 13859 (n 145); Select Committee on Artificial Intelligence of the National Science & Technology Council, *The National Artificial Intelligence Research and Development Strategic Plan: 2019 Update* (Report, June 2019); Fei-Fei Li and John Etchemendy, 'We Need a National Vision for AI', *Stanford University Human-Centered Artificial Intelligence* (Blog Post, 22 October 2019) <https://hai.stanford.edu/news/we-need-national-vision-ai>; Chris Bassler and Bryan Durkee, 'The Next Steps for the Pentagon's AI Hub', *Defense One* (online, 29 January 2021) <https://www.defenseone.com/ideas/2021/01/pentagons-ai-hub/171721/>.

[265] Daitian Li, Tony W Tong and Yangao Xiao, 'Is China Emerging as the Global Leader in AI?', *Harvard Business Review* (online, 18 February 2021) <https://hbr.org/2021/02/is-china-emerging-as-the-global-leader-in-ai>.

[266] See Katerina Sedova et al, *AI and the Future of Disinformation Campaigns: Part 1* (Report, Centre for Security and Emerging Technology, December 2021) 14.

[267] *2018 DoD AI Strategy Summary* (n 8) 7.

The NSCAI is a rarity in that it acknowledges the DoD's traditional contract law and practice pose challenges to commercial AI firms, affecting the DoD's attractiveness as a customer.[270] The NSCAI predicts the DoD will encounter challenges in numerous areas: start-ups deciding that doing business with the DoD is too difficult; the DoD's internal contracting, budgeting, and programming practices prevent promising prototypes from scaling; or the perception that working with the DoD prevents a commercial AI firm from competing for opportunities or investors urge against working with the DoD.[271] The NSCAI concluded that the DoD still operates at human speed when machine speed is necessary to achieve its national security goals.[272] Though the NSCAI explains that the government must undergo changes in business practices, organisational culture, and mindsets to adopt AI capabilities, each of these changes is required because the DoD needs commercial AI firms to develop and deliver capabilities to address defence application.[273] The legal mechanism through which the DoD leverages commercial innovation is contracts; thus, the law governing merits examination.

Just as commercial AI firms are atypical sellers in the defence context, the DoD is an atypical buyer in the commercial context. There is no bigger, more complex, better funded customer in the world.[274]  However, the most glaring difference between the DoD and any other customer is the DoD-specific requirements placed on potential vendors if they are to sell to the DoD.  Some of these requirements include: registering as a seller to the government[275]; certifying ownership of the company[276]; certifying that certain parts of their supply chain are not made in other countries[277]; certifying price and cost data in a special accounting system[278]; obtaining third-party certification of compliance with cybersecurity standards[279]; agreeing to mandatory clauses on issues ranging from drug-free workplaces[280] to training employees on identifying human trafficking[281]; and signing a contract governed by laws that apply exclusively to contracts with a single customer in the world.[282]

---

[268] *NSCAI Interim Report* (n 180) 18.
[269] See Baker (n 195) 173. The only practical way the United States can compel industry is under the Defense Production Act (DPA) — however, such authority can be challenged by industry in federal court and is viewed harshly by business critics as centralised government control of the economy. Thus, the DPA, though likely applicable to national security innovations in AI, has not been used to the limits of the law. Ibid. Additionally, using the DPA to compel commercial firms to develop AI may not be in the best interest of the DoD as it attempts to attract commercial partners.
[270] *NSCAI Interim Report* (n 180) 58.
[271] Ibid 60.
[272] *NSCAI Final Report* (n 4) 7.
[273] See ibid.

Although there are numerous factors that may affect a commercial AI firm's decision to compete for a DoD contract, the unique contract laws and procurement processes govern how the DoD engages with commercial AI firms.  The law and practice present obstacles and incentives alike for commercial firms.  It is in this context that the study focuses on the relationship between commercial AI firms and the DoD as the contract serves as both the conduit for the creation of the relationship and governs the dynamics between the parties.

5 *Synthesis of Topic I*

AI is a revolutionary technology, disrupting civilian and military domains alike.  The unique development of AI systems and their capability to aid humans in accomplishing complex tasks at incredible speed and scale make AI fundamentally different from other technologies. AI advancements are led by a sector of the commercial market that has not traditionally worked with the DoD; yet AI has been identified by the United States government as the technology necessary to compete in the current geopolitical and military landscape. There are several potential challenges the DoD must overcome to access AI innovation from the commercial sector. These obstacles include the diminished role the DoD plays in the AI market compared to other defence technology domains; the ethical, and perhaps business, concerns stemming from developing AI systems for military applications; and a contract law framework that appears unaligned with both the business practices of commercial AI firms and the development and deployment cycles for AI systems. The literature review in this topic reveals that while there are many possible challenges the DoD will face in attracting commercial AI firms, it is clear the contract law framework plays a significant role. Because the formal nexus between commercial AI firms and the DoD is manifested by a contract, this dissertation seeks to understand how the existing contract law frameworks available to the DoD align with the DoD's goals to access AI from the commercial market. The next topic below reviews the two primary contract law frameworks available to the DoD to engage with commercial AI firms — the traditional Federal Acquisition Regulations (FAR) and other transaction (OT) authority.

---

[274] The Department of Defense spent nearly $400 billion in contracts in Fiscal Year 2019.  Government Accountability Office, 'A Snapshot of Government-wide Contracting for FY 2019 (Infographic) (26 May 2020) <https://blog.gao.gov/2020/05/26/a-snapshot-of-government-wide-contracting-for-fy-2019-infographic/>.
[275] See 48 CFR § 52.204-7.
[276] See 48 CFR § 52.219-8.
[277] See 48 CFR § 52.225-1.
[278] See 48 CFR § 52.230-2.
[279] See *DFARS* §§ 252.204-7012, 7021.
[280] See 48 CFR § 52.223-6.
[281] See 48 CFR § 52.222-50.
[282] See generally 48 CFR and *DFARS*.

E *Topic II: Department of Defense Contract Law Frameworks*

The second topic of literature reviewed is defence acquisition, specifically contract law frameworks in the DoD.[283] The body of defence acquisition literature comprises of the law,

---

[283] The DoD's Defense Acquisition System — see Office of the Under Secretary of Defense for Acquisition and Sustainment, 'The Defence Acquisition System' (DoD Directive No 5000.01, rev ed 28 July 2022) — is composed of three parts that start with requirements development through the Joint Capabilities Integration and Development System (JCIDS) (see Joint Chiefs of Staff, 'Charter of the Joint Requirements Oversight Council and Implementation of the Joint Capabilities Integration and Development System' (Chairman of the Joint Chiefs of Staff Instruction 5123.01I, 30 October 2021)); programming and budgeting for requirements through the Planning, Programming, Budget and Execution (PPBE) (Department of Defense, 'The Planning, Programming, Budgeting,



policy, research, and practice relevant to the contracts that allows the DoD to engage with the commercial sector.  Through the contractual vehicles and procurement process, the DoD can obtain AI-enabled capabilities from commercial firms.  This topic is divided into three subtopics of literature: the purpose and goals of public procurement; acquisition reform efforts; and comparing the two predominant contract law frameworks available to the DoD, the traditional FAR-based procurement contracts and the non-traditional other transactions agreements.

### 1 Purpose and Goals of United States Public Procurement Law

Public procurement as a field of study is multi-disciplinary with research extending into scholarship on management, public administration, mathematics, finance, law, logistics management, information technology, and innovation.[284]  There are many goals of public procurement, some of which focus on the procurement itself, such as reducing cost, increasing quality, timeliness, risk management, increasing competition, and maintaining integrity and transparency.[285]  Additionally, public procurement serves to meet other policy goals, including social, economic, and international relations objectives.[286]  Because these goals are sometimes in conflict with one another, policy makers and public procurement professionals must decide which goals are most important to the procurement and make trade-offs of these goals.[287]

---

and Execution (PPBE) Process' (DoD Directive No 7045.14, rev ed, 29 August 2017)); and Defense Acquisition Process (Office of the Under Secretary of Defence for Acquisition and Sustainment, 'Operation of the Adaptive Acquisition Framework' (DoD Instruction 5000.02, rev ed, 8 June 2022)), program management to fulfil the requirement.  The requirements, budgeting, and acquisition components of the Defence Acquisition System are internal processes involving interplay between the various levels of bureaucracy within the DoD, the Executive Branch of the President, and the Legislative Branch of Congress.  The scope of this dissertation is limited to an assessment of the external driver of the system — contracts with the private sector.  However, it is important to note that contracting is just one component of the budgeting and acquisition framework.

[284] Anthony Flynn and Paul Davis, 'Theory in Public Procurement Research' (2014) 14(2) *Journal of Public Procurement* 139, 139–40.

[285] Khi V Thai, 'Public Procurement Re-Examined' (2001) 1(1) *Journal of Public Procurement* 9, 27.

[286] Ibid.

[287] See ibid.



Over twenty years ago, Khi Thai postulated that public procurement, despite a recorded history of nearly 5000 years and as one of the most important economic activities of government, lacks academic focus.[288]  Furthermore, notwithstanding the complexity of the American procurement system, and that federal procurement officers make on average a purchase every 0.31 seconds per each working day, there was not a set of common public procurement goals.[289] Thai also describes how the legal environment shapes public procurement, noting that the legal framework governs all business activities.[290]  This study is principally conducted from a legal perspective, with an emphasis on how the legislative process that enacts laws that are implemented by regulations and policy by the DoD interact with and influence the procurement practice.  A contract is the legal instrument that formalises the relationship between the DoD and commercial AI firm.[291]  Because the contract is a key mechanism that can be altered to affect the relationships between the variables in this study, the law governing contracts with the DoD is the primary field of research in this dissertation.

Steven Schooner's 'Desiderata' asks 'what does a government hope to achieve through its government procurement law?'[292]  He pointed out that while the federal government has been entering into contracts for as long as it has existed, there have been limited attempts to 'rationalize this phase of governmental activity in its relation to the functions of government and to the persons and firms with whom contracts are made.'[293]  Schooner provides his attempt to rationalise government contracts through discussion of nine goals of government procurement systems: competition, integrity, transparency, efficiency, customer satisfaction, best value, wealth distribution, risk avoidance, and uniformity.[294]  The first three are fundamental to the United States procurement system which, he argues, offers access to the best contractors, lowest prices, most advanced technology, favourable contract terms and conditions, and the highest quality goods.[295]  This access is attributed to the wide participation of potential competitors, the perception of fair treatment, and meaningful profit incentives.[296]  By maximising competition, the government leverages the power of the marketplace; competition is maximised by attracting contractor participation by instilling integrity and transparency into the system.[297]  However, other procurement goals, such as efficiency and customer satisfaction factor into the overarching framework.[298]

---

[288] Ibid 9, 11.
[289] Ibid 24, 27.
[290] Ibid 34.
[291] See Peter Kamminga, 'Rethinking Contract Design: Why Incorporating Non-Legal Drivers of Contractual Behavior in Contracts may Lead to Better Results in Complex Defence Systems Procurement' (2015) 15(2) *Journal of Public Procurement* 208, 210.
[292] Steven L Schooner, 'Desiderata: Objectives for a System of Government Contract Law' (2002) 2 Public Procurement Law Review 103, 103.
[293] Ibid 103 n 1, quoting J W Whelan and E C Pearson, 'Underlying Values in Government Contracts' (1962) 10 *Public Law* 298, 298.

Schooner explains that no procurement system can achieve all of these goals, and several of these goals are in conflict with each other.[299]  A procurement system incurs significant transactional, economic, and social costs when maximizing transparency, integrity, and competition.[300]  Although Schooner argues that these costs are an excellent investment,[301] it is worth considering whether the National Defense Strategy's goals in achieving technological advantage over its competitors is compatible with this trade-off.  Assuming the best method of accessing the widest pool of potential contractors is through maximizing competition, it follows that the perceptions and preferences of the commercial AI firms the DoD hopes to attract are critical factors in balancing these conflicting goals. The discussion of procurement goals is government-centric; fairness and transparency of a system promotes competition only if the pool of potential contractors perceives the system as fair and transparent. These goals must also balance the equation of incentives for the contractor, particularly in a market where the government is a relatively small customer as it is for AI-enabled technology.

---

[294] Schooner (n 292) 103.
[295] Ibid 104.
[296] Ibid.
[297] Ibid.
[298] Ibid 107.
[299] Ibid 110.
[300] Ibid.
[301] Ibid.

Often in discussions of public procurement objectives, the focus is on the public organisation, or, in the case of socioeconomic policies, on ensuring participation of discrete populations. Even in socioeconomic policies, the consideration is a trade-off of open competition for macroeconomic or social objectives, not whether the policies help ensure the military requirements are met timely, with the requisite quality.[302] The objectives of these policies are often tangential or unrelated to better leveraging emerging technology at speed and scale to meet the National Defense Strategy's objectives. The central mission of the DoD is to provide the military with the forces and equipment required to deter war and ensure the security of the United States.[303] Thus, the DoD may consider utilising contract law to make trade-offs of unrelated policy to maximise alignment with procurement goals like competition, efficiency, and customer satisfaction when it procures AI. Additionally, these goals rarely contemplate the perspective or preferences of the vendor. If maximising competition is the best approach to accomplishing its mission, the DoD should implement practices that align its procurement system with the preferences of potential contractors and make informed trade-offs of other procurement goals that conflict or detract from this goal. The next section discusses the history of acquisition reform, explaining how Congress has shaped contract law to balance the procurement goals, and leading into the discussion on the two legal frameworks that arose from those reforms.  The review of acquisition reform history shows that many of the problems the DoD is currently experiencing in attracting commercial engagement in advancing science and technology have existed for decades.  Moreover, this review contextualises congressional actions in passing legislative fixes to these problems and underscores its intent for the DoD to leverage alternative contract law.

---

[302] See 48 CFR Part 19.
[303] 'About', *U.S. Department of Defense* (Web Page) <https://www.defense.gov/About/>.

## 2 *Acquisition Reform Efforts*

The first post-World War II shake up to the acquisition system came as a reaction to the Soviet's successful launch of *Sputnik*.  When the first satellite was launched, 'a new element was injected into the urgency to proceed with acceleration of basic science and technology efforts.'[304] However, when the second Soviet satellite was successfully launched and was clearly far superior to the capabilities of the United States, it was viewed as a sign that the United States had indeed fallen behind technologically; the 'national prestige and security of the free world were of primary concern because of the psychological impact' of the Soviet victory in the space race.[305] This quickly led to Congressional action, and the National Aeronautics and Space Act of 1958[306] ('the Space Act') was enacted, creating the National Aeronautics and Space Administration (NASA).[307]  Intending to harmonise the goals of industry and government to better access private innovation, the Space Act provided authority for NASA 'to enter into and perform such contracts, leases, cooperative agreements or *other transactions* as may be necessary in the conduct of its work'.[308]  This was the first grant of other transaction authority in the federal government, which allowed NASA to enter into agreements that were not regulated by standard contract law and regulations.

---

[304] Wilson R Maltby, 'The National Aeronautics and Space Act of 1958 Patent Provisions' (1958) 27(1) *George Washington Law Review* 49, 50.
[305] Ibid 51.
[306] *National Aeronautics and Space Act of 1958*, Pub L No 85-568, 72 Stat 426 ('*Space Act of 1958*').
[307] *See* Maltby (n 304) 51.
[308] *Space Act of 1958* (n 306) § 203(b)(5) (emphasis added).

While the Space Act sparked the United States into action and resulted in significant technological advancements in a relatively short time frame, procurement reform since has had a mixed record at best. In the 1960s, in an effort to eliminate costs overruns, Defence Secretary Robert McNamara required formal source selection procedures, contractor performance evaluations, and fixed-price contracting.[309] These proved ineffective as the major programs of the time all experienced large cost overruns.[310] In 1962, the *Truth in Negotiations Act* ('TINA') sought to limit the opportunity for contractors to make excessive profits off of government contracts by requiring the contractor to provide detailed cost data to the government prior to contract negotiations.[311] In practice, TINA led to federal government-specific requirements that made it difficult for anyone seeking to do business with the DoD who was not already part of the defence industry.[312]

The next major wave of acquisition reform focused on competition. In 1984, the Competition in Contracting Act ('CICA') was passed to increase contracting competition, establishing full and open competition as the preferred method of soliciting and awarding contracts.[313] CICA established the legal norm for full and open competition as the rule,[314] although with significant exceptions.[315]

---

[309] Joseph Pegnato, 'Federal Procurement Reform: A Mixed Record at Best' (2020) 15 (Fall) *Journal of Contract Management* 35, 38.
[310] Ibid.
[311] Ibid.
[312] See Bill Greenwalt, 'Ike Was Wrong: The Military-Industrial-Congressional Complex Turns 60' *Breaking Defense* (online, 25 January 2021) <https://breakingdefense.com/2021/01/ike-was-wrong-the-military-industrial-congressional-complex-turns-60/>.
[313] See 41 USC § 253; Pegnato (n 309) 39.
[314] 48 CFR subpart 6.1 (Full and Open Competition).
[315] 48 CFR subparts 6.2 (Full and Open Competition After Exclusion of Sources), 6.3 (Other Than Full and Open Competition).



In 1986, the Packard Commission was appointed by Congress to assess the DoD procurement process.[316]  It found that the increasing layers of bureaucracy and regulations in the process was the reason weapon systems were taking too long to develop, overrun budgets, and incorporated old technology.[317]  Believing that the acquisition cycle of 10-15 years could be cut in half, the Packard Commission proposed an increased use of prototypes for development, the use of off-the-self products, and commercial-style competition to reduce costs.[318]  Soon after the Packard Commission finalised its report, the Defense Advanced Research Projects Agency ('DARPA') sought new ways to reduce bureaucratic 'red tape' and were given the first other transaction (OT) authority in the DoD.[319]  This new contract authority was not beholden to the FAR and other procurement laws, and enabled DARPA to act like a venture capital firm.[320]  However, this legal authority is parallel to the traditional contracting system; thus, DARPA sidestepped the procurement issues rather than addressed the problems of the system directly.

In the early 1990s, as the Cold War-era ended, the strategic impetus to develop technology for the DoD intersected with the information technology ('IT') age.  The *Federal Acquisition Streamlining Act of 1994* ('FASA') sought to simplify acquisition procedures, especially for low-cost contracts and commercial items.[321]  FASA recognised and addressed the difficulty commercial businesses had in working with the federal government.[322]  While FASA implemented many overarching reforms intended to encourage commercial business practices, the goals and promises of these supposedly commercial-friendly reforms have yet to be realised.[323] As part of the overall push to streamline processes and attract commercial business, DARPA's OT authority was extended to the rest of the DoD in 1994, extending the original science and technology authority to also carry out prototype projects that are directly relevant to weapons or weapon systems.[324]

---

[316] Government Accounting Office, *Defense Management: Status of Recommendations by Blue Ribbon Commission on Defense Management* (Report No NSIAD-89-19FS, 4 November 1988) 11 ('*Packard Commission*').
[317] Lopes (n 111) 16–7.
[318] *Packard Commission* (n 316) 2.  Many of the Packard Commission's findings and recommendations from 1988 are like the current issues plaguing the DoD acquisition system: prototypes have been used and tested far too little; high priority should be given to building and testing prototype systems before proceeding with full-scale development; streamlined procurement processes should be used: at 1–4.
[319] 10 USC § 4021 (2022) (originally codified as 10 USC § 2371).
[320] Sharon Weinberger, *The Imagineers of War* (Alfred A Knopf, 2017) 283.  In its first other transaction agreement, DARPA awarded $4 million to a start-up gallium arsenide firm that would go on to design an extremely fast, at the time, gallium arsenide microchip: at 284.
[321] See Pegnato (n 309) 39.
[322] Ibid.

Despite the attention by Congress to address commercial business concerns about an overly complex and burdensome procurement system, by the mid-1990s it was discovered that compliance with the many regulations added 18% to the overall cost of weapon systems delivered under DoD contracts.[325] In the early 2000s, the DoD focused on fighting in Afghanistan and Iraq and received increased funding from Congress. However, by 2009, Congress began to take note of poor cost and schedule performance of several high-profile weapon systems,[326] including the F-22 and F-35 advanced fighter aircraft. The Better Buying Power (BBP) Initiative was an internal DoD effort to increase efficiency and incentivise productivity and innovation in industry.[327] Both of these goals were expected to improve efficiency by reducing processes and bureaucracy.[328] While the BBP Initiative, in all three of its iterations made real attempts at addressing big problems, and helped realise some of FASA's goals of looking for best value rather than lowest price as the criteria for most source selections, progress was incremental.  Congress wanted to see faster, more robust reform.

---

[323] See Peter Levine and Bill Greenwalt, 'What the 809 Panel Didn't Quite Get Right', *Breaking Defense* (online, 4 April 2019) <https://breakingdefense.com/2019/04/what-the-809-panel-didnt-quite-get-right-greenwalt-levine/>.
[324] *National Defense Authorization Act for Fiscal Year 1994*, Pub L No 103-160, § 845, 107 Stat 1547, 1721–2 (1993).
[325] See Lopes (n 111) 7.
[326] See Susanna V Blume and Molly Parrish, *Make Good Choices, DoD: Optimizing Core Decision-making Processes for Great-Power Competition* (Report, Center for New American Security, 2019) 15, explaining that among the changes Congress made in response to poor cost and schedule performance came from the passage of the Weapons Systems Acquisition Reform Act which required the DoD to pursue competition with both prime and subcontractors throughout the program's lifecycle.
[327] Ibid.
[328] Ibid.

Starting in Fiscal Year 2016 and continuing through Fiscal Year 2018, Congress passed 247 acquisition reform provisions, an average of 82 provisions per year compared to an average of 47 such provisions in the preceding ten years.[329]  Overall, this active legislative period was aimed at developing more timely and efficient ways for the DoD to acquire goods and services, streamlining existing authorities and granting new authorities that would allow the DoD to rapidly prototype and field new programs.[330]  Congress explicitly stated its concern about innovation making its way into defence technology and dedicated several sections to attempt to boost defence-related innovation from outside the traditional defence industrial base.[331]

During these three years, Congress sought to simplify contracting for commercial items;[332] give the DoD more authority to negotiate for data rights,[333] and required major defence acquisition programs to be developed using 'a modular open system architecture approach to enable incremental development and enhance competition, innovation, and interoperability.'[334] The approach acknowledges the challenges defence acquisitions have in developing systems with different contractors with their own proprietary software code that makes interoperability between multiple systems challenging and expensive, if not impossible.  This requirement shows Congress's tacit understanding of the increasing reliance on software for weapons systems, an issue that becomes more complex with AI.

---

[329] Moshe Schwartz and Heidi M Peters, Congressional Research Service, *Acquisition Reform in the FY2016-FY2018 National Defense Authorization Acts (NDAAs)* (CRS Report No R45068, 19 January 2018) 1–2.
[330] See ibid.
[331] See ibid 2; *National Defense Authorization Act for Fiscal Year 2017*, Pub L No 114-328, § 884, 130 Stat 2000, 2318 (2016) ('*NDAA FY2017*').
[332] See Schwartz and Peters (n 329) 5-6.
[333] *NDAA FY2017* (n 331) § 809, 130 Stat 2266.
[334] Ibid § 805, 130 Stat 2252.



Among the many reforms Congress passed during this time, perhaps none has had a greater impact than the expansion of 'other transaction' authority. While discussed in greater detail below, each of the three years of National Defense Authorization Acts ('NDAAs') gave the DoD additional authority and flexibility to use OTs. The FY2016 NDAA expanded and codified OT authority for the DoD to use OT agreements to 'carry out prototype projects that are directly relevant to enhancing the mission effectiveness of military personnel and the supporting platforms, systems, components, or materials proposed to be acquired or developed by the Department of Defense, or to improvement of platforms, systems, components, or materials in use by the armed forces.'[335] The FY2018 NDAA contains eight sections aimed at expanding and improving the use of OTs,[336] including a preference for use of OT and experimental authority in the execution of science, technology, and prototyping programs.[337]

The most recent push for acquisition reform originated in the Fiscal Year 2016 NDAA, with the establishment of an advisory panel on streamlining and codifying acquisition regulations, known as the Section 809 Panel.[338] The Panel was made up of experts in acquisition and procurement policy charged with reviewing DoD acquisition regulations with an ultimate goal of maintaining defence technology advantage.[339] Accomplishing this goal, according to Congress, required the Panel to make recommendations to amend or repeal regulations to establish and administer buyer and seller relationships in the procurement system and improve the functioning of the acquisition system.[340]

---

[335] *National Defense Authorization Act for Fiscal Year 2016*, Pub L No 114-92, § 815, 129 Stat 726, 893 (2015) ('*NDAA FY2016*'); 10 USC § 4022(a)(1).
[336] Schwartz and Peters (n 329) 3.
[337] *National Defense Authorization Act for Fiscal Year 2018*, Pub L No 115-91, § 867, 131 Stat 1283, 1495 (2017) ('*NDAA FY2018*').
[338] *NDAA FY2016* (n 335) § 809, 129 Stat 889.
[339] Ibid.
[340] Ibid.



The Panel understood that, given the growing gap between DoD capabilities and the evolving threats it must prepare to meet, the acquisition system had to adopt a mission-first approach.[341]  This meant streamlining the acquisition system to 'manoeuvre inside the turn of the nation's near-peer competitors and nonstate actors, which are not bound by the same acquisition rules as the DoD.'[342]  While acknowledging that the DoD must promote competition, with transparency and integrity, the Panel noted that considerations of rapidly acquiring warfighting capability and delivery it to the military must take precedence over other public policy concerns.[343]

Concluding in 2019, the Section 809 Panel published five reports with 98 recommendations over approximately 2400 pages.  The overall effort was intended to simplify commercial buying,[344] and improve communication with industry.[345]  Although it is too soon to judge whether the Section 809 Panel was successful in meeting its goals, the response from commentators and industry has been mixed.  One critique pointed out the Panel missed an opportunity to review existing statutes and regulations and instead focused on more board issues such as the defence acquisition workforce and the congressional appropriations processes.[346]  However, many from industry praised the Panel's work.  One recommendation that has resonated with industry yet has not yet been implemented is Recommendation 81, 'Clarify and expand the authority to use Other Transaction agreements for production.'[347]  Named as one of

---

[341] Advisory Panel on Streamlining and Codifying Acquisition Regulations, *A Roadmap to the Section 809 Panel Reports* (February 2019) 1 ('*Section 809 Roadmap*').
[342] Ibid.
[343] Ibid.
[344] See ibid 6–7.
[345] See ibid 9–10.
[346] Levine and Greenwalt (n 323).
[347] *Acquisition Regulation Report Vol 3* (n 238) 440.  Disclaimer: this researcher was the principal author of Recommendation 81.



the top recommendations by the National Defense Industrial Association,[348] Recommendation 81

seeks to open the aperture beyond the limitations of the statute to allow for other transaction

agreements through production, helping to overcome the barriers of traditional contracting law

and practice.[349]

Despite decades of acquisition reform efforts, the delta between the DoD and commercial

industry, particularly in advanced technology sectors such as AI is a wide as ever.  As the

Section 809 Panel admitted, the DoD's system of procuring technology 'suffers from processes

and procedures that are obsolete, redundant, or unnecessary' and the DoD 'must work to move

quickly enough to keep pace with private-sector innovation.'[350]  This historical overview of

acquisition reform efforts has shown that progress is incremental, evolutionary and pendular.

Reforms often tinker with existing law and regulations, and then are engulfed with more

bureaucratic restrictions either internally from the DoD or externally by Congress.  Given the

importance of commercial technology to the DoD's strategic goals, and the unique position of

the DoD as a minor player in the commercial AI marketplace, perhaps a more revolutionary

approach is required.  In the following section, the two primary contracting methods used by the

DoD to acquire AI-enabled capabilities are compared and assessed.

---

[348] National Defense Industrial Association, '809 Panel Volume 3: Top 10 Recommendations' (14 May 2019) <https://www.ndia.org/media/sites/ndia/policy/documents/ndia_section_809_panel_v3_top_10_recommendations.ashx>.
[349] See *Acquisition Regulation Report Vol 3* (n 238) 444.  Legal and contracts commentators opined that the push for making the DoD's contracting more consistent with the commercial section is critical, though they fear the DoD's Cold War era style of bureaucracy could result in OTs looking more like traditional contracts: see 'Section 809 Panel Volume 3 Report: Seven More Key Takeaways', *Baker Tilly* (Article, 24 January 2019) <https://www.bakertilly.com/insights/section-809-panel-volume-3-additional-highlights-and-insights>.  See also Alex D Tomaszczuk et al, 'Section 809 Panel: 'The Commercialization of Government Contracting', *Pillsbury* (Article, 24 January 2019) <https://www.pillsburylaw.com/en/news-and-insights/section-809-dod-acquisiations.html>.
[350] *Acquisition Regulation Report Vol 3* (n 238) EX-4.





The authority of the DoD to contract is inherent in the United States Constitution. Article II, Section 1 provides '[t]he executive Power shall be vested in a President of the United States of America,'[351] and Section 2 adds that '[t]he President shall be Commander in Chief of the Army and Navy of the United States.'[352] This full grant of executive power authorises the President, or designee, to contract for supplying the armed forces. This concept of authority to contract was underscored by the United States Supreme Court, explaining that when acting in its proprietary capacity, the United States is bound by commercial law unless otherwise provided by statute or regulation.[353] Thus, while the DoD would have authority to enter into commercial contracts as a typical customer, it is bound by the federal statutes and regulations that pre-empt commercial law.[354]

There are several contract vehicles available under law for contracting officers to use to acquire AI-enabled capabilities. While there are a variety of contractual agreements that the DoD is authorised to use, each authority has its own unique purpose and rules. The body of these contractual agreements are grouped into two broad categories — those that are governed by the FAR and those that are not governed by the FAR.[355] Traditional procurement contracts are governed by the FAR and other procurement-related statutes, while the non-FAR contracts are governed by the specific statute providing the DoD with specific contract authority. Within the non-FAR contract vehicles, several permit the DoD to stimulate R&D in academia and the commercial sector,[356] while two authorities permit the DoD to acquire technology including AI systems and components for specific purposes.[357]  One of those authorities, other transaction authority to carry out prototype projects, presents an alternative legal framework to the FAR for buying AI systems from the commercial market.  The following section provides an overview of the respective legal frameworks and compares the salient attributes of FAR-based procurement contracts and other transaction agreements in the context of acquiring AI-enabled capabilities.

---

[351] *United States Constitution* art II § 1.
[352] *United States Constitution* art II § 2.
[353] *Cooke v United States*, 91 US 389, 398 (1875), holding that if the government 'comes down from its position of sovereignty, and enters the domain of commerce, it submits itself to the same laws that govern individuals there'.
[354] See *The Floyd Acceptances*, 74 US 666, 680 (1868).
[355] 'Contracting Cone', *Defense Acquisition University* (Web Page) <https://aaf.dau.edu/aaf/contracting-cone/>.

4 *Characteristics and Attributes of FAR Contracts*

As discussed in Chapter I, the *Federal Acquisition Regulation* ('FAR')[358] is the primary set of procurement rules the DoD, NASA, and the civilian executive agencies use that govern the formation and administration of contracts.[359]  The FAR System is the system of regulations that include the FAR and agency supplements such as the Defense FAR Supplement ('DFARS').[360] The vision of the FAR system is to timely deliver the best value product or service to the customer while maintaining the public's trust and fulfilling public policy directives.[361]  The statement of guiding principles for the FAR System states it will maximise the use of commercial products or services, minimise administrative costs, and promote competition.[362] The FAR encourages the acquisition team to exercise initiative and sound business judgment, and instructs that the contracting officer may assume that if a strategy, practice, or policy is in the best interest of the government, and is not addressed in the FAR nor prohibited by law, Executive Order, or other regulation, such initiative is authorised.[363]  However, given the FAR is over 2000 pages long,  DFARS is over 1300 pages, and other applicable procurement law (including statutes, regulations, Executive Orders, case law, and policy) is essentially unquantifiable, it is difficult to find meaningful capacity for such imaginative practice.

---

[356] See 10 USC § 4021 (Research Other Transactions); 15 USC § 3710a (Cooperative Research and Development Agreements); 15 USC § 3715 (Partnership Intermediary Agreements); 32 CFR § 37 (2003) (Technology Investment Agreements).
[357] See 10 USC § 4022 (Prototype Other Transactions); 10 USC § 4023 (Procurement for Experimental Purposes): section 4023 authorises the DoD to buy technology considered necessary for experimental purposes in the development of the best supplies that are needed for national defence.  This authority is available to the DoD for buying AI systems from the commercial market, though the statute is limited to experimentation, technical evaluation, assessment of operational utility, safety, and providing a residual operational capability; it does not authorise full scale production like section 4022.  Thus, this dissertation focuses on section 4022, other transactions for prototypes and the follow-on production clause, as a full alternative legal framework to the FAR.
[358] 48 CFR § 1.
[359] See 48 CFR § 1.103.
[360] 48 CFR § 1.301; *DFARS* § 201.1.
[361] 48 CFR § 1.102.
[362] Ibid.

While procurement law applies to the DoD, the FAR and DFARS do not, technically, apply to contractors.  However, the FAR and DFARS require the DoD to place conditions on who can compete for a contract, how a contract is awarded, and what clauses that do apply and bind the contractor are part of the contract.[364]  There are numerous requirements that must be met by any company seeking to work with the DoD.  Prospective contractors who wish to compete for a DoD contract opportunity are required to register in the System of Award Management,[365] and make approximately 30 representations and certifications on a variety of topics, such as company size, previous contracts, compliance with labour laws, and domestic origin of manufactured goods.[366]  Most of these certifications are required to be eligible for consideration in competition, so prospective contractors must navigate through the many DoD-specific and unique compliance regulations just for the opportunity to compete.  Spending the time and costs to comply with the many requirements does not guarantee funding from the DoD, so prospective contractors assume the risk of attempting to enter the defence market.

---

[363] Ibid.
[364] See 48 CFR Parts 6, 9, 15, 52.
[365] 48 CFR § 4.1102, requiring prospective contractors to be registered prior to award of a contract with limited exceptions such as small dollar awards, classified contracts, and contracts awarded overseas.
[366] 48 CFR § 52.204-8.



For contracts under the FAR, 'full and open competition in soliciting and awarding' contracts is the rule, with exceptions to reduce the pool of competitors to small or disadvantaged businesses, or to a single contractor in a sole-source award.[367] This policy is required by statutory law, namely CICA, and potential contractors may protest a solicitation or award that violates this statute.[368] However, while full and open competition appears the ideal paradigm for fairness to industry and value for the government, several barriers in the FAR and procurement practice may lead potential vendors to self-select out of the competitive pool. CICA does not require full and open competition of all potential contractors, just those who respond to the solicitation and are deemed responsible to perform the contract.[369] If the best potential contractors choose not to compete, or do not understand how to find opportunities to compete for DoD contracts, then full and open competition is an empty vessel. As many of the advancements in AI come from the commercial sector that is unaccustomed to defence contracts, the number of hurdles that must be cleared to compete for a FAR contract may pose barriers to entry for commercial AI firms.

One potential obstacle inherent in the FAR is how specifications, or contract requirements, are stated. Under the FAR, an 'acquisition begins at the point when agency needs are established and includes the description of requirements to satisfy agency needs, solicitation and selection of sources, award of contracts … and those technical and management functions directly related to the process of fulfilling agency needs by contract.'[370] Contracting officers are instructed to state requirements with respect to an acquisition in terms of functions to be performed, performance requirements, or essential physical characteristics.[371]  With AI, the advancement in capabilities is swift and the solution to problems is often unexpected.  It is possible the DoD does not fully understand what is possible in terms of available technology, or how a potential solution could be developed.  Under this rigid process, it is assumed the DoD already knows how to solve the problem and details how the contractor should perform.  With this pre-defined requirement generation, AI firms have little room to innovate.  Detailed performance specifics restrict machine learning capabilities to identify patterns or make predictions and limit the ability to iterate.  Machine learning helps solve complex problems when the solution evades humans; it is incongruous to require a specific outcome and method of achieving such outcome when contemplating an AI use case.

---

[367] 48 CFR § 6.101.
[368] 10 USC § 3201 et seq; 41 USC § 3301; see 48 CFR § 33.102.
[369] 48 CFR § 9.103(b).
[370] 48 CFR § 2.101.

Another problem that is connected to the requirements process under the FAR is it creates blind-spots for the DoD agency trying to find a way to meet its requirements.  While the agency is required to conduct market research to determine what is available[372] and must contract for a commercial item if one is available to meet its need,[373] the requirements process limits the scope and can exclude non-developmental items from consideration.  Agencies seeking to identify potential solutions often rely on requests for information that is posted to the government's market website.[374]  However, requests for information take time to respond and do not lead to a contract,[375] so there is little incentive for industry to engage, especially if companies are not actively searching for request for information that may apply to their technology.

---

[371] 48 CFR § 11.002(a)(2).
[372] 48 CFR § 10.001.
[373] 48 CFR § 12.101; 10 USC §§ 3452 et seq.
[374] 48 CFR § 15.201(c); General Services Administration, 'Contract Opportunities', *SAM.gov* (Web Page) <https://sam.gov/content/opportunities> (the government's point of entry for procurement solicitations).
[375] Ralph C Nash Jr, Karen R O'Brien-Debakey, and Steven L Schooner, *The Government Contracts Reference Book* (Wolter Kluwer, 4th ed, 2013) 429.



One famous example of how the Army relied on outdated processes and assumptions leading to an award of a developmental contract when a commercial solution existed is the Distributed Common Ground System-Army Increment 2 (DCGS-A2) contract. The Army, despite the legal requirement to determine whether commercial items could meet or be modified to meet its procurement needs,[376] failed to assess whether commercial options were available.[377] Palantir had developed a software platform that was marketed commercially for several years prior to the Army's solicitation, and enables agencies to integrate, visualise, and analyse large amounts of data from different sources in different formats — essentially the exact specifications required by DCGS-A2.[378] The Court of Appeals for the Federal Circuit affirmed the lower court's ruling that the Army was not justified in its decision to require developmental software when the Palantir platform was capable of modification for use.[379] Though not explicitly discussed in the court's decision, the news of the case highlighted one issue in particular: Palantir is a Silicon Valley-based start-up and the Army steered the contract towards traditional defence contractors to 'reinvent the wheel' at a higher cost.[380] As seen from the literature discussed above, cases such as *Palantir* can impact the commercial AI industry's perspective of the DoD as a customer.

The methodology for selecting contract awardees for FAR-based procurement contracts is also complex. There are different rules that apply to a commercial item[381] than to non-commercial items,[382] though defining what constitutes a commercial item under the FAR is challenging.[383]  The information required of competitors includes certified cost or pricing data,[384] and compliance with dozens of mandatory provisions.[385]

---

[376] 10 USC § 3453(c)(2).
[377] *Palantir USG Inc v United States*, 904 F 3d 980, 995 (Fed Cir, 2018).
[378] Ibid 985.
[379] Ibid 995.
[380] Ellen Mitchell, 'How Silicon Valley's Palantir Wired Washington', *Politico* (online, 14 August 2016) <https://www.politico.com/story/2016/08/palantir-defense-contracts-lobbyists-226969>.
[381] See 48 CFR Part 12.
[382] See 48 CFR Part 15.

The negotiated procurement process is very formal, restricting communication between potential contractor and the contracting agency,[386] and can take many months or years for the source selection authority to issue an award.  Even after the award is issued to a contractor, such award can be protested in three fora: with the agency,[387] at the Government Accountability Office (GAO),[388] or the Court of Federal Claims (COFC).[389]  Protests made in one forum does not preclude protests in the others, so even after the award is made, there could be additional months or years before contract performance and funding occurs.  In one of the most famous cases, the DoD's Joint Enterprise Defense Infrastructure (JEDI) cloud was intended to provide a DoD-wide platform to develop AI capabilities, and was originally awarded in October 2019 to Microsoft.[390]  The original award was delayed as the DoD had to overcome pre-award protests at GAO and COFC by Oracle alleging the specifications were unfair.[391]  After award, multiple losing bidders, including Amazon and Oracle, filed suit protesting the source selection with litigation continuing into 2021.[392]  Despite the importance of the cloud infrastructure to the DoD in building AI capabilities to compete with China, the length of the process led the DoD to abandon the contract, worth up to $10 billion, and recompete a new multiple award cloud

---

[383] Section 809 Panel (n 238) vol 3, 18-19.
[384] 48 CFR 15.403.
[385] See 48 CFR 52.301.
[386] See 48 CFR 15.201 and 15.306.
[387] 48 CFR 33.103.
[388] 48 CFR 33.104.
[389] 48 CFR 33.105.
[390] Andrew Eversden, 'Pentagon could Reassess Future of JEDI Cloud, Depending on Court Action' (29 January 2021) *C4ISRNET* <https://www.c4isrnet.com/battlefield-tech/it-networks/2021/01/29/pentagon-could-reassess-future-of-jedi-cloud-depending-on-court-action/>.
[391] Oracle America, Inc., B-416657 et al (14 November 2018); *Oracle America, Inc. v. United States*, 975 F 3d 1279, 1283 (Fed Cir, 2020) ('*Oracle*').
[392] *Oracle*, 975 F 3d 1279 (certiorari denied, 4 October 2021, 142 S Ct 68).



program that was finally awarded in December 2022 to Amazon, Google, Microsoft, and Oracle.[393] Although protests serve an important function in ensuring a fair and transparent process, the length between opening the competition and funding the contract for DoD projects represent a risk to commercial AI firms that may be unable to bear the opportunity costs of waiting so long, not to mention the concern that the DoD will miss out on acquiring technology necessary for it to carry out its mission.

While the complexity and length of the FAR process represent potential barriers that commercial AI firms must overcome to secure a contract award with the DoD, the DoD's management of intellectual property (IP) involved in the contract may prove even more challenging. Under the DFARS, items, components, or processes developed under DoD contract and funded by the government require the contractor to grant unlimited rights in the data or software developed.[394] Mixed funding from the government and private funds result in government purpose rights, allowing the government to release to competitors for future procurement.[395] Because machine learning models are trained and developed using data — potentially from government, proprietary, or third-party sources — the assignment of technical data and software license rights in FAR contracts may be too rigid and extensive to attract AI firms.[396]

Even senior acquisition officials in the DoD find the FAR to be monolithic, often explaining the way the DoD does business requires 'hacks' to stay relevant to an increasingly

---

independent and technologically advanced commercial industry.[397]   Will Roper, the former

Assistant Secretary of the Air Force for Acquisition, Technology and Logistics, assessed:

> The period in which the FAR was created is very different than today.  The government was still the central driver of the technology in this nation.  We represented most of the research and development in the nation during the height of the Cold War.  I don't think the original drafters of the FAR would've imagined a complete 180.  So we must hack the system to be relevant.[398]

The FAR represents a traditional way to procure goods and services, and, arguably,

works well for traditional hardware and services.  As discussed in Chapter I, there are many

advantages to the government provided by the FAR: practitioners are better trained and more

comfortable using traditional procurement contracts; there is more predictability, oversight, and

auditability; competition is transparent, and contractors have access to multiple fora to dispute

government action.[399]   However, based on the literature reviewed in Topic I and Topic II, it

appears that traditional contracts are ill-suited for engaging and attracting commercial AI firms

and do not align with the development cycle.  An alternative to the traditional system exists in

other transaction (OT) authority.  Because OT authority is intended specifically to attract non-

traditional defence contractors, a review of the legal framework and attributes of OT agreements

helps identify whether they align better with the business and technology considerations of

commercial AI firms.

---

Unlike FAR-based contracts, OT agreements are not bound by procurement regulations and statutes.[400] Rather, the statutory authority provides the legal parameters for an OT agreement.[401] Defined in the negative, OT agreements are not procurement contracts, cooperative agreements or grants; however, OT agreements are legally binding contracts nonetheless, with an offer, acceptance, consideration, authority, legal purpose, and meeting of the minds.[402] A proper OT agreement for prototype contemplates four requirements: purpose, prototype, participation, and production. An OT agreement is authorised for the purpose of carrying out 'prototype projects that are directly relevant to enhancing mission effectiveness.'[403] The term 'prototype project' is defined as a project that 'addresses a proof of concept, model, or process, including a business process; reverse engineering to address obsolescence; pilot or novel application of commercial technologies for defence purposes; agile development activity, creation, design, development, demonstration of operational utility; or any combination of' the foregoing.[404] This is a very broad definition that arguably permits the use of an OT agreement for the development and deployment of any machine learning application.

The authority to enter into an OT agreement is limited to four pathways: at least one non-traditional defence contractor or non-profit research institution participates to a significant extent; all significant participants are small businesses or non-traditional defence contractors; at least one third of the total cost is paid out of funds provided by sources other than the federal government; or the senior procurement executive of the agency determines exceptional circumstances justify the use of a transaction that provides for innovation business arrangements or structures that would not be feasible or appropriate under a FAR-based procurement contract.[405] Although 'significant extent' is not defined by the statute, a plain meaning of the term is used. In practice, participation by a small business, non-traditional defence contractor or non-profit research institution can meet the requirement by playing a critical role in the project that would otherwise result in higher cost, delay, or decrease in effectiveness or efficiency without such participation. Contributions of intellectual property or labour can meet this requirement. These conditions are incredibly broad and are easily met for AI projects given most commercial AI firms qualify as a small business[406] or non-traditional defence contractor.[407]

---

[400] Government Accountability Office, *Federal Acquisitions: Use of "Other Transaction" Agreements Limited and Mostly for Research and Development Activities* (Report No GAO-16-209, January 2016) 1.

[401] See 10 USC § 4022; *Other Transactions Guide* (n 104): the Other Transactions Guide is a deliberately non-binding document issued by the DoD's Under Secretary of Acquisition and Sustainment intended to merely guide contract officials, not to regulate the statute.

[402] *Other Transaction Guide* (n 104) 38; 10 USC § 4022(e).

[403] 10 USC § 4022(a)(1).

[404] *Other Transaction Guide* (n 104) 31.

The statute differs significantly from the FAR, providing flexibility that can lead to expeditious award and subsequent fielding of the prototype. While the award of OT agreements can be protested, review is limited to protests alleging that the agency is improperly exercising the authority, thus the award and solicitation for the award of an OT agreement is not reviewed by the Government Accountability Office.[408] The DoD can award agreements to multiple offerors from the competitor, increasing the DoD's chance at obtaining a successful prototype.[409]

Finally, the authority to enter into a transaction for a prototype project also permits the DoD to award a follow-on transaction or contract for production of a successfully completed project.[410]  An award of a follow-on production transaction requires that the award of the transaction for prototype project used competitive procedures for the selection of parties for participation in the transaction.[411]  The competitive procedures required for the follow-on production award are not those required by CICA; rather the statute states only that competitive procedures be used 'to the maximum extent practicable.'[412]  In practice, opportunities that are publicly advertised meet this requirement and flexibility exists for the agency to tailor the form of competition to the project.[413]

---

[405] 10 USC § 4022(d)(1)(A)–(D).
[406] 15 USC § 632.
[407] 10 USC § 3014: 'an entity that is not currently performing and has not performed, for at least the one-year period preceding the solicitation of sources by the Department of Defense for the procurement or transaction, any contract or subcontract for the Department of Defense that is subject to full coverage under the cost accounting standards'. Full coverage of cost accounting standards is unlikely to apply to most commercial AI firms as such standards to not apply to contracts for the acquisition of a commercial product or service: see 41 USC § 1502. Moreover, full coverage under the cost accounting standards only applies to firms that receive a single covered contract award of $50 million or more: *Other Transaction Guide* (n 104) 30.
[408] *Blade Strategies, LLC* (Matter No B-416752, 24 September 2018) 2.
[409] *Other Transaction Guide* (n 104) 14–5.

This capability to award a follow-on transaction or contract without further competition can permit the DoD to rapidly scale and field new technology and overcome the 'Valley of Death,' the gap between initial funding of early-stage technology pilot and scalable program can result in the end of many potentially worthwhile projects.[414]  The follow-on production authority can help bridge the Valley of Death and avoid forcing commercial firms that successfully complete a prototype project to choose between waiting years to compete and receive a follow-on contract and leaving the DoD market altogether to pursue strictly commercial ventures.[415]

The permissive statutory language and reluctance by the DoD to issue binding regulations governing OTs are by design.  Congress has expressed its pleasure in the successful use of OT agreements to support acquisition speed and innovation and explicitly stated the statute is written in an intentionally broad manner.[416]  Congress has urged the DoD to interpret the authority in the most flexible and broad manner possible and indicated that it is willing to tolerate more risk in the use of OT authority in order to pursue innovation and speed the development and fielding of critical new capabilities.[417]  The DoD, in issuing the Other Transaction Guide ('OT Guide'), made it clear that the document is not formal policy and encouraged practitioners to assume that a strategy, practice, or procedure is permitted if it is in the best interest of the DoD and not prohibited by law.[418]  As the OT Guide explains, 'the OT authorities were created to give DoD the flexibility necessary to adopt and incorporate business practices that reflect commercial industry standards and best practices into its award instruments.'[419]  In short, OTs are inherently flexible instruments that are intended to leverage best practices from industry to gain 'access to state-of-the-art technology solutions from traditional and non-traditional defence contractors, through a multitude of potential teaming arrangements tailored to the particular project and the needs of the participants.'[420]

---

[410] 10 USC § 4022(f).
[411] 10 USC § 4022(f).
[412] 10 USC § 4022(b)(2).
[413] See *Other Transaction Guide* (n 104) 5.
[414] See Anthony Davis and Tom Ballenger, 'Bridging the "Valley of Death"' (2017) (January–February) *Defence AT&L* 13.
[415] See Committee on Armed Services, United States Senate, *Report to Accompany S. 1519* (Senate Report No 115-125, 10 July 2017) 190–1 ('Senate Report No 115-125').
[416] Ibid.

In comparing the purposes of traditional procurement contracts with OT agreements, the FAR itself indicates it may not be an appropriate contract vehicle for procuring AI applications. Unlike OT agreements that are intended to complete a prototype project, FAR contracts are intended to buy products and services.[421]  By definition, such procurements have known requirements for which the work or methods can be precisely described in advance.[422]  The FAR distinguishes between contracts with known requirements and contracts focused on more nebulous problems where it is difficult to judge the probability of success or required effort.[423] Given the complexity of developing, integrating, and deploying AI, it is likely most efforts involving AI fall into this latter group.  OT authority provides the flexibility to contract for these challenging problems and allow for iterative and modular efforts that better align with the development and operations cycles for AI applications.[424]  In contrast to the FAR, the OT Guide explains that the most important part of the DoD's planning activities is not specifying how the contractor must perform but rather defining the problem, area of need, or capability gap.  This problem statement is agnostic to the technical approach, schedule, cost, or even industry, yet clearly articulates the area of need to allow for innovation trade space for a wide range of solutions.[425]

---

[417] Ibid.
[418] *Other Transaction Guide* (n 104) 3.
[419] Ibid 4.
[420] Ibid.
[421] 48 CFR § 1.102.
[422] See 48 CFR § 2.101.
[423] 48 CFR § 35.002.

Moreover, the lack of regulations and requirements permit OT agreements the flexibility to accommodate many different business arrangements and are fully negotiable.[426]  This freedom from the procurement regulations permits the DoD to enter into agreements with commercial firms much in the same way as another business or consumer would.[427]  The ability to freely negotiate and communicate throughout the evaluation process allows all parties the gain a better understanding of goals and intent, reflected in the lack of mandatory terms and conditions that would otherwise apply to FAR contracts and the bespoke source selection methods.[428]  The OT Guide encourages full negotiation on intellectual property terms, modifications and disputes in terms that make use of best practices in the private sector, and make common sense based on the nature of the prototype project.[429]  Although the ability to negotiate each term can lead to fewer clauses that are unique to the government, and lead to mutually beneficial terms, the process can take significant time and resources, and is less predictable than negotiations for FAR contracts that are generally limited to price and schedule.[430]  An example of this unpredictability was reported by the DoD Inspector General in 2021.[431]  One common method used by the DoD to award OT agreements is through a consortium manager, a third party contracted to manage multiple awards for prototype agreements.[432]  However, an audit found that DoD contracting personnel did not consistently award OT agreements in accordance with applicable laws, citing the lack of guidance and training as the cause.[433]  Additionally, because there were no requirements for tracking and reporting important information about the contractors performing work on the prototype projects, the DoD does not have direct oversight of the projects it is funding.[434]  Thus, while the flexibility to negotiate contract clauses is a valuable attribute of OT authority, the DoD could be hindered in making important funding decisions if governance clauses are not included in the OT agreement.  Some of these clauses are included in traditional contracts because they are required by the FAR and are thus non-negotiable; the lack of legal requirement for their inclusion in an OT agreement may enable potential contractors to require additional fees for compliance.

---

[424] See *Other Transaction Guide* (n 104) 16–7.
[425] See ibid 9.
[426] See ibid 4, 17.
[427] See ibid.
[428] See ibid 14–9.
[429] Ibid 15–6.
[430] See ibid 14–5.

While not explicitly describing AI, Congress recognised OT agreements should be the preferred method of contracting for science and technology.[435]  In the report accompanying that legislative preference, the Senate expressed frustration with the DoD for underutilising its OT authority, assigning blame to senior leaders, contracting professionals, and lawyers for narrowing interpreting what was intended as broad authority to leverage OTs for innovate projects.[436]

Two experts on contracting for emerging technology have weighed in on the advantages and drawbacks to using OT agreements to attract non-traditional defence contractors.  First,

---

[431] Inspector General, US Department of Defense, *Audit of Other Transactions Awarded Through Consortiums* (Report No DODIG-2021-077, 21 April 2021) 1.
[432] See ibid 1–3.
[433] Ibid 19.
[434] Ibid 27.
[435] *NDAA FY2018* (n 337) § 867, 131 Stat 1495.
[436] Senate Report Number 115-125 (n 415) 189.



Richard Dunn, former general counsel of DARPA, sees OT authority as an alternative procurement system to the FAR.[437] Claiming the traditional system 'costs-too much' and 'takes-too-long,' and discourages companies from participating in the government marketplace, Dunn advances the idea that OT authority allows for more innovative business arrangements and contracting that the DoD can leverage 'to advance science and technology, engage a wider industrial base,' and 'deliver new capabilities at the speed of relevance.'[438] Dunn advocates OT authority should be used as a parallel acquisition system with the FAR, rather than as a last resort.[439] He points out that in as early as 1994, the DoD concluded software, space systems, mature jet engines, and other technologies should be procured in a fully commercial manner.[440] Despite this conclusion and Congress's directive to establish a preference for OT agreements for technology, the DoD continues to focus its contracting education and training on the traditional procurement system and limits the use of OT authority.[441] Although the DoD spending on OT agreements has dramatically increased from about $700 million in 2015 to $7.7 billion in 2019,[442] OT agreements still represent a small fraction of the DoD's annual contract obligations of $282 billion to $402 billion in those same years, respectively.[443] Thus, the high water mark for DoD spending on OT agreements is still less than two percent of total contract funding, and just 18 percent of the DoD's total research and development funding.[444]

Following Richard Dunn in serving as general counsel of DARPA, Crane Lopes also found the DoD's underutilisation of OT authority puzzling.[445] In his assessment on the institutional factors limiting the use of OTs by DoD, Lopes conducted a review of OT literature to develop a conceptual framework to understand why OTs, despite their advantages in contracting for science and technology, are not more widely used in the DoD.[446] Lopes identified several advantages and potential disadvantages to using OTs to solve problems of traditional procurement. One advantage highlighted in literature is the ability for the DoD to negotiate terms and conditions to overcome barriers that prevent commercial firms from competing for opportunities in the defence marketplace.[447] However, some critics cite this ability to negotiate as a disadvantage for the government as it makes OT agreements more complicated than FAR contracts that have extensive mandatory clauses.[448] Others have pointed out that OT agreements are preferable to traditional procurement contracts because they are not subject to many regulations that apply to the FAR. This too presents a double-edged sword, as the lack of administrative safeguards, training, and metrics to measure success make OTs risky for the DoD.[449] Finally, while many cite the intent of OT authority is to attract non-traditional contractors, critics have considered this authority a failure because only a small portion of OT awards have gone directly to non-traditional contractors, with the majority of contract dollars going to traditional contractors.[450] Lopes concluded that OTs are more advantageous than FAR contracts for the DoD in attracting non-traditional firms to develop and field technology faster, yet there are institutional barriers that limit the risk-adverse bureaucracy that incentivises

compliance with regulations at the expense of innovation.[451]  The lack of training for awarding OT agreements and the institutional bias towards traditional procurement contracts appear to explain each argument used against OT authority.[452]

Lopes' comprehensive work in understanding the internal institutional factors that limit the DoD from using OTs more widely serves as a foundation for this research.  However, his research did not address whether an increased use of OT authority would serve to better attract commercial firms to work with the DoD.  This research builds upon the literature and aims at understanding the unexamined question that follows if the DoD accepts Lopes' recommendations to use OTs more widely to field advanced technology capabilities[453] from the opposite contract party's perspective: if the DoD used OTs more widely, would it be a more attractive customer to commercial AI firms?  Understanding whether the attributes and practice of OT agreements can better attract the commercial AI industry to support the DoD in meeting the objectives of its national defence strategy by delivering AI-enabled performance 'at the speed of relevance'[454] can influence future acquisition reform efforts as well as a re-examination of the training for contract officials.

6 *Synthesis of Defence Acquisition Law and Literature*

The review of the two legal frameworks demonstrates there are some critical differences between FAR contracts and OT agreements.  The distinction becomes more pronounced when assessing the alignment with the way AI systems are developed and deployed.  Figure 6 below highlights how the choice of contract law affects the key attributes of a contract and how well those attributes align with the technical considerations in the AI lifecycle.

---

[451] See ibid 611–3.
[452] Ibid 117.
[453] Ibid 653.
[454] See ibid.



*Figure 6: AI Business and Technical Alignment Comparison between the FAR and OT Authority*



| Contract Attribute | FAR | OT | AI Considerations |
|---|---|---|---|
| **Competition** | • Default full and open competition subject to protest on the solicitation and award<br>• Governed by CICA | • Fair and transparent<br>• Competitive procedures used to the maximum extent practicable if awarding follow-on award | • Many commercial AI firms are unfamiliar with DoD contracting procedures |
| **Solicitation** | • Request for proposals to fulfill requirements based on detailed specifications<br>• Announced on the government's website | • Problem statement, area or interest, or capability gap<br>• Announced through means expected to maximise exposure of the problem set to relevant technology providers | • AI systems are developed by understanding the problem and seeking relevant data first, and then determine the appropriate training method, algorithm, and features for the application[455] |
| **Complexity** | • Dozens of government-specific conditions, certifications, and accounting systems<br>• Foreign process for many commercial firms | • Intended to be streamlined and consistent with commercial best practices | • Many commercial AI firms are unfamiliar with DoD contracting procedures |
| **Communication** | • Limited to before solicitation is published and during designated period during source selection[456] | • Encouraged throughout problem statement development through performance with industry and academia | • It is critical for the developer to understand the problem to be solved as well as how it will be used |
| **Flexibility** | • Generally rigid<br>• Modifications must be within scope of specifications or subject to further protest under CICA[457] | • Inherently flexible<br>• Understands that the project may yield outcomes that surprise participants<br>• Directs iterative process[458] | • The AI lifecycle is iterative and never finished as new data changes underlying software code |
| **Terms & Conditions** | • Dozens of required non-negotiable terms and conditions of performance | • Fully negotiable | • Many commercial AI firms are unfamiliar with DoD contracting procedures |
| **Intellectual Property** | • Standard clauses provide government with unlimited, government purpose, or limited rights depending on whether development occurred under government contract<br>• Commercial license and negotiated license may be available to the extent permitted by law[459] | • Fully negotiable<br>• DoD can negotiate for rights above or below legal limitations applicable to FAR[460] | • AI is developed through finding data relevant to the application which is used to train the machine learning algorithm to produce a model capable of making inferences on new data; the data may come from a variety of sources, including government provided data as well as proprietary or open data |
| **Speed** | • Negotiation of terms is generally quick as negotiable terms are limited<br>• Pre-solicitation and source evaluation/ selection can take significant time<br>• Solicitation and award can be protested | • Negotiation of terms can be lengthy<br>• Pre-solicitation and source evaluation/ selection are unencumbered by FAR requirements and protest jurisdiction is limited so prototype award can be made swiftly<br>• Follow-on production agreement can be awarded with no additional competition cycles | • Advancements in the state of the art in AI occur extremely quickly<br>• Many commercial AI firms are unaccustomed to lengthy procurement times |

While not every OT agreement is free from concerns from both the DoD and industry, the law appears to align with the business and technical considerations better than FAR-governed contracts.

The DoD's two primary contract law options for AI-enabled capabilities provide contract practitioners a range of options to consider. The choice of contract vehicle implies a choice of legal framework, presenting questions of how each aligns the DoD's national security requirements with commercial AI firms' business preferences and the technical considerations of developing AI for defence applications. While both Dunn and Lopes point to the intent of Congress and legal flexibility as reasons for the DoD to leverage its OT authority to better engage with commercial industry, many contracting officers in the DoD still use the traditional FAR framework for AI system procurement. Critics of OT authority are concerned that the lack of regulation can lead to waste and abuse and argue OT agreements carry inherent risk as they lack the safeguards built into FAR contracts.[461] However, the literature on the DoD's procurement system does not directly answer the research question: why do commercial AI firms decide to contract with the DoD and whether contract attributes influence that decision? A robust study from the AI industry's perspective is elusive. Without examination of the perspectives, preferences, and motivations of the commercial AI industry, legal, policy, and practice decisions regarding contracts for AI-enabled capabilities are made in the dark.

This literature review uncovered no research that clearly explains how the two legal frameworks align the DoD's national security objectives in acquiring AI systems with defence applications with the business preferences and technical considerations of the firms providing those capabilities. There is a lack of resources informing the development of a research design and methodology for such a study. Moreover, there is a gap in existing literature explaining what factors and contract attributes are perceived as attractive to commercial AI firms or how those factors and attributes affect the business calculus of a firm deciding whether to compete for a DoD contract opportunity. Finally, although Congress, expert commentators, and the legal review above indicate that OT authority is better suited to contract for emerging technology and engaging with non-traditional contractors, the authority is nevertheless used sparingly compared to the traditional procurement system. Thus, understanding the theoretical underpinnings of how public procurement law and policy can attract contractors can assist in carrying out this new study.

---

[461] See L Elaine Halchin, Congressional Research Service, *Other Transaction (OT) Authority* (CRS Report for Congress No RL34760, rev ed, 15 July 2011) 24; Amey (n 120).

This research seeks to understand why commercial AI firms choose to work with the DoD. To help understand the concept of customer attractiveness and to develop the framework for understanding the research question, the following section reviews existing literature on social exchange theory.

### E *Topic III: Social Exchange Theory*

The third topic of literature reviewed is social exchange theory, with a focus on the concept of customer attractiveness. This broad theory, grounded in the fields of economics, sociology and psychology, is used as the theoretical lens to examine how the DoD is perceived as a customer by commercial AI firms. Social exchange theory connects both contract and non-business factors to the perceived attractiveness of a customer, focusing on the initiation and maintenance of a relationship. Given the complex procurement system as presented in the previous literature topic, the use of a theoretical tool is useful to assess how the distinctive



attributes of FAR contracts and OT agreements may affect commercial AI firms' perspectives on working with the DoD.

The concept of 'customer attractiveness'[462] can inform the DoD's efforts to understand how best to attract non-traditional companies to access emerging technology and act faster. The phenomenon of customer attractiveness is derived from sociology, psychology, and economics research based on the social exchange theory.[463] Customer attractiveness, though studied within other broad theoretical frameworks, is well-situated in social exchange theory because the core issues contemplated by the theory include questions of relationship initiation, termination and continuation of the relationship.[464] Social exchange theory posits that people 'choose between alternative potential associates and courses of action by evaluating the experiences or expected experiences with each in terms of a preference ranking and then selecting the best alternative.'[465] This process of choosing between alternatives is grounded in the person's perception of the attractiveness of the potential relationship or course of action, forming the construct of how rewarding a choice will compare to the alternatives.[466] In the customer-supplier context, the customer should seek to maximise its attractiveness to suppliers.[467] Research in social exchange theory in the commercial context indicates that firms who effectively attract suppliers to compete for contract opportunities and satisfy those suppliers during contract performance can access, develop, and utilise strategic resources to gain competitive advantages.[468]

---

[462] See Lambe, Wittman and Spekman (n 14) 12–3.

[463] Holger Schiele, Richard Calvi, and Michael Gibbert, 'Customer Attractiveness, Supplier Satisfaction and Preferred Customer Status: Introduction, Definitions and an Overarching Framework' (2012) 41 *Industrial Marketing Management* 1178, 1180.

[464] Ibid 1179–80.

[465] Andreas Herbert Glas, 'Preferential Treatment from the Defense Industry for the Military' (2017) 1 *Journal of Defense Analytics and Logistics* 96, 100.

[466] Ibid.

[467] Ibid.

[468] Holger Schiele et al, 'The Impact of Customer Attractiveness and Supplier Satisfaction on Becoming a Preferred Customer' (2016) 54 *Industrial Marketing Management* 129, 137.



This research uses social exchange theory research, both as a theoretical lens and to discover the drivers of how commercial AI firms perceive the attractiveness of the military as a customer.  The social exchange theory literature is used to build a conceptual framework that can examine how contract law and procurement practice of the DoD attracts or, conversely, repels commercial AI firms.  Understanding the preferences, opinions, and perceptions of commercial firms can inform the DoD how to optimise its attractiveness through alignment of its engagement with industry and choice of contract law.  This insight can uncover best practices in attracting and working with innovative firms to develop AI systems.

A supplier's perception of customer attractiveness can be determined by three main areas: value creation, the interaction process, and the emotional response of wanting to get closer to the buyer.[469]  The first consideration requires the buyer to maximise the potential value of an interaction with the supplier to increase attractiveness.[470]  Factors that can positively affect the potential value creation by the buyer-supplier interaction include profit, volume, innovation development and market access.[471]  The interaction process itself can also affect customer attractiveness.[472]  A collaborative process, which requires mutual trust of the parties and commitment to the common goal, is important for successful buyer-supplier relationships.[473]  When comparing the two legal frameworks, OT authority appears better suited to enable collaboration than the regulation-driven FAR contracting paradigm which epitomises a transactional relationship.  Interestingly, customer attractiveness, despite its economic

---

foundation, is also influenced by emotions and reflects the irrational part of the decision making.[474] This element is reflected in the drive of firms to work for the DoD out of a sense of personal fulfillment through work on challenging, meaningful projects. Like the public-private partnerships during the space race, firms can leverage the prestige of working on 'moonshot' problems for commercial gain in other markets. Thus, the seemingly non-economic factors can nonetheless lead to economic success thereby increasing the attractiveness of the opportunity. The DoD can distinguish itself from other potential buyers, like marketing firms, retailers, and banks, by connecting the mission to the requirement. Ultimately, in high demand industries such as AI, customers need to maximise their perceived attractiveness compared to other choices to work with the best, smartest suppliers.[475]

A review of existing literature outlined drivers of customer attractiveness that help to create a research framework for understanding how a supplier, or industry, may perceive the attractiveness of a customer in accordance with social exchange theory.[476] Such drivers include: economic, resource and social factors; familiarity, similarity, compatibility and knowledge of alternatives; expected value; trust; dependence and autonomy; communication; ethical behaviour and fairness; corporate image; and relational fit.[477]

---

[474] Ibid. This concept of actors behaving in a way that is inconsistent with utility theory, which is focused on maximising profits, and act in not entirely rational manner, is studied in the field of behavioural economics: see Richard H Thaler, *Misbehaving: The Making of Behavioral Economics* (Norton, 2016) 29–30. Thaler's theory started with the mission statement: 'Build descriptive economic models that accurately portray human behaviour': at 30. This research seeks to understand commercial AI firms' behaviour and then align contract law with that behaviour to better attract those firms to engaging with the DoD and supporting national security.

[475] Hüttinger, Schiele and Veldman (n 469) 1197–8.

[476] See Lisa Hüttinger, Holger Schiele and Dennis Schröer, 'Exploring the Antecedents of Preferential Customer Treatment by Suppliers: A Mixed Methods Approach' (2014) 19 *Supply Chain Management: An International Journal* 697, 699–700 ('Preferential Customer Treatment').

[477] Ibid.



Other researchers have identified that profits are an important contributor to the value a supplier places on a contract, so buyers that pay fair prices or buy large quantities are preferred.[478]  However, Jean Nollet explains that non-monetary factors also contribute to the attraction of a customer.[479]  Such factors include the possibility for suppliers to maintain their autonomy and power when dealing with buyers; the affect the relationship has on their reputation; the buyer's trustworthiness and fairness; timely information sharing; and effective and harmonious interpersonal relationships between the seller and buyer.[480]  Buyers can improve their attractiveness by participating in events with sellers such as trade shows, communicating in a consistent manner, and highlighting what distinguishes the buyer from other potential customers.[481]

Becoming an attractive customer increases the expected value of the initial exchange between the seller and buyer, making the seller more likely to accept the exchange.[482]  However, attraction is very subjective.[483]  Thus, the buyer must understand the seller's perception of value and align its action accordingly by using the appropriate tactics.[484]  In the DoD, the tactics employed to attract commercial AI firms will need to comply with its contract law and regulations.  The DoD must understand commercial AI industry's perceptions and align its engagement and contract negotiation with the values of the industry.

---

[478] Jean Nollet, Claudia Rebolledo and Victoria Popel, 'Becoming a Preferred Customer One Step at a Time' (2012) 41 *Industrial Marketing Management* 1186, 1189.
[479] Ibid.
[480] Ibid.
[481] Ibid.
[482] Ibid 1188.
[483] Ibid.
[484] Ibid.



Research on buyer-seller collaboration reveals a relationship dynamic where the buyer and seller hold differing perceptions of collaboration.[485] In interviews with manufacturing suppliers, the suppliers wanted to be involved earlier in the buyer's operational processes and were eager to take part in innovation projects as well as in setting a shared strategy.[486] The research found suppliers are more open for collaboration at different levels than buyers are,[487] indicating there is opportunity for buyers to build relationships with suppliers throughout the buying and performance process, specifically in innovative applications. This research supports the position that 'buyers need to enhance their pro-relationship behaviour in terms of listening to the voice of suppliers and develop better strategies for managing buyer-supplier relationships.'[488] As discussed above, the development of AI systems strongly aligns with a collaborative process beginning at the start of the system's lifecycle.

One study by Andreas Herbert Glas applies the concept of customer attractiveness to the defence logistics research context.[489] While research in social exchange theory and customer attractiveness predominantly focus on industry, Glas's transfer to the defence sector is novel.[490] Glas builds upon prior works by Hüttinger to identify the challenge militaries face today where they are dependent on suppliers who are in demand by other customers.[491] In the military logistics context, military customers buy from suppliers who often supply to other militaries, as well as civilian business to business or business to consumer markets.[492] As Glas warns, when there is rapid technological change, limited military research and development budgets force militaries to compete in highly innovative civilian markets.[493] Declaring an end of the times of 'the customer is always right,' he argues defence customers must make doing business attractive for suppliers.[494] This concept applies to the DoD acquiring AI systems, where the supplier — with many options to select customers and dictate terms — must be attracted to the customer.

---

[485] Agnieszka Blonska, 'To Buy or Not to Buy: Empirical Studies on Buyer-Supplier Collaboration' (PhD Thesis, Maastricht University, 2010) 74.
[486] Ibid.
[487] Ibid.
[488] Ibid.
[489] Glas (n 465) 96.
[490] See ibid 97.
[491] Ibid.
[492] Ibid.

Glas starts with Hüttinger's model to explore the customer attractiveness model using the factors of trust, commitment and comparative customer perceptions.[495] Applying this model to military customers, he tests hypotheses that the perception of comparative customers, commitment, and trust positively affect customer attractiveness which in turn positively affects the way suppliers treat such customers.[496] Glas surveyed 93 managers of German defence suppliers, finding commitment and trust have a significant positive influence on customer attractiveness.[497]

While Glas's research shows the German defence industry values trust and commitment of customers, the sample population is likely very different than this research's target population of commercial AI firms in the United States. Additionally, Glas's study is limited to testing customer attractiveness factors without exploring how factors such as contract law and procurement practice, ethics, and other motivations affect perceptions of customer attractiveness.

Social exchange theory provides a lens in which to understand why suppliers choose to conduct business with customers and how they make that decision; using social exchange theory as a theoretical lens can help explain the interaction between the commercial AI industry and the DoD. In the competitive, high-demand market for AI innovation, the seller is better described as the customer of potential buyers, whereas the buyers compete for the seller's finite resources. Sellers must decide how to allocate those resources, including time, effort, and expertise, to maximise revenue that can be reinvested into scaling their product or service, or to further developing and advancing their technology. Based on the literature review, there are likely many drivers that motivate commercial AI firms to compete for DoD contracts. In the AI revolution, demand is universal, and supply is limited.

---

[493] Ibid.
[494] See ibid 98–100.
[495] Ibid 101.
[496] Ibid 101–2.
[497] Ibid 109.

This research uses social exchange theory to frame the examination of how commercial AI firms perceive the attractiveness of the DoD as a customer. This research seeks to understand the DoD's relative attractiveness compared to alternative customers. The different attributes of FAR contracts and OT agreements are assessed from the perspective of the commercial AI firms — the better the attributes align with this population's preferences, whether for business, technical, or other reasons, the more attractive the contract opportunity. Other potential drivers, such as ethics, cultural fit and desire to support the DoD's mission are also examined. The relative strength of the various drivers can be assessed by examining the perceptions, opinions, and preferences of commercial AI firms on various contract attributes, customer characteristics, and ethics. The insight provided by collecting this data can identify attractive contract attributes available to the DoD. This knowledge can help the DoD decide which legal framework offers a better fit for attractive commercial AI firms and focus efforts at acquisition reform to support the DoD's national security mission.

F *Limitations of the Theoretical Framework*

Although this research draws on social exchange theory and customer attractiveness literature to help explain the preferences of commercial AI firms and develop a framework for



collecting and analysing data, there are limitations to using a theoretical framework. Because this research uses the framework in a novel application, there is a risk that an overreliance of the theory can oversimplify the problem and lead to excessively reductionist conclusions, potentially overlooking important variables and dimensions.[498] To mitigate this risk, social exchange theory is used as a framework to develop a more specific theory that can explain what contract attributes affect the perceived attractiveness of the DoD in the commercial AI market.

G  *Synthesis of DoD Contract Attributes through the Lens of a Social Exchange Theory*

As discussed, the FAR and OT frameworks, while both available to the DoD for acquiring AI-enabled capabilities, are fundamentally different; the former is a highly regulated model intended for filling specific requirements while the latter is an inherently flexible model focused on iterative prototyping and experimentation to solve a problem. Procurement contracts governed by the FAR balance policy objectives such as fairness and transparency with oversight and saving costs. However, these policy considerations can result in delay to the government filling its requirement and create barriers to contractors not specialising in defence contracts. OT agreements are flexible contracts that encourage public-private collaboration and offer the opportunity to freely negotiate terms and conditions to lower barriers to private businesses. However, OT agreements can be unpredictable as they are comparatively rare, and the law and practice are less settled. Moreover, without the predefined terms and conditions required by the FAR, OT agreements require drafting and negotiating terms with industry, requiring skillsets that are not widely taught to government contract attorneys and procurement officials.

---

[498] See Linda Dale Bloomberg and Marie Volpe, *Completing Your Qualitative Dissertation* (Sage, 4th ed, 2019) 168, explaining that although use of a theoretical framework in qualitative research provides a meaningful way of seeing, thinking, and understanding, and provides the ability to organise and focus a study, the framework can be too reductionist and deterministic, forcing the researcher to place data into predetermined categories.



There is scant research into the connections of these attributes to the way commercial AI firms conduct business. An understanding of the drivers impacting the business calculus of commercial AI firms is necessary for the DoD to develop best practices in acquiring AI systems. Figure 7 provides a synthesis of the literature reviewed. This synthesis highlights the differences between the two contract law frameworks and presents testable hypotheses of what commercial AI firms would prefer based on the concepts of social exchange theory as well as the lifecycle for designing, developing, testing, and deploying AI systems. This table offers an integrated review of the three major literature topics, adding concepts from social exchange theory to the table comparing the attributes of FAR contracts to OT agreements in the context of AI development.

Figure 7: Synthesis of Literature Review Topics



| Contract Attribute | FAR | OT | AI Considerations | Social Exchange Theory |
|---|---|---|---|---|
| Competition | • Default full and open competition subject to protest on the solicitation and award<br>• Governed by CICA | • Fair and transparent<br>• Competitive procedures used to the maximum extent practicable if awarding follow-on award | • Many commercial AI firms are unfamiliar with DoD contracting procedures | • Customers must maximise their perceived attractiveness compared to other choices to work with the best suppliers |
| Solicitation | • Request for proposals to fulfill requirements based on detailed specifications<br>• Announced on the government's website | • Problem statement, area or interest, or capability gap<br>• Announced through means expected to maximise exposure of the problem set to relevant technology providers | • AI systems are developed by understanding the problem and seeking relevant data first, and then determine the appropriate training method, algorithm, and features for the application | • Suppliers want to be involved earlier in the buyer's operational processes and were eager to take part in innovation projects |
| Complexity | • Dozens of government-specific conditions, certifications, and accounting systems<br>• Foreign process for many commercial firms | • Intended to be streamlined and consistent with commercial best practices | • Many commercial AI firms are unfamiliar with DoD contracting procedures | • Military customers must engage and attract potential suppliers in competitive commercial markets by making the relationship advantageous to the supplier<br>• Complex, customer-specific terms and conditions can make a selling relationship more costly than alternatives |
| Communication | • Limited to before solicitation is published and during designated period during source selection | • Encouraged throughout problem statement development through performance with industry and academia | • It is critical for the developer to understand the problem to be solved as well as how it will be used | • A collaborative process, which requires mutual trust of the parties is important for successful buyer-supplier relationships |
| Flexibility | • Generally rigid<br>• Modifications must be within scope of specifications or subject to further protest under CICA | • Inherently flexible<br>• Understands that the project may yield outcomes that surprise participants<br>• Directs iterative process | • The AI lifecycle is iterative and never finished as new data changes underlying software code | • Human factors are crucial components as they influence trust and commitment between parties, and communication, collaboration, and trust are important drivers of customer attraction |
| Terms & Conditions | • Dozens of required non-negotiable terms and conditions of performance | • Fully negotiable | • Many commercial AI firms are unfamiliar with DoD contracting procedures | • Burdensome, costly, rigid, complex, and mandatory terms and conditions placed upon the supplier are negative indicators of attractiveness |



| Intellectual Property | • Standard clauses provide government with unlimited, government purpose, or limited rights depending on whether development occurred under government contract<br>• Commercial license and negotiated license may be available to the extent permitted by law[499] | • Fully negotiable<br>• DoD can negotiate for rights above or below legal limitations applicable to FAR[500] | • AI is developed through finding data relevant to the application which is used to train the machine learning algorithm to produce a model capable of making inferences on new data; the data may come from a variety of sources, including government provided data as well as proprietary or open data | • Drivers of customer attractiveness include economic and resource factors; familiarity, similarity, compatibility, knowledge of alternatives; expected value; trust; and fairness |
|---|---|---|---|---|
| **Speed** | • Negotiation of terms is generally quick as negotiable terms are limited<br>• Pre-solicitation and source evaluation/ selection can take significant time<br>• Solicitation and award can be protested | • Negotiation of terms can be lengthy<br>• Pre-solicitation and source evaluation/ selection are unencumbered by FAR requirements and protest jurisdiction is limited so prototype award can be made swiftly<br>• Follow-on production agreement can be awarded with no additional competition cycles | • Advancements in the state of the art in AI occur extremely quickly<br>• Many commercial AI firms are unaccustomed to lengthy procurement times | • Military customers must engage and attract potential suppliers in competitive commercial markets by making the relationship advantageous to the supplier<br>• Lengthy and costly processes that are inconsistent with the seller's business model can make a selling relationship more costly than alternatives |

This table illustrates there are clear relationships between the contract attributes of FAR contracts and OT agreements to considerations regarding the development of AI systems and social exchange theory. From these relationships, hypotheses about commercial AI firms' perceptions and preferences are generated.

---

[499] DFARS § 252.227-7013.
[500] See *Other Transaction Guide* (n 104) 49.



- **Competition hypothesis:** Commercial AI firms prefer competing for awards without significant barriers to entry and are attracted to opportunities that require limited resources to compete.

- **Solicitation hypothesis:** Commercial AI firms prefer to assist their customers explore and solve the problem and are more attracted to opportunities that permit creativity and innovation rather than following strict specifications of how to perform a task.

- **Complexity hypothesis:** Commercial AI firms prefer straightforward processes that align with their commercial business models and technical approaches.

- **Communication hypothesis:** Commercial AI firms prefer the open-ended ability to communicate and collaborate with the end-user.

- **Flexibility hypothesis:** Commercial AI firms prefer flexibility in contract negotiation and performance, especially to experiment and iterate in developing an AI system.

- **Terms and conditions hypothesis:** Commercial AI firms prefer the ability to negotiate terms and conditions and avoid mandatory clauses unique to specific buyers.

- **Intellectual property hypothesis:** Commercial AI firms prefer negotiating intellectual property terms that they understand and are attracted to opportunities that fairly compensate their efforts in developing, ensure adequate protections and account for the unique way AI applications are developed.

These hypotheses are tested in the surveys and interviews and assessed in the analysis discussed in Chapters IV and V. Assessing how these attributes attract or discourage commercial AI firms



to compete for DoD contracts develops a more comprehensive understanding of the relative merits of each contract law framework.

<p style="text-align: center;">H <em>Conclusion</em></p>

In this chapter, Topic I reviewed literature at the nexus of AI and national security, examining how AI differs from other technologies and why that matters to the DoD. This topic addressed potential challenges the DoD may encounter as it attempts to leverage commercial AI innovation, finding the unique contract law governing the relationship between commercial AI firms and the DoD is a significant factor. Topic II reviewed the two parallel contract law frameworks available to the DoD in buying commercial AI systems and examined the challenges and benefits of each. Topic III reviewed social exchange theory to build a conceptual framework for this research to help explain how attractive the DoD is to commercial AI firms, and whether the DoD's contract law and practice impact that perception. The synthesis of the literature reviewed indicates that there are theoretical and pragmatic national security advantages to developing a contract law framework that enables the DoD to leverage the advancements to AI-enabled technologies from commercial AI firms. By advancing the DoD's understanding of the commercial sector's preferences and technological considerations, the DoD can align its contract law and practice to optimise its attractiveness to commercial AI firms. The following chapter discusses the research design and methodology used in this research.





A *Introduction*

Exploring an explanation of why commercial AI firms decide to contract with the DoD requires data on the perceptions of those firms which is not available in existing literature. This original research seeks to obtain data on the perceptions of commercial AI firms to help answer the research question. Insight into the factors that affect how attractive the DoD is as a customer from the perspective of commercial AI firms will lead to a better understanding of the alignment between current contract law and procurement practice and business preferences in the commercial AI industry. With that understanding, contract lawyers in the DoD can advise on the choice of legal framework, negotiate mutually beneficial contracts, and legislators and policy makers can better focus acquisition reform efforts to optimise the contract law for acquiring state of the art AI systems for defence applications.

B *Overview of Information Needed*

To answer the research question, an understanding of what factors influence commercial AI firms in their business decisions is required. Due to the lack of existing research on how commercial AI firms perceive the DoD as a customer, this research seeks to identify those factors by conducting cross-sectional exploratory survey research.[501] The survey targets are a hard to identify, and harder to quantify, population of commercial AI firms that sell defence-relevant AI technology and applications. The survey is used to determine factors that could influence the perception of the DoD as a customer. Factors that could influence the perception of the DoD as a customer could influence the decision to pursue a contract with the DoD. The questions attempt to elicit an understanding of how commercial AI firms compare the DoD to commercial customers. The survey also attempts to understand whether commercial AI firms prefer certain contract attributes over others; this could assist determining how well the FAR or OT authority align with commercial preference. Surveys, especially a cross-sectional, exploratory study, have limitations; survey results provide a snapshot of views of the survey participants and cannot be generalised to larger populations.[502] Despite efforts to mitigate bias and capture a sample that is representative of the population, it is unknown whether the sample represents the population.[503] Thus, the survey results should be viewed as informative of the sample only. Yet, the data is helpful to develop hypotheses and theories. The surveys were followed by in-depth interviews of a purposeful sample of business leaders at commercial AI firms. The interviewees gave personal accounts of their opinions, perceptions, and preferences regarding customer selection, contract attributes and the DoD. In combination with the survey findings, the interviews are used to develop an understanding of the research question.

---

[501] The surveys and interviews were approved by the University of Adelaide Office of Research Ethics, Compliance and Integrity (Ethics Approval No H-2020-037) and conducted in accordance with the National Statement on Ethical Conduct in Human Research 2007 (Updated 2018).

Demographic information was collected from databases such as Crunchbase, LinkedIn, and Govshop to understand the composition of the commercial AI industry and identify the target population. This information was used to recruit participants for the survey study. In the survey, participants provided demographic information relating to their firm's experience in competing for DoD contracts, the volume of business the DoD represents compared to the commercial market, location, size, funding sources, status as a traditional or non-traditional defence contractor, and whether the firm was veteran-owned. While the demographic information collected focused primarily on the firm as an entity, the demographic data collected on the individual participants included knowledge of the business decision making (this was a requirement to participating in this study), and their position in the firm, and experience contracting with the DoD.

---

[502] See Erin Ruel, William Edward Wagner III and Brian Joseph Gillespie, *The Practice of Survey Research: Theory and Applications* (SAGE, 2016) 125 (explaining that only a random or probabilistic sample allows statistically significant findings about the sample to be inferred to the population from which it is drawn). As discussed in this chapter, a non-random sample was surveyed because the target population (commercial AI firms that develop or deploy defence-relevant AI applications) is hard to locate or quantify, making a random sample impossible: at 149–57 (discussing methods for sampling hard to find populations and the tradeoffs of using nonrandom sampling).
[503] Ibid 149.

Perceptual information was collected in the survey and follow-on interviews. The focus of this research is to collect data on the commercial AI industry's perception of the DoD as a customer and what factors attract firms to contract opportunities. Thus, perceptual information is the most critical to this study. Information gathered included perceptions on how unique attributes of AI, both as a technology and as a commodity, impact the business model; how experience working for or with the DoD affects their decision to compete for contract opportunities; how the size and funding of their firm factors into their firm's decision making; how specific contract attributes affect whether their firm decides to submit a proposal for a contract opportunity with the DoD; and how much weight business and non-business factors are given in their firm's decision to engage in a contract opportunity. While perceptual information is critical to this study, perceptions are not facts — they are only what the participants perceive as facts.[504] Thus, the perceptions collected in this research 'are neither right nor wrong; they tell the story of what participants believe to be true.'[505]

Theoretical information was collected from the various literature sources described in Chapter II. This information provided a framework supporting the research methodology, theories that helped form the research question and development of the conceptual framework, and support for data interpretation, analysis, and recommendations and conclusions.[506]

---

[504] See Bloomberg and Volpe (n 498) 188, explaining that perceptions are rooted in long-held assumptions and one's own view of the world or frame of reference.
[505] Ibid.

## C *Research Design Overview*

This study is designed in the grounded theory tradition.  Grounded theory is a method of inquiry that aims to 'inductively generate theory about a particular behavioural phenomenon that is grounded in, or emerges from, the data.'[507]  The goal of grounded theory research is for the researcher to generate or discover a theory of processes, actions, or interactions that are grounded in the views of the participants, all of whom have personally experienced the process, action, or interaction.[508]  The developed theory can explain the perception and views of the participants or provide a theoretical framework for further research.[509]  The theory is based on multiple stages of data collection and the refinement of abstract categories of information to generate meaning and of building consensus to explain phenomena as they are experienced by the research participants.[510]  Because the 'epistemological premise of grounded theory assumes that the theoretical knowledge to be gained through research cannot be presupposed,' the methodological approach 'regards knowledge production as something that can be gained only through an inductive process.'[511]  In this study, data is collected from the research participants, each of whom are employed at a commercial firm that provides AI-enabled capabilities, either through a product or service.  Each participant has personal knowledge of how their firms make business decisions.  Through mixed methods of both quantitative and qualitative research, the experiences, perspectives, and opinions of the participants are collected to generate a consensus to explain the commercial AI industry's perception of the DoD as customer. Open coding and open-ended formal interviews are used to develop an explanation for the perception and generate a theory of how the DoD can be an attractive customer to the commercial AI industry. This theoretical framework can be used to examine other organisations and procurement systems to identify best practices in attracting commercial AI firms to compete for contract opportunities.

---

[506] See ibid 188–9.
[507] Ibid 96.
[508] Ibid.
[509] Ibid.
[510] Ibid 96–7.
[511] Ibid. 97.

This study is conducted in the pragmatist research paradigm. Pragmatism is an epistemological approach that is concerned with practical application and workable solutions to research problems.[512] Pragmatists typically employ multiple methods, both quantitative and qualitative, which can be combined in creative ways to gain a greater understanding of the research problem.[513] This study employs mixed methods to develop a greater understanding of why commercial AI firms decide to work with the DoD. By gaining insight directly from business leaders at commercial AI firms, this study seeks to generate a theory that explains what factors make the DoD an attractive customer. This theory can assist government contract lawyers advise on the choice of legal framework that optimising alignment of the DoD's defence objectives with commercial AI firm business and technical considerations.

### D *Overview of Methodology*

The research methodology follows a mixed-methods explanatory sequential plan.[514] This method was chosen because there is little empirical data providing insight into the reasons commercial AI firms decide to work with the DoD.[515] Explanatory sequential mixed methods design allows for quantitative research, typically through a survey instrument, to collect data to understand the phenomenon studied, and then gain additional insight to help explain the survey responses through interviews.[516]  The research began with quantitative research in the form of a survey of commercial AI firms.  The survey results were analysed and used to develop interview questions.  The qualitative research in the form of interviews of a purposeful sample of the survey participants help to explain and expand upon the survey findings.[517]

---

[512] Ibid 44.
[513] Ibid.
[514] Creswell and Creswell (n 157) 221.
[515] See ibid 221–2 (explaining that explanatory sequential design is an appropriate methodology in research areas new to qualitative approaches).

Prior to the survey research, qualitative methods in the form of a literature review were conducted, as discussed in Chapter II, to frame the study, develop research questions and identify hypotheses to be examined.  The complex research design was necessitated by the novelty of the research question and is consistent with the pragmatic worldview.[518]  The two data sources derived from different methodologies support and corroborate each other in data and methodological triangulation, illuminating different facets of the commercial AI firms' experiences, perspectives, and opinions.[519]  Viewing the data through the lens of social exchange theory also helps build a stronger foundation for understanding the rationale and motivation of the participants' perceptions and contract preferences.[520]  The research methodology process map is depicted in Figure 8 below.[521]

*Figure 8: Research Methodology Process Map*

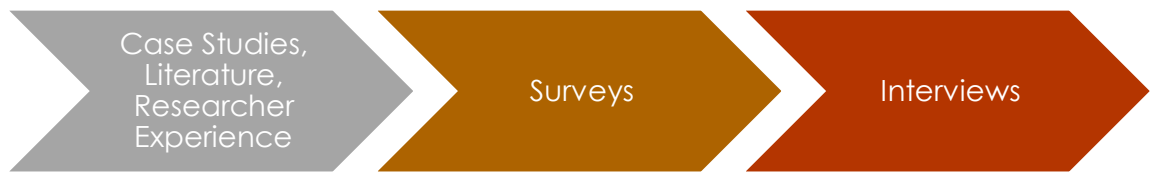

While the two qualitative phases of this mixed methods design are critical to understanding the research problem, the interview data is the focus of this research. The research question is a qualitative inquiry, searching for an understanding not only of what factors impact commercial AI firms' decision to work with the DoD, but why those firms ultimately make that decision. This focus on qualitative data is atypical in the explanatory sequential design.[522] However, this decision was intentional as it was determined a survey was the best tool for gathering previously unknown data about how the DoD is perceived by the sample population before conducting interviews to understand those opinions. Although the survey data made possible the identification of themes and helped build an understanding into the research questions, the qualitative interviews are critical in interpreting and validating the survey data, and ultimately addressing the question of why commercial AI firms choose to work with the DoD and whether the DoD can leverage its choice of contract law to influence that decision. This question cannot be answered by the survey data alone. Accordingly, Figure 9 reflects greater emphasis on the qualitative data.[523]

---

[522] Ibid 238.
[523] Adapted from ibid 235–6.



*Figure 9: Research Design Notation*

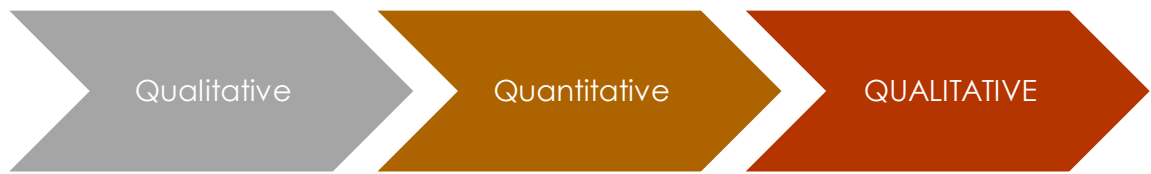

The expected outcome from this research is to gain insight into the impressions, perspectives, preferences, and opinions about contracting, generally, and the DoD as a customer. This new insight leads to identification of themes, preferred contract attributes, as well as recommendations for aligning the contract law and procurement practice of the DoD with the business and technical considerations of commercial AI firms.

1 *Surveys*

The quantitative data derives from close-ended surveys conducted on persons with knowledge or authority to make business decisions on behalf of their commercial AI firm. Several demographic datapoints were collected to identify whether any of these variables impacted a participant's perceptions. Demographic data collected in the survey included: size of the firm; status as a traditional defence contractor; location of the firm; experience working with the military. These demographics are predictor variables that may impact perception of the DoD as a customer.[524] Predictor variables are variables used to predict an outcome of interest — in this case, the perceptions of the DoD as a customer and preferences on contract attributes — in a survey.[525]  Perception data collected in the survey included: preferences on intellectual property terms, ability to negotiate clauses, the commerciality of terms, communication with end-user, speed of award process and funding, access to databases, type of work involved (challenging, prestigious, disruptive), and cultural factors.  These perceptions are outcome variables, possibly the result of the predictor variables.[526]  However, the prediction value of the demographic data is not the intent of the survey; the primary intent is to gain understanding in commercial AI firms' perception of the DoD as a customer and preferences towards certain contract attributes.  The demographic data and any potential value in predicting customer attractiveness or contract preferences was assessed only in analysis the interviews to avoid confusing random correlation with causation.  The cross sections of the various demographic populations within the sample were too small to meaningful assess any predictive value in the demographic variables.  The focus of this study is the perception data, while the demographic data was intended to assist in selecting a purposeful sample for the subsequent interviews.

---

[524] Ibid 50–1, explaining that 'predictor variables,' also called antecedent variables, are analogous to independent variables in that they are hypothesised to affect outcomes in a study, but dissimilar because the researcher is not able to systematically manipulate a predictor variable. It is unknown whether demographic data, such as former experience working with the DoD or firm size, affects perception of the DoD as a customer, but this survey research is experimental and such variables may impact a firm's opinion on the DoD. While this research cannot assign commercial AI firms to control groups that conduct all its business with the DoD, it may measure such experience within the sample as a predictor variable: see ibid 50–1.

The goal of the survey in this research is to begin to understand what motivates commercial AI firms to decide to work with the DoD.  The literature, reviewed in Chapter II, indicated that commercial AI firms may perceive the DoD as a flawed customer that contracts in a system that is unaligned with the business and technological considerations of a commercial AI firm, but nevertheless offers a unique opportunity to work on challenging and meaningful projects that could be advantageous.  The literature resulted in several hypotheses; the survey was designed to contribute to the testing of those hypotheses.  However, this is exploratory research, and despite efforts to conduct a survey of a random sample representative of the population, such a sample proved elusive. Despite a sample size of 111 commercial AI firms participating in the survey, there is no way to be certain the sample represents the population.[527] Due to constraints in understanding the true target population's size and limitations on recruiting methods, it was impractical to generate a random sample of the population. Thus, generalisation of the findings beyond this sample is limited. Nonetheless, the survey results provide insight into the business calculus of a large sample of commercial AI firms developing defence relevant technology and serve as an approximation or heuristic of the population. These findings are used to identify prominent opinions that are further explored in the interviews. Finally, these results, along with the interviews, are used to develop a model that identifies the relative strength of preference of various contract attribute options and alignment with DoD contract law to assist contracting officials optimise customer attractiveness.

---

[525] Ibid 51.
[526] Ibid, explaining that outcome variables 'are variables that are considered outcomes or results of the predictor variables in survey method studies'.  It is recommended that experimental survey research aim to measure multiple outcome variables: ibid.

## 2 *Interviews*

The intent of qualitative research is to provide rich information about the context and setting.[528] In mixed methods, the integration of quantitative and qualitative research can provide a better understanding than either method in isolation.[529] The research was conducted using an explanatory sequential mixed methods design, involving a two-phase data collection: the first phase collected quantitative survey data and subsequent analysis; the second phase built upon the first phase with purposeful sampling of interviewees for qualitative data collection.[530]  The intent of the of the interviews is to help explain in more detail the quantitative results from the survey.[531]

---

[527] Ruel, Wagner and Gillespie (n 502) 149. The research had a goal of obtaining a sample size of at least eighty-four (84) survey participants to improve reliability. This size is calculated using a test family: exact, statistical test: correlation: bivariate normal model, a priori power analysis test performed on G* Power 3.1 software. With the input parameters of two tails, medium (.3) correlation p H1, alpha value error probability of .05, power of .8, and correlation p H0 of 0. Franz Faul et al, 'G* Power 3: A Flexible Statistical Power Analysis Program for the Social, Behavioral, and Biomedical Sciences' (2007) 39(2) *Behavior Research Methods* 175; Creswell and Creswell (n 157) 151–2.
[528] Bloomberg and Volpe (n 498) 186.
[529] Creswell and Creswell (n 157) 4, 216, explaining the 'integration of qualitative and quantitative data' in this research 'will yield additional insight beyond the information provided by either the quantitative or qualitative data alone'.

The qualitative research was comprised of semi-structured one-on-one interviews of key informants selected according to purposeful sampling derived from data collected in the quantitative research phase.[532]  The researcher used open-ended questions to contextualise the quantitative data and triangulate the findings with the theoretical framework.[533]  Data analysis involved coding, synthesis, categorisation and aggregation using qualitative computer software programs for assistance.[534]  Interviewees provided informed consent of the risks of their participation and the researcher requested the interviewees to not discuss confidential, sensitive or propriety information during the interview.[535]  Interviewees were informed they had the option to end the interview at any time, and confidentiality was assured.  No interviewee requested to end the interview at any point during or after the interview and all consented to being audio recorded.

E *Sample Population*

A critical step to conducting the survey was identifying the target population as there is no published list of firms that provide AI solutions with potential defence applications eligible to contract with the DoD.[536]  The target population of AI firms that provide solutions of possible relevance to the DoD was identified using Crunchbase Pro, LinkedIn Premium, and Govshop. These resources provided information on firms' geography, technology, size, funding, and business maturity, ensuring an understanding of the target population's dynamics and representation.  Executives, directors, program managers, and investors from firms that provide AI capabilities to the commercial market were identified and recruited to participate in an online survey containing a total of 73 questions.[537]  The invitations were submitted to firms that, by their public description, provide solutions that could be used by the military, ensuring that the participants' opinions were relevant.[538]  A total of 370 firms based in the United States were identified as potential members of the population based on their publicly available information on AI products or services with a possible defence application.[539]  Invitations to participate in the survey were submitted by email or LinkedIn messaging.[540]

---

[530] Ibid 221–2; Bloomberg and Volpe (n 498) 187 (defining purposeful sampling as a strategy to select research participants).  In this research, the strategy employed for selection was to find diverse perspectives from business leaders at commercial AI firms with varied experience that can contribute to the evolving theory.
[531] Creswell and Creswell (n 157) 222.
[532] Ibid 216.
[533] Ibid 215.
[534] See ibid 191–4.
[535] Ibid 184–5; see Appendix for details on research ethics.

---

[536] 'The target population represents all of those from whom the sample to be surveyed will be drawn,' comprising the entire collection of the population the research aims to study and ideally generalise: Ruel, Wagner and Gillespie (n 502) 16.

[537] The survey instrument was written and administered on SurveyMonkey web-based software.

[538] The AI industry is incredibly diverse.  Many companies focus on narrow AI solutions for specific sectors.  AI solutions marketed towards industries such as finance, social activities, sales, and employment were excluded from this study, as their AI work is focused primarily on search functions, advertisement placement, and financial trading — capabilities that are not particularly relevant to the DoD.  Firms with solutions focused on computer vision, autonomous robotics, natural language processing, data analytics, business processes (such as automated workflows), AI-enabled warfighting capabilities, predictive algorithms, etc. were the focus of this study as those solutions can be adopted for both commercial and military use cases.

[539] There may be many more firms that provide AI-enabled capabilities and solutions that could be relevant to the DoD.  Firms that were not listed or identified in the databases described above were omitted.  AI firms that appear to only work in non-defence industries, such as advertising or entertainment, were not invited to participate in this study.  All firms surveyed were American-owned and operated, with one exception of a foreign owned, but American headquartered firm.  With multiple layers of subjective judgment used in identifying the target population, it is likely the field of AI firms that make up the target population is even larger.  As other research efforts identified, because there is no single, objective definition of an 'AI firm,' it is challenging to precisely account for the entire population.  See Arnold, Rahkovsky and Huang (n 242) 2.

[540] Personal messages on LinkedIn yielded far more responses than emails.  Feedback from several participants indicated that LinkedIn has the advantage of providing the researcher's background, photograph, professional contacts, and education experience whereas an email, though more formal, reveals very limited information about the researcher and cannot be verified as easily as a LinkedIn profile examination.



Out of the 370 firms invited, 153 firms agreed to participate, with 111 completing the survey.[541] Each of the survey participants certified they had personal knowledge of how their firm made business decisions, and most were C-Suite executives, vice presidents, or directors of sales or business development. For companies that performed contracts with the DoD and had a separate division or employees dedicated to government contracts (about 37% of firms surveyed), the survey participant was often the head of a firm's public sector or defence sector division. Participation in the survey was confidential, though demographic information about each company was collected.

The survey participants come from various backgrounds and represent firms of all sizes, from start-ups to some of the world's largest companies. While most firms came from the tech hubs of Silicon Valley (32), the DC region (20), Boston (11), New York (9), Seattle (5) and Austin (5), dozens of firms from outside these areas were surveyed as well. Figure 10 provides a chart depicting the geographic locations of the firms that make up the sample.

---

[541] 17 participants did not clear the screening questions meaning they either did not consent to the terms and conditions of participation, did not work at a commercial AI firm within the last 12 months or lacked knowledge about how their company makes business decisions. An additional 25 did not complete the survey after clearing the screening questions. Data from those participants who did not complete the survey for any reason was not analysed in the study in accordance with the terms of participation and consent.



*Figure 10: Location of Firm*

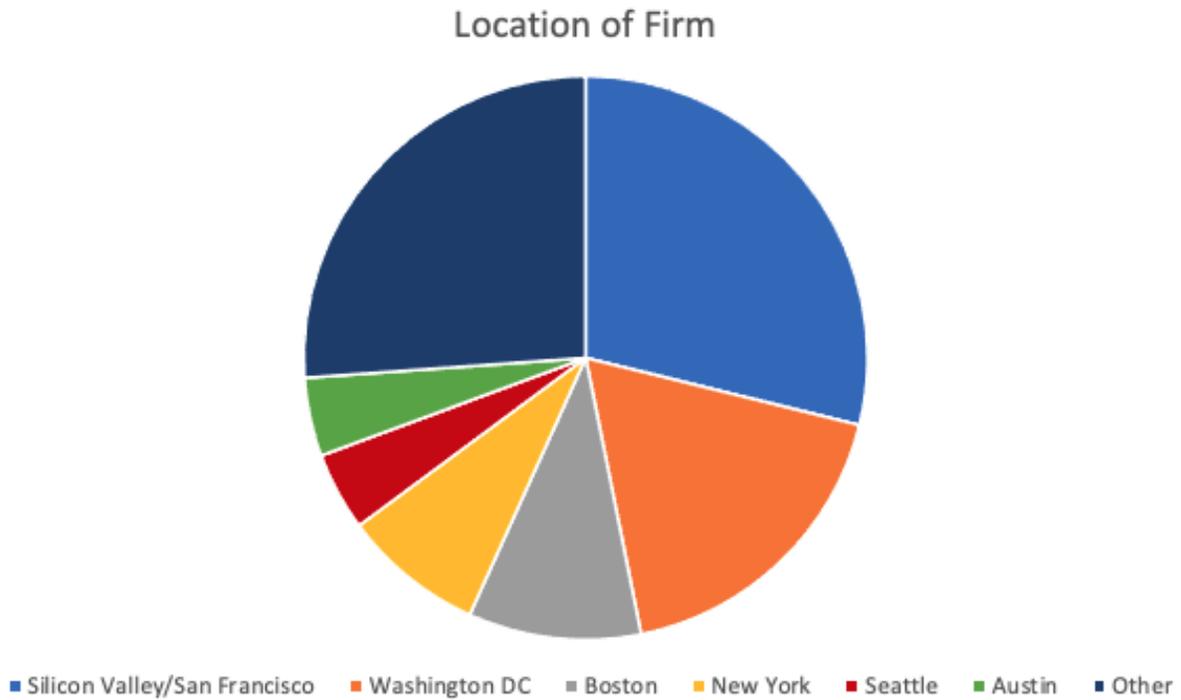

Firms surveyed were closely split between early-stage start-ups (founders, seed, angel funding) (49 start-ups) and firms with later-stage financing (Series A, B, C, D, E) (51 firms). The sample also includes 11 public companies. Approximately 84% of the firms surveyed would qualify as a small business for a DoD contract.[542] Figure 11 reflects the type of funding received by the firms.

---

[542] Based on size standards for NAICS code 541715 (up to 1000 employees): US Small Business Association Table of Small Business Size Standards Match to North American Industry Classification System Codes (effective 19 December 2022) <https://www.sba.gov/document/support-table-size-standards>. A review of FPDS data on DoD contracts for AI solutions revealed that the most common NAICS code used by the contracting agency was 541715 (Research and Development in the Physical, Engineering, and Life Sciences (except Nanotechnology and Biotechnology)).



*Figure 11: Level of Funding*

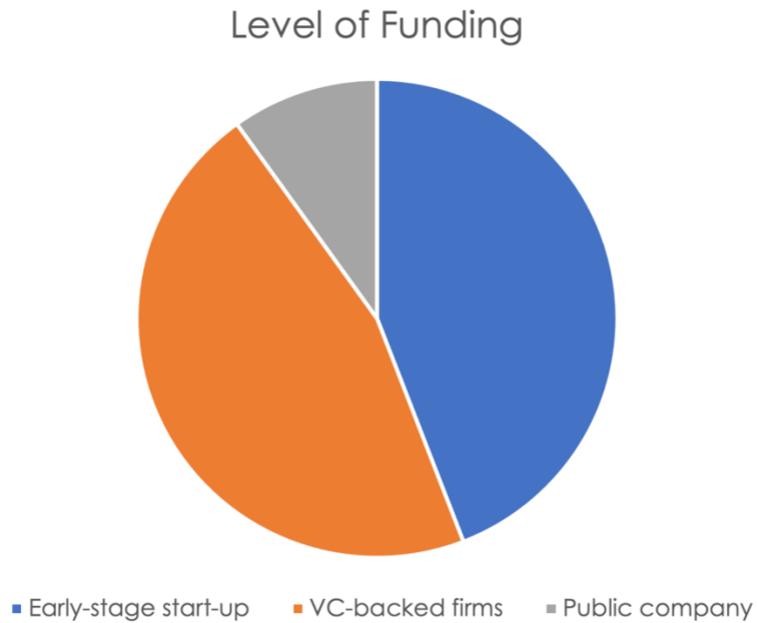

Of the companies surveyed, roughly 23% have never performed a contract with the DoD. Approximately half of the companies responded that contracts with the DoD represent less than 30% of their business revenue. One-quarter of the respondents stated that DoD contracts make up 70% or more of their total business earnings. Figure 12 reflects the relative share of total sales made up of DoD contracts, indicating how frequently a firm contracts with the DoD compared to other customers.



*Figure 12: Share of DoD Contracts to Firm's Total Sales (percentage)*

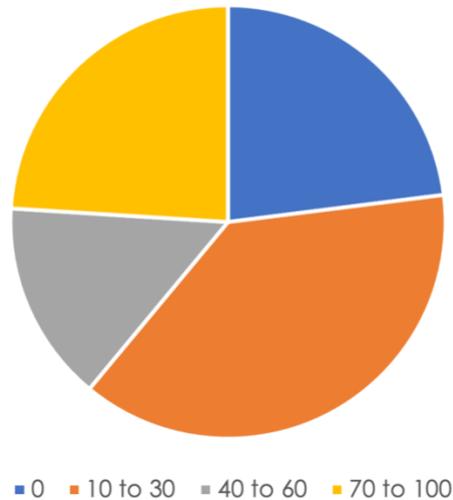

Thirty-four companies identified as traditional contractors, having performed a contract or subcontract subject to full coverage under the Cost Accounting Standards within the past year.[543] The other 77 companies are non-traditional defence contractors.[544] Eighteen companies have never competed for any government contract, while the other 93 companies have competed for at least one FAR, SBIR/STTR, or OTA contract at some point. Precisely one-third of the companies are members of at least one other transaction consortium.

While the survey sample consisted of volunteers from the known target population, the interviewees were purposefully sampled from the survey participants based on a variety of factors. The intent of using purposeful sampling is to gain insight and understanding of the

---

[543] See 10 USC § 3014, defining 'non-traditional defense contractor' with respect to a procurement or other transaction as 'an entity that is not currently performing and has not performed, for at least the one-year period preceding the solicitation of sources by the Department of Defense for the procurement or transaction, any contract or subcontract for the Department of Defense that is subject to full coverage under the cost accounting standards'.
[544] Ibid.



research question.[545]  In qualitative research, purposeful sampling is preferred to random

sampling to select information-rich cases which can shed light on the phenomenon studied.[546]

Out of the 111 survey participants, 15 were interviewed.  These 15 participants were

selected out of the 58 survey participants that volunteered to participate in an interview.

Consistent with grounded theory research, a theoretical sampling method was employed to select

the interviewees.[547]  The theory-based sample was selected on the basis that the individuals

interviewed could contribute to the evolving theory that the DoD's contract law and procurement

practice can affect its perceived attractiveness as a potential customer to commercial AI firms.

The interviewees were selected to maximise a range of experience in working with the DoD;

location, size, and funding of their company; and perspectives as reflected in their survey

responses.[548]  The interviewees included business decision makers, such as chief executive

officers, executive vice presidents, directors of sales, directors of business development, senior

managers, chief operating officers, and directors of federal or defence programs.[549]  The firms

represented ranged from start-ups of fewer than ten employees to some of the largest technology

firms in the world.  Firms ranged from self-funded to publicly traded corporations with the intent

of understanding the impact cash flow and source of funds has on the business calculus of a

commercial AI firm in deciding whether to compete for a DoD contract.[550]  They also varied in

experience with the DoD.  The spectrum of experience with the DoD was selected for the

interview sample: firms that had never considered working with the DoD; firms that attempted but have been unable to secure a DoD contract; firms that perform some DoD contracts that make up a small proportion of their business; dual-use firms with roughly equal commercial and defence portfolios; and defence contractors whose primary, if not only, customer is the DoD. This range of experience presented a cross-section of diversity found in the survey population. Firms from Silicon Valley, Seattle, Boston, New York, Austin, Denver, and Washington were interviewed to understand how the location of the company may affect the perception of working with the DoD. The type of technology sold by the firms was also considered in selecting the sample. Firms that provide enterprise autonomous hardware systems, machine learning data analytics, cybersecurity, natural language processing, and computer vision applications are included in the sample. While each of these areas of AI are relevant and sought after by the DoD, some have greater utility than others in the commercial market. Within the commercial market, some AI applications are intended for business-to-business sales while others are aimed at consumers. The interviewees represent vertical and horizontal AI firms. Vertical AI firms solve a very specific customer need while horizontal AI firms can more readily adjust to different use cases.[551]

---

[545] Bloomberg and Volpe (n 498) 186.

[546] Ibid.

[547] See ibid 187.

[548] Andrew S Bowne, 'Innovations: Making the Pentagon an Even More Attractive Customer for AI Upstarts' (February 2021) *Contract Management* 76, 78.

[549] Ibid.

[550] Ibid.

The number of interviews conducted was determined based on several factors. Scheduling, conducting, transcribing, and analysing interviews is extremely labour intensive and time consuming, so the number of interviews was limited by practical considerations and the researcher's finite resources. The number was not predetermined. The researcher originally scheduled eight interviews, and then added an additional seven to gather additional insights and

perspectives.  Although the researcher had access to many other potential interviewees, the decision to end the interview phase was made once saturation of the themes was reached.[552] Over the course of fifteen interviews, it became clear that several themes predominated the discussion about the motivations, preferences, opinions, and perceptions of commercial AI firms about the DoD as a potential customer, detailed in Chapter IV.  Figure 13 reflects anonymised descriptions of the fifteen interviewees.

*Figure 13: Interviewees*

| Size (Number of Employees) | Funding | Percentage of Business from DoD Contracts (Number/Value) |
|---|---|---|
| 101-1000 | Early Stage | 30/30 |
| 1-100 | Founders | 20/20 |
| 1-100 | Early Stage | 30/60 |
| 1-100 | Founders | 0/0 |
| 101-1000 | Late Stage | 0/0 |
| 1-100 | Late Stage | 10/30 |
| 1001+ | Public | 10/10 |
| 1-100 | Founders | 10/20 |
| 1-100 | Founders | 10/10 |
| 1-100 | Public | 100/100 |
| 1-100 | Late Stage | 30/30 |
| 1-100 | Early Stage | 20/20 |
| 1-100 | Late Stage | 10/10 |
| 1-100 | Founders | 0/0 |
| 1001+ | Public | 10/10 |

To ensure anonymity of all interviewees, the size of the firms is given in wide ranges and the location of the firms is not included in the table.  The interviewees come from the main sources of AI technology in the United States.  The geographic breakdown of the sample includes six from the Silicon Valley/San Francisco area; three from Washington, DC; two from Boston; and one each from New York, Austin, Denver, and Seattle.  The position descriptions are also presented separately to ensure anonymity.  The interviewees were all senior employees within

---

[552] Creswell and Creswell (n 157) 186, explaining 'saturation,' the approach of collecting data until fresh data no longer sparks new insights or reveals new properties, is a concept employed in grounded theory research to determine an adequate sample size.



their firms with experience and knowledge in how their firms make business decisions. Eleven

of the interviewees were C-Suite executives, of which seven were Chief Executive Officers. The

remaining four interviewees held senior manager or director level positions.

### F *Data Collection: Methods, Analysis, and Synthesis*

After providing demographic data, the survey participants were asked to give their

overall assessment of the DoD as a potential customer currently and how that assessment has

changed in the past year. The participants then rated statements about various issues related to

the perceived attractiveness of contract attributes and practices and their opinions on the

importance of common contract issues on a Likert scale.[553] This method provides insight into

the commercial AI industry's opinions, perceptions, and preferences on the DoD as a customer.

The survey consisted of affirmative statements that participants were able to assess their

agreement with the statement. Ten questions presented contract attributes and business

motivations that participants were asked to rate in terms of importance on a five-point Likert

scale. These questions addressed the views of the participants on concepts of communication,

collaboration, intellectual property, negotiation, timeline, flexibility, commercial marketability,

and contribution to national security.

Following these questions, the participants were asked to indicate their agreement with

42 affirmative statements on a seven-point Likert scale. These statements were intentionally mixed in terms of positive, negative, and neutral perception of the DoD as a customer to reduce the likelihood of bias and repeated responses. An example of a positively worded statement is: 'The [DoD's] contracting process is transparent.' Agreement with this statement indicates a positive perception of the DoD while disagreement indicates a negative perception of the DoD. Of the 42 statements, 26 were positively worded. An example of a negatively worded statement is: 'My company prefers the commercial contracting process to the [defence] contracting process.' Agreement with this statement indicates a negative perception of the DoD while disagreement indicates a positive perception of the DoD relative to the commercial contracting process. Seven negatively worded statements were mixed with the positively worded statements. Additionally, nine neutral statements were mixed in with the rest of the statements. An example of a neutral statement is: 'Our company prefers to submit proposed solutions in writing over direct, face to face interactions, such as pitches.' This statement tests preferences of commercial AI firms on buying practices that exist in both defence and commercial contracts, so agreement or disagreement is neutral as to the perception of the DoD. However, the degree of agreement provides insight into the preferences of evaluation and award tools that are available under existing law to DoD contracting officers that may attract commercial AI firms. Finally, the last three questions were open-ended with potential responses also given. These questions asked participants to identify factors that influence their firm's decision to not engage with the DoD; contract attributes that would prevent their firm from working with the DoD; and factors that influence their firm's decision to engage with the DoD.

---

[553] A Likert scale survey question asks participants to indicate their agreement or disagreement with a statement by scoring their response along a range from 'strongly agree' to 'strongly disagree,' with each response option on the scale associated with a numerical score: Ruel, Wagner and Gillespie (n 502) 59. In this study, a 5-point scale (testing opinions on the importance of various contract attributes) and a 7-point scale (testing opinions, preferences, and perceptions of customer attractiveness generally and specifically as related to the DoD as a customer) were used. The 7-point numerical scores were: 1, strongly agree; 2, agree; 3, slightly agree; 4, neither agree nor disagree; 5, slightly disagree; 6, disagree; 7, strongly disagree. Thus, an average score of 3 indicates the sample slightly agreed with the statement. Statements were intentionally mixed, with agreement of one statement indicating a positive association with the DoD and agreement of another statement indicating a negative association with the DoD.

The statements were written with the intent to gauge the perceptions of commercial AI firms regarding a variety of factors relevant to the research question. Factors identified in social exchange theory, such as collaboration, communication, culture, profit, expected success,



autonomy, and community approval were examined to test the applicability of customer attractiveness in the context of commercial AI firms viewing the DoD as a customer. The opinions regarding specific contract attributes that are common in either FAR contracts or OT agreements were examined to understand what attributes attracted firms and which ones were problematic. These questions provided insight into preferred contracting practices and allowed the researcher to analyse the alignment of current law and practice with those preferences as discussed in Chapters IV and V. Participants were also asked about their opinions on the commercial market compared to the defence market in terms of customer attractiveness and contract law and practice. The print version of the survey questions is provided in Appendix C.

After the surveys were complete, the researcher analysed the data using SurveyMonkey, Excel, and Tableau to derive greater insight. The findings are discussed in Chapter IV. The findings were used to refine the conceptual framework, research questions, and develop a theory that helps to explain why commercial AI firms choose to engage with the DoD to support national security. The findings also lead to refining the interview questions. To get rich qualitative data from the interviewees, the interviews were semi-structured which allowed the interviewees to present their own ideas about what affects their perception of the DoD as a customer and why their firms decide to work with the DoD. Interviews began with a confirmation of consent and started with a question about the business calculus of their firm when deciding whether to compete for a business opportunity. This question led to discussions about the primary drivers of a commercial AI firm in their business decisions and what they look for in a contract and customer. Follow-up questions were asked depending on the responses leading to open-ended discussions driven primarily by the interviewee. This interview style reduced the risk of including researcher bias in the findings, although it is impossible to



eliminate the researcher's views from interviews altogether.[554]  Interview questions were asked to gather perceptions, opinions, and attitudes of the purposeful sample with the goal to gain a better understanding of the following topics related to the research question:

1. Perception of attributes of the DoD as a customer

2. Factors affecting business decision to compete for DoD contract

3. Advantages of DoD contracts in both traditional procurement contracts and OT agreements

4. Disadvantages of DoD contracts in both traditional procurement contracts and OT agreements

5. Comparison of DoD to other customers

6. Factors that can be changed to make the DoD a more attractive customer to commercial AI firms

Using verbatim transcriptions and MaxQDA software, the interviews were analysed, and codes were developed in an iterative process throughout the data collection phase.  In the grounded theory tradition, codes were developed with the goal of understanding the meaning of the data and its potential for theorising.[555]  Thematic codes were first developed prior to the initial interview based on the survey findings and evolved as more data was collected that explained the survey findings.  After analysis of the survey findings, the researcher began the interview phase with five themes, including: attraction to the DoD; barriers to entry in the defence market; engagement practices within the DoD procurement system; ethics; intellectual property.  During the interviews, the researcher added one additional theme to include problems within the DoD's own organisational bureaucracy as a determinate of its relationship with the

commercial AI industry.  The emergent codes and sub-codes are discussed in detail in Chapter IV.

<div align="center">G <em>Issues of Trustworthiness</em></div>

Because this research focuses on the perceptions of business leaders within the commercial AI industry, this research is limited to what the participants believe are right and true.  However, as noted earlier, perceptions are not factual, and are limited by the participants own personal experience and worldview.  Because the survey and interviews asked for perceptive, opinions, and preferences regarding specific contract attributes unique to the DoD, a lack of experience or understanding of DoD contracts could affect the trustworthiness of the data on specific questions.  There are several issues that can impact the validity of the data that are possible in this research.  While efforts were made to recruit the population of commercial AI firms that sell defence-relevant technology, the extent of the target population is unknown and subjective measures were applied in identifying members, which may have excluded members. The respondents and participants, all of whom volunteered, may not be a representative sample of the commercial AI industry.  Although the demographics indicate a rich cross-section of the population within the sample, sampling bias and overfitting may exist in the quantitative data. Sampling bias refers to research that draws conclusions from a set of data that is not representative of the population you are trying to understand.[556]  It is common when participants volunteer for a study.  Overfitting is a problem where the creation of a model that is overly tailored to the data found in the model and not representative of the general trend.[557]  Mindful that both fallacies are difficult to detect, these findings are representative only to the sample and

are not necessarily representative of the general population of defence-relevant commercial AI firms.

Recognising these potential pitfalls, efforts were made to address issues of credibility, dependability, confirmability, and transferability.[558] To confirm the participant's perceptions were reflected correctly in the researcher's portrayal of them, the researcher would paraphrase within the interview and seek clarification.[559] This immediate confirmation was audio-recorded and transcribed. The perceptions are presented with descriptions to ensure readers can adequately understand the research.[560] Triangulation of data sources, inherent in the mixed methods research conducted, also improves the credibility. The researcher was careful to explain their employment by the DoD and personal experience to each participant in both the survey and again prior to the interview. While it was explained that the DoD did not sponsor or direct this research, the connection almost certainly had some impact on the process, the participants' comfort in discussing their perceptions of the DoD, and the researcher's decisions. To ensure dependability of data, survey data and transcripts are preserved, and the processes and procedures employed by the researcher to answer the research question are detailed in this dissertation.[561] To ensure confirmability, the researcher presents the rationale for all methodological, theoretical, and analytic choices.[562] To ensure transferability, the researcher outlines the rationale for sampling so readers can form their own opinions of the quality of research and relevance of the researcher's interpretations.[563]

---

[558] See Bloomberg and Volpe (n 498) 202–3.
[559] See ibid 203.
[560] See ibid.
[561] See ibid 204.
[562] See ibid 205.
[563] See ibid.



## H *Limitations and Delimitations*

There are several limitations inherent in mixed methods research that are present in this study.[564] Because there is no known definitive population of commercial AI firms that sell defence relevant AI-enabled capabilities, it is impossible to know whether the research sample is representative of the overall industry. Generalisation of this data to a larger or different population is therefore problematic. However, efforts were made to identify every firm listed across multiple databases that identified as a United States-based commercial AI firm that appeared to provide defence relevant technology. Each firm that appeared to meet these criteria was invited to participate in this study.

Delimiting factors include the choice of research problem, research questions, time, population selection, methods of data collection and analysis, theoretical perspectives, and alternative theories that were not adopted.[565] These factors impact this research — perhaps significantly.[566] The sample selected was based on the interest generated by the survey and explanations of the research purpose within the known target population. The number of interviews conducted was limited to fifteen as data saturation was achieved, though additional interviews may have led to insight that was not obtained in this research.

## I *Conclusion*

The choices of methodology in conducting this research are all tied to gaining a better understanding of the research question. The explanatory sequential design allowed the researcher to obtain unprecedented insight into perceptions of the commercial AI industry. This design resulted in the collection of a large dataset of quantitative data with empirical findings

---

[564] See ibid 207.
[565] See ibid 207–8.
[566] See ibid.



and an explanation through perspective data obtained through in-depth interviews.  The findings,

analysis, and synthesis of the surveys and interviews are detailed in the following chapter.





A *Introduction*

This chapter reports the findings of the survey and interviews conducted to collect data on the perceptions, opinions, and preferences of commercial AI firms. Through developing an understanding of the business and technical considerations made by business leaders at commercial AI firms, the DoD can evaluate the alignment of its contract law framework with buying AI systems. The data collection process aimed at testing hypotheses generated from the literature on the unique attributes of AI, it's nexus to national security, and the two government contract law frameworks. Theory and concepts from social exchange literature were used to analyse the data and develop an explanatory theory of how contract law attributes affect customer attractiveness.

This chapter presents analysis of the findings and concludes with a synthesis that integrates the findings from the survey, interviews, and literature. The following chapter provides a synthesis of the findings, explaining how contract law in the DoD can serve to attract commercial AI firms, setting up the theoretical contributions and recommendations for future research.

B *Survey Findings and Analysis*

The intent of the survey was to identify what factors influence the decision of commercial AI firms on whether to contract with the DoD. As discussed in Chapter III, the survey was deliberately scoped to explore possible explanations of why commercial AI firms contract or decide to not contract with the DoD. Additionally, the survey was designed to uncover the prevalence of preferences or perceptions held by commercial AI firms about the DoD as a



customer.  Accordingly, while many of the survey questions yielded notable findings that contribute to understanding what contract attributes are perceived as valuable to commercial AI firms and how the contract law can affect the attractiveness of the DoD, other questions provide minimal insight.[567]  The discussion below describes the results of the survey that advance the understanding of the research question.  The full survey results are provided in Appendix D.

The percentages of respondents choosing a particular response, the mean response score,[568] and the standard deviation[569] are reported.  The mean is an average measure of central tendency and indicates the average response score of the sample.[570]  The standard deviation is a measure of the amount of dispersion about the mean; this measure indicates whether there is variation within the sample on a specific variable.[571]  With the frequency of responses (indicated in percentile of the sample), mean (the average ordinal score of the Likert scale), and standard deviation (the measure of agreement or disagreement of the participants within a particular survey question), the survey results can be interpreted.  The results of the survey are presented below.[572]

---

[567] For example, several questions called for opinions on specific contract attributes that required the respondent to possess familiarity and understanding of some of the DoD-specific terminology and processes.  Review of the survey responses and the demographics of the respondents indicated that such questions yielded potentially misleading or confusing results.  16.22% of the respondents never competed for a defence contract, and 23.42% of the sample never performed any kind of defence contract, and an additional 26.13% of the sample indicated that contracts with the DoD amounted to only about 10% of their business.  With nearly a quarter of the sample having no experience in performing a defence contract, survey questions that required specialised knowledge of the DoD contracting process were categorically removed from the analysis because of concerns about the validity of the question.  These questions included calls for opinions on cost accounting standards, request for proposals, prime contractors, government point of entry, dispute resolution, and the difference between FAR contracts and OT agreements.  Many of these questions had a majority of responses indicating a neutral opinion ('somewhat agree', 'somewhat disagree', or 'neither agree nor disagree').  This type of response was highly unusual for questions that did not require specialised knowledge of the DoD contracting process.  For example, the survey asked for respondents to indicate their agreement with the statement 'The DoD is more fair and timely in resolving disputes and issues than in the commercial marketplace.'  This calls for experience in both commercial and defence markets on dispute resolution.  The responses indicated a neutral opinion from the sample (49.54% stated they 'neither agreed nor disagreed' with an additional 15.6% indication either slight agreement or disagreement).  Another example is the statement 'if given a choice, my company would prefer a FAR contract over an Other Transaction Agreement.'  While this question is directly related to the research question, the requirement for specialised knowledge of the difference between the two types of contracts lacks foundation and thus, potentially lacks validity.  Nearly half (47.27%) of the respondents indicated they neither agreed nor disagreed with the statement.  While 21.82% of respondents indicated they strongly disagreed with the statement (a total of 39.09% indicated some level of disagreement with the statement compared to only 13.64% indicating some level of agreement), it is tempting to draw the conclusion that commercial AI firms are more likely to prefer OT agreements to FAR contracts; however, it is unknown whether the respondents, with limited knowledge of either contract possessed the understanding required to provide a valid response.  As such, findings of these questions are excluded from this analysis and included in the complete survey results presented in Appendix D.  The interviews permitted a better assessment of the interviewees' experience and understanding of these DoD-unique terms; thus, conclusions were informed by interviews.

[568] The mean is defined as the arithmetic centre of a variable's distribution.  Ruel, Wagner and Gillespie (n 502) 240.

[569] The standard deviation is the measurement of variation about the mean; it is calculated as the square root of the variance, which is calculated by subtracting the mean of the sample distribution from the mean of each sample, summing the resulting differences over the total number of samples, and dividing by the total number of samples drawn minus one: at 131–2.

[570] Ibid 226.

## 1 *General Opinion of the DoD as a Customer*

The first component of the survey asked commercial AI firms to assess their opinion of the DoD as a customer or potential customer, acknowledging that some participants have never worked with the DoD before completing this survey.  Figure 14 shows the opinions the respondents have of the DoD as a customer.

---

[571] See ibid.

[572] As discussed in Chapter III, the survey was administered to a non-probabilistic sample.  Accordingly, while demographic data was collected, comparing two or more populations (such as traditional defence contractors compared to non-traditional defence contractors, or start-ups compared with public corporations) was avoided.  The dilution of the sample would negatively impact generalization and reliability.  Thus, the findings and analysis of the survey responses reflect the entire sample.  However, as discussed in Chapter III, the demographic data and the differences within the data was considered in selecting interviewees from the sample.



*Figure 14: General Opinion of the DoD as a Customer*

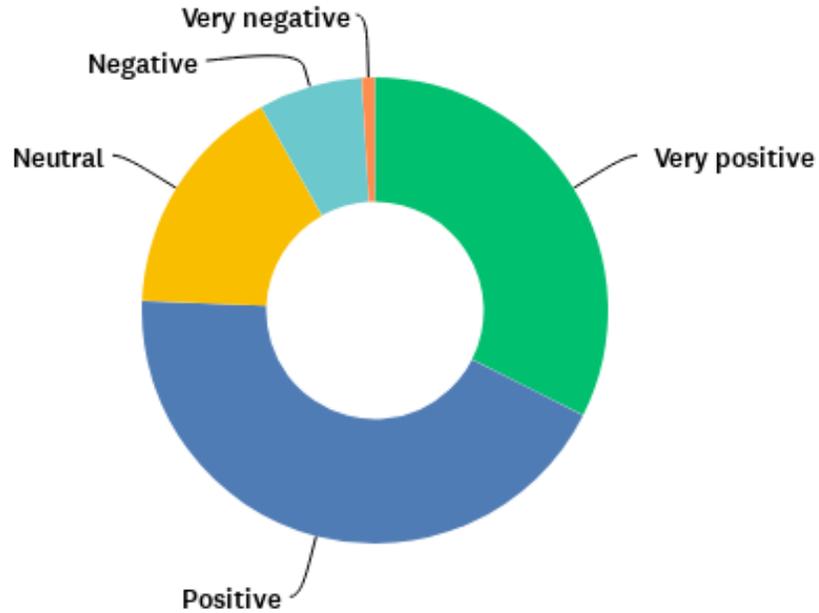

| Question | Very positive (1) | Positive (2) | Neutral (3) | Negative (4) | Very negative (5) | Mean | Standard Deviation |
|---|---|---|---|---|---|---|---|
| How would you describe your overall opinion of the DoD as a (potential) customer? | 32.43% | 43.24 | 16.22 | 7.21 | 0.90 | 2.01 | 0.93 |

Approximately three-quarters of the sample expressed either a very positive or positive (32.43% and 43.24%, respectively) opinion of the DoD as a customer or potential customer. In contrast, very few had negative opinions of the DoD (7.21% had a negative opinion and only one out of the 111 survey participants, 0.9%, had a very negative opinion of the DoD). This finding indicates that commercial AI firms tend to view the DoD as an attractive customer. The following sections help explain this general opinion and reveal a paradox — although commercial AI firms generally view the DoD as an attractive customer, DoD contract law and procurement practice is generally viewed negatively.



## 2 *Importance of Contract Attributes*

The second category of survey questions assessed the importance of contract attributes — the presence or absence of a contract term or procurement practice which would affect the attractiveness of a contract opportunity — that commercial AI firms may place on contracts and customer relationships. The survey instructed participants to 'rate the level of importance your company places on the following factors.' The choice of responses was provided on a five-point Likert scale. The available responses consisted of 'Extremely important,' 'Very important,' 'Somewhat important,' 'Not so important,' and 'Not at all important.' These responses are coded as 1 through 5, respectively, with averages closer to either pole indicating stronger opinions on the relative importance of a factor, whereas averages closer to 3 indicate either lack of consensus (if the standard deviation is wider) or neutral opinion (if the standard deviation is narrower). Figures 15-20 show the survey responses indicating the perceived importance of contract attributes, presented in order from most to least important based on the mean responses.



*Figure 15: Importance of collaboration with the customer about the problem*

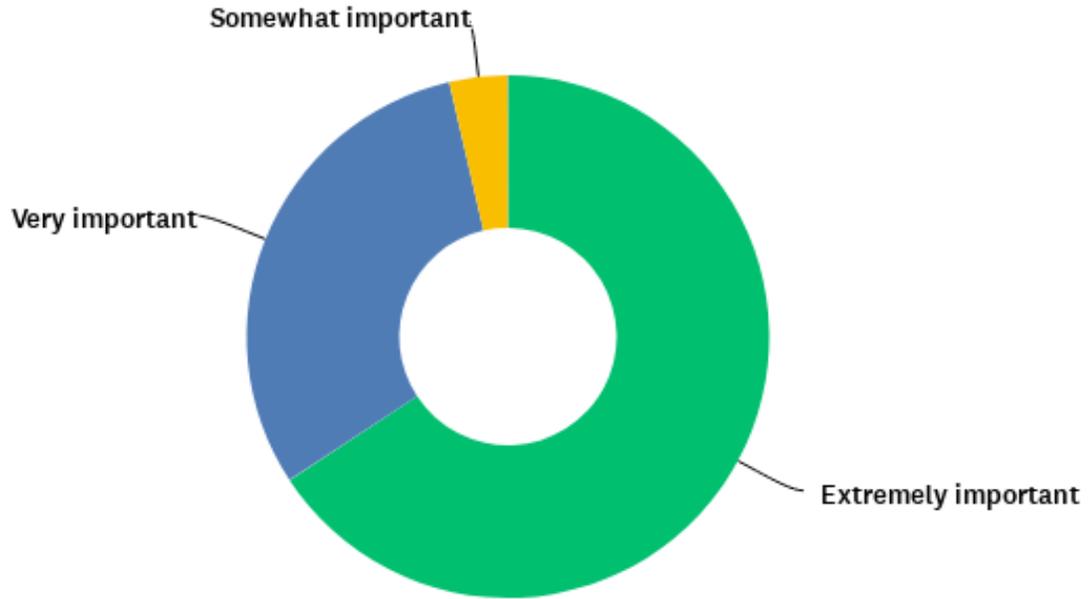

| Question | Extremely Important (1) | Very Important (2) | Somewhat Important (3) | Not so Important (4) | Not at all Important (5) | Mean | Standard Deviation |
|---|---|---|---|---|---|---|---|
| Ability to collaborate with the customer about its problem/need and potential solutions before contract award | 65.77 | 30.63 | 3.6 | 0 | 0 | 1.38 | 0.55 |

The respondents were unanimous in their opinion that collaboration with the customer about defining the problem and identifying potential solutions before contract award was important (65.77% stated collaboration was extremely important, 30.63% responded it was very important, while 3.6% admitted it was at least somewhat important). This finding corroborates the literature from all three topics discussed in Chapter II.[573] The AI development literature explained that AI applications require intimate knowledge of the problem and the data.[574] The

developer needs to understand the problem as well as the data that will train the model — this information requires the developer to work alongside the end-user.[575]  The contract law literature identifies a fissure between the two legal frameworks with respect to collaboration between the commercial AI firm and DoD customer.[576]  In FAR competitions, the problem to be solved is stated as a requirement, where the customer instructs the contractors how to perform.[577]  In competitions for OT agreements, the bidding contractors offer solutions to the customer's problems and are encouraged to communicate with end-users to understand the problem and discuss approaches to solving it.[578]  This opinion is consistent with a fundamental concept of social exchange theory, that a collaborative process, which requires mutual trust and commitment towards a common goal, is critical to the success of a buyer-supplier relationship.[579]  The hypothesis that commercial AI firms prefer collaborative relationships when developing the solution is supported by this finding, and corroborates the social exchange theory claim that buyers need to enhance their pro-relationship behaviour vis-à-vis suppliers, especially in innovative endeavours.[580]

---

[575] Ibid.

[576] *Other Transaction Guide* (n 104) 8, explaining the most important part of the planning activities is defining the problem to clearly articulate to offerors. Cf 48 CFR § 11.002, requiring acquisition officials state requirements in terms of functions to be performed, performance required, or essential physical characteristics, presuming the agency already knows the solution and how to perform the task and describe the product required to solve the problem.

[577] 48 CFR § 11.002.

[578] See *Other Transaction Guide* (n 104) 16–7.

[579] See Hüttinger, Schiele and Veldman (n 469) 1197.

[580] Blonska (n 485) 74.



*Figure 16: Importance of communication and collaboration with the end-user*

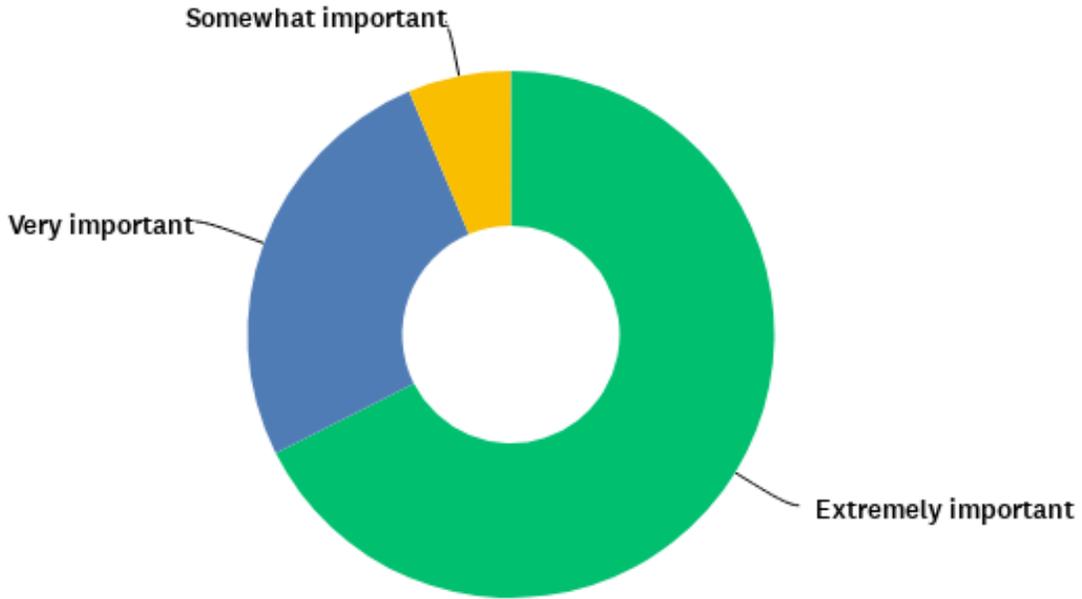

| Question | Extremely Important (1) | Very Important (2) | Somewhat Important (3) | Not so Important (4) | Not at all Important (5) | Mean | Standard Deviation |
|---|---|---|---|---|---|---|---|
| The ability to communicate and collaborate with the end-user of the product/service during contract performance | 67.57 | 26.13 | 6.31 | 0 | 0 | 1.39 | 0.6 |

Like the previous question about the importance of collaboration, all the respondents stated the ability to communicate with the end-user during contract performance is an important contract attribute. Over two-thirds of the sample indicated communication and collaboration was 'extremely important.' Just as the sample found collaboration before contract award to be important, the same finding applies to collaboration during contract performance. As described above, this finding is consistent with the literature on AI development as it is critical for the developer to understand the problem, the nature of the relevant data required to describe, predict,



or recommend an outcome, and how the AI system will be used.[581]  Additionally, this finding is consistent with social exchange theory that highlights the importance of building trust through shared goals to develop a successful relationship.[582]  Thus, the survey data supports the hypothesis that commercial AI firms prefer the open-ended ability to communicate and collaborate with the end-user.

*Figure 17: Importance of swift contract award and funding*

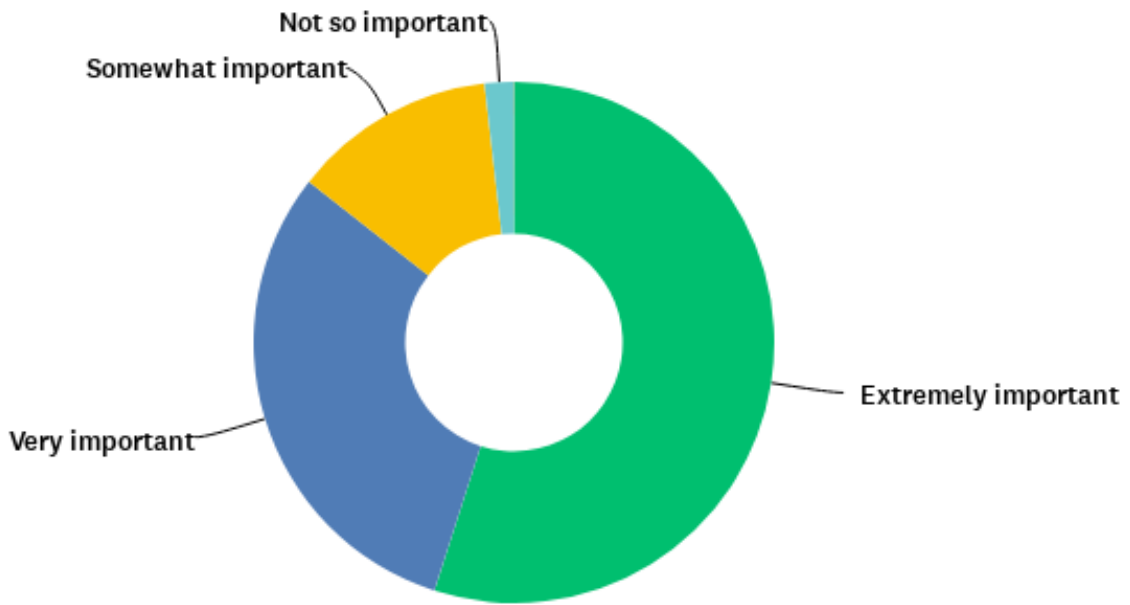

| Question | Extremely Important (1) | Very Important (2) | Somewhat Important (3) | Not so Important (4) | Not at all Important (5) | Mean | Standard Deviation |
|---|---|---|---|---|---|---|---|
| Timeline of a contract award and funding | 54.95 | 30.63 | 12.61 | 1.8 | 0 | 1.61 | 0.77 |

The timeliness of contract award and funding was viewed as important by all but 1.8% of the sample.  Most respondents stated the timeline for funding was 'extremely important' with an additional 30.63% finding it 'very important.'  The literature on the national security context

demonstrated the timeliness of acquiring and fielding new capabilities is a major concern for the DoD as China can acquire AI systems much faster.[583]  Additionally, advancements in the state of the art in AI occur extremely quickly and many commercial AI firms are used to rapid funding from commercial customers.  The survey data supports the hypothesis that timely contract award and funding are important contract attributes to commercial AI firms.[584]

Timeliness of contract award can affect the perceived attractiveness of the DoD as a customer, as the DoD is competing with commercial customers for AI capabilities.  Social exchange theory posits that in competitive markets, the buyer must make the relationship advantageous to the supplier.[585]  Thus, the DoD's ability to award contracts as fast as commercial buyers may affect the relative attractiveness of the DoD as a customer.  The different contract law frameworks in the DoD can affect the timeliness of contract award.  Although OT agreements can take time to negotiate terms and conditions, the streamlined solicitation and competition processes can mitigate the impact of time spent in crafting the contract.  The FAR competition process is typically longer due to CICA and greater exposure to protests; however, the negotiations typically focus on bargaining for the price and schedule of performance,[586] while most of the clauses are pulled directly from the FAR and DFARS.[587]  Thus, understanding the value of negotiation to commercial AI firms can provide insight into whether the trade-off to time is worthwhile.  Figure 18 demonstrates that negotiation, specifically on the IP rights, is an important contract attribute to commercial AI firms.

---

[583] See *NSCAI Final Report* (n 4) 61–2.
[584] See above Chapter II(G).
[585] See Glas (n 465) 97.
[586] 48 CFR § 15.405(b), stating the 'contracting officer's primary concern is the overall price the Government will actually pay'.
[587] See 48 CFR Part 52 and *DFARS* Part 252.



*Figure 18: Importance of negotiating intellectual property rights*

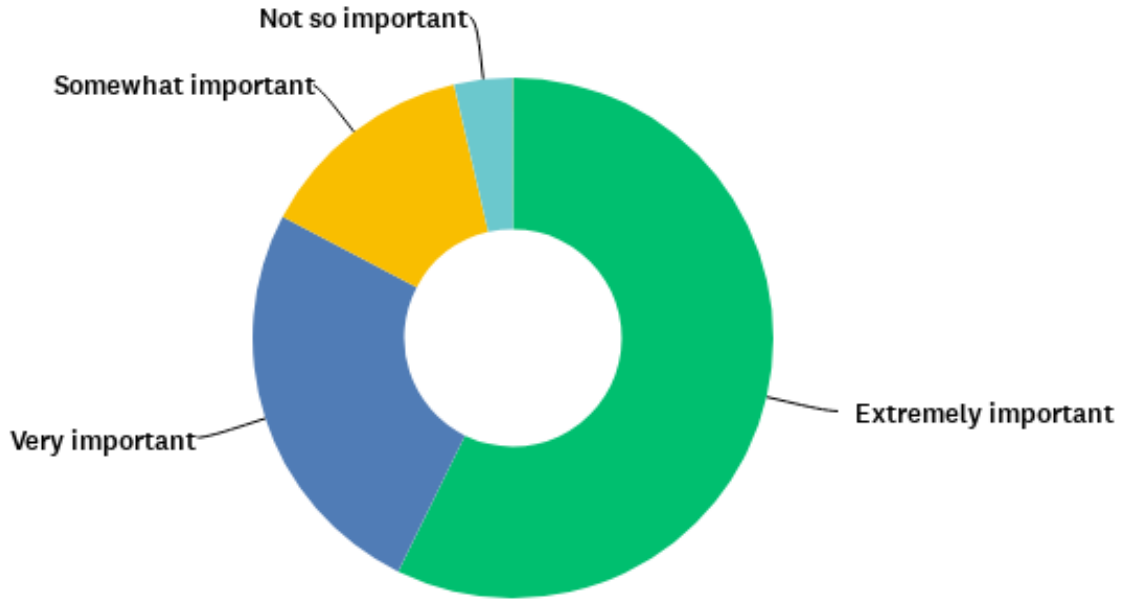

| Question | Extremely Important (1) | Very Important (2) | Somewhat Important (3) | Not so Important (4) | Not at all Important (5) | Mean | Standard Deviation |
|---|---|---|---|---|---|---|---|
| Ability to negotiate intellectual property rights | 57.27 | 25.45 | 13.64 | 3.64 | 0 | 1.64 | 0.85 |

Most (57.27%) of the sample stated the ability to negotiate IP rights is extremely important. All but 3.64% of the respondents believed negotiating IP rights is at least somewhat important (25.45% indicated it is 'very important' and 13.64% states it is 'somewhat important'). This finding supports the hypothesis that commercial AI firms prefer negotiating IP terms and are attracted to opportunities that ensure adequate protections of their innovation. The ability to negotiate terms aligns with the social exchange theory research that demonstrates relationships founded on trust and fairness are attractive.[588]

---

[588] Lambe, Wittman and Spekman (n 14) 16.



The development of AI systems makes IP negotiation especially important as the success of an application depends upon the data used to train the model. This construct challenges the DFARS paradigm for assigning data rights, as the question of when software is developed changes with each training iteration.[589] Thus, the ability to negotiate IP rights afforded by OT authority can attract commercial AI firms who appear to strongly value this contract attribute.

*Figure 19: Importance of flexibility to iterate*

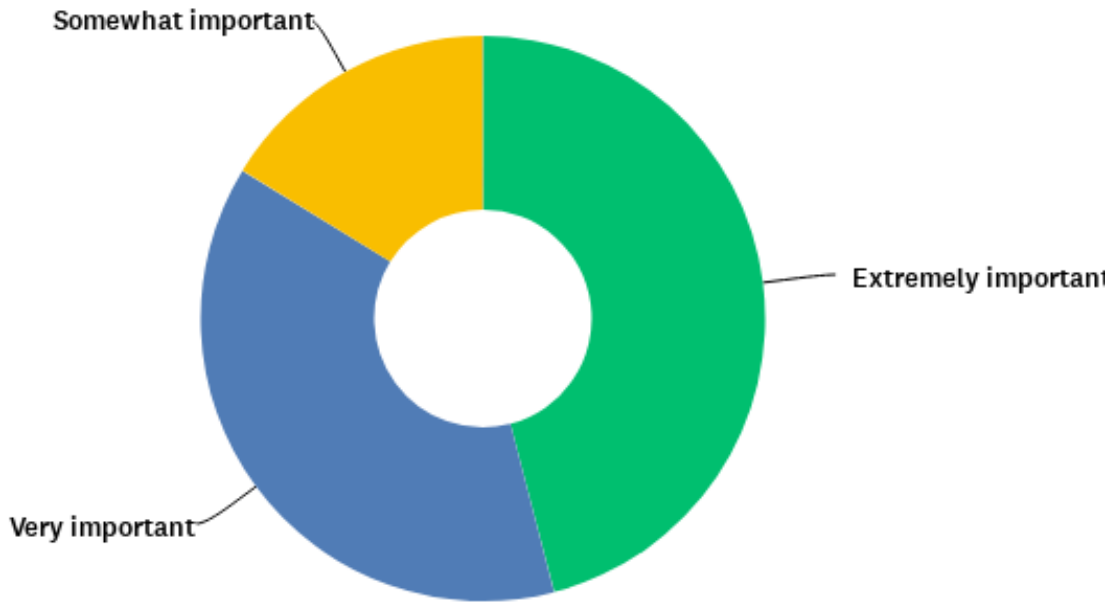

| Question | Extremely Important (1) | Very Important (2) | Somewhat Important (3) | Not so Important (4) | Not at all Important (5) | Mean | Standard Deviation |
|---|---|---|---|---|---|---|---|
| The flexibility to iterate the product/solution during the performance of the contract | 45.95 | 37.84 | 16.22 | 0 | 0 | 1.7 | 0.73 |

All respondents indicated that the flexibility to develop a solution through iteration was at least somewhat important (45.95% stated flexibility to iterate is 'extremely important', while 37.

---

[589] See Bowne and McMartin (n 155) 9.



84% stated it is 'very important' and the remaining 16.22% stated it is 'somewhat important').

The survey data supports the AI literature that suggests iteration, experimentation, and

prototyping are critical to the develop of AI systems.[590]  This finding supports the hypothesis that

commercial AI firms prefer flexibility in contracts that permit experimentation and iteration in

developing AI systems.[591]  The choice of law can greatly impact the flexibility to iterate during

contract performance.  The flexibility to develop AI systems through experimentation is limited

in FAR contracts, unless such iterations were drafted into the scope of the contract.[592]  Given the

FAR contracts start with a defined requirement and contain detailed performance specifications,

such flexibility is unlikely.[593]  OT agreements, however, are inherently flexible, and the OT

Guide directs the prototype project is developed in an iterative process.[594]

---

[590] See Malone, Rus and Laubacher (n 185) 6.
[591] See above Chapter II(G).
[592] 48 CFR § 43.201.
[593] 48 CFR § 2.101; 48 CFR § 11.002.
[594] See *Other Transaction Guide* (n 104) 49.



*Figure 20: Importance of contributing to national security*

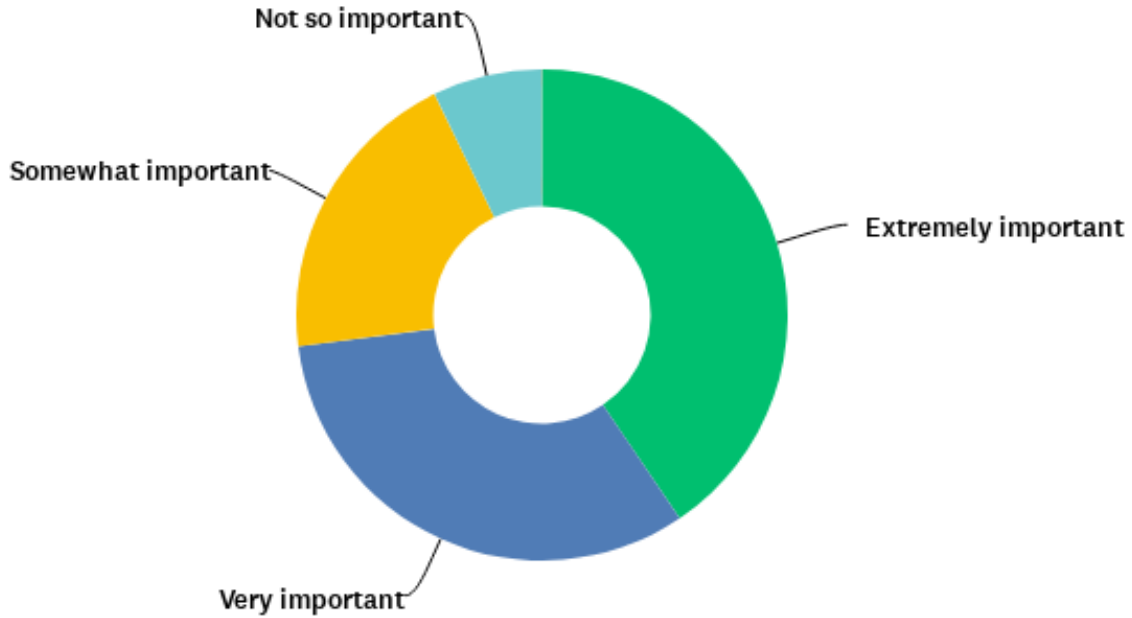

| Question | Extremely Important (1) | Very Important (2) | Somewhat Important (3) | Not so Important (4) | Not at all Important (5) | Mean | Standard Deviation |
|---|---|---|---|---|---|---|---|
| Contribution to national security | 40.54 | 32.43 | 19.82 | 7.21 | 0 | 1.94 | 0.94 |

Although several respondents thought that contributing to national security was 'not so important,' nearly three-quarters of the sample stated that such contributions were 'extremely important' (40.54%) or 'very important' (32.43%) to their firm's decision to contract with a customer. This opinion reveals the old defence innovation paradigm from the space race and Cold War experience where Silicon Valley firms believed advancing the state of the art in technology will support national security and earn them profits may persist.[595] Accordingly, this finding calls into question the validity of the claims that commercial AI firms are unwilling to support the DoD.

---

[595] Brose (n 174) 45.



Collectively, these findings indicate that contract law and practice that focus on the relationship through collaboration, communication, negotiation, and are considerate of the business concerns such as timeliness, IP protection, and flexibility to iterate on a prototype solution are likely to be viewed as attractive to commercial AI firms.

## 3 *Perceptions, Opinions, and Preferences*

The third component of the survey asked participants to assess the degree of agreement with affirmative statements regarding the perceptions, opinions, and preferences about choosing projects and customers, offering a means of comparing the DoD with the commercial customers. Responses were based on a seven-point Likert scale, with available selections consisting of 'Strongly agree,' 'Agree,' 'Somewhat agree,' 'Neither agree nor disagree,' 'Somewhat disagree,' 'Disagree,' and 'Strongly disagree.' These responses are coded as 1 through 7, respectively, with averages closer to either pole indicating stronger perceptions relating to the statement, whereas averages closer to 4 indicating indicate either a lack of consensus (if the standard deviation is wider) or neutral opinion (if the standard deviation is narrower).

The types of statements presented in the survey are grouped by themes. While there is overlap of the themes with some statements fitting in two or more thematic groups, the themes surveyed include perceptions from commercial AI firms regarding: potential barriers to entry in the defence market; a comparison between the DoD and commercial market using concepts from social exchange theory; and preferences on contract law and procurement practice. The survey data is presented below.

### a. *Barriers to Entry*

The survey responses in Figures 21 and 22 present the sample's perception of whether the DoD's procurement process presents barriers to entry — challenges to accessing the defence



market, competing for, and winning a contract. The analysis of both questions follows Figure 22.

*Figure 21: The DoD's procurement process is simple and straightforward*

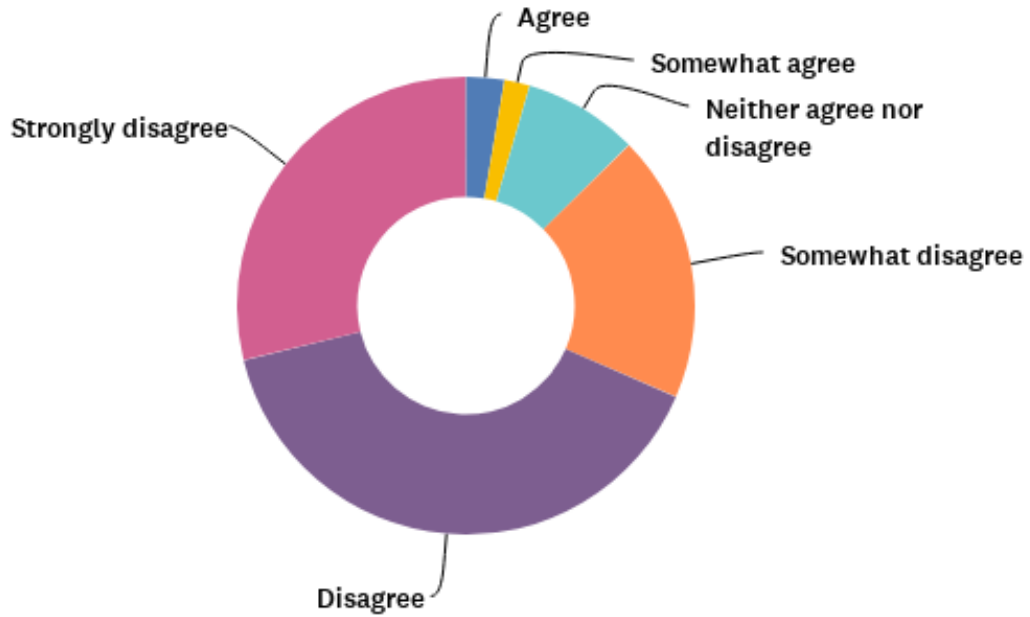

| Question | Strongly agree (1) | Agree (2) | Somewhat agree (3) | Neither agree nor disagree (4) | Somewhat disagree (5) | Disagree (6) | Strongly disagree (7) | Mean | Standard Deviation |
|---|---|---|---|---|---|---|---|---|---|
| The process of competing for and performing a contract with the Department of Defense is simple and straightforward. | 0 | 2.7 | 1.8 | 8.11 | 18.92 | 39.64 | 28.83 | 5.77 | 1.16 |



*Figure 22: The barriers to entry are low*

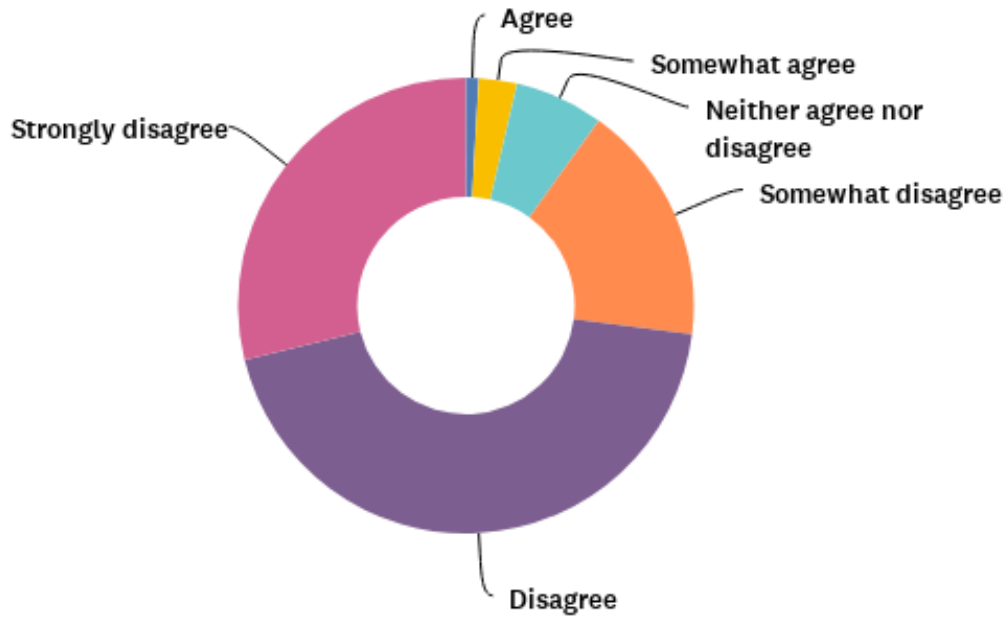

| Question | Strongly agree (1) | Agree (2) | Somewhat agree (3) | Neither agree nor disagree (4) | Somewhat disagree (5) | Disagree (6) | Strongly disagree (7) | Mean | Standard Deviation |
|---|---|---|---|---|---|---|---|---|---|
| The barriers to entry in the government defence marketplace are low and easily overcome. | 0 | 0.9 | 2.7 | 6.31 | 17.21 | 44.14 | 28.83 | 5.87 | 1.04 |

The findings for these questions demonstrate the sample did not agree that the DoD's contract process was simple and straightforward. Nor did the sample agree that the barriers to entry into the defence market are low and easily overcome. Very few respondents 'agreed' or 'somewhat agreed' with the statements (only 4.5% agreed or somewhat agreed that the contract process was simple and straightforward while only 3.6% agreed or somewhat agreed the barriers to entry are low and easily overcome). These findings lead to the inference that commercial AI firms perceive securing a contract with the DoD is challenging. If this inference is true, it is likely that some commercial AI firms will avoid competing for a DoD contract, thus limiting the



DoD's access to commercial innovation. Social exchange theory suggests that buyers must maximise their attractiveness by ensuring the suppliers perceive the relationship to be more advantageous than the alternative.[596] Therefore, the DoD should simplify the contract process and lower barriers to entry to increase its relative attractiveness. Although the hurdles creating these high barriers to entry were designed to further legitimate government interests, such as preventing fraud and waste,[597] these barriers may limit competition from new entrants if they perceive the barriers too high to reasonably expect earning a contract.[598] OT authority permits the DoD to mimic commercial business practices, streamline procurement, excludes regulations, and institutes a preference towards awarding agreements to non-traditional defence contractors.[599]

    b. *Comparison with Commercial Market Using Social Exchange Theory Concepts*

    A social exchange concept known as 'comparison of alternatives' refers to a phenomenon in which the more attractive option is more likely to result in a buyer-seller relationship than the less attractive option.[600] Drivers of attraction include economic, resource and social factors: familiarity, similarity, compatibility, knowledge of alternatives, expected value, trust, communication and relational fit.[601] As commercial AI firms, unlike vendors for defence-specific technologies, have access to a commercial market, the DoD represents an alternative.

---

[596] See Hüttinger, Schiele and Veldman (n 469) 1197–8.
[597] See Government Accountability Office, *Contract Management: DOD Vulnerabilities to Contracting Fraud, Waste, and Abuse* (Report No GAO-06-838R, 7 July 2006), reporting that the DoD's use of new contracting techniques, such as interagency acquisitions or multi-award contracts, and lack of oversight were causes of fraud, waste, and abuse, leading to increased audit authority and oversight to slow down the procurement process.
[598] See Government Accountability Office, *DOD is Taking Steps to Address Challenges Faced by Certain Companies* (Report No GAO-17-644, 20 July 2017) 8, reporting that while the DoD's acquisition environment is driven by laws that provide transparency, fairness, and socioeconomic goals, companies expressed frustration to GAO with the complexity of DoD's acquisition process, specifically the time, cost, and risk associated with competing for and performing a contract.
[599] See 10 USC § 4022(d).
[600] See Hüttinger, Schiele and Schröer (n 476) 700.
[601] See ibid.



The survey responses in Figures 23-27 present the sample's perception of the DoD as a customer using concepts from social exchange theory research.

*Figure 23: Competing for a contract with the DoD is more time consuming and difficult than it is with a commercial customer*

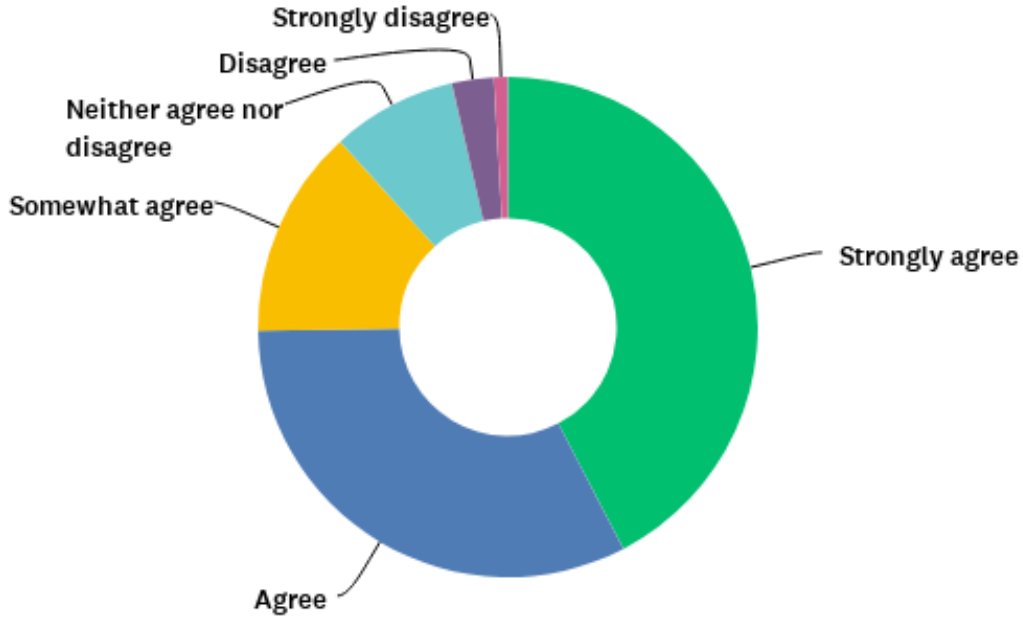

| Question | Strongly agree (1) | Agree (2) | Somewhat agree (3) | Neither agree nor disagree (4) | Somewhat disagree (5) | Disagree (6) | Strongly disagree (7) | Mean | SD |
|---|---|---|---|---|---|---|---|---|---|
| Competing for a contract with the Department of Defense is more time consuming and difficult than it is with a commercial customer. | 42.34 | 32.43 | 13.51 | 8.11 | 0 | 2.7 | 0.9 | 2.03 | 1.24 |

Unfortunately for the DoD, most respondents believe that competing for a contract with the DoD is more time consuming and difficult than it is with a commercial customer (only 3.6% of the sample disagreed or strongly disagreed with this statement). This finding substantiates the NSCAI's fear that the DoD will encounter challenges in attracting commercial AI firms that may



decide doing business with the DoD is too difficult.[602]  Despite the decades of acquisition reform

efforts from FASA[603] to the Section 809 Panel[604] that sought to make contracting with the DoD

easier for commercial companies, this finding indicates that the perception of commercial AI

firms is that it is still difficult.  However, OT competition procedures are streamlined and can

mirror business practices.  Without CICA and various other regulations that govern FAR contract

competitions, OT agreements can appear less difficult to commercial AI firms.  Social exchange

theory holds that buyers must maximise their perceived attractiveness compared to alternatives,

thus, streamlining the competition process can better attract suppliers.

---

[602] *NSCAI Final Report* (n 4) 7.
[603] Pegnato (n 309) 39.
[604] *Acquisition Regulation Report Vol 3* (n 238) 1.



*Figure 24: Performing a contract with the DoD is more costly than a commercial contract*

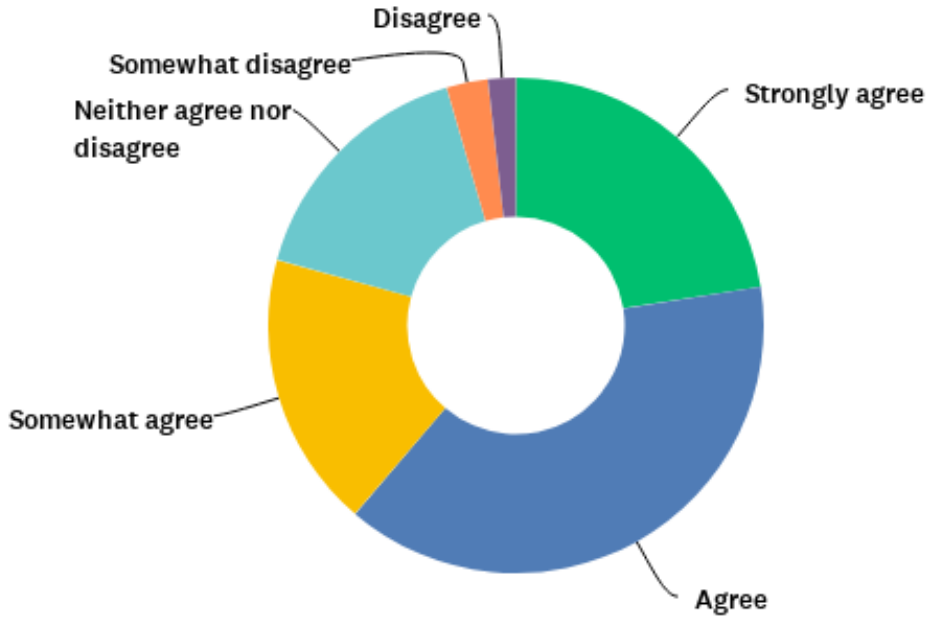

| Question | Strongly agree (1) | Agree (2) | Somewhat agree (3) | Neither agree nor disagree (4) | Somewhat disagree (5) | Disagree (6) | Strongly disagree (7) | Mean | SD |
|---|---|---|---|---|---|---|---|---|---|
| All else being equal, performing a contract with the Department of Defense is more costly than a contract in the commercial marketplace. | 22.52 | 38.74 | 18.02 | 16.22 | 2.7 | 1.8 | 0 | 2.43 | 1.18 |

Most respondents believe performing a DoD contract is more costly than in the

commercial marketplace (61.26% of respondents agree or strongly agree with the statement,

totalling 79.28% when adding those that somewhat agree). As social exchange theory posits

suppliers choose between alternative buyers based on evaluating the expected value and

experience derived from the relationship,[605] the DoD should ensure that contract compliance

---

[605] See Glas (n 465) 100.



costs are minimal.  Studies have found that compliance with the FAR requirements can add

significant costs borne by the contractor.[606]  Many of those regulatory compliance costs, such as

the accounting systems, do not apply to OT agreements unless included through negotiation.[607]

*Figure 25: Performing a contract with the DoD is more difficult than with a commercial customer*

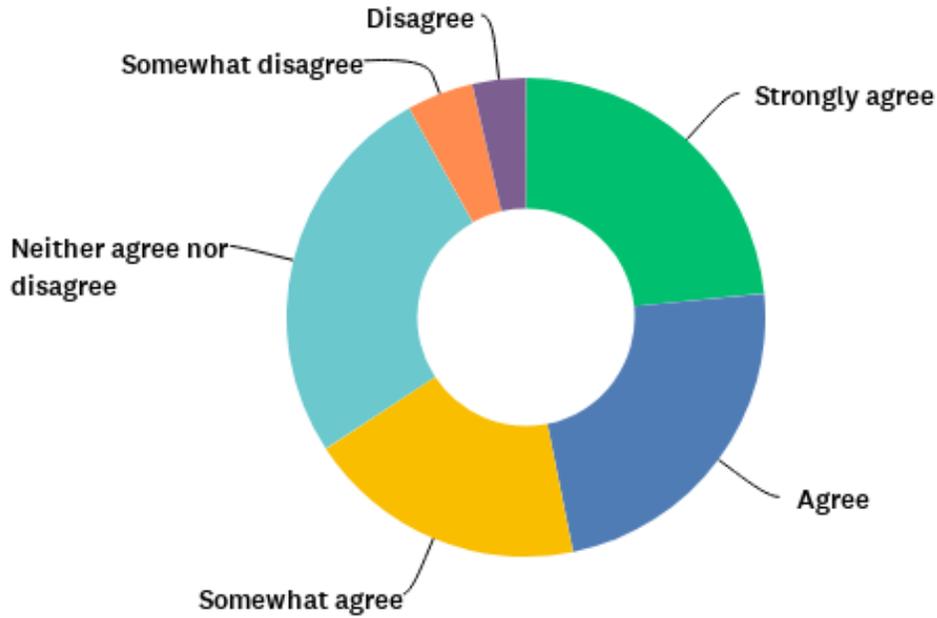

| Question | Strongly agree (1) | Agree (2) | Somewhat agree (3) | Neither agree nor disagree (4) | Somewhat disagree (5) | Disagree (6) | Strongly disagree (7) | Mean | SD |
|---|---|---|---|---|---|---|---|---|---|
| Performing a contract with the Department of Defense is more difficult than it is with a commercial customer. | 23.42 | 23.42 | 18.92 | 26.13 | 4.5 | 3.6 | 0 | 2.76 | 1.37 |

Although only 46.84% of the respondents strongly agreed or agreed that performing a

DoD contract is more difficult than in the commercial marketplace, only 3.6% disagreed (no one

strongly disagreed).  Nonetheless, when considering potential customers, many commercial AI firms perceive the DoD to be more difficult to work with than a commercial customer.

*Figure 26: Preference for commercial contracting process to defence contracting process*

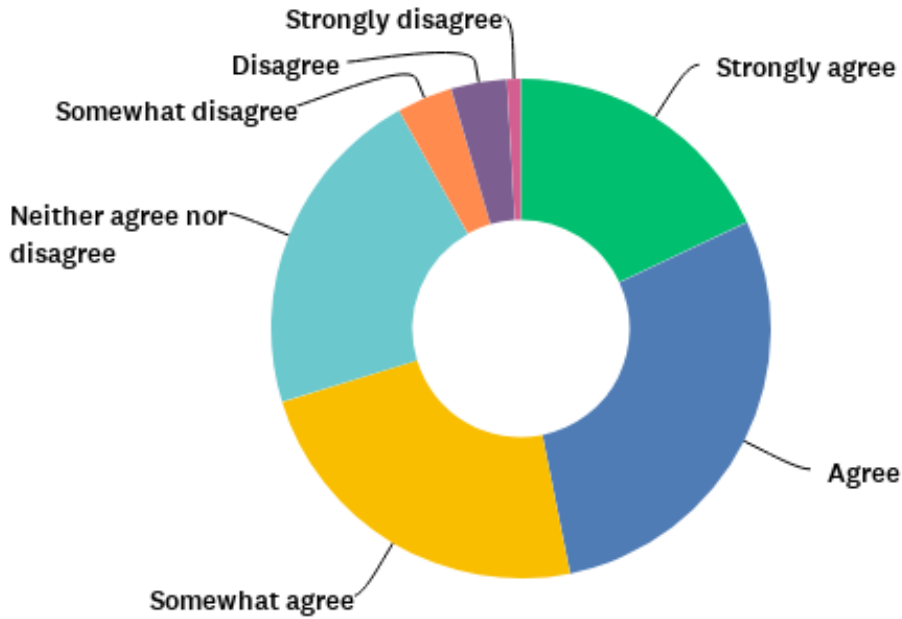

| Question | Strongly agree (1) | Agree (2) | Somewhat agree (3) | Neither agree nor disagree (4) | Somewhat disagree (5) | Disagree (6) | Strongly disagree (7) | Mean | SD |
|---|---|---|---|---|---|---|---|---|---|
| My company prefers the commercial contracting process to the defence contracting process. | 18.02 | 28.83 | 23.42 | 21.62 | 3.6 | 3.6 | 0.9 | 2.78 | 1.34 |

Most respondents prefer the commercial contracting process to the defence contracting process (70.27% indicated some level of agreement with 46.85% stated they strongly agreed or agreed and 23.42% that somewhat agreed), while only 8.1% of respondents indicated some level of disagreement with the statement.  This finding indicates very few prefer the defence contracting process to the commercial practice.  As discussed above, the DoD is unlike commercial buyers and have numerous procurement objections that necessitate some departure



from the commercial contracting process.[608]  However, this finding underscores the concern that DoD's procurement objectives may impede its effort to attract commercial AI firms.[609]  This is an acute challenge in the context of buying AI innovation when commercial vendors have many alternative buyers that conduct business in a relatively more attractive manner,[610] and the ability of the DoD to attract and leverage cutting-edge technologies is critical to its national security.[611]

*Figure 27: Prototyping is a precursor to integration and scaling new AI use cases*

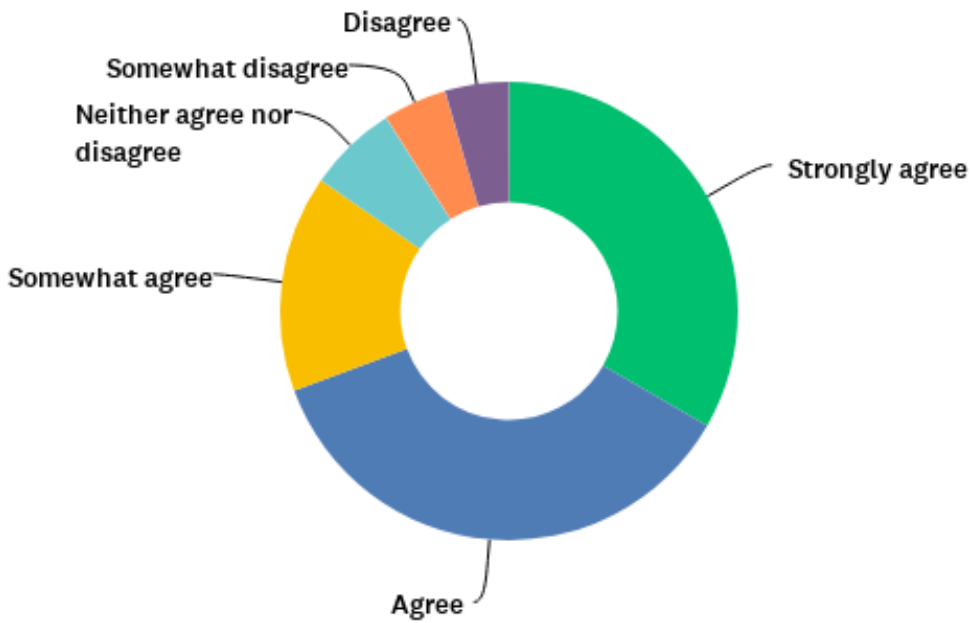

| Question | Strongly agree (1) | Agree (2) | Somewhat agree (3) | Neither agree nor disagree (4) | Somewhat disagree (5) | Disagree (6) | Strongly disagree (7) | Mean | SD |
|---|---|---|---|---|---|---|---|---|---|
| Starting with a prototype, pilot, minimum viable product, or proof of concept is an essential step before integrating or scaling any new use case. | 33.33 | 36.04 | 15.32 | 6.31 | 4.5 | 4.5 | 0 | 2.26 | 1.34 |

Most respondents believe that starting with a prototype, pilot, minimum viable product, or proof of concept is an essential step before integrating or scaling any new use case (33.33% strongly agree; 36.04% agree; 15.32% somewhat agree). This finding supports the claim that developing or integrating an AI capability meets the statutory requirement to use an OT agreement.[612] Moreover, it indicates that flexibility to iterate and experiment during contract performance is a practical advantage for AI development. Thus, FAR contracts may prove too restrictive for integrating and scaling a new AI application. Finally, social exchange theory underscores the importance of compatibility in relationship,[613] so contract law that aligns with the technical model for development is likely more attractive than an incompatible legal framework.

The data collected from this theme demonstrate commercial AI firms view the DoD contracting process as challenging and appear to prefer the commercial contracting practice. Social exchange theory provides many factors that can predict the attractiveness of a customer; factors that a supplier perceives as contributing to value creation by interacting with the potential customer, such as profit, innovation development, or ease of process, increase attractiveness.[614] That there appears to be a consensus among commercial AI firms that competing for a contract with the DoD is more difficult and time consuming than with a commercial customer, that performing a contract with the DoD is more difficult and costly than with a commercial customer, and very few companies prefer the DoD contracting process over the commercial process indicates the DoD is not optimising its attractiveness to commercial AI firms. If commercial AI firms do not see the value of competing for a DoD contract compared to alternative commercial opportunities, the DoD will struggle to leverage commercial AI innovation. Unlike industrial-era manufacturing where the DoD could readily obtain supplies it needed as the largest, and often only, customer, whether the DoD is able to attract innovative AI firms to supply cutting-edge AI systems will impact how the DoD projects power in near future conflict.[615]

---

[612] 10 USC § 4022(a).
[613] Hüttinger, Schiele and Schröer (n 476) 700.
[614] Hüttinger, Schiele and Veldman (n 469) 1197.

        c. *Ethical Use of AI*

The survey also sought to understand the perception of commercial AI firms on the role ethical concerns play in forming opinions about the DoD as a customer. While the *Project Maven* case study suggests the ethical concerns relating to providing AI-enabled capabilities to the battlefield can be divisive, the literature review indicated that incident may not be illustrative of the commercial AI industry's opinions on supplying the DoD with AI technologies. The survey responses in Figures 28 - 30 present the sample's responses to statements concerning the ethics in the context of AI development and deployment.

---

[615] See above nn 76–88 and accompanying text.



*Figure 28: Importance of ethics in developing and leveraging AI capabilities*

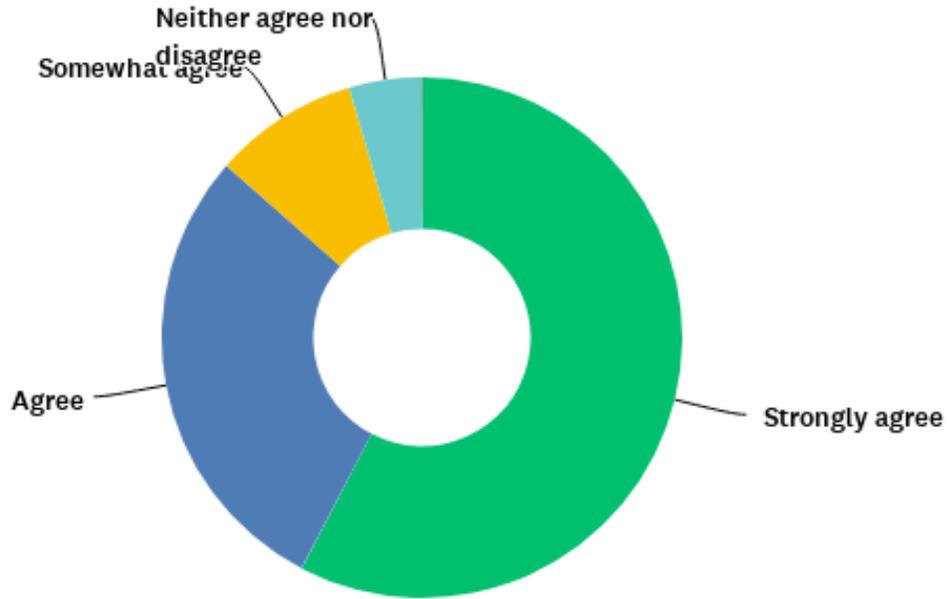

| Question | Strongly agree (1) | Agree (2) | Somewhat agree (3) | Neither agree nor disagree (4) | Somewhat disagree (5) | Disagree (6) | Strongly disagree (7) | Mean | SD |
|---|---|---|---|---|---|---|---|---|---|
| Ethics in developing and leveraging artificial intelligence capabilities is important to my company. | 57.66 | 28.83 | 9.01 | 4.5 | 0 | 0 | 0 | 1.6 | 0.83 |

No respondent disagreed with the statement 'ethics in developing and leveraging AI capabilities is important to my company.' While this finding makes clear that ethics is a concept that is important in commercial AI development, the limitation of survey questions is the lack of context provided. Thus, little can be inferred by this question alone. Nonetheless, the question is explored further in the interviews and covered in the analysis of the interviews later in this chapter.

The analysis for Figures 29 and 30 follows Figure 30.



*Figure 29: Trust in the DoD to use AI ethically*

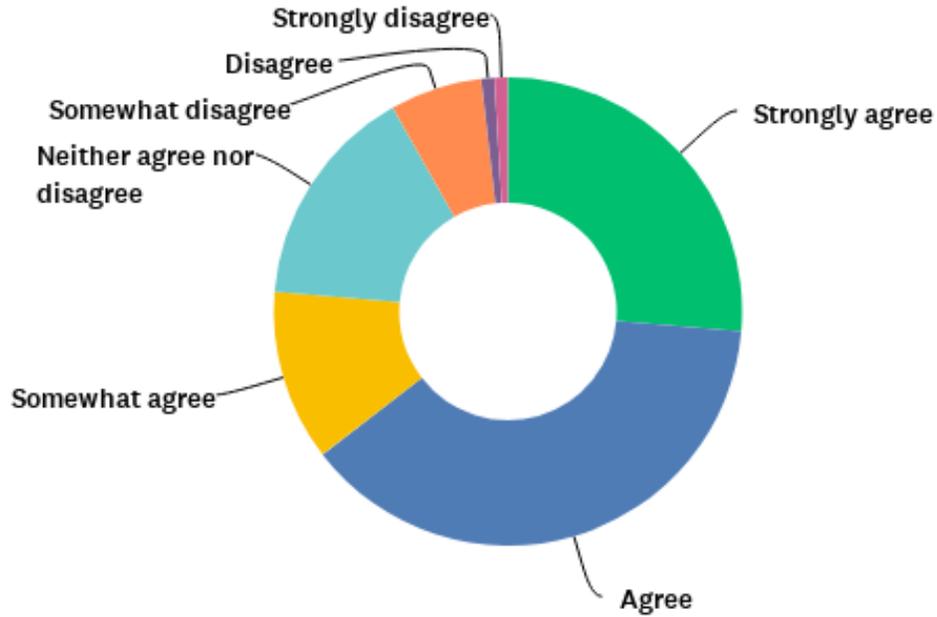

| Question | Strongly agree (1) | Agree (2) | Somewhat agree (3) | Neither agree nor disagree (4) | Somewhat disagree (5) | Disagree (6) | Strongly disagree (7) | Mean | SD |
|---|---|---|---|---|---|---|---|---|---|
| Our company trusts the Department of Defense to use artificial intelligence ethically. | 26.36 | 38.18 | 11.82 | 15.45 | 6.36 | 0.91 | 0.91 | 2.44 | 1.32 |



*Figure 30: Comfort with the DoD using their AI technology for lethal purposes*

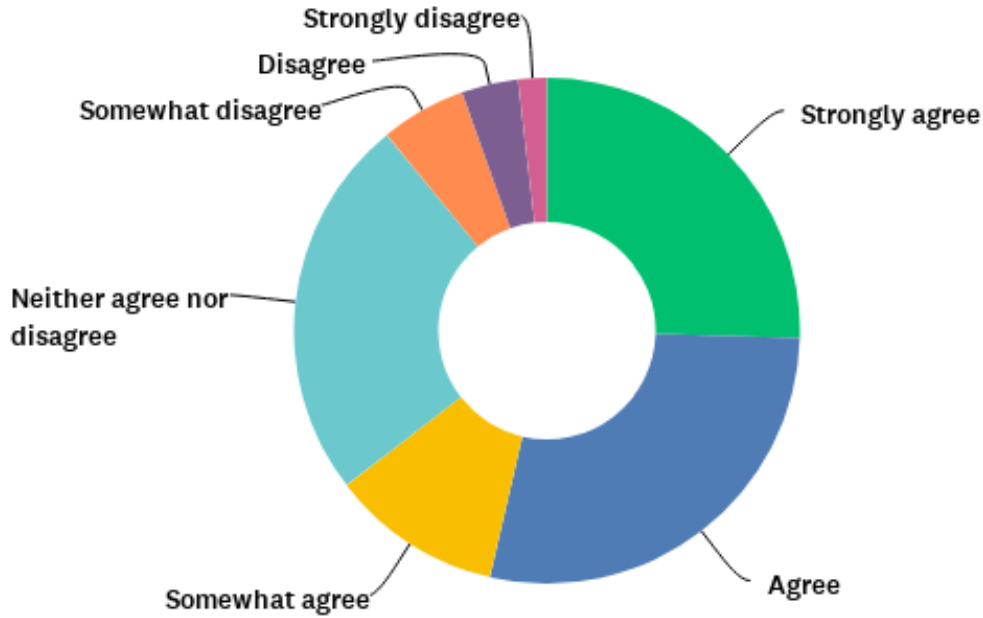

| Question | Strongly agree (1) | Agree (2) | Somewhat agree (3) | Neither agree nor disagree (4) | Somewhat disagree (5) | Disagree (6) | Strongly disagree (7) | Mean | SD |
|---|---|---|---|---|---|---|---|---|---|
| My company is comfortable that the Department of Defense may use our product/service for lethal purposes. | 25.45 | 28.18 | 10.91 | 24.55 | 5.45 | 3.64 | 1.82 | 2.75 | 1.52 |

Most of the sample (64.54% strongly agreed or agreed as opposed to only 1.82% that disagreed or strongly disagreed) trusts the DoD to use AI ethically. While the level of trust commercial AI firms place in the DoD decreased when using AI for lethal purposes, nearly 65% of participants (included those that strongly agreed, agreed, or somewhat agreed) indicated they were comfortable with the DoD using AI for lethal purposes (compared with 11% of the sample that indicated they somewhat disagreed, disagreed, or strongly disagreed). Overall, these findings are positive for the DoD. Additionally, the stated concerns that ultimately resulted with Google ending its participation in *Project Maven* are not necessarily representative of all



commercial AI firms.  However, as stated above, these findings lack context — the term 'ethics' is abstract and subjective interpretations of the concept can impact the respondent's choice.  How the DoD uses AI towards lethal effect may positively or negatively impact the respondents' opinions.  The respondents may or may not be familiar with current DoD policy recognises that decisions about the potential use of lethal force require appropriate human control.[616]  While these nuances are more thoroughly discussed in the interview analysis, the literature, survey, and interviews indicate that the ethical concerns regarding military use of AI appear less impactful to a commercial AI firm's decision to contract with the DoD than concerns about the difficulty and cost of performing a contract.

      d.   *Perceptions of the DoD as a Customer*

The next theme examined the perceptions of the DoD as a customer.  How commercial AI firms perceive work with the DoD will affect their business goals is important to gaining an understanding of what motivates firms to engage with the DoD or stay clear of defence contracts. The survey responses in Figures 31 - 35 present the sample's perceptions of the DoD as a customer.

---

[616] DoD Directive 3000.09 (n 19) 14; see Ryseff et al (n 81) 18–9, reporting that a survey of technologists from Silicon Valley, the defence industrial base, and top computer schools indicated a significantly higher level of comfort with the military using AI supervised by soldiers than fully autonomous systems.



*Figure 31: The DoD is an attractive customer*

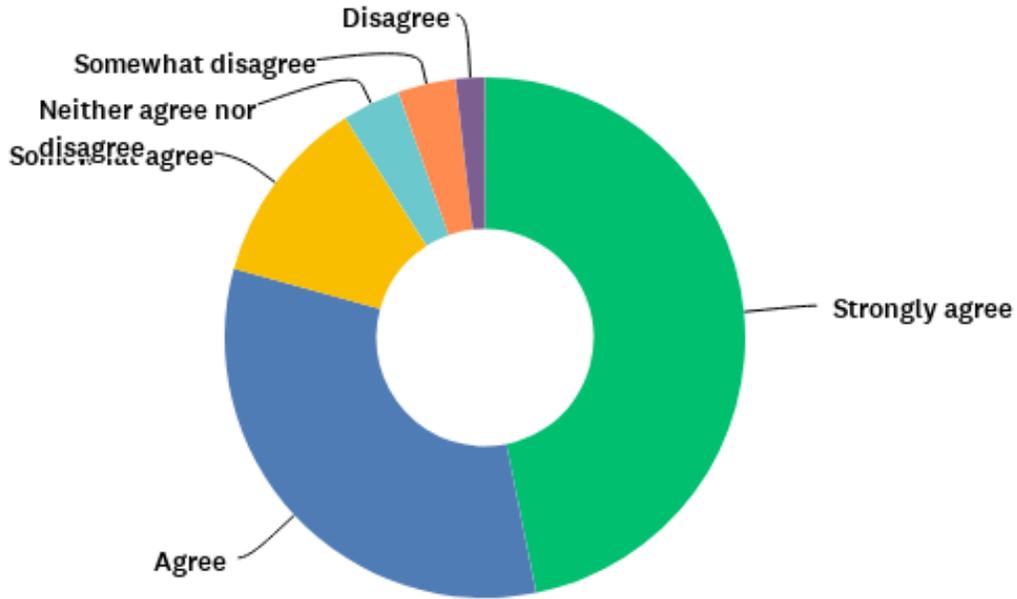

| Question | Strongly agree (1) | Agree (2) | Somewhat agree (3) | Neither agree nor disagree (4) | Somewhat disagree (5) | Disagree (6) | Strongly disagree (7) | Mean | SD |
|---|---|---|---|---|---|---|---|---|---|
| The Department of Defense is an attractive customer to my company. | 46.85 | 32.43 | 11.71 | 3.6 | 3.6 | 1.8 | 0 | 1.9 | 1.15 |

Most respondents view the DoD as an attractive customer (79.28% strongly agree or agree compared to only 1.8% that disagreed). This finding corroborates the results from Figure 14, above, that indicated the sample held a generally positive opinion of the DoD. While this finding is positive for the DoD, the other findings indicate the DoD could improve its attractiveness to commercial AI firms further if it improved how such firms perceive its unique contract law and practice.



*Figure 32: The DoD's contracting process is transparent*

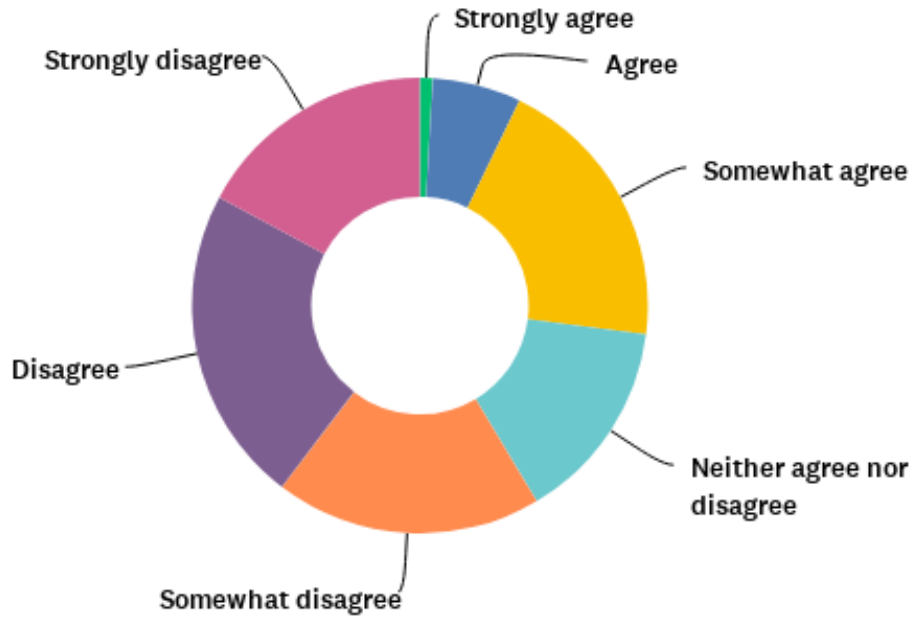

| Question | Strongly agree (1) | Agree (2) | Somewhat agree (3) | Neither agree nor disagree (4) | Somewhat disagree (5) | Disagree (6) | Strongly disagree (7) | Mean | SD |
|---|---|---|---|---|---|---|---|---|---|
| The Department of Defense's contracting process is transparent. | 0.9 | 6.31 | 19.82 | 14.41 | 18.92 | 22.52 | 17.12 | 4.8 | 1.59 |

Despite transparency being a goal of public procurement and the rationale for full and open competition and bid protest law,[617] commercial AI firms do not appear to perceive the DoD's contracting process as transparent. While most respondents were relatively neutral (53.15% responded they somewhat agreed, neither agreed nor disagreed, or somewhat disagreed with the statement), 39.64% disagreed or strongly disagreed compared to 7.21% who agreed or strongly agreed. This finding indicates that commercial AI firms are more likely to view the DoD's contracting process as opaque rather than transparent. Such perception can affect the

DoD's attractiveness, as social exchange theory research demonstrates that trust and commitment to a common goal are precursors to successful relationships.[618]

*Figure 33: The DoD uses easy to understand terms and conditions*

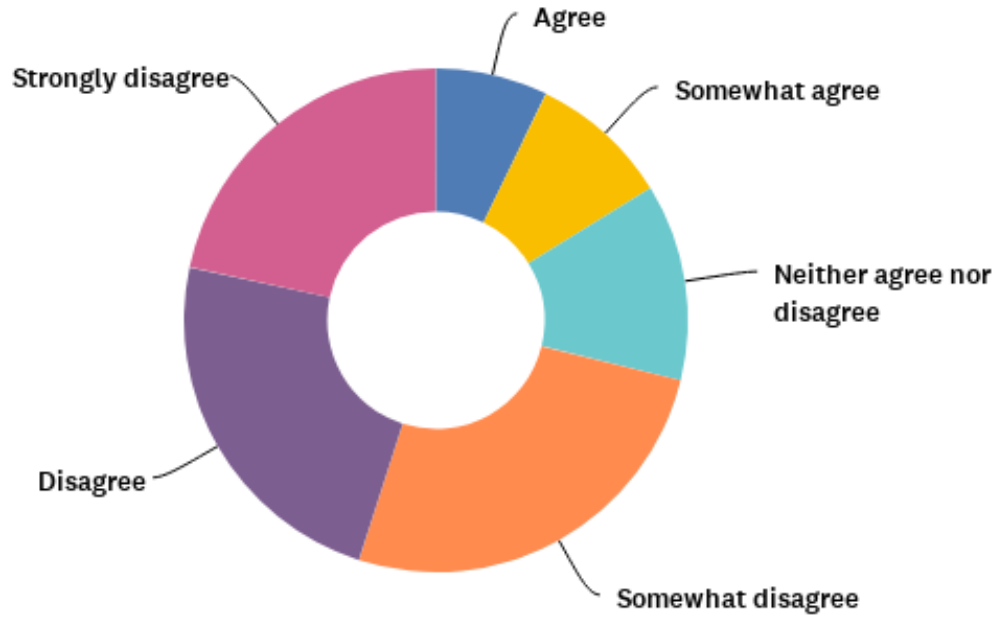

| Question | Strongly agree (1) | Agree (2) | Somewhat agree (3) | Neither agree nor disagree (4) | Somewhat disagree (5) | Disagree (6) | Strongly disagree (7) | Mean | SD |
|---|---|---|---|---|---|---|---|---|---|
| The Department of Defense uses easy to understand terms and conditions. | 0 | 7.21 | 9.01 | 12.61 | 26.13 | 23.42 | 21.62 | 5.14 | 1.49 |

More respondents disagreed that the DoD uses easy to understand terms and conditions than those that agreed (71.17% indicated some level of disagreement while only 16.22% responded that they agreed or somewhat agreed with the statement). As many commercial AI firms are unaccustomed with the DoD's contract law, the many unique clauses required by the FAR and DFARS may contribute to this finding. Thus, negotiating terms and conditions can

help commercial AI firms work with the DoD in a manner that is easier to understand.  Because mutual trust in the buyer-supplier relationship is important to successful endeavours,[619] the DoD can leverage OT authority to align terms and conditions with commercial contract practices.

*Figure 34: Concern that contracting with the DoD will harm business*

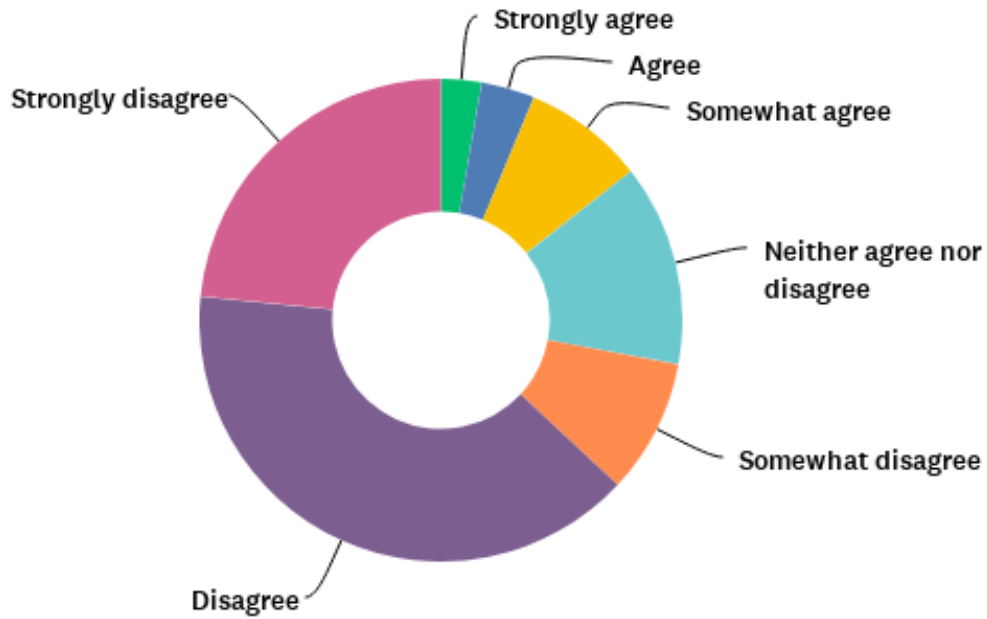

| Question | Strongly agree (1) | Agree (2) | Somewhat agree (3) | Neither agree nor disagree (4) | Somewhat disagree (5) | Disagree (6) | Strongly disagree (7) | Mean | SD |
|---|---|---|---|---|---|---|---|---|---|
| My company is concerned that contracting with the Department of Defense may negatively impact our business. | 2.7 | 3.6 | 8.11 | 13.51 | 9.01 | 39.64 | 23.42 | 5.35 | 1.56 |

According to social exchange theory research, how a prospective customer affects value creation, including profit, volume, innovation development, and market access factors into a supplier's perception of attraction.[620]  Fortunately for the DoD, far more respondents disagreed than agreed with the assertation that their company is concerned that contracting with the DoD may negatively impact their business (63.06% disagreed or strongly disagreed, while only 6.3% agreed or strongly agreed with the statement).

---


[619] Ibid.
[620] Ibid.

*Figure 35: Concern that contracting with the DoD will harm image*

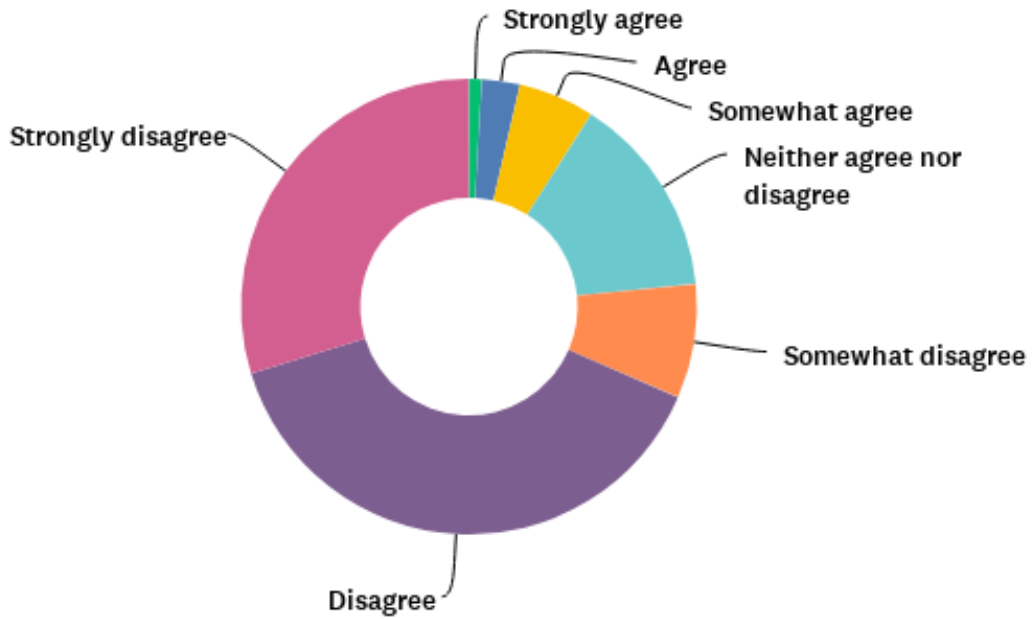

| Question | Strongly agree (1) | Agree (2) | Somewhat agree (3) | Neither agree nor disagree (4) | Somewhat disagree (5) | Disagree (6) | Strongly disagree (7) | Mean | SD |
|---|---|---|---|---|---|---|---|---|---|
| Our company's image will be negatively affected by performing a contract with the Department of Defense. | 0.9 | 2.7 | 5.41 | 14.41 | 8.11 | 38.74 | 29.73 | 5.61 | 1.4 |

Most respondents disagreed that contracting with the DoD will negatively affect their company's image (68.47% disagreed or strongly disagreed, while only 3.6% agreed or strongly agreed with the statement). The findings from this theme suggest that most commercial AI firms do not believe working with the DoD will negatively affect their business or reputation. However, the DoD's contract law and practice are generally not perceived as transparent to



commercial AI firms.  This perception likely negative impacts mutual trust and therefore the overall attractiveness of the DoD as a customer.  Despite the inhibitors of trust and attractiveness relating to the contract practice, this survey found most commercial AI firms perceive the DoD as an attractive customer.  Improving the contract attributes to better instil trust in the procurement process may further positively affect how the DoD is perceived as a customer.

e.  *Preferences on Contract Law and Procurement Practice*

Understanding the preferences of commercial AI firms on contract attributes and procurement practices can help the DoD increase its attractiveness to the industry.  All phases of the procurement process can be adapted under the FAR and OT statute, so aligning the practice of engagement, solicitation, evaluation, selection, and administration of contracts with the preferences of commercial AI firms is legally permissible.  However, there is a lack of empirical data identifying what contract attributes and procurement practices are preferred by commercial AI firms.  Figures 36 - 39 present the findings from the survey that provide insight into these preferences.



*Figure 36: Preference of creating solutions over following specifications*

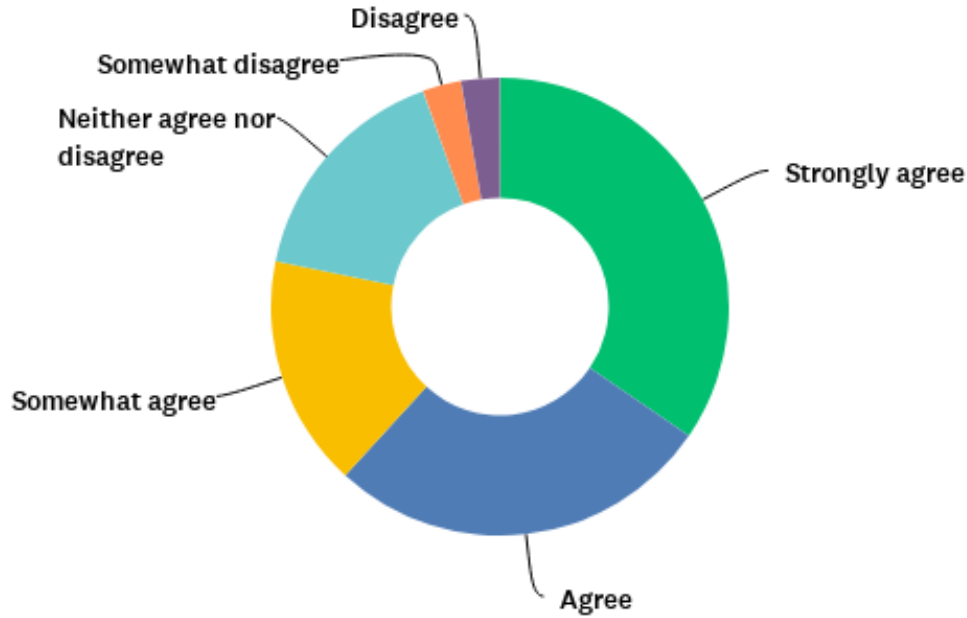

| Question | Strongly agree (1) | Agree (2) | Somewhat agree (3) | Neither agree nor disagree (4) | Somewhat disagree (5) | Disagree (6) | Strongly disagree (7) | Mean | SD |
|---|---|---|---|---|---|---|---|---|---|
| When choosing a new project, my company prefers the opportunity to come up with solutions to problems rather than follow pre-set specifications written by the customer. | 34.55 | 27.27 | 16.36 | 16.36 | 2.73 | 2.73 | 0 | 2.34 | 1.32 |

Most respondents (61.82% strongly agreed or agreed with the statement) favoured projects that permitted their firm to innovate solutions rather than follow customer-set specifications. Only 5.46% of the respondents indicated any disagreement (2.73% somewhat disagreed and 2.73% disagreed) with the statement, indicating very few respondents prefer following pre-set specifications written by the customer. This finding aligns with the literature about the technological considerations of developing AI applications which tend to require



exploration and experimentation with the data to train a model can describe, predict, or recommend the desired output.[621]  It also corresponds to empirical studies about customer attractiveness that found suppliers want to collaborate with the buyer and take part in innovation projects.[622]  Because OT agreements typically start with a problem statement rather than predetermined performance specifications, this finding indicates the sample generally prefers to work on projects that permit suppliers to problem solve.

*Figure 37: Terms and conditions*

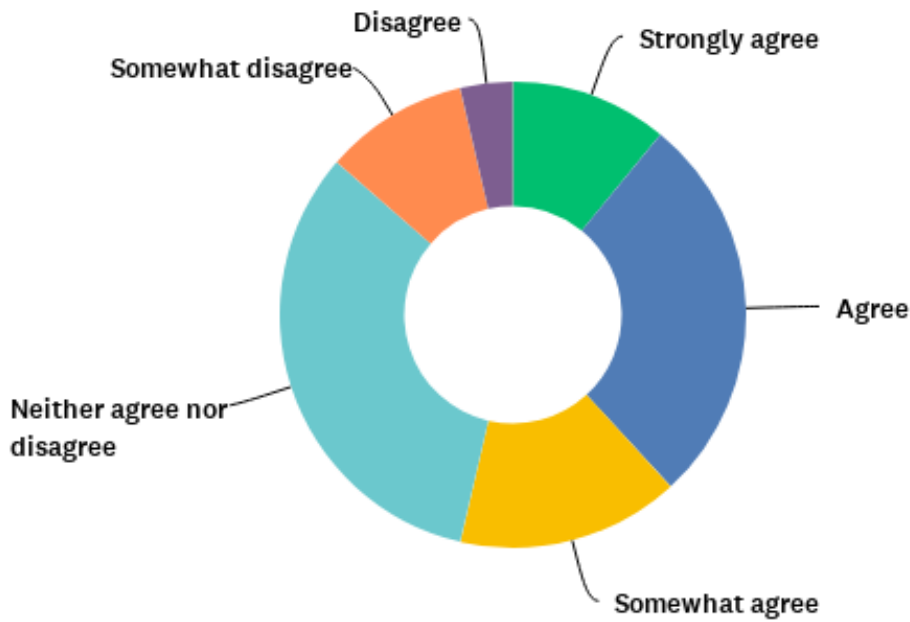

| Question | Strongly agree (1) | Agree (2) | Somewhat agree (3) | Neither agree nor disagree (4) | Somewhat disagree (5) | Disagree (6) | Strongly disagree (7) | Mean | SD |
|---|---|---|---|---|---|---|---|---|---|
| If the contract terms and process were identical, the military would be a more attractive customer than a commercial buyer. | 10.91 | 27.27 | 15.45 | 32.73 | 10 | 3.64 | 0 | 3.15 | 1.32 |

Although most respondents had a largely neutral opinion (58.18% indicated they somewhat agreed, neither agreed nor disagreed, or somewhat disagree), 38.18% of respondents either strongly agreed or agreed that if the contract terms and process were identical, the military would be a more attractive customer than a commercial buyer (compared to on 3.64% that disagreed). This finding indicates that the difference for some commercial AI firms between the DoD and commercial customers is the terms and conditions. For those companies, it appears that the DoD is a more attractive customer than the commercial alternative, but for the terms and conditions that apply to the respective contracts. While the terms and conditions in FAR contracts are typically mandated,[623] they are negotiable in OT agreements, with agreements officers encouraged to adopt commercial business practices.[624] Thus, OT agreements can leverage this flexibility and better align with commercial AI firms' preferences.

*Figure 38: Fixed requirements compared with agile and iterative development*

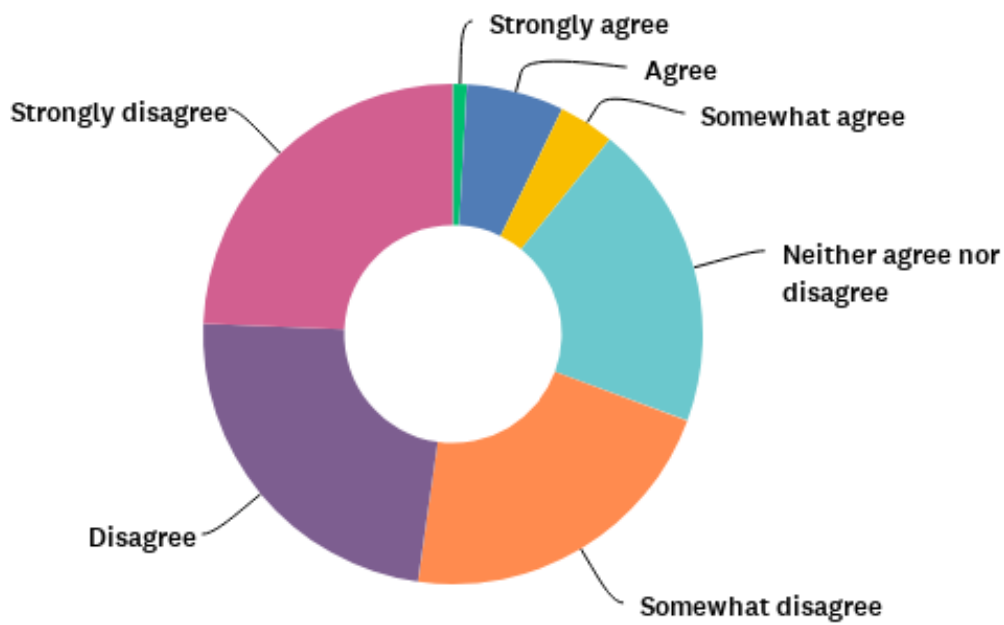

| Question | Strongly agree (1) | Agree (2) | Somewhat agree (3) | Neither agree nor disagree (4) | Somewhat disagree (5) | Disagree (6) | Strongly disagree (7) | Mean | SD |
|---|---|---|---|---|---|---|---|---|---|
| Fixed requirements and milestones are preferable to agile and iterative steps when developing and deploying our product/service. | 0.9 | 6.31 | 3.6 | 19.82 | 21.62 | 23.42 | 24.32 | 5.23 | 1.49 |

Nearly half of the respondents indicated they disagreed or strongly disagreed with the statement that fixed requirements and milestones are preferable to agile and iteratives steps when developing and deployment their AI solutions (when adding those that somewhat disagreed, the percentage rises to 69.36% compared to only 10.81% that indicated any agreement with the statement). This finding suggests that commercial AI firms are more likely to prefer agile and iterative steps to fixed milestones when developing and deploying their product or service. This finding is consistent with the literature reviewed about the AI development lifecycle.[625] OT agreements permit flexibility throughout development of the prototype and the OT Guide encourages utilising agile and iterative processes.[626] However, FAR contracts are bounded by the scope of the contract, which can increase performance and cost predictability for the competitors and government, but also limit the flexibility during contract performance.[627]

---

*Figure 39: Commercial contracting practice compared with DoD contracting practice*

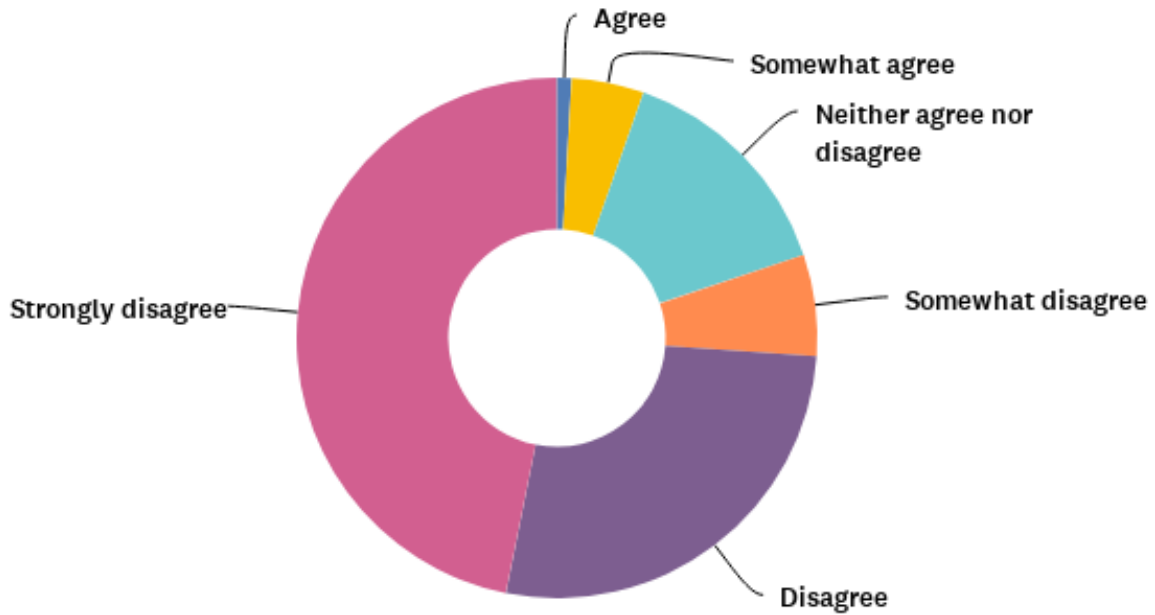

| Question | Strongly agree (1) | Agree (2) | Somewhat agree (3) | Neither agree nor disagree (4) | Somewhat disagree (5) | Disagree (6) | Strongly disagree (7) | Mean | SD |
|---|---|---|---|---|---|---|---|---|---|
| The commercial marketplace should model its contracting practices on the Department of Defense's procurement system. | 0 | 0.9 | 4.5 | 14.41 | 6.31 | 27.03 | 46.85 | 5.95 | 1.29 |

When compared to the commercial marketplace, the DoD's procurement system appears to pose obstacles to attracting commercial AI firms. The question of whether the commercial marketplace should model its contracting practices on the DoD's system elicited the strongest response in the survey (based on percentage of strongly agreed or disagreed as well as the mean closest to the pole on the Likert scale). About three-quarters (73.88% of respondents) disagreed or strongly disagreed that the commercial marketplace should model its contracting practices on the DoD's system compared to just one out of 111 respondents that agreed that the commercial



market should adopt DoD practices.  This finding appears to be an indictment on the DoD's contracting practices, indicating that while the DoD is viewed as an attractive customer, few commercial AI firms prefer the way the DoD contracts compared to commercial business practices.

4 *Conclusion*

Collectively, the survey findings show many commercial AI firms are willing to work with the DoD.  However, commercial AI firms have specific concerns about the DoD's contract law and procurement practices.  These concerns appear to limit the DoD's attractiveness, especially as the sample indicates a preference towards commercial contracting practices over the DoD's traditional contract law framework.  The interviews of fifteen survey participants provide additional context to explain why certain contract attributes attract commercial AI firms and how the DoD's contract law impacts the decision to contract with the DoD.

C *Interview Findings and Analysis*

The survey served to develop a broad understanding about the perceptions and opinions commercial AI firms have about doing business with the DoD and whether they preferred any contract law attributes over alternatives.  The interviews developed the understanding of why commercial AI firms hold those views on the DoD.  The interviews were semi-structured, allowing the interviewee to discuss whatever they thought was relevant to their firm's decision making when considering whether to pursue a DoD contract.  The following topics guided the discussions:

- Perception of attributes of the DoD as a customer

- Factors affecting the decision to contract with the DoD



- Advantages of DoD contracts in both traditional procurement contracts and OT agreements

- Disadvantages of DoD contracts in both traditional procurement contracts and OT agreements

- Comparison of DoD to other customers

- Factors that can be changed to make the DoD a more attractive customer to commercial AI firms

Throughout the interviews, several main themes were discussed frequently by the interviewees.  These themes were identified, examined, and coded, with the coded data placed in categories.  MaxQDA, a software tool for conducting qualitative research, was used to organise interview transcripts and code the data.  The main themes that appeared over the course of conducting interviews are represented in Figure 40.



*Figure 40: Interview Data Codes*

| Emergent Code | Sub-Codes |
|---|---|
| Attraction to DoD | • Profits/scalability<br>• Innovative contract practices (OT agreements and commercial solution openings)<br>• Supporting national security and democratic values<br>• Culture and fit<br>• Trust<br>• Work on challenging problems<br>• Perception in community<br>• Mission-oriented<br>• Professionalism |
| Barriers to Entry | • Difficult to understand process<br>• Length of process<br>• Start-up unique problems<br>• Cost<br>• Regulatory hurdles<br>• Perception of Traditional Contractors<br>• Unique terms |
| DoD Internal Problems | • Technical knowledge gap<br>• Budget/fiscal law<br>• Modernisation priorities |
| Engagement | • Solicitation/requirements<br>• Communication with end-users<br>• Collaboration<br>• Prototype/Pilot/Demonstration<br>• Agile methodology |
| Ethics | • National security concerns<br>• Concern about public perception<br>• Trust |
| Intellectual Property | • Complexity of DoD IP law<br>• Importance to AI firms<br>• Importance to DoD<br>• Trust<br>• Negotiation |

Each interviewee discussed each theme reflected in the code table, so they are presented as equally weighted codes and are listed in the table alphabetically. The sub-codes are based on order of prevalence across all interviews with priority of order representing the most discussion and consensus within the sample. The findings for each of these codes are presented below without reference to the literature reviewed. These findings are then interpreted and integrated with the literature and survey findings in the synthesis section of this chapter.



*1 Attraction to DoD*

Of the 15 interviewees, each indicated that their firm viewed the DoD as a flawed, yet attractive customer. While the flaws are detailed below, the reasons that commercial AI firms found the DoD an attractive customer focused on the impact a DoD contract could have on its business. The primary reason the DoD was named an attractive customer was the direct business revenue and profit stream generated by a DoD contract. While revenue and profit were not necessarily the number one reason the DoD is attractive for each interviewee, it was the only reason discussed by each interviewee. Although almost all interviewees agreed that scaling a product is easier in the commercial market, those with experience in the defence market understood that if their company was able to advance to a defence program, the opportunity for revenue and growth is huge. Firms with limited experience with the DoD recognise the opportunity afforded by a defence contract but perceive that opportunity is limited to traditional defence contractors.

Although direct business was important to the interviewees, there was a strong agreement that the DoD was uniquely attractive in the indirect impact it can have on its revenue and profit generation. Both start-ups and large corporations explained that being a DoD supplier equates to credibility in other markets. One interviewee from a large enterprise solutions corporation explained how DoD contracts boost commercial reputation more than commercial contracts.

> If we're comparing two equivalent dollar whale contracts, [one contract] to a global retail company, or [one contract] to the DoD, the DoD is worth more to us than the commercial retailer. And the reason is because at the end of the day, when it comes to the products that my company makes, the stamp of approval from the DoD is a credential towards security and trust. If the DoD can trust us with their ... workloads, you can trust us with your backend web application or your credit card processing or whatever else. So that's why it's worth it to us, because it's going to drive accretive revenue in our [commercial products].



This notion that the DoD as a customer builds credibility in the commercial market was echoed by start-ups. One CEO rationalised that profit margins on DoD contracts can be very low to still be incredibly valuable because the upside to the contract is in the marketing. The CEO explained:

> I can use [the DoD contract] then to finance the commercial side. And then I know on the commercial side, I can make ten times multiple of [the DoD contract] knowing someone thinks that they're getting tech that the DoD is finding useful.

However, some of the interviewees acknowledged that the reputation a company can earn through performing contracts for the DoD is not always positive when transitioning to the commercial market.

The interviewees explained that recurring and steady revenue streams were important to their business and the DoD can be a very good customer in that regard. Though smaller, newer companies expressed the recurring revenue is vital to keeping their business afloat, larger, more established companies were able to gamble for big payout contracts, though they also see the DoD as a customer that can provide stability with long-term, steady contracts. However, start-ups, were wary of defence contracts that offered relatively small dollar amounts for limited work with the promise of scale after successful performance. One CEO of a Washington, DC-based start-up commented that the companies with a business model focused on getting small grants from the DoD on occasion will not add to the DoD's capabilities in AI. They opined that 'at the end of the day, it's DoD that killed them. Absent DoD recognising the harm it's doing to itself as well as to others, and then finding a comprehensive solution' the small awards by DoD



organisations will ultimately end some commercial AI firms and lead to problems for the DoD in building capabilities with AI.[628]

One factor that makes the DoD an attractive customer is the mission-focused personnel. This factor is unique to the military when compared with other potential commercial AI customers.  Though a lack of technical sophistication causes some problems, overall, the mission and service-oriented military makes for an attractive customer to commercial AI firms.  A CEO from an Austin based start-up new to DoD contracts explained:

> [The military personnel we work with] are just decent people that are very professional. They want to improve things, and it's just a pleasure to speak to them, because what's typical in [industry], everybody's always kind of [angry] and … all the decisions are based around what the bonus check is going to be like.  So, this is, to us, a very refreshing thing where there is just a lot of mission-oriented folks.

Another factor that the interviewees cited as an attractive factor of working with the DoD was the ability to contribute to national security.  Several interviewees explained that it is important that the United States has the best weapon systems and all interviewees acknowledged there is a significant role played by AI and the commercial industry in ensuring that the military remains competitive.  One executive at a large Silicon Valley-based firm remarked that 'there's a sense of a noble endeavour to support the military and the efforts of DoD…we're supporting an awesome mission, and we're proud to help support the missions of the US military.'  Another executive at a Boston-based company commented that they have brought technology solutions to the DoD for nearly two decades and expressed that 'it brings me great satisfaction to be able to enable, facilitate, and give capabilities.'  Several interviewees pointed out the perception that the DoD has a knowledge gap about what AI can do and they believe helping show the DoD what AI can and cannot do is a service to the country.  One director at a Silicon Valley firm just starting to work on defence contracts commented that work with the DoD is 'more fun, it's more interesting…there's a little bit more than just a sale,' which is rewarding.  This sentiment — that work with the DoD is interesting and rewarding — reflects the concept of attraction in social exchange theory.[629]  Because the nature of the work the DoD requires is often interesting, meaningful, and supports national security, commercial AI firms that are attracted to those opportunities appear to find the DoD an attractive customer.  The DoD can leverage the nature of the problems that require AI solutions to increase its perceived customer attractiveness.[630]

---

[628] See Samuel Hammond and Gabriella Rodriguez, 'America's Flagship Program for Innovative Small Businesses is Broken' *The Hill* (9 September 2022) <https://thehill.com/opinion/congress-blog/3636756-americas-flagship-program-for-innovative-small-businesses-is-broken/> (arguing the Small Business Innovation Research (SBIR) program brings short term revenue to small innovative companies but typically yield dead end research with little success at bringing capabilities to the DoD); Michèle Flournoy and Gabriella Chefitz, 'Sharpening the U.S. Military's Edge: Critical Steps for the Next Administration', *Center for a New American Security* (13 July 2020) <https://www.cnas.org/publications/commentary/sharpening-the-u-s-militarys-edge-critical-steps-for-the-next-administration> (explaining that the DoD struggles to bridge the gap between successful research or prototype and scalable capability because the small-dollar amount awards are often not followed with a production contract until a year or more, leaving the innovative company without funding and the DoD without the capability it needs).

One interviewee from a large company with extensive defence experience explained their opinion on Google's actions regarding Project Maven, stating:

> Everybody wants national security; not a lot of people want war.  And so products that can lead to performance improvement or efficiencies for national security stakeholders…those are perceived as a natural good because it's a tool as opposed to the other end of the spectrum of weapons that have a uniquely negative connotation.

Another acknowledged that there are plenty of Americans who would negatively perceive work with the DoD, but many in the commercial AI industry recognise and appreciate the opportunities afforded to Americans is due to a strong defence posture.  They explained that the commercial AI industry understands the importance of strong national security well due to the constant cyber-attacks, disinformation, and IP theft by competitor states that goes on at AI firms. However, a CEO of a Silicon Valley start-up explained that investors largely are unmoved by national security interests and are motivated by more direct profit generation than defence contributions. Another interviewee from a large company with extensive defence experience explained that their company is willing to provide the military with commercial technology that can be applied by the military, but the military unique applications were up to the DoD. The view that commercial AI firms were concerned about providing AI capabilities to be used for lethal purposes for fear that such business would lead to a negative public opinion about their company was echoed by several interviewees. Thus, the concerns appear to be less to do with ethical concerns than business concerns.

---

[629] See Nollet, Rebolledo and Popel (n 478) 1189.
[630] See Hüttinger, Schiele and Veldman (n 469) 1197–8, explaining that in high demand industries, customers must maximise their perceived attractiveness compared to other choices to work with the best suppliers; Nollet, Rebolledo and Popel (n 478) 1189, explaining that buyers can improve their attractiveness by highlighting what distinguishes them from other potential buyers.

Of the 15 interviewees, three were veterans and two others had worked for the DoD as civilians. The sense of responsibility of the commercial AI firms in supporting national security was not absolute but shared by most interviewees. The noticeable difference between the interviewees who had no personal ties to the DoD and those that did was the interviewees with a personal connection to the DoD indicated more frustration with the contracting process.

Another driver cited by many of the interviewees was the ability to work on challenging problems with the DoD. One executive at a Washington, DC-based firm with extensive experience working with the DoD admitted that there are many reasons why the DoD is not an attractive customer to commercial AI firms due to the many barriers presented by the contract law and procurement process. However, what the DoD has that makes it an attractive customer to commercial AI firms, despite the various problems with contracting, is big, exciting challenges. The interviewee described a conversation with a DoD procurement official who



stated, 'everybody wants to do business with the DoD because we're big and we spend the most

money, and we're the best customer.' The interviewee stated they thought:

> None of those things are true anymore…that's 1950s talk. So the last thing that DoD has
> going for it is cool problems. And the good news is that people who do new stuff are drawn
> to cool problems, so that's a good feature to have, but if DoD doesn't regain some of those
> original things, they're going to lose.

An executive at a small start-up added that they want to tackle very tough challenges, and the

DoD still has plenty of them, particularly in applying AI to defence-specific problems. Another

executive at a traditional defence contractor stated that the problems the DoD has are interesting,

adding that:

> There are very few places where you can go in industry where you are solving a problem
> as big as we solve. That's a big deal. You can almost get addicted to the heart of the
> mission, and we find that a lot of our people could take more money to go into commercial,
> but they stay with us for the problems and it's the same thing with us. There are things
> about commercial that might be more attractive, but the problem set and mission keeps us
> here. Those are definitely the things that go into our calculus of why we work with the
> DoD.

One factor that made the DoD an attractive customer from the perspective of the

interviewees was the trustworthiness of the DoD when compared to the commercial market. A

large majority of interviewees cited the values, ethical behaviour, and trustworthiness of the DoD

as a driver to compete for a contract with the DoD. Though firms with little to no experience

with the DoD recognised the procurement process burdens new entrants, the competition is

perceived as fair. While there were significant concerns about the DoD's IP law, most

interviewees explained that they believed the DoD would not steal IP, which was reported as a

common fear in the commercial market. On the issue of trust, the interviewees described the

DoD as honest, genuine, authentic, and straightforward. These descriptors contrasted



significantly with how commercial customers, particularly in business-to-business transactions, were perceived by the interviewees.

Though there are many factors that make the DoD attractive in the eyes of commercial AI firms, none of the interviewees found the way the DoD conducts business attractive. As reported below, the interviewees found the DoD's contract law framework and procurement practice as a factor limiting their engagement with the DoD.

## 2 *Barriers to Entry*

Anything that the interviewees perceived as an obstacle to accessing the defence market or required more effort, time or cost than the commercial market was coded as a barrier to entry. The primary subcodes in this category include the complexity of the procurement process, length of time to award, cost, regulations and unique terminology. The interviewees that came from start-ups also had perceptions of traditional contractors and investors that were categorised as a perceived barrier to entry.

One executive from an Austin-based start-up that is a new entrant into the defence market succinctly explained the defence procurement process: 'It seems to be surprisingly difficult… I don't know how a company that didn't have [a defence contracts expert] hand-holding them could possibly succeed.' They explained their first experience attempting to win a DoD contract from the Air Force:

> The structure of what the Air Force and DoD in general looks like is very opaque to people who don't have that experience. It's hard to understand who's in charge. There's no single point of contact, and there was nobody who was trying to find us the right place to go to. There's the customer discovery issue of, how do you figure out who might be interested in your technology and why, and what kind of modifications might be necessary. And that is a problem. We have generally applicable technology for a lot of [use cases], but it was really hard to figure who the right point of contact is, and without a consultant from a venture fund, I don't know how a company would ever find the right place to go.



The technology developed by this firm is potentially industry-disrupting and clearly relevant to a solving a persistent problem in the DoD. However, the firm found the contracting process so opaque and challenging that an executive concluded, 'we got very lucky, and I don't really know how anybody would do it without that kind of luck.'

One of the primary reasons cited by the interviewees on why they believe the DoD has barriers to entry is that the contract law that applies to the DoD differs so significantly from commercial contract law. One executive from a large Silicon Valley firm with extensive experience in business to business as well as defence sales opined that the DoD's system is different because it attempts to be fair and just. However, as the interviewee pointed out, the effect of certain laws intended to promote fairness creates perverse incentives for industry. They explained that small business set asides[631] influence the way a company structures itself so that the company gains access to opportunities with restricted competition. The perception from several of the interviewees was that the DoD contracting law and process do not attract competition, particularly from commercial AI firms. According to several interviewees, the perception is that the pool of competitors for what the DoD calls full and open competition is largely made up of traditional defence contractors and the winner is the contractor that may not be the best, but simply the best at following the myriad regulations.

Almost every interviewee voiced frustration that the process to compete for and win a contract was slow, onerous, and costly. Many lamented the unique laws that apply only to the DoD and have no relation to commercial contracts for the same technology. This difference affects every interviewee's firm as they had to choose to hire specialised experts in DoD contracting or, as two interviewees' firms decided, forego DoD opportunities altogether. The inability of the DoD to lower the burdens its contract process places on commercial AI firms keeps innovative technology out of the DoD.  As one director at a Silicon Valley firm explained, 'it's a running joke how difficult it is to do business with the Federal Government and how difficult it is for a small or new business to do or to break into that field to the point where most people won't even do it.'  The perception that working with the Government is too difficult for small businesses that some may not even try negatively affects the DoD's ability to attract commercial AI firms, reducing competition and access to innovation.[632]

---

[631] 'Small business set asides' refer to the requirement that certain contracts under a dollar threshold must restrict competition to small or socioeconomically disadvantaged companies. See 48 CFR Part 19.

With AI, where much of the DoD's requirements can be met with commercial solutions, commercial AI firms can sell similar — possibly identical — products or services to the commercial market and DoD.  However, the contracts and the procurement process are drastically different.  The terminology used is described by many of the interviewees as foreign and the licensing terms are very different.  For firms that are unfamiliar with the contracting process and law, these differences create uncertainty, which many interviewees view as high risk to their companies.  A CEO at a Silicon Valley firm that has never worked with the DoD before explained what they thought was necessary to attract other Silicon Valley firms:

> I think that the only way that [the DoD is] going to get Silicon Valley-type start-ups to start engaging is if there's a really proactive push to both lower the barrier and to demystify the barrier associated with start-ups, working more or less as they do today.  That means finding a way to do it such that they don't have to know what 'FAR' is and all the other acronyms.

A director at another Silicon Valley firm that has just started to perform on a DoD contract explained that many high-tech firms have no idea how different and complex working with the DoD is until they try it.  The same interviewee expressed that the DoD is as clueless about working with start-ups as start-ups are about working with the DoD.

---

[632] See Glas (n 465) 98–100.



The length of time a DoD contract takes from announcement to award and funding was perceived as unacceptably long to nearly all interviewees. The length of time required to compete for and get awarded a contract with the DoD negates many of the advantages of working for the DoD. This was reported by start-ups who expressed the runway, or amount of funding required to operate their business, was too short to wait more than six months for a contact. Large, established companies reported that the length of time to award represented opportunity costs that are lost. Enterprise companies reported that it often takes several years for the DoD to award contracts on a major program. By the time the DoD awards the contract, the technology is obsolete, such is the rate of technological advancement in AI. A director at a tech giant explained on a recent contract that took several years to award after the proposal was written that 'by the time that this contract actually gets executed, in many cases, the things that we promised, we wouldn't even want to deliver because they are legacy hardware.' This predicament creates problems on both sides of the contract — a customer receives outdated technology too late to use and a vendor must take on additional costs to provide and sustain obsolete products. Thus, although the rationale may differ between start-ups and established companies, commercial AI firms see the length of time it takes the DoD to award a contract as a factor in deciding whether to compete for a contract opportunity.

A director at a large Silicon Valley firm with substantial commercial and defence business expressed the length of time the DoD needs to award a contract presents a huge obstacle. They explained the perception of AI firms when they realise how long it takes to get a DoD contract:

> The thing that's confounding about [the length of time] is, [the DoD is] losing the opportunity, the advantages of Silicon Valley. You know that there's a huge opportunity because of the funding available in DoD. And then AI companies go and it's like,



'[contracting with the DoD] is not like watching paint dry, it's like building pyramids in Egypt, one block at a time.'

Another interviewee drove the point: 'we filed a proposal sometime around two years ago. We're still waiting on a decision.' This interviewee opined that the DoD is inherently risk-adverse as 'the system is designed to punish people whenever they make a mistake, but there is no reward for making smart decisions — there is no incentive to move fast.'

Adding to the length of the process is the ability of losing offerors to protest the award. While the DoD perceives this process as vital to fairness and transparency, commercial AI firms were quick to note that protests do not exist in the commercial market. Bid protests can cost the winning offeror a fortune and further delay contract execution. One interviewee stated the number one legal reform they would like to see is reducing the options for losing offerors to engage in post-award litigation.

Along with the length of time, most interviewees had negative perceptions of the accounting requirements for DoD contracts. Many interviewees stated their companies would not bid on a contract if it required cost accounting standards. They were only interested in firm, fixed price contracts or time and materials contracts. The cost accounting standards in the DoD were perceived as unduly burdensome. One interviewee with experience in both the DoD and commercial market explained 'the accounting standards and system that the government levies on us really hurts us from competing in the commercial market. A lot of companies set up a federal business and a commercial business, really just to divide the accounting systems — they are so complex.' Because of the many unique requirements, several interviewees in charge of start-ups stated you can only pick one market to focus on, commercial or defence, because learning the rules for defence contracts is too difficult to manage both markets.



Another barrier to entry described by the interviewees was the cost associated with contract compliance.  The length of time it takes to compete for a contract for the DoD adds to the cost, in terms of real money used to prepare a complex proposal, opportunity cost from eschewing other potential contracts during the competition, and the lack of revenue earned over the time it takes for the DoD to award a contract.  Because competition is often robust for DoD contracts, the perceived likelihood of winning a DoD contract compared to a commercial contract was lower overall amongst the interviewees.  Additionally, the mandatory FAR clauses were perceived largely as waste.  This added expense, many of which were described as 'zero-value add' by interviewees, limit the amount of funding the commercial AI firms have available for actual innovation.  Much of the revenue generated from a DoD contract goes to expensive compliance programs such as accounting standards, various unique labour law requirements, and other overhead and administrative costs that are specific to DoD contracts.  One director at a Silicon Valley-based start-up new to the defence market expressed frustration with how time-consuming and costly compliance and certification to obtain Authority to Operate (ATO) on the DoD's IT system.  They explained that because the barriers to entry, just to qualify to compete for, let alone win, a contract are so high, especially on highly technical projects that include most contracts involving AI, only a small number of firms can compete.  The barriers are such that once a company clears those barriers, they want to protect those barriers so that they are guaranteed access to DoD contracts that the commercial competition cannot compete.  The level of complexity and cost of making a relatively minor mistake, given the razor thin margins of which source selection decisions are made, require firms to hire specialised experts to assist them for one customer that requires sellers to play by a completely unique set of buyer-centric rules.



The differences between commercial contracts and DoD contracts were perceived by the interviewees as immense. One area where the difference was most pronounced was the terminology used by the DoD. The language used, and reliance on acronyms, made the DoD a very insular customer, detached from industry standards. Most interviewees acknowledged that the DoD's communication with industry lacks clarity. One executive at a Silicon Valley-based start-up offered that 'one of the biggest things [DoD] could do is try to figure out how to write some of these [requests for information and requests for proposals] in the language of the people they want to respond.' Another executive at an Austin-based start-up believed that the DoD would be more effective at engaging with and attracting commercial AI firms if they broke down the language barrier and used industry terminology rather than DoD-specific jargon. One CEO from Silicon Valley stated simply, 'it's just a completely foreign language.' The organisations that have adopted industry terminology, like DIU, were more likely to be described as attractive by the interviewees than organisations that used defence-specific jargon and acronyms to communicate with industry.

Although the interviewees were in strong agreement that traditional contracting methods, such as FAR-governed procurement contracts, were vastly different than the commercial market and presented unique barriers, most interviewees expressed OT agreements were better able to attract commercial AI firms. OT agreements were perceived by the interviewees as faster than the FAR. Many preferred OT agreements to traditional contracts specifically because most of the costly and cumbersome regulations that plague traditional procurement contracts do not apply. A director at a large firm with extensive defence experience indicated that 'if we didn't have OTs,' they wouldn't compete for a DoD contract because FAR contracts 'are just too long of a process. It's too protracted. It's too much [unnecessary requirements] and hoops you have



to jump through.'  The customer-centric model of contracting in a highly competitive field such as AI appears to negatively affect the perception of the DoD as a customer.[633]

An executive at a start-up pointed out that the communication possible during the source selection process is much more open with an OT than a FAR contract, referring to the 'blackout periods' under FAR Part 15 that limits communication between the offerors and the DoD between submission of offers to contract award.[634]  They also remarked that the practice of down-selecting the pool of competitors in an OT saves losing offerors time and money by knowing sooner that they are not going to win the contract while it permits demonstrations of the down-select pool that allows less-established firms to showcase their abilities against more established firms.  Because past performance is such a heavily weighted factor in traditional procurement source selection, this OT characteristic permits small commercial AI firms to have a chance at securing a contract that otherwise would be out of reach.

According to several interviewees, in theory and in law, OTs should be much more like commercial contracting than the traditional procurement contracts.[635]  A caveat was that several interviewees experienced the DoD inserting FAR requirements and processes into OT agreements, stripping the flexibility and advantages that make OT agreements attractive contract vehicles for AI companies.  Several interviewees mentioned the DoD insisted on DFARS clauses in OT agreements they negotiated.  Others, even with extensive defence contracting experience, were unaware that negotiating IP clauses is permitted under OT authority.  Thus, while OT agreements were viewed as a better option, many saw the DoD's practice of limiting negotiation and inserting FAR requirements into OTs as problematic.

---

[633] See Glas (n 465) 98–100, explaining when there is rapid technological change, militaries must compete in highly innovative commercial markets, declaring an end of 'the customer is always right' paradigm in business.
[634] See 48 CFR § 15.306.
[635] The law for prototype OT agreements exempts the contract from most procurement laws and regulations that apply to traditional procurement contracts in the DoD: see 10 USC § 4022; *NDAA FY2016* (n 335) § 815, 129 Stat 893.  The legislative history of 10 USC § 4022 indicates the intent of Congress was that the DoD should use OT authority to rapidly acquire information technology systems to ensure the DoD will 'remain competitive in the commercial marketplace': Committee on Armed Services, House of Representatives, *Report on H.R. 1735 Together with Dissenting Views* (House Report No 114-102, 5 May 2015) 202, § 853 ('HR 114-102').  Congress saw OT agreements as 'an effective tool for research and development contracts,' due to the ability to 'tailor the contracting language and thus eliminating many aspects of the [FAR] that may not be pertinent' in the contract: HR 114-102 (n 635) 202.  Congress stated OT authority 'can make [the DoD] attractive to firms and organizations that do not usually participate in government contracting due to the typical overhead burden and "one size fits all" rules': House of Representatives, *Conference Report to Accompany H.R. 1735* (House Report No 114-270, 29 September 2015) 702 ('HR 114-270').  Additionally, it stated 'expanded use of OT agreements will support Department of Defense efforts to access new sources of technical innovation, such as Silicon Valley start-up companies and small commercial firms': HR 114-270 (n 635) 702.  The exemption from traditional procurement law, regulations, and policy, and the flexibility of contracting inherent in the OT authorities permit the DoD to form contractual relationships that are more in line with business contracts.

The interviewees were nearly unanimous in their belief that OT agreements were better suited than FAR contracts to attract commercial AI firms, citing the flexibility, speed, limited regulatory burdens, and ability to negotiate terms and conditions as favourable business conditions as well as compatible with AI system development.[636]  As one executive at a Silicon Valley firm explained, flexibility and iteration are critical to developing an AI-enabled solution. They explained the requirement generation under the FAR is too brittle and limited for AI development, while an effective OT agreement can be flexible enough to leverage commercial AI solutions.  Another executive from a Boston-based start-up added that the ability to prototype, test, and then iterate is critical to an AI firm to understand the DoD problem and to better serve the customer, noting that no one can draft detailed specifications for a machine learning application as typical for a FAR contract without first developing the model through an iterative process.  A director at a large corporation opined that as AI development continues, OT agreements are more appropriate given the flexibility of the contract, though once AI is required at a large enough scale and there is no longer a need to develop the AI for each use case, a FAR contract with a traditional commercial software licence may be more effective. Overall, the opinion of the interviewees regarding OT agreements was that they were closer to the commercial process and more consistent with the way AI is developed than traditional FAR-regulated contracts. As one executive at a Boston-based firm pointed out, OTs demonstrate that 'the DoD is trying to make it easier for little companies like ours to actually get contracts.'

---

[636] Although several interviewees did not know the distinction between FAR contracts and OT agreements, all interviewees explained their business preferred contracts that permitted frequent communication with the customer, collaborating with the end user on finding a solution to their problem, flexibility to experiment and iterate as the solution is developed, speed of process, and IP protections.  As discussed throughout Chapter V, these preferences align with OT agreements and are uncommon or legally limited in FAR contracts.  For those interviewees that had experience in providing AI technologies to the DoD under both FAR and OT authority, each preferred OT authority.

The complexity, length of time, and cost associated with DoD contracts present significant barriers to entry from the commercial AI industry's perspective. Some of these problems described above stem from the law and regulations. However, as many interviewees explained, many of these legal challenges can be overcome with existing legal pathways, such as OT authority, and mitigated with better communication and engagement. The DoD's failure to leverage the existing law can cause frustration and lead to a negative perception of the DoD as a customer. As one executive warned, if a commercial AI firm has a bad experience with the DoD, 'now they'll go tell all their friends, they'll tell their investors, they'll tell their advisors.'

3 *DoD Internal Problems*

The vast DoD bureaucracy played a role in the perception of several of the interviewees. A few of the interviewees that had extensive experience working with or for the DoD prior to their work at commercial AI firms were aware of the problems the DoD faces internally that can limit its efforts to attract commercial AI firms. While outside the scope of this study, it is relevant that the law, regulations and policy over the planning, programming, budgeting, and execution of acquisitions were viewed as factors that inhibit the DoD from leveraging commercial innovation.

The lack of technical knowledge and sophistication with AI at high levels of leadership in the DoD and Congress also represented concerns for several the interviewees. Many of the



interviewees stated a perception that the lack of technical knowledge within the DoD results in some contracts being awarded to firms, not with the best technical solutions, but with the best marketing. As one executive at a Silicon Valley firm explained, without the ability to evaluate potential solutions, the DoD is limited in recognising novel and innovative solutions. Several interviewees emphasised the unique nature and development of AI and their perception that many in the DoD do not fully understand how AI is developed and what it can and cannot do. Without understanding the technology, the DoD appears to view AI as a service, while most of the interviewees remarked that AI is better viewed as a tool that can be leveraged. This distinction affects the procurement strategy.

The DoD's training, education, talent management, recruitment, and budget programs are all factors that the interviewees described as potential barriers to the DoD from executing its national defence strategy. The perceived knowledge gap also affects how commercial AI firms view the DoD as a customer because they impact how the DoD awards contracts for AI technologies.

4 *Engagement*

Engagement between the DoD and commercial AI firms is an area where most of the interviewees believed was insufficient overall. Engagement was a term used to describe the act or process used by the DoD to reach commercial AI firms and notify them of opportunities for contracts. Industry engagement, communications, publication and solicitation of requirements and problem statements were discussed by the interviewees.

Traditional procurement contracts were viewed as very formal, and communication was restricted according to nearly all interviewees. Typical DoD engagement with commercial AI firms was described by interviewees as 'old school,' 'monolithic,' 'boring,'



and did little to attract firms, especially start-ups, that hadn't previously worked with the DoD. However, some best practices have emerged.  Two interviewees explained that the opportunity to demonstrate their technology via competition for a prototype OT or a commercial solutions opening,[637] rather than simply writing about it, led to contracts with the DoD that they would not have competed for had the process been the traditional request for proposals solicitation method. Another executive offered that the way the DoD asks for written proposals does not work for evaluating AI, and that a better way to engage commercial AI firms and select the best solutions is to request demonstrations and pilot versions of the AI-enabled capability, adding that demonstrations and pilots are how industry buys AI.

Other interviewees agreed and stated the general perception that streamlined competition procedures with shorter evaluation timelines focused on innovative commercial solutions are preferred to the traditional detailed performance requirements and lengthy and complex competition procedures.  The streamlined processes, interviewees explained, were better aligned with commercial practices, and makes sense for machine learning applications that are challenging to describe as a specific requirement.  One director at a Silicon Valley firm explained the DoD pushes out opportunities to firms to fulfil its requirements.  Those requirements are often described in detail by the DoD.  This practice may have worked when the DoD had technical expertise to draft the specifications for cutting edge technology but is problematic for machine learning which typically requires experimentation before the model is developed.

Compounding the problem is that many commercial AI firms, given the demand for AI-enabled technology, attract customers through their advertisements and do not typically respond to requests for proposals.  One interviewee described this situation as a fisherman casting a line into a pool with no fish, only other fishermen.  Both customer and seller are using incompatible solution finding and business capture practices.  As one executive pointed out, the DoD needs to be more active in identifying, engaging with, and pursuing commercial AI firms as the passive model of posting a request for proposals is not going to yield competition.  One executive opined that because there is no clear path for AI firms with potential military applications to bring their ideas to the DoD, their 'guess is that there's interesting stuff happening with potential defence applications that the DoD has no idea about.'

---

[637] Commercial solution opening is a competitive process to obtain solutions or new capabilities to provide technological advances.  10 USC § 3458.

The interviewees had a clear preference for more open-ended solicitations.  While several interviewees acknowledged that a traditional request for proposal that described the work or product in detailed specifications was appropriate in instances that the DoD already knew the best solution, nearly all agreed that contracts that called for developing or adopting AI technology for new use cases was incompatible with detailed specifications.  Open-ended solicitations align with AI development and encourage companies to provide solutions to DoD problems.  These methods contrast with the traditional methods used by the DoD where it is common for the buyer to determine the solution and then look at offerors to suggest a price for the work or product that will solve the problem, which, according to the interviewees, is nearly unheard of in the commercial market.  A CEO at an established firm that just recently began work on its first defence contract explained there is no innovation in responding to the traditional solicitation.  With AI, they added, 'nobody can define a way or define a solution today within the area of AI' so narrowing the requirement to a set of specifications limits innovation, presumes the DoD already knows the solution and restricts competition.



Communication and collaboration with the end-user were described by every interviewee as not only critical to understanding the problem to develop the solution, but also as important in building relationships with their customers. Additionally, as AI systems are developed in an agile method, with multiple iterations and constant feedback and evaluation cycles, communication and collaboration is perceived by the interviewees as critical. Despite how important the ability to hold meaningful conversations with the end-user and collaborating to find the right solution is to develop and deploy AI-enabled capabilities, the interviewees did not believe the DoD valued this relationship. One CEO with experience working with commercial customers and the DoD compared the two:

> In a commercial contract, you know whether you're doing something good or bad very quickly. You're touching base with the customer on a regular basis. With the DoD, I get the sense that they're not available for us to meet every week or every two weeks — it's on a monthly basis. Right now, I'm trying to adjust my agile development cycle to be a monthly agile development cycle, which is contrary to being agile.

This concern not only affects the DoD's attractiveness as it lacks the collaboration, communication, and flexibility sought by commercial AI firms,[638] it indicates the DoD's contracting practice is unaligned with AI development best practices.[639]

Another executive from a firm with significant DoD contracts explained that without the ability to collaborate with the end user, the relationship is transactional. Several interviewees agreed that commercial AI firms are not attracted by transactional relationships and much prefer collaborative relationships where they are involved in developing the customer's solution alongside the customer. A CEO at a start-up explained that in the DoD, it is common for the buyer to be a distinct organisation from the user and those two organisations do not always communicate with each other.

---

[638] See above Chapter IV(B)(2) and (3).
[639] See Gadepally (n 20) 2; *NSCAI Interim Report* (n 180) 56.



The interviewees were also in close agreement that the DoD is too risk adverse compared to the commercial market. The overarching perception was the DoD was incapable of thinking like an investor, so they were always going to pursue more proven and mature technologies instead of the disruptive innovations it says it needs to remain competitive. Several interviewees described one key difference between investors and the DoD is that in the commercial market, bold ideas are funded with the intent of discovering what works and what does not. If a disruptive technology emerges, more funding occurs to accelerate and scale the technology; if an idea fails, it does so quickly so the funding can be freed up to reinvest in the proven technology or more new ideas. Rapid prototyping is required to 'fail fast' and ensure the finite resources are properly allocated to support the advancement of promising technologies.

One interviewee opined that the DoD could better attract innovative firms, learn what current technology is capable of, and reduce risk by awarding small, short duration contracts to multiple awardees to build pilots and then down-select to the one or two contractors with the best solution. The overwhelming majority of interviewees concurred there is value offered to technology evaluation in pilots, demonstrations, and prototypes. All interviewees believed that these methods were better than written proposals for both buyer and seller, especially for new AI applications. Pilots and demonstrations were cited as the primary method that industry buys AI. As one interviewee explained, the DoD should try to leverage that comfort in process and adopt industry best practices.

Another interviewee explained that for machine learning applications, the demonstration would be very effective at showcasing capabilities directly relevant to DoD use cases if the DoD provided its own dataset to the competitors. Speed and accuracy can be observed in real-time. This interviewee described how their firm, a new entrant into the defence market, was able to



showcase their product for an agency that provided a synthetic dataset and beat out a major defence prime contractor for an award. They believed the relatively new firm would never have been awarded the contract based on a white paper or proposal and only decided to compete when they discovered a demonstration with a dataset would be the method of evaluation.

Overall, engagement — the process of how the DoD begins a relationship with a commercial AI firm before a contract is signed — is an area that the interviewees saw as an opportunity for the DoD to improve. However, best practices were identified by the interviewees, specifically streamlined competition procedures through pilot or demonstrations, matching the way AI innovation is procured in the commercial market. Additionally, open communication with actual users of the technology, developing a collaborative relationship to solve problems together, and frequent interactions from advertising opportunities to evaluations of technical demonstrations were universally noted best practices in contracting for AI-enabled capabilities.

5 *Ethics*

Ethics was an issue that was discussed by nearly all interviewees. However, no interviewee cited the ethical considerations stemming from the DoD's use of AI for military activities as an absolute 'deal breaker' for their firm. Several interviewees expressed that their firms, and their beliefs that other firms in the industry, typically have no concerns providing AI-enabled capabilities to the DoD. However, they distinguished technologies that are dual-use with technologies that are specifically intended for use in combat. One interviewee from a large firm explained:

> Where we would probably draw the line is when we start getting asked to build things of a uniquely military capability. If [the DoD is] just asking for us to produce cars, that's fine. If they're asking us to produce tanks, that's probably a different thing. Building language translation services, computer vision, all that stuff, we can do all that, you have to go then



train your model to do what you want. Building ruggedised appliances to do machine learning at the edge, sure, we can do that. You want to figure out what application you want to run on that and it's going to fly on a drone and add a certain computer vision model on it, that's up to you.

Other interviewees expressed their firms were more than willing to support the DoD with full understanding that their technology may be used to lethal effect. One CEO explained they believed other firms in the AI industry held 'an unreasonable aversion dealing with the DoD,' adding that 'at the end of the day, we're on the same team—of course I'm going to want the military to have the best weapon systems.' Another interviewee stated their firm was comfortable providing their technology to the DoD because 'nobody is a fan of war, but if we're going to have war, we'd prefer the United States to have the best armed forces.'

Although the interviewees universally used the term 'ethics,' it appeared that the term as a concept lacked a cohesive meaning amongst the interviewees. There is no industry standard for the ethical development and deployment of AI systems, so the perceptions provided by the interviewees were subjective. Nonetheless, several interviewees offered definitions of what they meant by ethics in the context of military use of AI systems. Most contextualised the ethical use of AI systems on a spectrum with commercial applications on one end and lethal effects on the opposite end. Thus, the more benign the impact, the less concern the firm would have over ethics. However, several interviewees explained that applications on the harmful end of the spectrum could be mitigated with humans in control of the AI system. While several interviewees from start-ups held their firms would not be comfortable providing technology that would be used for lethal means even with human control, the majority, including large tech firms and start-ups, acknowledged that their AI systems could be useful in a warfighting context, and they would be comfortable working with the DoD on a contract that could result in lethal effect.



A few of the interviewees connected the concept of ethics with their firm's commerciality. One interviewee voiced a concern that if AI is used in a lethal manner, either directly or indirectly (such as in an intelligence or surveillance role through a computer vision model), there is an expectation that the commercial market will become off-limits.[640] Others candidly discussed that their firms were cognizant of how developing AI capabilities for the DoD could affect their commercial business, for better or for worse, and made decisions based on how they perceived the commercial market would react.

When asked specifically about the DoD's ethical principles for AI, the interviewees expressed their opinion that the DoD is framing the ethical issues in the development, deployment, and use of AI as well as any organisation. While not all interviewees were familiar with the DoD's ethical principles of responsible AI use specifically, all identified best practices within the AI industry, such as mitigating unintended bias, emphasis on model safety and reliability, transparency and auditability, and governance (the ability to disengage or deactivate systems that exhibit unintended behaviour).[641] Several interviewees acknowledged that the DoD, unlike commercial customers of their AI systems, had ethical principles enforced by legal obligations, such as the Law of Armed Conflict.[642] These obligations ensured such principles were not adopted merely out of self-interest to create an appearance of ethical behaviour.

---

[640] Although several interviewees expressed that commercial AI firms may draw the line at providing their technology to the DoD if it is used for lethal purposes was prevalent among the interviewees, most of the sample stated that *their* firms would have no issue with the technology being used in any manner that the DoD required. It appears the impact of a relatively small group of Google employees, and the subsequent media response, do not represent the views of the industry. As Google and others have since competed for and performed DoD contracts, developing and delivering AI tools to the armed forces, the concern about ethics appears to be a consideration for AI firms, but likely not one that will significantly limit the DoD's access to commercial technology.

[641] These recommended best practices are consistent with the DoD's ethical principles. See Department of Defense, 'DOD Adopts Ethical Principles for Artificial Intelligence' (Media Release, 24 February 2020) <https://www.defense.gov/News/Releases/Release/Article/2091996/dod-adopts-ethical-principles-for-artificial-intelligence/>.

[642] As discussed in this section, ethics, though a consideration for commercial AI firms deciding whether to work with the DoD, is not as significant of a factor as the DoD's contract law and procurement practice in that calculus. However, this point made by several of the interviewees requires some additional explanation. The interviewees



Overall, the issue of ethics was described as a factor that commercial AI firms consider when determining whether to compete for a contract with the DoD, though for nearly all firms, it was much less of a factor than business factors (and for some firms, it was described as a business factor) and rarely came up in discussions about why their firm may choose to avoid working with the DoD.  Perceptions of the DoD's contract law appear to outweigh the ethical concerns in providing AI-enabled capabilities to the military.  Thus, while ethics was a factor in the perceived attractiveness of the DoD as a customer, and the industry's understanding that its interest is mostly aligned with the United States' national security interests make the issue of ethics a neutral or even positive attribute rather than a problem for the DoD to overcome.

## 6 Intellectual Property

One of the dominant concerns expressed by the interviewees was how their intellectual property is protected in a contract.  Out of the 15 interviews, IP was the most prevalent coded

---

correctly identified the legal obligations all members of the armed forces in the DoD must adhere to, known collectively as the Law of Armed Conflict (LOAC).  In the United States, such legal obligations are rooted in customary international law, treaty law, domestic law, and policy.  International legal obligations include the concept of command responsibility, where commanders are criminally responsible for war crimes committed by their subordinates if they knew, or had reason to know, that the subordinates were about to commit or were committing such crimes and did not take all necessary and reasonable measures in their power to prevent their commission: *Protocol Additional to the Geneva Conventions of 12 August 1949, and relating to the Protection of Victims of International Armed Conflicts*, signed 8 June 1977, 1125 UNTS 3 (entered into force 7 December 1978) art 87 ('Protocol I').  One such war crime is the indiscriminate attack of the civilian population and civilian objects: at art 51.  US law requires commanders to give lawful orders and adhere to LOAC: 10 USC § 892.  Additionally, the rules of engagement may limit the use of force beyond international legal obligations: see J Ashley Roach, 'Rules of Engagement' (1983) (January–February) 36(1) *Naval War College Review* 46, 46–9, explaining rules of engagement can serve to limit the use of force beyond the international legal obligations for policy or military reasons.  Finally, US policy requires that all systems be designed to allow commanders and operators to exercise appropriate levels of human judgment over the use of force: DoD Directive No 3000.09 (n 19).  Although application of these obligations in the context of AI and autonomous weapon systems is relatively novel and potentially untested, such obligations continue to exist.  Thus, AI systems used by the DoD must adhere to international and domestic legal obligations, including command responsibility and the prohibition against indiscriminate attack, as well as other policy restrictions.  The DoD's ethical AI principles augment AI-specific guidance on implementing those legal obligations within the acquisition and deployment of AI-enabled weapon systems.  However, the DoD's ethical principles, while consistent with the legal obligations, serve to reinforce normative standards, and provide broad guidance to commanders and operators deciding when and how to use AI systems in combat, as well as how acquisition officials design, develop, test and evaluate the systems to ensure such systems are reliable and governable throughout their lifecycle.  As such, the ethical principles may restrict the use of AI systems further than the law requires and may require further implementing guidance as new AI capabilities emerge.



topic.  Every interviewee discussed IP as a consideration for their firm on any contract. However, most interviewees were more comfortable with and favoured the commercial market's IP practice, particularly in how the commercial framework can cope with the idiosyncrasies of AI development and deployment.  Most of the interviewees described IP as 'the most important' consideration, explaining their IP was their firm's 'differentiator' or 'competitive edge.'  Many of the interviewees explained that their IP was their most important asset and set up their business model with protecting IP as the primary goal.

The primary concerns regarding IP and defence contracts include the following: lack of understanding of the unique IP licensing framework in defence contracts; how the government uses contractor data; ability to negotiate terms; and whether a DoD contract can yield a return on investment with the revenue expected from licensing their technology to the government. Several interviewees from start-ups expressed that the DoD IP law and practice is so foreign to their firms that it had to hire expensive experts to help them understand the data rights clauses. One CEO stated that their firm avoided working with the DoD specifically because DoD contracts are so different from the commercial market that there are 'pitfalls you don't even think about,' concluding that the risk of inadvertently licensing a trade secret was too great.  Other interviewees stated that IP is so important to their firms and the risk to their business poised by leaking their IP is so great with the DoD's IP practice that the contract opportunity offered by the DoD must 'be a pretty big pot of gold' to make the opportunity attractive enough for a firm to spend the resources necessary to learn and understand the unique IP law and practice well enough to accept the risk.

Interviewees from larger firms also stated they were uncomfortable with the DoD IP laws based on how different they are from the commercial licensing practice.  The DFARS framework



for data rights hinges on whether the development of the product occurred under a government-funded contract.[643]  However, as some interviewees pointed out, machine learning programs are developed when the algorithm is trained with data.  While new algorithms developed under a contract are relatively clear to determine the government's rights under the DFARS, interviewees explained that even their commercial-off-the-shelf models will evolve based on the input of the training data, often provided by the government.  In commercial contracts, software licensing is generally more specific as to rights of the license holder.  The fact that the DoD treats commercial computer software and non-commercial computer software differently[644] adds to the confusion.

The vast difference in law and practice means the firms must decide whether to enter into a deal with the DoD they do not fully understand (at least not without investing in personnel with specific experience on DoD licensing rights) or do not believe will yield a return on investment. Several interviewees stated that they would not enter into a DoD contract that required their firm to agree the DoD would receive 'unlimited rights' or 'government purpose rights,'[645] with one interviewee explaining that those clauses 'would absolutely be a deal breaker... [unlimited rights] would be extremely problematic and not worth it for our company.'  Another interviewee explained the question of IP rights is 'a contracting question, so it depends a lot on the money,' adding that small contracts with the DoD that asserts rights inconsistent with commercial licensing is not worth their firm's efforts.

The DFARS terms determine IP rights very differently than the commercial market, as the determining factor of how rights are assigned under the DFARS is whether the creation of the technology was funded by the government.  One interviewee explained that the default process of data rights assignment under traditional procurement contracts appeared to be 'a lack of sophistication on the buyer's part…[that] can be viewed as somebody's sort of playing coy, and trying to create a window where they'll have rights on something they shouldn't have rights on, which creates a significant distrust.'  However, firms expressed that negotiating IP terms and license rights would be attractive.  Because IP is viewed as critical, if the DoD insists on terms that firms perceive as overreach, they may decide to avoid working with the DoD altogether, as a few interviewees warned.  Thus, the DoD should take care to consider how much IP it needs, align its contracting strategy with those needs, and be prepared to negotiate.[646]

---

[643] *DFARS* §§ 227.7103-4(a), 227.7203-4(a).
[644] See *DFARS* § 252.227-7014; 48 CFR § 52.227-19; 48 CFR § 12.212.
[645] See *DFARS* § 252.227-7013(a)(13), (16).

A large majority of the interviewees discussed the importance of IP in the context of developing AI and deploying machine learning to DoD use cases.  These interviewees explained that as important as IP is generally to the technology industry, IP is particularly critical in AI development, where the product is the IP, and the decision to grant IP rights to the government may be 'existential' for their company.  Additionally, as several interviewees explained, machine learning presents additional complexity that renders traditional data rights provisions clumsy, if not obsolete.  As a commercial AI firm builds a machine learning algorithm, the DoD supplies the relevant dataset.  While the firm owns the IP to the algorithm, the DoD owns their data.  However, as the algorithm trains on the dataset, it learns and adapts, thereby changing and improving upon the untrained model.  The dataset becomes more useful with the output as insight into patterns or predictions becomes possible. Thus, both the commercial AI firm and DoD gain IP that they did not have outside of contract performance. While the interviewees explained that commercial practices commonly offer the customer a license for the software and output for a limited duration and number of users, the DoD could assert unlimited or government purpose rights into the output or trained model that was developed with government funding under the contract. The vast majority of the interviewees were uncomfortable working with any customer that would have the ability to use their IP in the manner that these data rights afford.[647] Although most interviewees acknowledged that they understood the government has different requirements than commercial customers and may require different or additional rights to the software and technical data, they explained the contracting officers typically did not understand what the requirement was for and asked for more rights than reasonably necessary.[648]

---

[646] The DoD's IP strategy should be part of its acquisition strategy at the outset of the contracting process.  While the interviewees indicated a perception the DoD overreaches for IP rights, in some cases, the DoD may require additional license rights to access and use code to ensure the ability to integrate the code in existing platforms, ensure cybersecurity, or maintain the software.  Additionally, the DoD may require access to a model's training data to ensure risks of bias or responsibility are mitigated.  If that is the case, the interviewees generally accepted that such requirements could change their firm's opposition of providing data rights to the DoD, though they acknowledged that such license would require fair compensation and special protections from release to third parties.

Some interviewees discussed other ways that AI and traditional IP, even in a commercial law context, are incompatible. Because of the immense importance of IP to AI, many commercial AI firms will not file patents on inventions, because patents make public an AI firm's 'business advantage.' Rather, many firms will keep their innovation protected as a trade secret. Because traditional procurement contracts must comply with the Bayh-Dole Act,[649] contractors may, under certain circumstances, be legally obligated to apply for a patent on inventions developed under DoD contract. This requirement caused one interviewee to express their firm will not perform any development for the government as the release of any of their IP will erase their firm's value in the eyes of potential investors.

---

[647] Unlimited rights authorize the government to 'to use, modify, reproduce, perform, display, release, or disclose technical data in whole or in part, in any manner and for any purpose whatsoever, and to have or authorize others to do so': *DFARS* § 252.227-7013(a)(16). For computer software, unlimited rights mean 'rights to use, modify, reproduce, release, perform, display, or disclose computer software or computer software documentation in whole or in part, in any manner and for any purpose whatsoever, and to have or authorize others to do so': *DFARS* § 252.227-7014(a)(16). Government purpose rights, which the DoD is entitled to when it funds any aspect of the development of the software, including providing data at its own expense, provides the same rights as the 'unlimited rights' clause within the government, but the release and disclosure is limited to outside the government for government purposes: *DFARS* § 252.227-7014(a)(12).
[648] While DoD policy under DoDI 5010.44 and guidance in *DFARS* Part 227 direct the DoD to only acquire the technical data and computer software rights necessary to satisfy agency needs, defence officials may attempt to protect themselves from uncertainty and hopes of avoiding vendor-lock by requiring too many rights. Eric Lofgren, 'GMU Playbook: Striking the Balance with Intellectual Property', *Acquisition Talk* (Blog Post, 11 November 2021) <https://acquisitiontalk.com/2021/11/gmu-playbook-striking-the-balance-with-intellectual-property/>.
[649] *Patent Rights in Inventions Made with Federal Assistance*, 35 USC Chapter 18 (2018).

Beyond the perception of overreach, most interviewees indicated that the traditional IP clauses are too confusing, some even stating that for emerging technology firms, 'the burden to learn the government IP system, including everything that comes along with it, that is just a hill too high.' The overwhelming majority of interviewees agreed that negotiated IP terms and clauses, as is possible in an OT agreement, align with industry norms and are more likely to provide the flexibility required in AI deals than the DFARS licensing framework.

Several interviewees pointed out that AI procurement in the commercial market typically consists of two negotiations, one for the product or service, and another for the license agreement, each with various pricing and terms. This commercial practice differed significantly from their experience in contracting with the DoD where contracting officers presume rights to data are included with the procurement of the product or service. One executive described what they believed would be a best practice by negotiating IP terms in commercial terms rather than DFARS terms, though expressed frustration that this practice remained elusive in their dealings with the DoD. As the interviewee explained:

> If I could find a government customer who is sophisticated enough to negotiate the elements of the standard clauses, then that's something I would really consider, because now I know I have a sophisticated party on the other side, they understand that we were actually negotiating terms of a license… and understanding that we can really customise [the license agreement]. Now I can talk about revocable, non-revocable and exclusive, non-exclusive. What time do you get this piece? There's just an infinite number of options for us to come up with a really great business deal. I've never found that in the DoD.



The consensus of the interviewees on IP concerns for AI companies looking to work with the DoD was surmised by one executive from a traditional defence contractor:

> IP is my number one concern…because the DoD struggles with IP. For a commercial business, IP is everything. My company is IP. What do people want to buy from us? Not software in a box. They want to buy access through a license or through a subscription, access to our data pool, and how we filter and leverage and how we apply algorithms to the data pool to give them the displayed data that they want for their use. So IP is everything. Honestly, you could take away people and facilities and everything from a company, and all the value would still be there, because it's the data pool and how we're able to access the data pool. When I see a defence requirement, and I've seen two of them recently that said, 'If you're not going to give us government purpose rights on day one, don't bid.' Or 'If you're not going to give us unlimited rights on day one, don't bid.' Hey, easy, that's a no bid. I don't need to bid that… If there's an opportunity to negotiate something different, then that would open that requirement back to a potential bid. Because now it's, 'Tell me exactly what data rights you want and when you want them, and maybe we can work something up.'

While the DFARS clauses and typical IP practice in the DoD were nearly universally derided by the interviewees, some interviewees expressed there are some positives about working with the DoD regarding IP. As important as the interviewees believe their IP is to their firms, protecting IP is of paramount importance. A few interviewees applauded the recent trend of the DoD to secure data rights and software in escrow. This practice protects the contractor's IP, but ensures the government has access to the IP if required for sustainment of an obsolete program or if the contractor goes out of business before the end of the program's lifecycle. Some interviewees expressed that they trust the DoD to protect their IP more than they trust their commercial customers, especially in a business-to-business transaction. Though this trust is grounded in their trust of the personnel involved in the DoD transaction rather than in the contract itself, there is alignment in the interviewees' perception of the DoD as a trustworthy



customer across the themes discussed during the interviews.  This alignment indicates a major finding that the DoD is attractive to commercial AI firms because it is trustworthy.





The major findings from the survey and interviews provide insight to help understand the primary research question: why do commercial AI firms decide to contract with the DoD? The survey was designed to determine whether commercial AI firms perceived the DoD as an attractive customer, testing the hypotheses developed from social exchange theory research.[650] The research indicated that suppliers in competitive markets can decide which customers to work with; these alternatives require customers to maximise their attractiveness.[651] By increasing the expected value of the exchange, customers can attract the best suppliers — however, attraction is subjective so the customer must understand the supplier's values and align its action accordingly.[652] The surveys identified these perceptions and demonstrated the subjectivity of suppliers' attraction to various contract and buyer attributes. However, several themes and prevalent preferences emerged from the survey data. The subsequent interviews helped explain the survey data and provided a description of the perception, opinions, and preferences of commercial AI firms regarding the DoD as a customer. This section deconstructs the findings from the previous section and provides an interpretation of what the findings mean and synthesis of the integrated data. Returning to the sub-questions advanced in Chapter I, this synthesis provides narrative explanations drawn from the data analysis. This will enable the DoD to assess its choice of contract law framework to optimise its attractiveness to commercial AI firms.

## A *Analysis of Major Thematic Findings*

The surveys and interviews provided new data that illuminated several themes about the DoD's contract law and procurement practice. With a better understanding of the perspective of commercial AI firms regarding the DoD as a customer, these themes can be used to identify and leverage best practices in contract law and procurement practice to better align with commercial AI firms' preferences in contracting. The survey data provided an overview of the sample's opinions, revealing what perceptions and preferences the participants held regarding contracts and the DoD, including a comparison of DoD contract law and practice with commercial practices. The subsequent interviews developed the understanding of why the participants held those views. The integration of the surveys, interviews, and literature reveal findings relevant to the research questions. These major findings are discussed below.

---

[650] See above Chapter II(G).
[651] See Glas (n 465) 96–102.
[652] See Nollet, Rebolledo and Popel (n 478) 1188.

*Major Finding #1: The contract formation process is an important consideration for most commercial AI firms in determining whether to contract with the DoD.*

From concerns about timeliness of the contract award to protecting intellectual property, commercial AI firms view the contracting and procurement process as a major factor in their decision to contract with the DoD. Most commercial AI firms find this decision complex. While the DoD can offer large contracts and they can result in challenging and exciting work, there are no guarantees of success. Whether the contracting process is timely, enables collaboration, and permits flexibility that accounts for both the unique nature of AI and the firm's business model are key indicators of an attractive contract. The ability to negotiate terms and conditions, especially IP licensing, as well as the ability to collaborate with end-users towards developing a solution all play an important role in the decision to contract with the DoD for many commercial AI firms.

The formation of a relationship requires attraction, which can be determined through the perceived value creation, interaction process, and the emotional response towards working with



the buyer.[653]  Commercial AI firms have alternative options to the DoD.  Thus, the DoD should

optimise its customer attractiveness by aligning its contract law and practice with preferred

contract attributes like streamlined, easy to understand competition processes, robust

communication with the end-user, flexibility to collaborate and experiment towards developing a

solution, and ability to negotiate IP protections.

*Major Finding #2: The DoD's unique contract law and procurement practices pose significant*

*barriers to entry on commercial AI firms.*

The traditional FAR-based procurement practice appears to be incompatible with AI

development.  The FAR requires the acquisition to start at the requirement for a need, with the

solution drafted not by the developer, but by the customer.  This practice is contrary to

commercial AI firms' preferences to working with the end-users to solve problems.  Machine

learning is best used for problems where the solution is not yet known, such as making

predictions based on patterns within the data.  Detailed specifications presume knowledge of the

best solution.  Additionally, the unique terminology and required clauses that often differ from

commercial contracting practices are perceived as significant hurdles to new entrants to the

defence market.[654]  Complex accounting systems, customer-focused IP terms, and highly

technical proposal requirements often require firms to hire specialist accountants, lawyers, and

consultants to prepare a proposal and perform a contract for the DoD.  The complexity of process

and regulatory burdens results in additional overhead costs that do not apply to commercial

contracts.  As such, commercial AI firms must consider the difficulty and expense of competing for and performance of a contract when making the choice between a DoD contract and a commercial contract.

---

[653] See Hüttinger, Schiele and Veldman (n 469) 1197–8.
[654] See generally 48 CFR Part 52.

This finding supports the hypotheses that commercial AI firms prefer competing for contracts unencumbered by significant barriers to entry and which require limited resources to compete, follow a straightforward process, and permit the ability to negotiate terms and conditions.[655]  The value creation and interaction process affect the perception of customer attractiveness.[656]  This research suggests the DoD would appear more attractive to commercial AI firms if it lowered the barriers to entry on commercial AI firms, especially as AI innovation is largely occurring outside the traditional defence industrial base.

*Major Finding #3: The barriers to entry posed by DoD contract law and practice impacts competition for AI systems.*

Despite the DoD's policy of maximising competition, and the legal requirements under CICA for traditional procurement contracts, the complexity of the procurement process was frequently cited as a factor in influencing their firm's decision to not engage with the DoD.  The unique contract terms in DoD contracts, numerous regulatory requirements, and lengthy and confusing performance specifications are costly and risky for AI firms.  The value of the contract award is diminished by hiring expertise specific to the defence procurement process and obtaining certifications to ensure compliance with the numerous regulations applying to defence contractors.

---

[655] See above Chapter II(G).
[656] See Hüttinger, Schiele and Veldman (n 469) 1197–8.



One reason a customer is perceived to be unattractive is if the process to work with the customer is too complex, costly, or slow.[657]  These attributes were associated with the DoD's traditional FAR-based approach that is requirements-driven rather than problem-driven. Although CICA was designed to keep procurement fair and transparent, the complexity and extensive timeline for contracts awarded under the FAR are largely incompatible with the high-speed schedules that start-ups must achieve to secure additional funding.  Furthermore, due to the rapid rate of technology advancement for AI applications, a lengthy contract process can make the original solicitation of requirements obsolete by the time the contract is awarded.  This phenomenon affects the DoD, which is left with old technology on a new contract, as well as the vendor, which may have to support multiple versions of software — a legacy program for the DoD and a newer version for its commercial customers.

*Major Finding #4: Traditional DoD procurement contracts are incompatible with the unique characteristics of AI.*

Developing and deploying AI-enabled programs rarely permits plug-in solutions; AI developers state it is essential to start with a prototype, pilot, minimum viable product, or proof of concept before integrating or scaling any new use case.[658]  New use cases require shared data, trust, collaboration, and flexibility to experiment and iterate, with a constant cycle of testing and evaluation.  It is unrealistic to start an AI project with defined specifications drafted by the DoD.  Because commercial AI firms often tie their business model to their IP, the ability to negotiate IP clauses and protect their proprietary information is critically important.  Additionally, the advancement of technology far out-paces traditional procurement lead times. Finally, many commercial AI firms have limited experience working with the DoD and its complex procurement rules and practices.

---

[657] See Hüttinger, Schiele and Schröer (n 476) 699–700.
[658] See above Figure 27: Prototyping is a precursor to integration and scaling new AI use cases; nn 197–204 and accompanying text.

Under the FAR, requirements are drafted and published to industry via the Government point of entry.[659] Once the requirements are released, communication between the DoD and potential contractors is limited,[660] and further restricted after acceptance of offers.[661] After award, performance of the contract must adhere to the specifications originally drafted and modifications must be performed in the scope of the original contract, and only if agreed to by the contracting officer.[662] Given the level of specificity required to execute and perform a FAR-based contract, the DoD must be very certain of exactly what is needed. Such specificity limits the potential of AI development and, according to many interviewees and survey respondents, is less attractive to commercial AI firms.

As hypothesised, OT agreements appear better aligned with the technical considerations of the AI lifecycle as well as industry preferences.[663] Moreover, commercial AI firms prefer flexibility in contract negotiation and performance, especially to experiment and iterate in developing an AI system.[664]

*Major Finding #5: Commercial AI firms prefer flexible, negotiation-driven, collaborative contracts to rigid, regulation-driven contracts.*

---

[659] See 48 CFR 5.003.
[660] 48 CFR § 15.201.
[661] 48 CFR § 15.306(b).
[662] See 48 CFR § 43.102.
[663] See above Chapter II(G).
[664] Ibid.



The research participants strongly prefer collaborative efforts with their customers than more transactional, rule-based contracts. While social exchange theory research on industries, such as manufacturing or logistics, has found suppliers prefer collaborative relationships,[665] this research is the first to use social exchange theory to understand attractiveness in the context of AI development. This finding suggests the unique way AI applications are developed and deployed make this technology inherently and exceptionally collaborative. Most interviewees saw the relationships and collaboration between their firms and the DoD as motivating to their employees. Contracts that did not include collaboration with the DoD, either in formation of the contract or in performance, were viewed less favourably than those contracts that transcended the transaction. Both the survey and interviews strongly indicated that developing and deploying AI capabilities required strong collaboration and communication between the firm and customer. To truly understand the problems, and ensure the AI model was valid and effective, frequent communication and sharing of data is required. Trust is a concept that was frequently mentioned by the interviewees as necessary for a successful relationship. Purely transactional relationships are limited in their utility for AI applications.

Because traditional procurement contracts begin with a statement of work that meets the pre-established requirements, the prospective contractors are not part of the process of understanding the problem and developing a solution. During this phase, AI firms prefer to discuss the problem with the end-user, analyse the dataset and architecture, negotiate data rights and licenses, and plan how to structure, train, and model the data. Traditional specifications in a request for proposal are generally more directive than collaborative during this process. According to the research participants, open topics, such as commercial solutions openings or request for prototype proposals, are generally better suited for commercial AI firms to collaborate with the end-users and develop a bespoke approach and solution to the problem the DoD is interested in solving. Additionally, contracts and selection processes that permit down-selecting options, iterative prototypes and experimentation were favoured over opaque source selection methods with limited, or non-existent, communication between the potential contractor and end-user for months at a time. The interviewees explained this preference is a result of the advantages collaboration and communication provide for both the business and the technology as AI is developed with human and machine interfacing, and the developers can better train the algorithms through meaningful and consistent interaction with the end-user.

---

[665] See Glas (n 465) 96.

*Major Finding #6: IP is a critical contract consideration for commercial AI firms.*

In the interviews, the participants universally agreed that IP was critically important to their firm, many claiming IP was their most important asset. Several interviewees remarked that the decision to give certain data rights is 'existential' to their firm's continued business. The interviewees opined that the nature of the development of AI models and methods presents unique challenges in protecting IP. Nearly all interviewees believed IP clauses needed to be specifically tailored and specially negotiated. There is a belief amongst the AI industry that current IP law is incompatible with AI as the input of data, possibly owned by the customer, third party, open source, or combination thereof, and use of that data to develop the model are more complex and nuanced than traditional IP constructs. Thus, care to negotiate case-specific IP agreements is required, though challenging and potentially risky to all parties to the contract.

This finding is consistent with social exchange theory research which describe drivers of customer attractiveness as economic and resource factors that include compatibility, trust, and



fairness.[666]  Additionally, this finding supports the hypothesis that commercial AI firms prefer negotiating intellectual property terms that they understand and are attracted to opportunities that fairly compensate their efforts in developing, ensure adequate protections and account for the unique way AI applications are developed.  The interviewees explained their concerns about IP licensing centred on the following questions: whether the licensing framework was compatible with the complexity of the AI development lifecycle; whether they could trust the DoD to protect their valuable IP; and whether they would receive fair compensation for the value of their IP. Because the DFARS licensing framework appears too blunt to account for the nuances of data sharing and collaborative development, OT agreements appear well suited to align with the both the technical considerations and industry preferences to better attract commercial AI firms.

*Major Finding #7: Most commercial AI firms view the DoD as an attractive customer.*

Commercial AI firms largely view the DoD as a potential customer that can positively impact their business.  Most firms surveyed indicated they viewed the DoD favourably, even when compared with commercial customers.  Very few survey respondents reported that the DoD was not an attractive customer to their company.  Explanations for this perception offered by the interviewees and inferred from the survey data include the view that the DoD is a large and influential customer, is trustworthy, and offers the opportunity to work on meaningful and challenging projects.  Many firms acknowledged their decision to engage with the DoD was due in part to work on challenging projects, contributing to the United States' national security,[667] and the potential revenue that can be generated as a direct or indirect result of performing a DoD contract.[668]

---

[666] See Hüttinger, Schiele and Veldman (n 469) 1197–8.
[667] The desire to work on challenging projects or contribute in a meaningful way, including to national security, was discussed by many of the interviewees as a distinguishing characteristic of the DoD compared to alternative customers.  The attraction to such opportunities can be explained by the emotional response of the supplier.  Ibid.

Customer attractiveness can occur through the presence of myriad factors.[669] A relationship can be formed if the supplier perceives the buyer can maximise the value of the relationship; however, the relationship succeeds if value is created, the interaction is collaborative, and the relationship provokes an emotional response.[670] Thus, while profit generation is important to commercial AI firms, so is the ability to collaborate with the military on developing AI solutions that contribute to national security.

However, the research sample indicated that although it perceives the DoD as an attractive customer, such perception is despite its largely negative perception of the DoD's contract law and practice. This finding indicates the DoD can optimise its attractiveness by aligning its contract law requirements with commercial preferences and AI considerations.

### B *Synthesis of Major Thematic Findings*

Collectively, the findings from the surveys and interviews help develop a holistic understanding of the research problem. Although most of the research sample perceived the DoD as an attractive customer, most identified significant problems with the DoD's contract law and practice. The findings suggest that traditional contract law is perceived as too burdensome for some commercial AI firms to overcome, which could result in the DoD losing access to innovation and competition for contracts of AI-enabled capabilities. However, alternative contract law, such as other transaction agreements and commercial solutions openings, align well with commercial AI firms' preferences as well as the AI development and deployment lifecycle. The synthesis of these findings reveals that other transaction authority provides a legal framework that is better able to attract commercial AI firms than FAR-based contracts which often legally require the presence of contract attributes that negatively affect the DoD's perceived attractiveness.

---

[668] Several interviewees explained the profit margins on a DoD contract may not favourably compare to commercial contracts but providing technology to the DoD can be leveraged as a selling point for business and consumer applications due to the DoD's credibility as a technology user.
[669] See Nollet, Rebolledo and Popel (n 478) 1188–9.
[670] See Hüttinger, Schiele and Veldman (n 469) 1197–8.

The synthesis considers the collective findings through the lens of social exchange theory and the legal constructs underlying the issues presented in the data. Using social exchange theory as a theoretical lens, this research hypothesised that commercial AI firms would seek to enter into new associations and maintain old ones because they expect doing so will be rewarding, both economically and socially.[671] The concept of customer attractiveness reflects the social exchange theory premise that expected outcomes of alternative exchanges are compared; the more resources a party must expend to be involved in an exchange, the less valuable that relationship becomes.[672] Exchanges that are perceived to be cooperative, where both parties expect to engage in behaviours that benefit mutual goals, positively influence attractiveness.[673] Thus, this research hypothesised commercial AI firms would be attracted to exchanges that were expected to require limited resources, such as cost, time or effort, and contribute to achieving mutual goals, such as collaboration, communication, and flexibility to innovate.[674] The research findings, described above in Chapter IV, indicate social exchange theory can explain commercial AI firms' perceptions about the DoD as a customer and contract and exchange preferences.

---

[671] Lambe, Wittman and Spekman (n 14) 6 (explaining that while monetary advantages are important in exchanges, the most important benefits involve social rewards, such as emotional satisfaction, values, pursuit of personal advantage, social approval and respect).
[672] See ibid 8.
[673] See ibid 23.
[674] See above Chapter II(G).



The findings provide an understanding of what attracts commercial AI firms to the DoD. This understanding combined with an analysis of how the two contract law frameworks available to the DoD acquiring AI-enabled capabilities align with those preferences suggests that OT authority can better attract commercial AI firms than traditional contracting practice. This insight allows for the identification of best practices for the DoD to leverage existing law to attract commercial AI firms and reveal areas that could benefit from directed reform.

This research pursued a better understanding of why commercial AI firms decide to contract with the DoD. This question is important given the role advancements in AI-enabled technologies will play in the United States' defence strategy in its competition with China, and the DoD's reliance on commercial industry to push the envelope of innovation to meet its strategic objectives. The findings identified above indicated that commercial AI firms decide to contract with the DoD because many of those firms perceive the DoD to be an attractive customer. This attraction is due to the perceived value of a DoD contract, the expectation that performing a DoD contract will provide direct (profit) and indirect (reputation in commercial market, ability to work with mission-oriented and trustworthy people, and ability to work on challenging and meaningful projects that may contribute to national security) benefits to their firm. However, at least some commercial AI firms decide against contracting with the DoD. This research indicates a significant reason for this decision is the serious concern that the DoD's contract law and practice do not align with their firm's preferences or technological considerations. As discussed above, these conclusions are consistent with social exchange theory. However, as will be discussed in Chapter VI, this research identified findings specific to AI development in the defence context and may not be generally applied to all defence



contracting contexts. The emerging theoretical framework can help DoD contract officials make informed decisions on their choice of law and practice to optimise customer attractiveness.

The survey and interviews were designed to gain insight into the research questions through testing hypotheses developed through the course of the literature review. The questions and hypotheses tested by this research are detailed below, with a narrative answer to the sub-questions based on the integration of the research findings with the literature.

- *Does existing contract law applicable to the DoD align with acquiring AI-enabled technologies from commercial firms?*

- *Which legal framework best aligns with the development and deployment of AI systems in the DoD?*

The DoD can contract for AI capabilities under two contract law frameworks — the FAR and OT authority. Based on this research, it appears that OT authority aligns well with acquiring AI-enabled technologies from commercial firms, whereas the traditional FAR model is less aligned with the preferences and considerations of commercial AI firms. As hypothesised, the flexibility, focus on solving a problem rather than meet pre-set requirements, and freedom from many time-consuming and costly regulations make OT agreements comparatively more attractive than FAR contracts. Additionally, the ability to freely negotiate IP licenses and iteratively develop a model through a collaborative relationship make OT authority better aligned with the technological considerations of development and deployment of AI systems.

The following sub-questions relate to how the DoD's choice of law impacts the availability of contract attributes. The synthesis of the findings and integration with the law explain that the research participants indicated preferences towards certain contract attributes;



whether such attributes are present in a contract with the DoD may depend on the choice of law applied to the effort.

- *What contract attributes do commercial AI firms prefer? Why?*

- *What unique characteristics of AI development and deployment affect the formation and performance of a contract for the DoD?*

- *How does the choice law affect the DoD's ability to contract for AI-enabled capabilities?*

This research revealed the presence or absence of contract attributes affect how commercial AI firms perceive the attractiveness of the DoD. According to some of the interviewees, some contract attributes may affect a commercial AI firm's decision to compete for a contract with the DoD. If the contract opportunity appears to be at the end of a lengthy process, such as writing a lengthy proposal with extensive source selection criteria and the possibility of the award getting delayed by protests, some commercial AI firms may avoid competition.

As hypothesised in Chapter II, commercial AI firms prefer certain contract attributes. The DoD's choice of law often impacts whether these preferred attributes are available during the contract lifecycle of competition, negotiation, and performance. The findings indicate a clear preference for streamlined competition without significant barriers to entry. While competition under the FAR is governed by CICA and typically involves written proposals with lengthy deliberations, OT agreements can adopt best practices from commercial contracts, permitting competition through demonstrations or down selects that are preferred by commercial AI firms. The research participants expressed a preference for opportunities that permit creativity and innovation rather than following strict specifications of how to perform tasks. As discussed,



FAR contracts typically define specifications that instruct the contractor how to meet the requirement.  Not only did the participants indicate this method was less attractive than exploring the problem with the customer, but such restrictive specifications also make little sense when developing AI systems.  However, OT authority permits open-ended problem statements that invite innovative solutions that can arise through communication and collaboration with the end-user.

The research participants preferred straightforward processes that align with their commercial business models and technical approaches.  They also prefer flexibility in contract negotiation and performance.  Finally, the research participants strongly prefer the ability to negotiate terms and conditions, especially IP licenses.  This preference is explained by the value commercial AI firms place on their IP and the unique technical considerations that require a nimble approach to assigning license rights.  The DFARS licensing framework was described by many interviewees as too rigid or incompatible with machine learning as the data pipeline used to train the model makes the assessment of when development occurred challenging.  Moreover, the Bayh-Dole Act that applies to FAR contracts requires the developer to patent inventions developed during performance of a government contract.[675]  Several interviewees expressed their firm's insistence on keeping their models trade secrets to protect the value of their IP.  The OT Guide encourages the DoD to negotiate IP licenses and neither the DFARS nor Bayh-Dole Act apply to OT agreements.[676]  Although such negotiation requires skill from both parties, OT authority permits the parties to collaborate towards mutual goals.

---

[675] 35 USC § 202.  While the developer retains title to the patent, the Government receives a royalty-free, irrevocable, worldwide license to practise the invention: *DFARS* § 252.227-7038.  If the developer fails to or chooses not to apply to patent the invention, the Government can claim title, limiting the ability of defence contractors from holding trade secrets: *DFARS* § 252.227-7038.  Although this deviation from the Bayh-Dole Act is legally permissible, it may be unwise for the government to encourage trade secret practice, especially if the subject invention is funded by the government.
[676] *Other Transaction Guide* (n 104) 49.



The research participants indicated a clear preference for collaboration over a strict business transaction and favoured experimentation and iteration over fixed requirements. The interviewees universally described collaborating with the customer about its problems before the contract award as important. The flexibility to iterate the solution during the performance of the contract was also unanimously acclaimed as important. Collaboration and iteration are technically achievable in traditional procurement contracts,[677] though require planning and are limited when the contract contains fixed requirements that presume the DoD already knows how to solve the given problem (which is antithetical to the reason for leveraging machine learning).[678] However, OT authority was given to the DoD by Congress specifically to build prototypes before production, and incentivising collaborative relationships and flexible contract terms to attract non-traditional defence contractors. Experimentation and iteration are legal requirements for prototype projects under OT authority.[679]

If the DoD implements best practices in streamlining the procurement process and developing collaborative efforts between the DoD and industry, it will increase its customer attractiveness, enabling it to better leverage the commercial AI innovation. Because this research suggests commercial AI firms strongly prefer commercial contracting practices yet still view the DoD as an attractive customer, the DoD can become an even more attractive customer by utilising more commercial contracting practices.  Using social exchange theory concepts as guidance, the DoD can develop a contracting process to optimise its customer attractiveness. Developing a relationship through collaboration in refining the problem to be solved is preferred to a transactional approach of requiring the performance of a set of pre-written tasks.[680] Collaborating to refine a problem as recommended in the OT Guide rather than dictating performance specifications for a preconceived solution would be more attractive to commercial AI firms.[681]  Selecting sources through interactive methods, such as challenges and demonstrations, would be more attractive than evaluating lengthy written proposals.[682] Regulations and requirements are negative indicators of customer attractiveness whereas open communication, flexibility and willingness to collaborate on solutions are positive indicators of attractiveness.[683]  Holding robust negotiations on terms and conditions, even to mirror commercial contracts or develop contract-specific clauses, would be preferred over limited discussions on price, schedule, and technical requirements based on existing contract clauses and regulations.[684]

---

[677] See 48 CFR Part 12 (provides requirements for using commercial terms to the maximum extent practicable when procuring commercial items), Part 13.5 (Simplified Acquisition Procedures permit streamlined processes when buying goods or services at small contract amounts), and § 39.103 (modular contracting allows for the acquisition of a system of information technology to be divided into several smaller acquisition increments); but see Government Accountability Office, *Agile Assessment Guide: Best Practices for Agile Adoption and Implementation* (Document No GAO-20-590G, September 2020) 17–22, explaining the challenges the DoD had in implementing agile methodologies for software development, in part due to the contracting officers requiring structured tasks and performance checks. These challenges were described by the interviewees as discussed in Chapter IV(C)(4): one interviewee explained: 'In a commercial contract, … [y]ou're touching base with the customer on a regular basis. With the DoD, … it's on a monthly basis…I'm trying to adjust my agile development cycle to be a monthly agile development cycle, which is contrary to being agile'.
[678] See Malone, Rus and Laubacher (n 185) 17–8.
[679] 10 USC § 4022(a).

Because the OT statutes authorise the DoD to conduct business like a business, the DoD can legally adopt commercial contracting practices to become a more attractive customer.  The change in practice and policy requires no change to existing law provided the DoD leverages OT authority to contract for AI technologies.  Aligning the FAR with commercial preferences and technological considerations requires additional legislative and regulation reform.

---

[680] Hüttinger, Schiele and Schröer (n 476) 700; Blonska (n 485) 74.
[681] Compare *Other Transaction Guide* (n 104) 11, with 48 CFR § 2.101.
[682] Compare *Other Transaction Guide* (n 104) 17, with 48 CFR § 15.3.
[683] Hüttinger, Schiele and Schröer (n 476) 700.
[684] Compare *Other Transaction Guide* (n 104) 17–21, with 48 CFR § 15.306.



## C *Conclusion*

The findings demonstrate that the DoD's contract law and procurement practice play a significant role in how commercial AI firms view the attractiveness of the DoD as a customer.[685] Traditional DoD contract law and regulations present significant barriers to commercial AI firms considering contracting with the DoD. Moreover, the unique attributes of the AI lifecycle and architecture present novel issues for contract formation that require collaboration and flexibility between developer and end-user. Nevertheless, there are reasons for optimism for DoD acquisition professionals. Alternative contract law processes exist in the form of other transaction authority that the DoD can leverage to better align DoD contracts with commercial AI firm preferences and improve collaboration between the AI system developer and end-user. By understanding the AI industry's perceptions, preferences, and opinions, the DoD can better identify best practices available under existing law and target areas for further reform.

While efforts were made to seek out a large and representative sample of the commercial AI industry that currently focuses on DoD-relevant technology, the findings are limited to the sample. Additionally, these findings are specific to parameters that should be considered when interpreting this research. The first is contract specificity. These findings reflect the sample's opinions of the DoD which has unique contract law and institutional attributes that may not exist or may function differently in other procurement systems. The second is national specificity. These findings indicate opinions of working with the DoD are influenced by internal preferences to support the national security strategy and are driven by a sense of purpose which may or may not exist in other countries. Finally, these findings are limited to the commercial AI industry. The findings here reflect opinions of working with the DoD relating to unique business and product development conditions found in the AI industry that may or may not be drivers in other technology sectors.

---

[685] Although it is frequently claimed that commercial AI firms are reluctant to work with the DoD, this study found no significant empirical evidence to support that claim.

These limitations notwithstanding, the research findings can assist DoD lawyers, procurement officials, policymakers, and legislators in developing best practices to attract commercial AI firms. Many of the attributes valued in a customer relationship are permitted under existing contract law. The DoD can leverage other transaction authority to create bespoke relational and collaborative efforts that start with a focus on the problem rather than a requirement. The lack of mandatory clauses and emphasis on communication and negotiation can reduce the number of DoD-unique terms and conditions and eliminate jargon that create high barriers to entry for commercial AI firms. Though OT agreements are never guaranteed to result in a contract award faster than a FAR contract — and can take months to negotiate — much of the competition and source selection processes that slow contract award under the FAR can be streamlined under OT authority. Additionally, the minimisation of compliance costs that accompany traditional FAR-based contracts reduces the cost of performance for OT agreements. These advantages can make the DoD a more attractive customer to commercial AI firms.

This research presents insight into the perceptions, preferences, and opinions of commercial AI firms that the DoD can leverage to better shape its engagement and collaboration, thus improving its mission capabilities through innovative AI solutions. Consistent with social exchange theory and the concept of customer attractiveness, commercial AI firms prefer a collaborative and relational partner to a transactional customer. OT agreements can align with the preferences and technological considerations to attract commercial AI firms that may otherwise avoid contracting with the DoD due to concerns about the DoD's traditional contract law and procurement processes. This alternative contract law framework allows the contract to



lower the barriers to entry for firms new to the defence market.  Even firms that are accustomed to FAR-based contracts acknowledge that open-ended problems that can be experimented with in an iterative and flexible way with the end-user — attributes inherent in OT agreements — are ideal for developing AI capabilities.  However, soliciting the development of an AI capability under OT authority is not a panacea to the challenges the DoD must overcome in acquiring AI capabilities for defence applications; this relatively unbounded and untested authority must be used wisely.  The contracting team, specifically the lawyers, must be sufficiently educated and trained in both OT authority and the science of AI to truly exploit the opportunities this alternative contract law framework offers.  As OT agreements are unbounded by regulation, this training and knowledge sharing of best practices is critical to mitigate exposure to unknown risks.[686]  A DoD contracting profession that understands the underlying technology of AI systems, the mindset of commercial AI firms, and the needs of the armed forces can have a transformative effect on the perception of the DoD as a customer.  Access to commercial AI innovations can help the DoD compete and defend in the AI era.

---

[686] William J Weinig, 'Other Transaction Authority: Saint or Sinner for Defense Acquisition?' (2019) 26(2) *Defense Acquisition Research Journal* 107, 109, explaining that OT authority is inherently risky, however, there is no record of defence official abusing the authority and the benefits, including speed, flexibility, and ability to attract commercial vendors, outweigh the risks.



CHAPTER VI: OPTIMAL BUYER THEORY AND RECOMMENDATIONS

The DoD, multiple Presidential administrations, Congress, and expert strategists agree that the DoD must leverage AI to meet its strategic goals.[687]  Despite this agreement, rhetoric and policies have not yielded the desired effect as the NSCAI concluded 'America is not prepared to defend or compete in the AI era.'[688]  Unlike other technology revolutions in military affairs, the DoD is a relative outsider; it needs to engage the commercial AI industry as *a* customer, not *the* customer.[689]  Moreover, much of the advancements in AI are coming from start-ups and non-traditional defence contractors.[690]  It is widely accepted that traditional DoD contracting is slow, costly, and burdensome to contractors.[691]  However, pre-existing research of the commercial AI industry's perception of the DoD as a customer is limited.  This research helps answer how the DoD can better attract commercial AI firms by developing an understanding of such firms' preferences, perceptions, and opinions about contract attributes and customer selection.  By aligning its contract law and practice with commercial preferences and the technological considerations of developing AI capabilities, the DoD can attract and leverage commercial AI innovation.

---

[687] See *NSCAI Final Report* (n 4) 159163.

[688] Ibid 1.

[689] See Gian Gentile et al, *A History of the Third Offset, 2014–2018* (RAND, 2021) 3.

[690] See *NSCAI Final Report* (n 4) 65; Mark Sullivan, 'Silicon Valley Wants to Power the U.S. War Machine', *Fast Company* (online, 1 November 2021) <https://www.fastcompany.com/90686262/silicon-valley-wants-to-power-the-u-s-war-machine>.

[691] See *NSCAI Final Report* (n 4) 65.  A recent report concludes the DoD is vastly different from a typical technology firm, exclaiming '[i]f you were to design an organization to be the exact opposite of a tech start-up, the end result would look a lot like the DoD': Melissa Flagg and Jack Corrigan, *Ending Innovation Tourism: Rethinking the U.S. Military's Approach to Emerging Technology Adoption* (Policy Brief, Center for Security and Emerging Technology, July 2021) 4 ('*Ending Innovation Tourism*').  The report explains that while technology firms 'strive to be freewheeling, fast-moving, and disruptive, the military is rigid, regimented, and risk averse' and the DoD's 'technology acquisition process is no different': *Ending Innovation Tourism* (n 691) 4.  Although this research is consistent with those conclusions, the findings from surveying and interviewing commercial AI firms about their opinions on the DoD as a customer leads to a more optimistic conclusion.



This mixed methods research collected empirical data that indicated the DoD is viewed as an attractive customer to much of the commercial AI industry.  However, the attraction is diminished by a major flaw: many of the DoD's contract and procurement practices are unpopular and potentially incompatible with the AI development process.  The rigid and transactional nature of traditional DoD contracts may bring in some commercial AI firms intrigued by the nature of the work with the DoD, but some contract attributes may drive firms away.  While traditional contract law applies 'one-size-fits-all' rules that focus on government oversight, securing IP rights, and transparency to the public, the legal framework lacks the efficiency, flexibility, trust and collaboration sought by innovative firms.[692]  This process moves too slowly compared with the speed of AI development, causing some talented engineers to leave companies engaged with the DoD or start-ups going bankrupt waiting for defence contracts.[693]  By not adapting the contracting process to the business and technological realities of commercial AI firms, the DoD risks losing out on AI advancements that could prove critical to meeting its national security objectives.[694]  The interviews with executives and business leaders explain why the DoD's traditional contract law framework is largely unpopular with commercial AI firms.  However, this research discovered a consensus opinion that the DoD can become more attractive to commercial AI firms by leveraging best practices and flexible, collaborative contracts.

This research establishes the DoD already possesses the legal tools it needs to better attract commercial AI firms.  However, the research reveals that education and training is required to maximise the DoD's ability to leverage those legal tools in a manner that both attract commercial interest and meets the DoD's objectives in leveraging AI innovation. DoD officials should receive education not only on these existing legal authorities, but also on engaging with businesses, the unique complexities and disruptive nature of AI technology, and negotiating terms and conditions outside the FAR framework. The DoD can learn from the data collected from commercial AI firms and use these insights to develop a set of practices which will form a successful relationship that better aligns with commercial AI firms' contract preferences and meets the needs of the DoD.

---

[692] See Weinig (n 686) 109–17.
[693] Heikkila (n 252).
[694] See ibid.

Contracting for AI technologies for defence applications is complex, and the research participants held many different views about the DoD as a customer, and the attractiveness of various contract attributes. Thus, it appears there is no single solution to better attract AI firms to work with the DoD. Nonetheless, of the two contract law frameworks available to the DoD to buy AI capabilities, OT authority appears better suited to the task than the FAR. The DoD and its contracting professionals will need to balance its procurement objectives and decide which objectives will enable it to fairly attract commercial AI innovation at a reasonable cost to the government while ensuring its national security mission remains paramount.

This research identifies clear preferences for certain contract law attributes and procurement practices held by commercial AI firms. The unique nature of AI fits well with the flexible, iterative process achievable through existing law in the form of OT authority. Thus, leveraging this alternative contract law framework for buying AI technologies can lead to improved perceptions of the DoD as a customer by the commercial AI industry. However, while OT agreements are inherently flexible and do not require many of the regulations found in traditional FAR contracts, some contract conditions are similar and remain important to the DoD. Likewise, the presence or absence of certain contract conditions can cause a commercial AI firm



to decide against contracting with the DoD.  The DoD can overcome the negative characteristics of its contract law and procurement practice by adopting more attractive best practices.  These include reducing the timeline to funding; increasing and improving communication throughout the contract formation process; collaborating on terms and conditions to promote mutual trust and commitment; and increasing flexibility in performance in consideration of the unique agility and fluidity of developing and deploying AI applications for new use cases.  The attractiveness of these contract attributes is a manifestation of social exchange theory — the presence of contract attributes that demonstrate the buyer is willing to enter into a relationship built on trust and collaboration to achieve mutual goals make that buyer more attractive to suppliers than a contract that does not align with the supplier's goals or diminishes the importance of the interaction.[695]

The research indicates the perception of the DoD as a customer is negatively affected by many current contract law and procurement practices.  However, the data collected helped identify which contract law and procurement practices align with commercial AI firms' preferences.  These best practices, all of which are currently authorised under existing law, include framing requirements as open-ended problems rather than predetermined specifications; providing opportunities to make oral pitches and demonstrate capabilities rather than write technical, burdensome proposals; engaging in open communication between the buyer and end-user throughout the contract lifecycle; providing the ability to freely negotiate contract terms and conditions, especially IP clauses; permitting flexibility in contract performance to enable agile and iterative development, testing and evaluation, and deployment of AI applications.  These best practices are the product of evaluating the connection of this research's findings to the

---

[695] See above nn 469–484 and accompanying text.



literature on AI technology and its role in national security, defence contract law, and social

exchange theory.  This process not only identified practical recommendations, but it also led to

the development of an overarching theory that can help defence procurement officials and

lawyers leverage the law to better attract commercial AI firms in the formation of a contract.

<p align="center">A <i>Optimal Buyer Theory</i></p>

According to the research data, the attractiveness of a defence organization to

commercial AI firms is increased when the contracting process is transparent, flexible to enable

experimentation and iteration, and communication and collaboration is high.  Conversely, the

attractiveness of a defence organization to commercial AI firms is decreased when the

contracting process is opaque, rigid, and communication and collaboration is low.  The length of

time the process takes to funding execution, the amount of defence-unique regulations required,

and the cost of compliance are negative attributes to the DoD's attractiveness as a customer.  The

access to a large market, revenue, and inference of credibility associated with providing

technology for the armed forces are positive attributes to the DoD's attractiveness as a customer.

Likewise, the offer of exciting and challenging work and the ability to support national security

attract commercial AI firms.

The DoD can maximise the perceived attractiveness of its contact opportunities by

making a series of decisions.  If the goal is to attract the greatest quality of commercial AI firms,

then decisions should be made to maximise the likelihood that commercial AI firms compete for

a given contract opportunity.[696]  The optimal buyer theory posits that collaborative efforts,

characterised by transparent communication, flexibility to experiment and iterate, negotiated terms that benefit all parties, and purpose-oriented efforts are most attractive to commercial AI firms and are best aligned with developing and deploying AI-enabled technologies. The optimal buyer is not transactional; it is a partner and collaborator engaged in a joint venture where the goals are mutual. This requires trust and trust is built not on the paper of a contract but in the relationships built through communication and empathy. Thus, while the choice of contract law framework impacts the attractiveness of the DOD to commercial AI firms, the practice and engagement must also align with business models and AI development best practices to optimise the DoD's ability to leverage the commercial market.

---

[696] Social exchange theory research indicates that firms who effectively attract suppliers to compete for contract opportunities and satisfy those suppliers during contract performance can access, develop, and utilize strategic resources to gain competitive advantages.  Schiele et al (n 468) 137.

By optimising its attractiveness to commercial AI firms, the DoD can increase competition and ensure access to defence relevant AI applications. The decision tree below depicts the optimal buyer theory, derived from social exchange theory in the context of the DoD contracting for AI capabilities, on how the contract attributes affect the perceived customer attractiveness.[697]

---

[697] A decision tree is a representation of a function that maps a vector of attribute values to a single output value, or 'decision.' Decision trees are a type of classification algorithm used in machine learning: see Russell and Norvig (n 18) 657–65.



*Figure 41: Optimal Buyer Theory (Perceived Customer Attractiveness Model)*

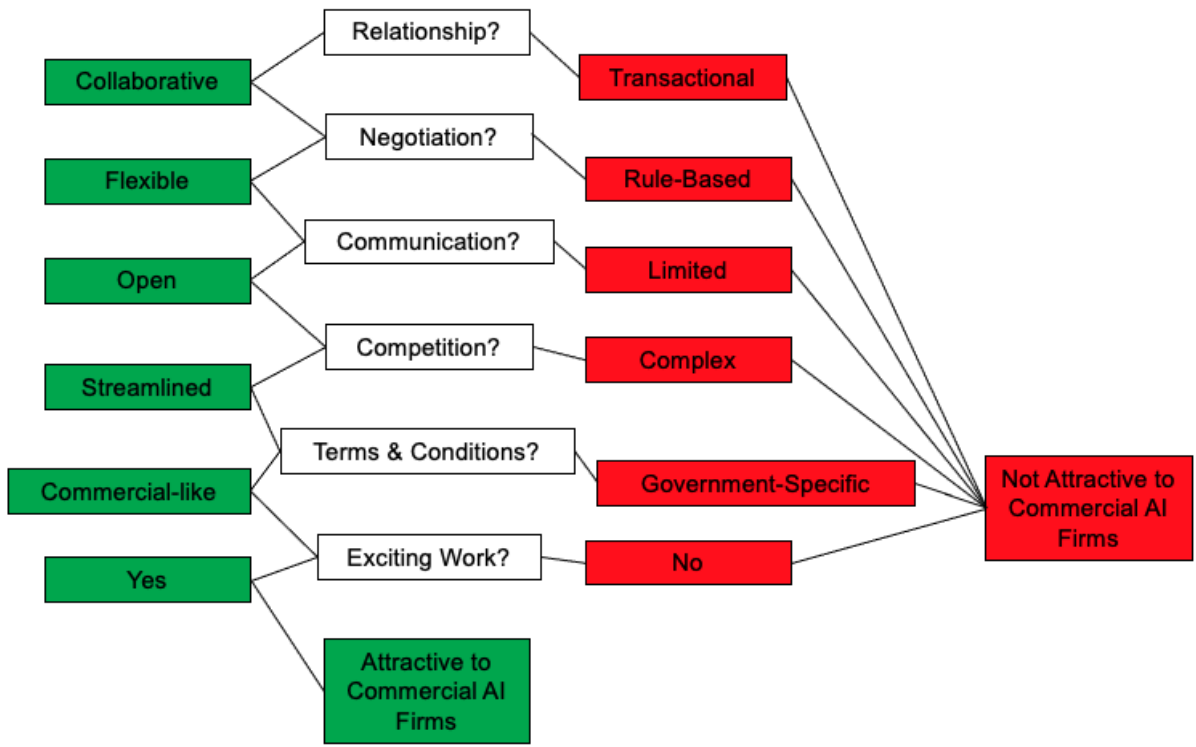

Based on customer attractiveness literature, and grounded in the research data, the decision nodes represent the available classes or contract attributes that the data represents. Each split represents a decision that affects the contract's attractiveness to a generalisation of the commercial AI firms that participated in this research.

The **first split** is based on the nature of the relationship formed by the contract, with relationships that are more transactional being identified as not attractive to most commercial AI firms. Those that are more collaborative are classed as potentially attractive, although the attractiveness of a collaborative relationship is dependent on additional contract attributes.

The **second split** is based on whether the contract is flexible and negotiable, which is potentially attractive, or is rigid and driven by regulations, which is not attractive.



The **third split** is based on whether there is open and effective communication between the commercial AI firm and the end-user, which is attractive, or is limited in time or scope, which is not attractive.

The **fourth split** is based on the competition process.  Streamlined processes that have limited barriers to entry and permit evaluation and selection by demonstrations are attractive and lengthy evaluation cycles with high barriers to entry based on written proposals are not attractive to commercial AI firms.

The **fifth split** is based on whether the terms and conditions are unique government-centric terms, which is not attractive, or commercial-like contracting terms that make performance less complex and consistent with a commercial AI firm's business model.

The **sixth split** is based on whether the work is challenging, interesting, or important.

The most attractive contracts to commercial AI firms are classified on the left side of the decision tree.  DoD organizations can use this theory to structure their contracts to maximise their attractiveness to commercial AI firms.  By focusing on the formation and structure of the legal relationship as manifested in the contract to maximise the attractiveness to the seller, this theory helps explain how a buyer can optimise its ability to form relationships with the best vendors and developers of AI-enabled technology.  By focusing on the contract law options available and understanding how the seller values those options, the potential customer can make decisions throughout the procurement process and contract formation that will positively affect its attractiveness to commercial AI firms, thus becoming the 'optimal buyer'.

The optimal buyer theory, a specific and contextual application of social exchange theory grounded in commercial AI firms' opinions, perceptions, and preferences about the DoD as a customer, is operationalised in the decision tree below.  This decision tree depicts the optimal



procurement process and choice of contract law that results in the most attractive buyer — determining which procurement approach or legal authority to use can be viewed as a classification problem, where the 'output' is categorical.[698]  Here the categories are presented as decisions between traditional procurement contract law and alternative contract law.  The principal decision that affects all others is whether the FAR or the OT statutes should govern the contract; thus, this decision is the root node of this decision tree.

This decision tree aims to use the empirical evidence collected in this research to illustrate the general perceptions of commercial AI firms to help DoD procurement officials and contract lawyers understand the impact of their choices between various options in engaging with industry and the formation of the contract.  The model is simplistic in that it may overfit the limitations of this dataset, but it is a heuristic that illustrates the findings and resulting concepts and theory.[699]

While this model can be a useful tool to understand how the choice of law affects the overall attractiveness of a contract, this automated approach can result in overreliance on past data without factoring in the preferences of potential vendors for a particular contract.  Thus, procurement officials should understand the findings from this research indicate a general preference of the industry as a whole; communication, collaboration, and negotiation remain critical antecedents to optimising contract formation for AI capabilities.

The model offered here is intended to demonstrate how the optimal buyer theory is applied to forming a DoD contract for AI capabilities.  Though this model reflects the relative strength of the commercial AI industry's preference of one contracting attribute over an alternative, practitioners must still assess the goals of the DoD and potential vendors to choose the correct path. A critical feature of optimal buyer theory is that each contract and interaction between the DoD and a commercial AI firm requires the contracting officials to think through the complex and potentially competing priorities of the DoD:[700] the DoD needs to leverage commercial innovation to compete and defend;[701] many commercial AI firms lack experience with the DoD's unique contract law and procedures;[702] commercial AI firms have many potential buyers;[703] the DoD must balance its procurement objectives with the need to optimise its attractiveness to commercial AI firms;[704] the traditional FAR-based contract law framework is largely unpopular with commercial AI firms and potentially incompatible with the AI development lifecycle.[705]

---

[698] See ibid 657–8.

[699] Future research may develop a probabilistic sample, enabling the creation of statistical models to predict an optimal procurement strategy for a contract.  Machine learning can be a useful tool in creating such a model.

In the model, each leaf node corresponds to a decision variable by the contracting agency. The branches represent an outcome of the choice, which either opens or closes the options for the next node or choice.[706] Each choice impacts the attractiveness of the contract and therefore affects the attractiveness of the customer. Each branch represents the outcome of the test.[707] Each node is weighted based on the relative strength of the commercial AI industry's preference for that particular contract attribute or process.[708] Each leaf node represents a classification label of the decision taken by the DoD. Each decision by the DoD can affect the degree of desirability of the two different outcomes as perceived by commercial AI firms.

---

[700] Social exchange theory research suggests the value of the exchange changes with each interaction: Hüttinger, Schiele and Veldman (n 469) 1197. Optimal buyer theory posits the buyer must consider the impact each decision within an exchange for its effect on perceived customer attractiveness as well as how losing access to that supplier can affect the buyer's ability to meets its objectives.
[701] See *2018 NDS* (n 211) 3–4; *NSCAI Final Report* (n 4) 65–72.
[702] See *NSCAI Interim Report* (n 180) 60.
[703] See Arnold, Rahkovsky and Huang (n 242) 24.
[704] See *Section 809 Roadmap* (n 341) 1.
[705] See above Chapter IV.
[706] See Russell and Norvig (n 18) 657–8.
[707] Ibid.
[708] See above Chapter IV.

The research conducted for this study provides insight into how commercial AI firms prefer certain outcomes or are indifferent to a particular outcome. Strong preferences of an outcome are more likely to lead to a decision to compete or not compete for a contract opportunity. Although commercial AI firms may be indifferent to a particular decision, the decision made at one node can affect the availability of a preferred outcome at a subsequent decision node. The choice of law can affect whether there is a choice at all at several decision nodes.

While potential vendors may decide to compete for the contract opportunity even if the contracting agency makes the least preferred decision at each decision node, the optimal buyer theory posits the buyer should seek to maximise the attractiveness of a contract. Thus, the optimal buyer theory is a utility function where the maximum utility to the buyer is the ability to attract the best vendor and form a contractual relationship.[709] An attractive contract increases the likelihood that a commercial AI firm would compete for a contract, providing the DoD with access to better capabilities and the ability to negotiate the price. Below is the optimal buyer theory decision tree as applied to the DoD's contract law frameworks.

---

*Figure 42: Optimal Buyer Theory Conceptual Framework (DoD Contract Law)*

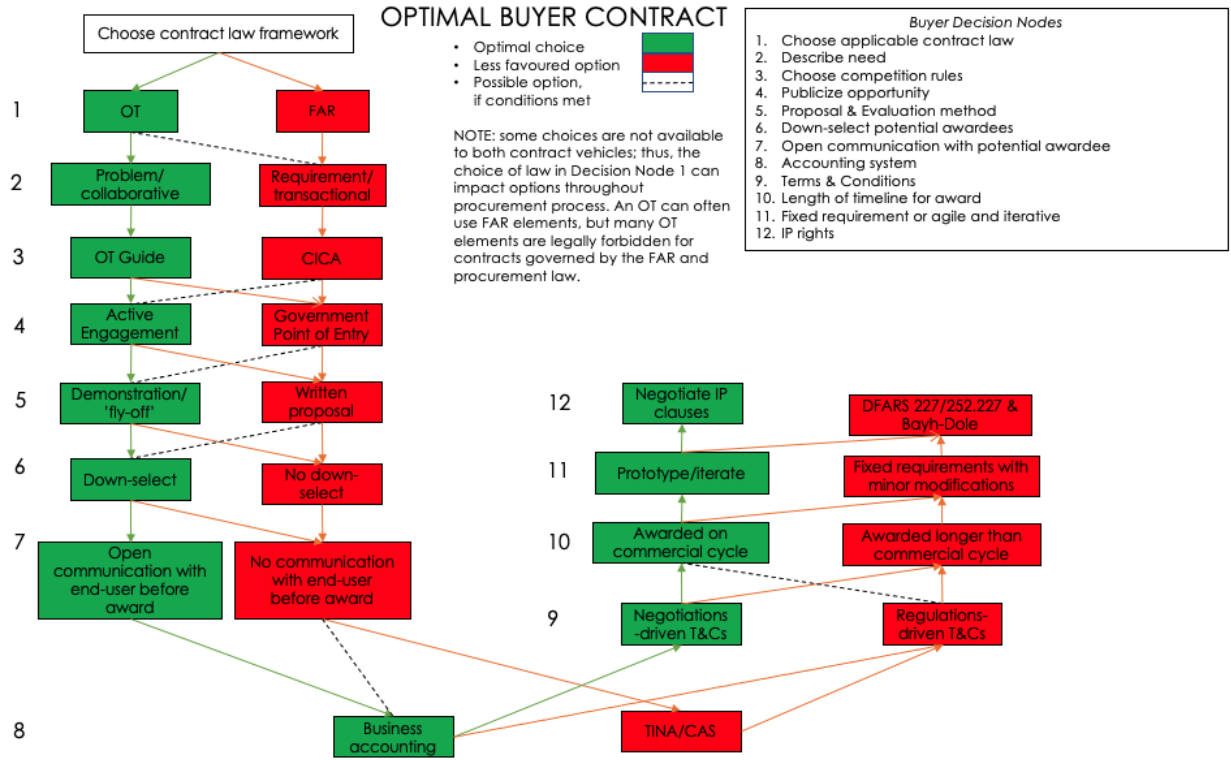

The nodes and branches of the decision tree are discussed below.

1.    The decision tree begins with the choice of contract law framework.  The primary acquisition contracts for AI capabilities are traditional procurement contracts governed by the FAR and OT for prototypes authorised by 10 USC § 4022.  Because the legal authorities for these two contract vehicles are vastly different, this choice impacts many subsequent choices.  Therefore, this initial decision on the legal framework that will apply to the contract formation is still the most consequential in predicting the attractiveness of the contract opportunity.

2.    The next choice is how the contracting organisation will describe its need.  Under the FAR, this process is detailed in FAR Part 11 and focuses on describing the requirement for goods or services.[710]  For an OT, the description of the need is focused on the problem, not the

---

[710] 48 CFR § 11.002.



solution.[711]  Although the FAR states that acquisition begins with the requirement,[712] contracts

under the FAR may use a commercial solution opening[713] or broad agency announcement[714] in

some instances to allow more flexibility in the stating the requirement.

3.      The next choice is the method of competition.  Under the FAR, competition is governed

by CICA.[715]  Under CICA, the competition includes the opportunity for potential vendors to

protest the specifications in the solicitation as well as the award.  Although the standard is 'full

and open' competition, many contracts are 'set aside' for only specific small business

interests.[716]  Competition for OTs can take many shapes provided contracts are competed to the

'maximum extent practicable.'[717]  FAR contracts are governed by CICA and cannot be competed

under the OT authority unless competitive procedures used to solicit an OT agreement or

contract meet the FAR and CICA requirements.[718]  While the FAR requirement is to maximise

full and open competition, if the barriers to entry are too high, the costs too great, and the

opportunities are unknown to potential contractors, it is unlikely traditional procurement

contracts will result in meaningful competition.  Even with the different standards for competition, OTs can enhance the level of competition on a contract if the engagement, publication, and request for offers is more attractive to the commercial AI industry.

---

[711] *Other Transaction Guide* (n 104) 11.

[712] 48 CFR § 2.101.

[713] Office of the Under Secretary of Defense for Acquisition and Sustainment, 'Class Deviation—Defense Commercial Solutions Opening Pilot Program', (Class Deviation No 2018-O0016, Department of Defense, 26 June 2018). This class deviation allows a contracting organisation to use competitive procedures other than those specified under FAR 6.102 which require specificity in stating the requirements.  This authority is limited to situations when 'meaningful approaches with varying technical or scientific approaches can be reasonable anticipated': at 2.

[714] 48 CFR § 35.016.  This authority is limited to basic and advanced research and is not appropriate for acquisition at scale.

[715] 48 CFR Part 6.

[716] See 48 CFR Subparts 6.2, 6.3 and Part 19.

[717] 10 USC § 4022(b)(2).

[718] See 48 CFR §§ 6.001, 6.101, 13.104, 13.106, 13.500 (requiring competitive procedures in accordance with CICA on negotiated procurements or simplified acquisitions unless statutory exception applies).  Because the competition requirements are undefined by OT authority, CICA competition requirements can be used to award either a FAR contract or OT agreement, though such process would limit the utility of the OT: see 10 USC § 4022(b)(2); *Other Transaction Guide* (n 104) 13–4.  However, a FAR contract cannot be awarded for an agreement solicited under OT authority as such award violates notice, solicitation, competition, and protest requirements: see 48 CFR §§ 5.101, 5.201, 6.1, 33.106.

4.      How the opportunity is publicised has an impact on what firms will compete for the contract.  FAR contract opportunities are required to be published on the Government Point of Entry.[719]  This research demonstrated that most firms that have performed a contract with the DoD previously will review such opportunities, most firms that have not yet performed a contract with the DoD will not see the publication and will not compete for the contract.  This limits the competition to firms that have worked with the DoD previously and excludes new firms.  While paid advertisements outside the Government Point of Entry are authorised in addition to this requirement,[720] in practice few FAR contract opportunities are posted outside the traditional medium.  In contrast, OTs can be published widely across social media platforms, reaching commercial AI firms where they are likely to see the opportunity.[721]

5.      Source selection methods drive the manner and timeline of the award.  Thus, the choice of source selection methodology affects the competition as complex, tedious methods that will take months or longer to award a contract are likely to limit the competitive sources to traditional defence contractors.  The procedures under FAR Part 15 are often onerous to the offeror and the review and selection procedures are largely opaque.[722]  In contrast, OT authority permits source selection processes more consistent with commercial practices and timelines.[723]  Additionally, through the flexibility of the OT source selection methods, commercial AI firms can be evaluated through demonstrations, challenges, or experiments that allow the agency to preview how the firm will solve a DoD problem using actual or synthetic DoD data.[724]

---

[719] 48 CFR § 5.003.
[720] 48 CFR § 5.101(b).
[721] Market research is an important step in finding the best pathways to publicise the opportunity, but beta.sam.gov is certainly not a viable method of attracting innovative companies.  Richard Dunn implores the DoD 'DO NOT, repeat, DO NOT just post your solicitation on beta.sam.gov and expect non-traditional companies to come to you. … If you want to attract new and innovative solutions, reach out … through multiple channels and shop your problem.'  *Guide to Other Transactions Authority* (Strategic Institute, 3rd ed, 2021) 45.
[722] See 48 CFR Subpart 15.3.
[723] *Other Transaction Guide* (n 104) 16.  However, some agencies award OT agreements as if FAR Part 15 applied, slowing down the process and limiting the utility of an OT for the DoD and the making the process less attractive to industry: at 39–40.

6.      While an option for both FAR and OT contracts,[725] down-selecting offerors was described by the interviewees as an attractive practice. DIU utilises down-selects during their OT source selection by conducting three phases: the first requires a solution brief of a short white paper or presentation; the second phase entails a pitch on the rough order magnitude, cost and schedule, and data rights; the third phase invites proposals and negotiation.[726] Each phase down-selects offerors. As such, not every competitor will have to submit full proposals, saving time and costs. While offerors prefer to secure the contract, if they are not going to earn the award, the interviewees strongly prefer to know as soon as possible that they are not competitive.

7.      Communication is a near-universally attractive contract formation attribute. The research participants indicated that communication throughout the contract formation and performance is necessary. Moreover, having an open dialogue with the end-users of the technology is critical. However, under the FAR, communication is restricted mainly to the contracting officer between receipt of proposals and contract award.[727] Although communication may be limited during an OT competition in the same manner as a FAR procurement, the OT authority permits more open engagement. It should focus on working with the competitors to analyse the problem. With AI problems, the end-user needs to define and refine the use case with the commercial AI firm. Logically, this should occur before contract award.

---

[724] See *Other Transaction Guide* (n 104) 16, explaining the flexibility inherent in 10 USC § 4022 leaves agencies free to create their own process to solicit and assess potential solutions provided it is fair and transparent.

[725] 48 CFR §§ 15.202, 15.306; *Other Transaction Guide* (n 104) 29.

[726] *Other Transaction Guide* (n 104) 29.

[727] 48 CFR § 15.201.



8.      Contract pricing though cost reimbursement,[728] particularly contracts covered by the Cost Accounting Standards (CAS),[729] require more complex, comprehensive, and costly accounting methods than are typically required in the commercial market.  According to several interviewees, some firms will not consider competing for a contract that requires CAS compliance due to the complexity and cost.  OT authority permits the use of business accounting standards as CAS and TINA do not apply to OT agreements unless negotiated by the parties.

9.      The ability to negotiate contract terms and conditions was one of the most attractive contract formation attributes described by the research participants.  Under the FAR, negotiation is limited to price, schedule, technical requirements, and other terms and conditions.[730] However, even contracts for commercial items have nearly sixty mandatory, non-negotiable clauses, while other FAR-based contracts have over a hundred mandatory clauses.[731]  Moreover, the FAR governs price, schedule, and technical requirements, with additional requirements and restrictions that limit the scope and freedom to negotiate.  Conversely, negotiating an OT can start with a virtually blank sheet of paper — all terms and conditions are negotiable.[732]

10.     The length of process it takes the DoD to announce a contract opportunity to contract award and payment is a factor that directly contributes to a commercial AI firm's decision to compete or not.  While it is a common myth that OT agreements are inherently faster to award than FAR-based contracts, that is not categorically true.[733]  OT authorities are more flexible than the FAR, so the source selection process can be streamlined.  However, as each clause may be negotiated, the process can take considerably longer than some FAR-based contracts that contain mostly boilerplate language.[734]  Regardless of the DoD's choice of law, it should be mindful of the consequences of moving too slow as the timeliness of contract award affects the attractiveness of the buyer as well as the ability of the DoD to acquire state of the art AI technologies.

---

[728] 48 CFR § 16.301-1.
[729] 48 CFR Part 30.
[730] 48 CFR § 15.306.
[731] See 48 CFR § 52.301.
[732] *Other Transaction Guide* (n 104) 40.
[733] See ibid 39–40.

11.     The research participants indicated a strong preference for experimenting, prototyping, and iterating the AI on a use case.  As machine learning is an inherently iterative process, training, testing, and evaluating is critical to developing and deploying a model.  This process requires contract flexibility.  Fixed milestones and waterfall methodology that is based on hardware systems development appear to be incompatible with the AI development lifecycle.  Because the FAR starts with a requirement with predefined technical specifications, it is challenging for the contracting officer and contractor to foresee the schedule and scope from the outset of contract formation.  As such, significant modifications may be required during contract performance.  If the modifications affect the schedule or scope enough, the contract must be re-competed as the originally competed work may not cover the change.[735]  However, OT agreements are not impeded by this problem because CICA and the FAR do not apply.  As such, OT agreements are well-suited to carrying out prototype projects in a manner that permits agile methodology development.[736]

12.     As discussed in Chapter IV, the ability to meaningfully negotiate IP clauses in a manner that reflects the value of the commercial AI firm's IP is highly valued.  Negotiation of IP clauses that recognise the unique issues AI poses to traditional legal constructs is necessary to attract commercial AI firms to compete for the contract and ensures the DoD obtains the appropriate data rights. DFARS Part 227 reflects traditional hardware-centric technical data that focuses on whether the technology was developed under government contract or non-contract funding.[737] When contracting for AI systems that use machine learning, the program is fundamentally changed by the data input; thus, determining when development occurred becomes a challenging question while using this construct. Alternatively, IP terms and conditions can be freely negotiated in an OT that mirrors commercial practice or develops novel and innovative ways to allocate rights to data.[738]

---

[734] See ibid.
[735] Such a change is called a 'cardinal change' which is a breach of contract and if the modification is ordered, the contractor may file a claim and the modification may be protested by other potential contractors.  Nash, O'Brien-Debakey and Schooner (n 375) 73–4, citing Ralph C Nash and Steven W Feldman, *Government Contract Changes* (West, 3rd ed., 2007-2012) vol 1 [4:2]-[4:9].
[736] 10 USC § 4022(a).

The decision to use a traditional procurement contract under the FAR will limit the entire procurement process and resulting contract to the procurement statutes, FAR and DFARS, unless a waiver or class deviation is granted by a senior procurement executive.[739] However, should the contracting officer choose to use OT authority as the contract law framework, only the OT statute and generally applicable law binds the contract, and FAR processes, terms and conditions are not required, though they may be used. This flexibility is useful, though as discussed in Chapter II, can result in unintended consequences. Thus, careful consideration and planning is required before choosing the contract vehicle. This model assists in demonstrating most commercial AI firms prefer procurement processes and contract terms that are only legally possible with an OT agreement.

The optimal buyer theory is specific to AI contracts as the characteristics of the industry, development, and funding are unique. However, this theory has general application with defence organizations buying emerging technologies where they do not control or dominate the market and must compete with other buyers.  These conditions present comparable issues to those studied in this research.[740]  The optimal buyer theory is likely applicable in conditions where, like here, the public organisation (particularly defence related) is a relatively small and late-adopting customer in the technology market and must adapt its contracting practices and engagement to become an attractive customer to innovative firms with several other potential customers vying for their services and resources.  Other fields where this theory could apply include hypersonic, space assets (launch vehicles and satellites), or microelectronics, all of which have private and public customers competing for limited resources that are increasingly important.[741]

---

[737] See *DFARS* Part 227, § 252.227-7014.
[738] See *Other Transaction Guide* (n 104) 50; Bowne and McMartin (n 155) 9.
[739] 48 CFR § 1.404.

## B *Recommendations*

Based on the findings and scope of this research, there is a need for additional action in this field.  Recognising the need for the DoD to procure the expertise and technology from AI firms,[742] this thesis concludes by offering recommendations for legal practice, policy, and further research.

---

[740] See Bharat Rao, Adam Jay Harrison and Bala Mulloth, *Defence Technological Innovation: Issues and Challenges in an Era of Converging Technologies* (Edward Elgar Publishing, Inc., 2020) 2; Kelley M Sayler, *Emerging Military Technologies: Background and Issues for Congress* (CRS Report No R46458, Congressional Research Service, 17 July 2020) 1, 15, 23, explaining that commercial research in areas such as lasers, quantum technology, AI, autonomous systems, and biotechnology has resulted in rapid advancement in these sectors that contribute to the growing importance of the commercial sector to the United States DoD, as well as raise concerns that the commercial nature of the development of such technologies can lead to competitors obtaining military applications more easily than when military technology was derived primarily from military funded research; Allen and Chan (n 7) 2.

[741] See Harald Andås, *Emerging Technology Trends for Defence and Security* (FFI Report No 20/01050, Norwegian Defence Research Establishment, 7 April 2020) 52, explaining the convergence of military and commercial use for advanced technologies in the electromagnetic spectrum is more contested and competitive due to commercial investment in advanced radiofrequency technology; Christopher Darby and Sarah Sewall, 'The Innovation Wars: America's Eroding Technological Advantage' (2021) 100(2) *Foreign Affairs* 142, 148-9.

[742] See Allen and Chan (n 7) 8; Andrew Ilachinski, 'Artificial Intelligence & Autonomy: Opportunities and Challenges' (Research Paper, CNA, October 2017) 14; Andrew S Bowne, 'Innovation Acquisition Practices in the Age of AI' [2019] 2019(1) *Army Lawyer* 74, 76.



**Recommendation 1: The DoD should Implement Best Practices in Contracting to Better Attract Commercial AI Firms**

As discussed above, social exchange theory is a useful framework to assess government contracts and their impact on the attractiveness of the customer. Through this research, a theory derived from social exchange theory but contextual to AI systems was created — the optimal buyer theory. This theory can help DoD procurement officials decide which legal authorities to leverage to maximise the potential attractiveness of a contract opportunity for AI-enabled capabilities. As many of the most attractive contract attributes are already legally authorised yet underutilised, the DoD can create policies that enable procurement practitioners to better align the procurement process with the commercial AI industry's preferences. These best practices include attributes that permit flexibility, negotiation, collaboration, communication, iteration, and engagement. Such a contract should minimise regulatory burdens to reduce complexity and cost that may prove too burdensome for many innovative commercial AI firms.

This recommendation is aligned with Congressional intent and the National Defense Strategy, both of which have stated the DoD shall rely upon the commercial technology and industrial base to meet the national security requirements.[743] Congress has repeatedly directed the DoD to reduce barriers to the use of commercial products and processes.[744] The DoD can better utilise the commercial technology and industrial base by establishing a preference for OT authority. In 2018, Congress directed the DoD to do just that, mandating the Secretary of Defense establish a preference for using other transaction authority under Sections 4021 (research OT), 4022 (prototype OT), and 4023 (procurement for experimental purposes) for science and technology.[745]

---

[743] See 10 USC § 4811(b)(1); *2018 NDS* (n 211) 3.
[744] 10 USC § 4811(b)(3).

There are several arguments made within the DoD and by commentators that the FAR should continue to be the standard regulatory framework despite Congress's persistence in directing the DoD to use the authorities it has provided to attract non-traditional contractors. The Defense Acquisition University (DAU) provides several common arguments against using OT authority in its description of prototype OT agreements.[746] DAU explains that the government assumes increased risk without the process, precedent, and protection of the FAR; the flexible negotiations that are attractive to industry can have long term negative impacts for the government; and the pursuit and execution of an OT agreement requires highly experienced and empowered staff.[747] Additionally, DAU states the lack of guidance, structure and processes can challenge and intimidate inexperienced staff.[748] The Project On Government Oversight takes a strong stance against OT agreements, claiming that the government is at a disadvantage whenever it negotiates with industry outside of the vast array of procurement laws that apply to traditional procurement contracts that serve as a safety net to prevent fraud, waste and abuse.[749]

---

[745] *NDAA FY2018* (n 337) § 867, 131 Stat 1495. While this research has focused on prototype OTs as the preferred legal framework for acquiring commercial AI capabilities, 10 USC § 4023 (procurement for experimental purposes) is also available provided the service secretary considers the purchase 'necessary for experimental or test purposes in the development of the best supplies that are needed for the national defense': 10 USC § 4023(a). This authority applies to nine technological domains, such as signal, transportation, space-flight, and telecommunications, including parts and accessories, and designs thereof: 10 USC § 4023(a). Arguably, AI is present in each of the domains. This authority is even more flexible than prototype OTs and should be leveraged to quickly (there is no competition requirements) buy commercial datasets, models, software, and hardware to carry out technical evaluation, experimentation, assessment of operational utility, or provide a residual operational capability: see 10 USC § 4023(b).
[746] 'Contracting Cone: Prototype OTs', *Defense Acquisition University* (Web Page) <https://aaf.dau.edu/aaf/contracting-cone/ot/prototype/>.
[747] Ibid.
[748] Ibid.
[749] See Amey (n 120). The Project On Government Oversight article overreaches in claiming that OT agreements have 'no oversight' and asserting that the Procurement Integrity Act does not apply. Ibid. These claims are false; the OT statute specifically provides audit authority for the Comptroller General and that the Procurement Integrity Act applies in the same manner it would for a traditional procurement contract. 10 USC § 4022(c) and (h).



Lopes found that DoD employees are risk-averse to try something new like OT agreements, especially because nearly all of their training was on traditional procurement contracts.[750] However, these challenges can be mitigated with robust education and development of OT experts within the DoD.[751]

### Recommendation 2: The DoD should Prioritise Education and Training in Other Transactions, Commercial Contracts, and Artificial Intelligence

The DoD workforce must be better educated in OT authority and AI to leverage the best practices in attracting commercial AI firms. Congress has directed the DoD to focus on training in both these areas.[752] Nonetheless, the DoD has been slow to meet these requirements. One recent study examined the educational opportunities provided to the DoD contracting community on OTs compared with traditional procurement training and concluded the limited training available lacks practical focus and reaches only segments of the workforce.[753] Likewise, the National Security Commission on AI concluded the DoD lacks a digital infrastructure and recommended the DoD develop a technical backbone through systematic training and education.[754]

The DoD should follow these recommendations. Based on the interviews conducted for this research, it is clear the commercial AI industry perceives the DoD's collective lack of education and comfort in awarding OTs over FAR contracts results in the overuse of traditional procurement frameworks that are slow, rigid, and costly. For contracting offices that do award

---

[750] Lopes (n 111) 603.
[751] See ibid 621–3.
[752] Congress directed the DoD to establish minimum levels and requirements for continuous and experiential learning for management, technical, and contracting personnel on the award and administration of other transactions and other innovative forms of contracting. *NDAA FY2018* (n 337) § 863. Additionally, Congress directed the DoD to develop an education strategy for servicemembers in relevant occupational fields to develop knowledge in AI and the impact of AI on military strategy and doctrine: *National Defense Authorization Act for Fiscal Year 2020*, Pub L No 116-92, § 256, 133 Stat 1290 (2019).
[753] Soloway, Knudson and Wroble (n 126) 40.
[754] *NSCAI Final Report* (n 4) 63–66.



OTs, the interviewees report that most offices rely on FAR contract terms and conditions, eroding the advantages provided by the inherent flexibility in OT authorities. This finding is supported by research into the perception of DoD procurement experts that admit there is a lack of OT expertise that discourages the DoD from using OTs.[755] Many commentators and researchers have argued that the DoD should provide more rigorous training and education to influence the greater use of OTs.[756] While OTs are arguably available to any AI contract, an understanding of how AI is developed and deployed is required to articulate the use case as a prototype and unlock the OT authority.[757] Without understanding that each AI use case requires some amount of testing and integration, procurement officials may believe OT authority does not cover a commercial-off-the-shelf program. Thus, education in AI and OT authorities is required to maximise OTs' use for AI contracts. Increased use of this authority will 'normalise' the practice as viable alternative to the FAR. The DoD's contract attorneys must play an important role in developing an expertise in all legal pathways to acquire commercial AI capabilities to ensure program managers, contracting officers, and resource managers are comfortable using authorities outside the traditional pathways.

---

[755] See Lopes (n 111) 621–26.

[756] See, eg, ibid; Soloway, Knudson and Wroble (n 126) 45.

[757] The development and deployment of defence-relevant AI and the model training methodologies both make iterative training and human-machine teaming critical steps in the deployment of an AI-enabled capability: Gadepally et al (n 20) 2, 'recent AI advances of relevance to the DoD have largely been in fields where a human is either in- or on-the-loop. The phase of human-machine teaming is critical in connecting the data and algorithms to the end user and in providing the mission users with useful and relevant insight. Human-machine teaming is the phase in which knowledge can be turned into actionable intelligence or insight by effectively utilising human and machine resources as appropriate.'; see Yipeng Hu et al, 'The Challenges of Deploying Artificial Intelligence Models in a Rapidly Evolving Pandemic' (2020) 2 *Nature Machine Intelligence* 298, 299, explaining that an AI model trained with data from one use case or multiple use cases do not necessarily translate to a new use case, thus rigorous design, analysis and validation of the model is required before generalising data or translating a model to a new use case. Additionally, AI is not a stand-alone 'technology that can be implemented as an individual function.' Wiebke Reim, Josef Åström and Oliver Eriksson, 'Implementation of Artificial Intelligence (AI): A Roadmap for Business Model Innovation' [2020] 2020(1) *AI* 180, 182. Accordingly, as each use case will require training, iteration, and human-machine teaming, each case use will meet the definition of a 'prototype project': see *Other Transaction Guide* (n 104) 31; 10 USC § 4022.



This research shows there is a perceived lack of understanding in the DoD of both OTs and AI. This evidence lends support to the calls to expand educational opportunities in the DoD. A better-informed workforce is necessary to creating a contracting practice that is more attractive to commercial AI firms.

### Recommendation 3: Congress should Broaden the DoD's Authority to Use Other Transactions

While OT authority is already broad and arguably applicable to any contract for developing, integrating, or acquiring AI technologies and capabilities, there are statutory requirements that may limit the DoD's ability to award OT agreements in circumstances where it would benefit from the flexibility of OT agreements. An OT awarded under Section 4022 authority must be to carry out a 'prototype project' that is directly relevant to enhancing the mission effectiveness of DoD personnel or improving platforms, systems, components, or materials proposed to be acquired or developed or used by the armed forces.[758] The statute defines 'prototype project' as any project that addresses — '(A) a proof of concept, model, or process, including a business process; (B) reverse engineering to address obsolescence; (C) a pilot or novel application of commercial technologies for defense purposes; (D) agile development activity; (E) the creation, design, development, or demonstration of operational utility; or (F) any combination' thereof.[759] AI, machine learning models, and autonomous systems, whether through development or piloting a commercial technology for defence purposes fit within this broad definition, unlocking the option to leverage this statute for most, if not all, AI contracts. Nonetheless, the requirement

---

can reduce the perceived availability of the authority and result in procurement officials opting for the FAR when OTs would be a more attractive option.[760]

The advantage of expanding the authority is the resulting increase in experience and comfort level amongst DoD contracting offices, leading to the development of best practices that can better attract potential contractors.[761]  Opening OT authority beyond prototypes is not new.  The Space Act of 1958 authorised NASA to enter into other transaction agreements the same way it could enter into contracts, leases, or cooperative agreements — when 'necessary in the conduct of its work and on such terms as it may deem appropriate.'[762]

Another limitation of the current OT authority is found in the ability to award a follow-on production award.[763]  In order to award a production OT, a prototype OT must be competitively awarded and the participants in the transaction successfully completed the prototype project provided for in the OT agreement.[764]  Under the statute, the participants that were awarded the prototype OT are the only parties eligible for award of a follow-on production OT.[765]  These limitations were discussed by the Section 809 Panel in its recommendation to Congress that it expand the authority to use OTs for production.[766]  The Panel argued that these current statutory limitations frustrate the national security efforts to achieve rapid fielding of new capabilities.[767]

---

[760] Expanded education in AI and OT authority as recommended above could help mitigate this issue.

[761] See Lopes (n 111) 399.

[762] *Space Act of 1958* (n 306) (codified in 51 USC § 20113(e)).  NASA's use of Space Act Agreements to encourage commercial innovation in space research and development can serve as a case study for the role the DoD's OT authority can have in advancing defence-relevant AI.  'The impact of Space Act Agreements on the development of the commercial space industry has been profound, enabling SpaceX to do what no other private enterprise has done before…. This new approach has numerous benefits.  It facilitates rapid development in terms of money and contracts[.] Perhaps just as importantly, it allows commercial companies to tap into the vast technical wisdom of an institution that has 50 years of experience in space.  It provides legitimacy to new commercial companies, which allows them to more effectively attract capital investment': Chad Anderson, 'Rethinking Public-Private Space Travel' (2013) 29 *Space Policy* 266, 268–9; see also Mattia Olivari, Claire Jolly and Marit Undseth, 'Space Technology Transfers and Their Commercialisation' (Policy Paper No 116, OECD, July 2021) 26, explaining that Space Act Agreements benefit start-ups by providing access to government infrastructure and services.

[763] 10 USC § 4022(f).

[764] Ibid.

[765] Ibid.

One scenario where this limitation can present obstacles to the rapid fielding of AI capabilities is when transitioning basic and advanced research, often performed at government laboratories or academic institutions, to applied research and scaling delivery, often performed by industry.[768]  In that scenario, the commercial firm would likely be precluded from entering into an OT and therefore limited to traditional procurement methods.  The Section 809 Panel recommended removing these obstacles and provided promulgating language to change the statute.[769]  To ensure wide use of OT authority and maximise the DoD's attractiveness as a customer to commercial AI firms, Congress should enact legislation that mirrors NASA's OT authority to award flexible contracts as necessary and enact the Section 809 Panel's recommendation to expand the ability to award a production OT to parties that were not participants on the prototype project.

### Recommendation 4: The United States Government should Conduct Research on the Impact the Planning, Programming and Budget Systems have on the DoD's Ability to Attract Commercial AI Firms

This research focused on understanding why commercial AI firms decide to compete for DoD contracts.  Accordingly, the procurement process and contracts that serve as the legal embodiment of the relationship were the subject of the research.  While the contract focuses on how the DoD interacts externally, internal factors, such as Congressional appropriations and prioritization of acquisition programs that compete for the budget, represent factors worthy of additional research.  Several perceptive interviewees acknowledged that the focus on sustaining legacy systems limit the DoD's ability to fund new technology. Although these internal factors were outside the scope of the research question, further study on the impact these internal structures have on the DoD's ability to field commercial AI capabilities is recommended.[770]

---

[766] See *Acquisition Regulation Report Vol 3* (n 238) vol 3, 440–47.
[767] Ibid 444.
[768] See *NSCAI Final Report* (n 4) 187.
[769] See *Acquisition Regulation Report Vol 3* (n 238) vol 3, 447.

One possible action that could address an internal problem (requirements generation) and an external problem (engaging commercial AI firms on national security issues) is to create a viable pathway for commercial AI firms to present ideas for funding. From a requirements generation perspective, perhaps the traditional model of the government determining its own needs is causing a lag in decision making and development of military-specific technology that can be rapidly fielded. One way to reduce the timeline of the national defence apparatus to field emerging technology is for industry to anticipate future defence needs and collaborate with the government. There are limited processes for industry to submit ideas to the DoD, and such processes appear to struggle in convincing the DoD that a truly disruptive technology is potentially feasible and perhaps capable of providing a national security advantage in a novel, if not revolutionary manner.[771]

As several interviewees shared, the time in a firm's lifecycle to develop an idea for disruptive technology is as an early-stage start-up. Venture capital firms look at the potential return on investment. The more disruptive the technology or business model is, the greater the potential for significant return on investment. It is not uncommon for a venture capital firm to essentially place bets on a hundred ideas in the hopes that one becomes a 'unicorn', the term for start-ups valued at over one billion dollars.[772] This provides the capital to place more bets. However, according to one of the interviewees, a CEO of a venture-backed start-up, the risk increases as the company matures because there are more stakeholders once a company receives multiple funding rounds. The DoD, funded by the taxpayer, and accountable to Congress, is inherently more risk averse than venture capitalists. Before funding innovative ideas, the DoD typically looks for more advanced technology readiness levels rather than potential for return on investment.[773] Most basic research into long-shot ideas is carried out by DoD organizations, such as DARPA, or Federally Funded Research and Development Centers (FFRDC) that report to the DoD.[774] Thus, the environment is not conducive for the DoD to obtain early access to disruptive innovation from industry.[775]

---

[770] Both House and Senate Armed Services Committees have expressed interest in commissioning a panel study on planning, programming, budgeting, and execution reform. Committee on Armed Services, House of Representatives, *Report on H.R. 4350 Together with Additional and Dissenting Views* (House Report No 117-118, 10 September 2021) § 1079; Committee on Armed Services, United States Senate, *Report to Accompany S. 2792* (Senate Report No 117-39, 21 September 2021) § 1002.

[771] See Pete Modigliani et al, 'Modernizing DoD Requirements: Enabling Speed, Agility, and Innovation' (Research Paper, MITRE Center for Technology and National Security, March 2020) 3; Thomas Holland, 'How the Army Ought to Write Requirements' (2017) (November–December) *Military Review* 100; Adam Jay Harrison, Bharat Rao and Bala Mulloth, 'Developing an Innovation-Based Ecosystem at the U.S. Department of Defence: Challenges and Opportunities' (2017) (May) *Defence Horizons* DH No 81:1-16, 3.

As previously discussed, the typical manner in which the DoD's requirements are met is by drafting specifications, publicising them for an open comment period, potentially incorporating the feedback, and then submitting a formal solicitation in the form of a request for proposals, whereby potential vendors can certify they can perform and submit an offer based on price.[776] This process assumes that the DoD already knows what its requirements are and what potential solutions could fulfil its requirements. Although open topics, broad agency announcements and commercial solutions openings provide more flexibility to the industry to present potential solutions that the DoD was unaware of, these processes generally result in the award of a contract for relatively mature technology.[777]  Several interviewees explained the perception of the commercial AI industry is that the DoD does not have the capacity for technology evaluation that venture capital or technology firms often employ, making it challenging for the DoD to recognise the merits of a proposed innovative solution.

---

[772] Aileen Lee, 'Welcome to the Unicorn Club: Learning from Billion-Dollar Startups', *TechCrunch* (online, 2 November 2013) <https://techcrunch.com/2013/11/02/welcome-to-the-unicorn-club >.
[773] See Government Accountability Office, *Technology Readiness Assessment Guide* (GAO-20-48G, January 2020) 24.
[774] *NSCAI Final Report* (n 4) 187.
[775] Ibid.
[776] See generally, 48 CFR Parts 6, 9.5, 11, 15.

Recognising that the source of AI innovation is often academia or industry, one option for the DoD is to collaborate with research universities to conduct early-stage research.[778]  Such a collaboration could focus on anticipating the future needs of the DoD.  In a collaborative effort, the DoD could provide briefings to university researchers on problems both current and expected.  Industry can use this insight to develop technology that is responsive to anticipated problems and needs.  One example where the DoD is collaborating with academia is the Department of the Air Force and Massachusetts Institute of Technology AI Accelerator.[779]  This collaboration produces fundamental research to advance the state of the art in machine learning with a focus on dual-use technologies that are ethically designed, developed, and deployed.[780] This collaborative effort could serve as the blueprint for future endeavours and scale to other industry and academia partners.

---

[777] See Karen Walker, 'Understanding the TRL Scale and SBIR/STTR Programs' *Arizona State University Knowledge Enterprise – Research Development* (Web Page, 4 June 2020) <https://funding.asu.edu/index.php/articles/understanding-trl-scale-and-sbirsttr-programs>.

[778] Such a collaboration could be achieved through a cooperative agreement or other transaction for research.  A cooperative agreement is a legal instrument reflecting a relationship between the United States Government and a non-federal entity when the principal purpose of the relationship is to carry out a public purpose of stimulation, such as AI research, rather than acquiring property or services and where substantial involvement is expected between the executive agency and the non-federal entity: 31 USC § 6305.  Other transaction authority for research permits the DoD to establish flexible partnerships with organisations, including academia, to conduct basic, advanced, and applied research: 10 USC § 4021.

[779] 'About', *DAF-MIT AI Accelerator* (Web Page) <aia.mit.edu/about/>.

[780] Ibid (the researcher is the chief legal counsel of the organisation).



**Recommendation 5: Future Researchers Should Use this Research Framework to Assess and Test the Applicability of Social Exchange Theory and Optimal Buyer Theory to Other Defence Procurement Systems**

This research focused on the procurement practices and two alternative legal regimes for awarding and structuring contracts in the DoD. The data collected came from commercial AI firms that were identified to provide defence-relevant applications. This scope limits the generalisability of the findings and theoretical contribution. However, this research provides a conceptual framework that can be used to test and compare other procurement systems or other technology sectors within one procurement system. Further research focused on different procurement systems or technology sectors can also contribute to the development of Social Exchange Theory as a lens to view public/private contractual relationships.

Defence procurement of AI is not exclusive to the DoD. Many militaries around the world are making efforts to leverage the power of AI in modernising their forces.[781] Given the degree that Five Eyes[782] members collaborate, interoperability of AI-enabled capabilities is likely desirable. Accordingly, the conceptual framework constructed in this research can be utilised to assess the attractiveness of the defence procurement systems in Australia, Canada, New Zealand, and the United Kingdom in future research.

Attraction is subjective, so the perspectives of the commercial AI industry may not match the microelectronics, hypersonic, or space launch industries. As this research found the development lifecycle of AI technologies is an important factor in assessing the alignment with contract law, other technologies may affect the alignment of industry preferences and development cycles with the DoD's contract law differently.  Additionally, the cultural and legal differences of procurement systems in different countries may impact the attractiveness of certain contract attributes.  Targeting specific industries to discover what factors are important to firms within those industries could result in a better understanding of that community.  Likewise, researching what factors are important to firms participating in other procurement systems will provide insight into the transferability of the optimal buyer theory to other countries.  Further research using this methodology can lead to the discovery of universal attractors, as well as an understanding of the idiosyncrasies of a particular industry or country that should be accounted for by public procurement officials.  Through a better understanding of its suppliers, defence procurement officers can better refine their contract practices in a more flexible, bespoke way, acknowledging that a one-size-fits-all approach to contracts may leave potential contractors out of the defence market.

---

[781] See Morgan et al (n 7) 5.
[782] See Office of the Director of National Intelligence, 'Five Eyes Intelligence Oversight and Review Council (FIORC)', *The National Counterintelligence and Security Center* (Web Page) <https://www.dni.gov/index.php/ncsc-how-we-work/217-about/organization/icig-pages/2660-icig-fiorc>; 'Defense Innovation Board's AI Principles Project', *Defense Innovation Board* (Web Page) <https://innovation.defense.gov/ai/>.

### Recommendation 6: The DoD should Collaborate with Industry, Academia, and International Partners to Develop an Ethical Framework for the Development and Deployment of Artificial Intelligence

Although the research findings show that the primary reason most commercial AI firms avoid working with the DoD is rooted in the business and legal practices used by the DoD, further research and policy are required on the topic of ethics.

The DoD's ethical principles are a baseline for the conversation, however, there must be a common understanding of how to implement those principles in the development and deployment of AI.  The definitions of terms such as 'fairness,' 'transparency,' 'explainability,' and 'responsibility' must be normalised.[783]  The subjectivity of these definitions should be mitigated by clear standards that apply to governments, academia, and industry.[784] How data is collected, stored, and used should be rigorously debated to ensure concepts like 'privacy' and 'bias' have understandable and consistent meanings. While many institutions employ ethicists and claim to use 'AI for good,'[785] the question remains: good for whom? While institutions in the international community may be quick to adopt terminology like 'responsible AI,'[786] it is important that robust debate and cooperation is facilitated to ensure a common understanding of such terms. Without a cross-cultural understanding, implementation of these standards in the commercial and governmental development and deployment of AI will be ineffective.

---

[783] See Felipe Thomaz et al, *Ethics for AI in Business* (Report, Saïd Business School, University of Oxford, 22 June 2021); Thilo Hagendorff, 'The Ethics of AI Ethics: An Evaluation of Guidelines' (2020) 30 *Minds and Machines* 99,

Whether the use case is for an automated workflow, predictive maintenance, or computer vision that aids in selecting targets, ethics is a critical aspect of AI development and deployment. The DoD currently enjoys the trust of most of the commercial AI industry, but one high profile mishap that can be traced to an AI system will likely change that perception. Without robust dialogue in the international community, and accepted standards that apply to industry and government alike, it is likely some commercial AI firms will avoid contracting with the DoD. The DoD should continue to lead efforts, alongside industry, academia, and the international community, to establish effective and actionable standards to build trust in this technology and mitigate the potential for misuse.

After reviewing the federal government's position in developing, adopting, and fielding AI, the National Security Commission on Artificial Intelligence reached what it characterised as an 'uncomfortable' conclusion: 'America is not prepared to defend or compete in the AI era.'[787] The United States Government cannot compete without the help of commercial industry.[788] The ability of Ukraine to leverage commercial AI innovation to exploit vulnerabilities in Russia's conventional systems further underscores, in a practical sense, the future of warfare requires application cutting edge technology to military use cases.[789] Understanding the perspectives of commercial AI firms can allow the DoD to become a more attractive customer to the industry. This understanding will allow contract practitioners to leverage the existing laws to optimise the procurement process in a manner that fits the unique attributes of AI development and deployment. An optimal contract in this context is one that is mutually attractive to commercial AI firms and beneficial for the DoD, both tactically in the contract terms and performance, and strategically by developing an alternative to the slow, costly, and burdensome procurement process that appears to drive away innovative AI firms.

Based on the perspectives of commercial AI firms discovered in this research, the DoD is far from optimal, yet it remains an attractive customer. However, this attraction is despite the DoD's contract law and practice, not because of it. The mission of the DoD and the benefits, both real and intangible, that commercial AI firms derive from supporting national security continue to attract innovative companies to defence contract opportunities, but the contract law and procurement practice frustrate and limit the number of firms willing to forego alternative commercial contracts that are comparatively less costly, complicated, difficult, time consuming, and risky.

---

[787] *NSCAI Final Report* (n 4) 1.
[788] Ibid.
[789] See Gregory C Allen, 'Across Drones, AI, and Space, Commercial Tech is Flexing Military Muscle in Ukraine' *Center for Strategic & International Studies* (online, 13 May 2022) <https://www.csis.org/analysis/across-drones-ai-and-space-commercial-tech-flexing-military-muscle-ukraine>.

After decades of tinkering with acquisition reform, Congress has provided a permissive legal environment that the DoD can leverage to align contract formation with commercial AI firms' preferences and technological considerations. Despite the legal authority to deviate from outdated procurement practices, the DoD must make drastic changes to the status quo if it is to succeed in its strategy to compete in the AI era. This research provides insight that demonstrates the importance of reflection and systematically seeking perception data from industry as the findings contribute to understanding why commercial AI firms decide to contract with the DoD. Using social exchange theory as a framework to understand what contract attributes commercial AI firms find attractive can inform procurement officials' acquisition strategies and choice of law.

Lawyers play a critical role in leading that change. The technical nature and complexity of the contract requirements and agility required to negotiate bespoke terms and conditions that are not only attractive to commercial AI firms but protect the DoD's interest dictate a comprehensive understanding of the law, critical thought, and business judgment. The optimal buyer theory offers a roadmap that DoD lawyers can use to maximise the attractiveness of a contract opportunity as already authorised by DoD contract law. Understanding the commercial AI industry's preferences allows the DoD to identify best practices in contracting. This research shows that commercial AI firms are attracted to contract opportunities that are expedient, free from unnecessary regulations, collaborative from the start, and flexible to accommodate the



unique considerations of developing and deploying AI-enabled capabilities.  By setting clear and actionable ethical standards and leveraging its legal authority, the DoD can become an even more attractive customer to the commercial AI industry.  If the DoD leverages its contract law to align with the business and technological considerations of commercial AI firms, it can better attract commercial AI firms to gain timely access to the AI innovation the United States needs to compete in the AI era.

一代人工智能发展规划的通知 (20 July 2017) <http://www.gov.cn/zhengce/content/2017-07/20/content_5211996.htm>]

Weinberger, Sharon, *The Imagineers of War* (Alfred A Knopf, 2017)

Weinig, William J, 'Other Transaction Authority: Saint or Sinner for Defense Acquisition?' (2019) 26(2) *Defense Acquisition Research Journal* 107

Weiss, Mitchell, and A J Steinlage, *Shield AI* (Harvard Business Review Case Study 9-819-062, 3 October 2018)

West, Darrell M, and John R Allen, *Turning Point* (Brookings, 2020)

Whelan, J W, and E C Pearson, 'Underlying Values in Government Contracts' (1962) 10 *Public Law* 298

Yukins, Christopher R, 'A Versatile Prism: Assessing Procurement Law Through the Principal-Agent Model' (Fall 2010) 40(1) *Public Contract Law Journal* 63

Jan Zawadzki, 'Vertical vs. Horizontal AI Startups', *Towards Data Science* (online, 21 June 2020) <https://towardsdatascience.com/vertical-vs-horizontal-ai-startups-e2bdec23aa16>

Zhang, Daniel, et al, *The AI Index 2022 Annual Report* (AI Index Steering Committee, Stanford Institute for Human-Centered AI, March 2022)

## B *Cases*

*Blade Strategies, LLC* (Matter No B-416752, 24 September 2018)

*Cooke v United States*, 91 US 389 (1875)

*The Floyd Acceptances*, 74 US 666 (1868)

Oracle America, Inc., B-416657 et al (14 November 2018)

*Oracle America, Inc. v. United States*, 975 F 3d 1279, 1283 (Fed Cir, 2020)

*Palantir USG Inc v United States*, 904 F 3d 980 (Fed Cir, 2018)

## C *Legislation and Executive Materials*

10 USC § 111 (2018)

10 USC § 892 (2018)

10 USCA § 3014 (2022)



10 USCA § 3201 et seq (2022)

10 USCA § 3453 (2022)

10 USCA § 3458 (2022)

10 USCA § 4021 (2022)

10 USCA § 4022 (2022)

10 USCA § 4023 (2022)

10 USCA § 4811 (2022)

15 USC § 632 (2018)

15 USC § 3710a (2018)

15 USC § 3715 (2018)

31 USC § 6305 (2018)

32 CFR § 37 (2003)

35 USC Chapter 18 (2018)

35 USC § 202 (2018)

41 USC § 253 (2018)

41 USC § 1502 (2018)

48 CFR § 1 et seq (2022)

50 USC § 4511 (2018)

51 USC § 20113 (2018)

Committee on Armed Services, House of Representatives, *Report on H.R. 1735 Together with Dissenting Views* (House Report No 114-102, 5 May 2015)

Committee on Armed Services, House of Representatives, *Report on H.R. 4350 Together with Additional and Dissenting Views* (House Report No 117-118, 10 September 2021)



Committee on Armed Services, United States Senate, *Report to Accompany S. 2792* (Senate Report No 117-39, 21 September 2021)

*DFARS* §§ 201 et seq

Executive Office of the President, *Maintaining American Leadership in Artificial Intelligence*, 89 Fed Reg 3967 (Executive Order No 13859, 11 February 2019)

House of Representatives, *Conference Report to Accompany H.R. 1735* (House Report No 114-270, 29 September 2015)

*National Aeronautics and Space Act of 1958*, Pub L No 85-568, 72 Stat 426 (1957)

*National Defense Authorization Act for Fiscal Year 1994*, Pub L No 103-160, 107 Stat 1547 (1993)

*National Defense Authorization Act for Fiscal Year 2016*, Pub L No 114-92, 129 Stat 726 (2015)

*National Defense Authorization Act for Fiscal Year 2017*, Pub L No 114-328, 130 Stat 2000 (2016)

*National Defense Authorization Act for Fiscal Year 2018*, Pub L No 115-91, 131 Stat 1283 (2017)

*National Defense Authorization Act for Fiscal Year 2020*, Pub L No 116-92, 133 Stat 1290 (2019)

*National Defense Authorization Act for Fiscal Year 2023*, Pub L No 117-263, 136 Stat 2395 (2022)

*United States Constitution*

## D *Treaties*

*Protocol Additional to the Geneva Conventions of 12 August 1949, and Relating to the Protection of Victims of International Armed Conflicts*, opened for signature 8 June 1977, 1125 UNTS 3 (entered into force 7 December 1978)

## E *Other*

'About', *DAF-MIT AI Accelerator* (Web Page) <aia.mit.edu/about/>

'About', *U.S. Department of Defense* (Web Page) <https://www.defense.gov/About/>

'Advancing AI for Everyone', *Google* (Web Page) <https://ai.google>

'AI for Good', *Microsoft* (Web Page) <https://www.microsoft.com/en-us/ai/ai-for-good>

APPENDIX B: HUMAN RESEARCH ETHICS APPROVAL



APPENDIX C: SURVEY QUESTIONS





This appendix presents the aggregated survey responses to the survey questions corresponding to Appendix C.  Chapter III discusses the methodology for conducting the survey. 111 respondents completed the survey.  Questions 1 and 2 were screening questions that are not included below.  Question 8 and 9 asked for the firm location and number of employees, respectively, while question 13 asked for the type of funding each company has received, are provided in Chapter III, Section E.  Chapter IV provides the findings of selected survey responses in greater detail.

Questions 2 – 7 and 10 – 12 and 14 develop an understanding of the sample demographics and experience working with the DoD.

| No. | Question | Yes | No |
|-----|----------|-----|-----|
| 3 | Has your company ever competed for or been awarded a Federal Acquisition Regulations (FAR) contract with the Department of Defense? | 53.15 | 46.85 |
| 4 | Has your company ever competed for or been given a Small Business Innovation Research (SBIR) or Small Business Technology Transfer (STTR) award with the Department of Defense? | 59.46 | 40.54 |
| 5 | Has your company ever competed for or been awarded an other transaction agreement (OTA) with the Department of Defense? | 41.44 | 58.56 |
| 6 | Is your company a member of any other transaction consortia? | 34.23 | 65.77 |
| 7 | In the past 12 months, has your company performed a contract or subcontract for the Department of Defense that is subject to full coverage under the Cost Accounting Standards (CAS)? | 30.63 | 69.37 |
| 12 | Does your company have a separate division or employee(s) focused on government contracts? | 36.94 | 63.06 |
| 14 | Is your company owned or operated by a military veteran? | 19.82 | 80.18 |

Questions 10 and 11 ask for the relative amount of business a respondent does with the DoD compared to commercial customers.



**Question 10.** Approximately what percentage of contracts (number) does your company perform for the Department of Defense compared to commercial customers?

| Percentage | Responses |
|---|---|
| 0 | 23.42 |
| 10 | 26.13 |
| 20 | 8.11 |
| 30 | 9.01 |
| 40 | 3.60 |
| 50 | 4.50 |
| 60 | 2.70 |
| 70 | 3.60 |
| 80 | 3.60 |
| 90 | 9.91 |
| 100 | 5.41 |

**Question 11.** Approximately what percentage of contracts (value) does your company perform for the Department of Defense compared to commercial customers?

| Percentage | Responses |
|---|---|
| 0 | 22.52 |
| 10 | 15.32 |
| 20 | 12.61 |
| 30 | 10.81 |
| 40 | 2.70 |
| 50 | 5.41 |
| 60 | 6.31 |
| 70 | 1.80 |
| 80 | 3.60 |
| 90 | 13.51 |
| 100 | 5.41 |

Questions 15 and 16 ask for a general opinion of the DoD as a customer.

| No. | Question | Very positive (1) | Positive (2) | Neutral (3) | Negative (4) | Very negative (5) | Mean | Standard Deviation |
|---|---|---|---|---|---|---|---|---|
| 15 | How would you describe your overall opinion of the DoD as a (potential) customer? | 32.43 | 43.24 | 16.22 | 7.21 | 0.90 | 2.01 | 0.93 |



| No. | Question | Much better (1) | Better (2) | About the same (3) | Worse (4) | Much worse (5) | Mean | Standard Deviation |
|-----|----------|-----------------|------------|--------------------|-----------|-----------------|------|--------------------|
| 16 | How has your perception of the DoD as a (potential) customer changed in the past 12 months? | 10.81 | 24.32 | 55.86 | 8.11 | 0.90 | 2.54 | 0.81 |

Questions 17 – 26 ask about the relative importance of contract attributes. The responses are on a 5-point Likert scale ranging from 'Extremely Important', coded as '1', to 'Not at all Important, coded as '5'. The mean is calculated by averaging the sum of the coded responses and dividing over the number of respondents. The lower (closer to 1) the mean is, the more important the contract attribute is to the sample as a whole. The standard deviation represents the amount of dispersion about the mean; this measure indicates whether there is variation within the sample on a specific variable. The greater the standard deviation, the greater the difference of opinions within the sample.

| No. | Question | Extremely Important (1) | Very Important (2) | Somewhat Important (3) | Not so Important (4) | Not at all Important (5) | Mean | Standard Deviation |
|-----|----------|-------------------------|--------------------|------------------------|----------------------|--------------------------|------|--------------------|
| 17 | Ability to negotiate intellectual property rights | 57.27 | 25.45 | 13.64 | 3.64 | 0 | 1.64 | 0.85 |
| 18 | Ability to negotiate (non-intellectual property) terms and conditions | 19.82 | 44.14 | 30.63 | 4.5 | 0.9 | 2.23 | 0.85 |
| 19 | Timeline of a contract award and funding | 54.95 | 30.63 | 12.61 | 1.8 | 0 | 1.61 | 0.77 |
| 20 | The ability to collaborate with the customer about its problem/need and potential solutions before contract award | 65.77 | 30.63 | 3.6 | 0 | 0 | 1.38 | 0.55 |



| No. | Question | Extremely Important (1) | Very Important (2) | Somewhat Important (3) | Not so Important (4) | Not at all Important (5) | Mean | Standard Deviation |
|---|---|---|---|---|---|---|---|---|
| 21 | Ability to communicate with the customer throughout the competition and negotiations prior to contract award | 52.25 | 33.33 | 11.71 | 2.7 | 0 | 1.65 | 0.79 |
| 22 | The flexibility to iterate the product/solution during the performance of the contract | 45.95 | 37.84 | 16.22 | 0 | 0 | 1.7 | 0.73 |
| 23 | The ability to communicate and collaborate with the end-user of the product/ service during contract performance | 67.57 | 26.13 | 6.31 | 0 | 0 | 1.39 | 0.6 |
| 24 | The potential for commercial market success (vice government market success) of the product/solution | 37.84 | 27.93 | 20.72 | 10.81 | 2.7 | 2.13 | 1.12 |
| 25 | The contribution of the product/solution to national security | 40.54 | 32.43 | 19.82 | 7.21 | 0 | 1.94 | 0.94 |
| 26 | The consistency of business practices utilized to compete and perform a contract in the commercial and government markets | 18.92 | 37.84 | 27.03 | 16.22 | 0 | 2.41 | 0.97 |

Questions 27 – 69 ask the respondents about their agreement with statements given that relate to various contract, business, and technology concerns. The responses are on a 7-point Likert scale ranging from 'Strongly Agree', coded as '1', to 'Strongly Disagree', coded as '7'. The mean is calculated by averaging the sum of the coded responses and dividing over the number of respondents. The lower (closer to 1) the mean is, the more strongly the sample agrees with the statement. The standard deviation represents the amount of dispersion about the mean;



this measure indicates whether there is variation within the sample on a specific variable. The greater the standard deviation, the greater the difference of opinions within the sample.

| No. | Question | Strongly agree (1) | Agree (2) | Somewhat agree (3) | Neither agree nor disagree (4) | Somewhat disagree (5) | Disagree (6) | Strongly disagree (7) | Mean | Standard Deviation |
|---|---|---|---|---|---|---|---|---|---|---|
| 27 | The barriers to entry in the government defense marketplace are low and easily overcome. | 0 | 0.9 | 2.7 | 6.31 | 17.21 | 44.14 | 28.83 | 5.87 | 1.04 |
| 28 | The process of competing for and performing a contract with the Department of Defense is simple and straightforward. | 0 | 2.7 | 1.8 | 8.11 | 18.92 | 39.64 | 28.83 | 5.77 | 1.16 |
| 29 | Earning a contract with the Department of Defense is more important to my company than a commercial contract of the same size. | 12.61 | 12.61 | 15.32 | 27.03 | 7.21 | 16.22 | 9.01 | 3.88 | 1.82 |
| 30 | All else being equal, my company's priority is maximizing profit. | 14.41 | 20.72 | 21.62 | 11.71 | 15.32 | 9.91 | 6.31 | 3.48 | 1.79 |
| 31 | Earning a contract with the Department of Defense is a good path for my company to maximize profit. | 18.92 | 23.42 | 29.73 | 14.41 | 5.41 | 6.31 | 1.8 | 2.9 | 1.48 |
| 32 | All else being equal, performing a contract with the Department of Defense is more costly than a contract in the commercial marketplace. | 22.52 | 38.74 | 18.02 | 16.22 | 2.7 | 1.8 | 0 | 2.43 | 1.18 |



| No. | Question | Strongly agree (1) | Agree (2) | Somewhat agree (3) | Neither agree nor disagree (4) | Somewhat disagree (5) | Disagree (6) | Strongly disagree (7) | Mean | Standard Deviation |
|---|---|---|---|---|---|---|---|---|---|---|
| 33 | All else being equal, performing a contract with the Department of Defense is more likely than a contract in the commercial marketplace to provide long term success for the company. | 7.21 | 21.62 | 22.52 | 23.42 | 11.71 | 9.91 | 3.6 | 3.55 | 1.54 |
| 34 | Performing a contract with the Department of Defense is more likely than a contract in the commercial marketplace to meet approval in the community. | 6.31 | 15.32 | 7.21 | 44.14 | 13.51 | 9.01 | 4.5 | 3.88 | 1.46 |
| 35 | Performing a contract with the Department of Defense is more likely than a contract in the commercial marketplace to provide our company with the most autonomy. | 0.9 | 7.21 | 8.11 | 36.94 | 25.23 | 18.02 | 3.6 | 4.47 | 1.24 |
| 36 | Performing a contract with the Department of Defense is more likely than a contract in the commercial marketplace to have a predictable outcome. | 2.7 | 18.92 | 21.62 | 27.93 | 17.12 | 7.21 | 4.5 | 3.77 | 1.42 |
| 37 | Our commercial customers are more trustworthy than the Department of Defense. | 0.91 | 6.36 | 8.18 | 47.27 | 12.73 | 19.09 | 5.45 | 4.44 | 1.27 |



| No. | Question | Strongly agree (1) | Agree (2) | Somewhat agree (3) | Neither agree nor disagree (4) | Somewhat disagree (5) | Disagree (6) | Strongly disagree (7) | Mean | Standard Deviation |
|---|---|---|---|---|---|---|---|---|---|---|
| 38 | The culture (beliefs, behavior, values) of the Department of Defense is consistent with the culture of my company. | 24.32 | 27.03 | 20.72 | 18.02 | 5.41 | 1.8 | 2.7 | 2.69 | 1.46 |
| 39 | Ethics in developing and leveraging artificial intelligence capabilities is important to my company. | 57.66 | 28.83 | 9.01 | 4.5 | 0 | 0 | 0 | 1.6 | 0.83 |
| 40 | Our company is concerned about the ethical use of our product/service by our customers. | 41.44 | 28.83 | 12.61 | 7.21 | 3.6 | 4.5 | 1.8 | 2.23 | 1.51 |
| 41 | Our company trusts the Department of Defense to use artificial intelligence ethically. | 26.36 | 38.18 | 11.82 | 15.45 | 6.36 | 0.91 | 0.91 | 2.44 | 1.32 |
| 42 | My company is comfortable that the Department of Defense may use our product/service for lethal purposes. | 25.45 | 28.18 | 10.91 | 24.55 | 5.45 | 3.64 | 1.82 | 2.75 | 1.52 |
| 43 | My company is concerned that contracting with the Department of Defense may negatively impact our business. | 2.7 | 3.6 | 8.11 | 13.51 | 9.01 | 39.64 | 23.42 | 5.35 | 1.56 |
| 44 | Access to large databases is critical to the success of our product/service. | 23.64 | 23.64 | 17.27 | 11.82 | 9.09 | 6.36 | 8.18 | 3.11 | 1.88 |



| No. | Question | Strongly agree (1) | Agree (2) | Somewhat agree (3) | Neither agree nor disagree (4) | Somewhat disagree (5) | Disagree (6) | Strongly disagree (7) | Mean | Standard Deviation |
|---|---|---|---|---|---|---|---|---|---|---|
| 45 | Access to large databases in the Department of Defense is a significant consideration in deciding to work with the military. | 9.09 | 14.55 | 20 | 24.55 | 6.36 | 12.73 | 12.73 | 3.94 | 1.82 |
| 46 | Our company is willing to adapt our business model to conform to the Department of Defense's procurement process and regulations. | 18.02 | 36.04 | 29.73 | 4.5 | 9.01 | 2.7 | 0 | 2.59 | 1.25 |
| 47 | Our company is willing and capable to comply with the Department of Defense's cost accounting standards. | 21.82 | 40 | 14.55 | 10 | 7.27 | 2.73 | 3.64 | 2.64 | 1.54 |
| 48 | Our company's image will be negatively affected by performing a contract with the Department of Defense. | 0.9 | 2.7 | 5.41 | 14.41 | 8.11 | 38.74 | 29.73 | 5.61 | 1.4 |
| 49 | Competing for a contract with the Department of Defense is more time consuming and difficult than it is with a commercial customer. | 42.34 | 32.43 | 13.51 | 8.11 | 0 | 2.7 | 0.9 | 2.03 | 1.24 |
| 50 | Performing a contract with the Department of Defense is more difficult than it is with a commercial customer. | 23.42 | 23.42 | 18.92 | 26.13 | 4.5 | 3.6 | 0 | 2.76 | 1.37 |



| No. | Question | Strongly agree (1) | Agree (2) | Somewhat agree (3) | Neither agree nor disagree (4) | Somewhat disagree (5) | Disagree (6) | Strongly disagree (7) | Mean | Standard Deviation |
|---|---|---|---|---|---|---|---|---|---|---|
| 51 | My company is comfortable working with a prime defense contractor. | 26.13 | 40.54 | 18.92 | 6.31 | 6.31 | 1.8 | 0 | 2.32 | 1.21 |
| 52 | Our company prefers to submit proposed solutions in writing over direct, face to face interactions, such as pitches. | 2.7 | 6.31 | 8.11 | 28.83 | 18.02 | 24.32 | 11.71 | 4.73 | 1.5 |
| 53 | My company is willing and able to write proposals in response to a FAR Part 15 Request for Proposals. | 22.52 | 35.14 | 11.71 | 18.92 | 8.11 | 2.7 | 0.9 | 2.67 | 1.43 |
| 54 | My company regularly reviews the Government Point of Entry (FedBizOps/ sam.gov) for opportunities to work with the Department of Defense. | 27.03 | 23.42 | 11.71 | 4.5 | 10.81 | 15.32 | 7.21 | 3.23 | 2.07 |
| 55 | When choosing a new project, my company prefers the opportunity to come up with solutions to problems rather than follow pre-set specifications written by the customer. | 34.55 | 27.27 | 16.36 | 16.36 | 2.73 | 2.73 | 0 | 2.34 | 1.32 |
| 56 | Fixed requirements and milestones are preferable to agile and iterative steps when developing and deploying our product/service. | 0.9 | 6.31 | 3.6 | 19.82 | 21.62 | 23.42 | 24.32 | 5.23 | 1.49 |



| No. | Question | Strongly agree (1) | Agree (2) | Somewhat agree (3) | Neither agree nor disagree (4) | Somewhat disagree (5) | Disagree (6) | Strongly disagree (7) | Mean | Standard Deviation |
|---|---|---|---|---|---|---|---|---|---|---|
| 57 | Starting with a prototype, pilot, minimum viable product, or proof of concept is an essential step before integrating or scaling any new use case. | 33.33 | 36.04 | 15.32 | 6.31 | 4.5 | 4.5 | 0 | 2.26 | 1.34 |
| 58 | It is the commercial industry's responsibility to engage with and support the Department of Defense's national security mission. | 17.12 | 23.42 | 18.92 | 27.03 | 9.01 | 1.8 | 2.7 | 3.04 | 1.46 |
| 59 | My company's business strategy can be the same for both the commercial and defense marketplaces. | 3.6 | 17.12 | 27.93 | 9.01 | 17.12 | 17.12 | 8.11 | 4.03 | 1.7 |
| 60 | Working exclusively on DoD contracts is a viable strategy for business growth. | 7.21 | 12.61 | 11.71 | 10.81 | 18.02 | 26.13 | 13.51 | 4.52 | 1.85 |
| 61 | My company prefers the commercial contracting process to the defense contracting process. | 18.02 | 28.83 | 23.42 | 21.62 | 3.6 | 3.6 | 0.9 | 2.78 | 1.34 |
| 62 | The Department of Defense is an attractive customer to my company. | 46.85 | 32.43 | 11.71 | 3.6 | 3.6 | 1.8 | 0 | 1.9 | 1.15 |



| No. | Question | Strongly agree (1) | Agree (2) | Somewhat agree (3) | Neither agree nor disagree (4) | Somewhat disagree (5) | Disagree (6) | Strongly disagree (7) | Mean | Standard Deviation |
|---|---|---|---|---|---|---|---|---|---|---|
| 63 | Foreign investment and sales are critical to my company's success. | 7.21 | 19.82 | 9.01 | 22.52 | 9.01 | 19.82 | 12.61 | 4.16 | 1.88 |
| 64 | If given the choice, my company would prefer a Federal Acquisition Regulations (FAR) contract over an Other Transaction Agreement (OTA). | 4.55 | 5.45 | 3.64 | 47.27 | 8.18 | 9.09 | 21.82 | 4.64 | 1.63 |
| 65 | The Department of Defense's contracting process is transparent. | 0.9 | 6.31 | 19.82 | 14.41 | 18.92 | 22.52 | 17.12 | 4.8 | 1.59 |
| 66 | The Department of Defense uses easy to understand terms and conditions. | 0 | 7.21 | 9.01 | 12.61 | 26.13 | 23.42 | 21.62 | 5.14 | 1.49 |
| 67 | The Department of Defense is more fair and timely in resolving disputes and issues than in the commercial marketplace. | 0 | 5.5 | 4.59 | 49.54 | 11.01 | 18.35 | 11.01 | 4.65 | 1.29 |
| 68 | If the contract terms and process were identical, the military would be a more attractive customer than a commercial buyer. | 10.91 | 27.27 | 15.45 | 32.73 | 10 | 3.64 | 0 | 3.15 | 1.32 |
| 69 | The commercial marketplace should model its contracting practices on the Department of Defense's procurement system. | 0 | 0.9 | 4.5 | 14.41 | 6.31 | 27.03 | 46.85 | 5.95 | 1.29 |



Questions 70 – 72 ask the respondents to select any answer choices from a selection or to write in responses if applicable considerations were not provided as a choice. The responses indicate the percentage of respondents that selected the answer choices, and many respondents selected more than one answer choice. These three questions were optional, so not all 111 respondents answered each of these questions. Thus, the percentages provided below reflect the percentage of respondents that selected a given answer out of the total respondents that answered that question (provided with each question).

**Question 70.** The following factors influence my company's decision to not engage with the Department of Defense. (104 answered)

| Answer Choice | Percentage |
|---|---|
| Lengthy process | 68.27 |
| Complexity of process | 59.62 |
| Lack of communication | 43.27 |
| Rigid process | 41.35 |
| Lack of security/facility clearance | 41.35 |
| Complexity of contract | 40.38 |
| Intellectual property concerns | 34.62 |
| Limited ability to negotiate | 31.73 |
| Uncertain budget | 31.73 |
| Inexperienced government workforce | 30.77 |
| Too difficult | 29.81 |
| Too costly | 23.08 |
| Export controls | 14.42 |
| Public perception | 11.54 |
| Mandatory non-business clauses (ie, drug-free workplace) | 8.65 |
| Ethical concerns | 7.69 |
| None of these factors | 0 |

The write-in responses are provided below (verbatim):

> *RFPs come out with extremely short turnaround times. If not embedded with customers to 'bake' contract, you feel disadvantaged. It is dominated by traditional primes who know the game inside and out which is frustrating for a primarily commercial company to translate and forces them to Innovation hubs (AFWERX, SOFWERX etc.) which is a bite sized approach and seems intended for those with concepts and not commercially dominating products.*

> *Difficult to find the right point of contacts*



*We actively engage the DOD, but there are barriers to entry. Hiring qualified people with clearances is a challenge. As a product company, we still need to provide services. The services are to support the product and prepare the data for results. We require collaboration with the teams trying to implement an AI/ML project. It's imperative to work with the data owners to understand use cases and define the problem to be solved. We see this as a mandatory element.*

*Large system integrator involvement with government agencies acts as a barrier to entry by blocking access to decision makers through their well-established relationships.*

*Hard to navigate opportunities.*

*The Fed will release RFPs simply to learn and there is no resulting award. So vendor teams spend time and effort (which has a direct cost) only to have used a complex process to educate government buyers and decision-makers. Scoring of vendor solutions and innovation of solutions can be handled very differently upstream from any RFP / RFI / BAA process. Further, the Prime at play has much to do with this aspect. Some Primes are nearly usury in their negotiations and practices.*

*We are engaged 100% as a company with the federal government through direct relationships or as subcontractors to major federal system integrators*

*Unclear what the DoD's need is.*

*Inability to talk to the customer ahead of time. Talking to the customer in a way where they can be open, honest and transparent is the most crucial item to ensure a contract works for both sides*

*While we would meet all standards, receiving all certifications is not an appropriate investment of time for the company.*

*We do engage with DOD, but the following reflects a pain point: The proposal process is too opaque, often against unclear requirements. Those writing solicitations should go to a requirements course*

*The requirement to be FedRAMP ATO prior to being awarded a contract.*

*Performing our SBIR had a material impact on our ability to hire. We lost 12 candidates when they were informed of the SBIR, their opinion was that there was no room for ethical deployment of AI technology to the DoD under any circumstance. Our SBIR also had a negative impact on our ability to raise funding from traditional venture investors. There is a disturbing trend in the San*



*Francisco/Bay Area AI ecosystem whereby many believe any work with the US DoD is 'unethical'. Our startup shut down in December 2019.*

*It is far easier, fun, and fulfilling for us to build crazy smart AI/ML algorithms and platforms than to bid on a government contract.*

*Our #1 issue in working on defense related projects has been that our employees have not wanted to be a part of it. Not only have multiple employees expressed that they would not want to be placed on a defense project, but also, they expressed concern that our business worked on them at all. Comments often said something like 'I want a strong military, but not with a president who is bragging about his nuclear arms button on twitter.'*

*Overall, working with the DOD has been pleasurable and a privilege. The biggest pain points: Unprofessionalism of contract managers. In some cases they don't even notify companies that they haven't been selected for award. In fact, in some cases they go completely radio silent on an entire cohort of SBIR Phase I awardees with regard to their Phase II proposals, and don't respond literally for months at a time. This is unacceptable. Also, changing success criteria in between cycles is unfair. Companies will be awarded a Phase I with a stated Phase II selection criteria, but during the time period of the Phase I contract, (the Air Force in this case) then changes the selection criteria and maximum budget amount for Phase II on a whim, in once example dropping the Phase II allowable budget from $750k to $500k. Overall, it is still worthwhile to work with the DOD, but it literally can force a company to go out of business. It can help companies succeed in the long run, but in the short run it may break them. Last thing, the concept of dual-use companies is flawed. It assumes that companies have already commercialized a technology, and now they will just repurpose it for the DOD. This isn't reality for most technology startups. Most tech startups are founded by people like graduate students, who discover or develop a fundamental technology. They will develop it for all uses in parallel, not spend 3-5 years to develop it for commercial markets and then sell it to the DOD. This concept was probably developed by people who have never actually built a company. If we really want to get technologies to the warfighter faster, and avoid tech loss to foreign countries, we need to acknowledge this reality.*

*Lead generation sucks for us. We have basically given up on SBIRs and BAAs where we don't know the customer already. The system has obviously been optimized for defense contractors. Moreover, its a lot more work for monies that are not as favorably viewed by investors who typically do not understand the defense market nor do they like the fact that we have to constantly replace revenue. In the commercial sector, the transactions are a lot easier (typically secured over a meeting and a few phone calls) and we get MRR which we don't have to replace next year. Plus the additional cost is against commercial processes that we are constantly refining. What I mean is that we cannot be really good at both and certainly could not be good in both markets in early stages. The two markets are so*



*vastly different that we have to pick one to focus on and almost ignore entirely the
other.*

**Question 71.** The following factors, if required as part of a contract, would prevent my company
from working with the Department of Defense. (100 answered)

| Answer Choice | Percentage |
|---|---|
| Unlimited rights in technical data/computer software | 55.00 |
| More than six months after submitting proposal to receive funding | 54.00 |
| Security/facility clearance | 36.00 |
| Government purpose rights in technical data/computer software | 28.00 |
| Mandatory (non-negotiable) contract clauses | 26.00 |
| Cost Accounting Standards | 20.00 |
| Export controls | 11.00 |
| Mandatory non-business clauses (ie, drug-free workplace) | 10.00 |
| Subject to auditing | 9.00 |
| Potential use on weapon system | 9.00 |

The write-in responses are provided below (verbatim):

*Government not providing facility access for projects*

*Getting an approved purchasing system requires a contract requiring an approved
purchasing system which results in a chicken or egg problem.*

*The company does not want to be dictated by a single customer. Our intention is to
grow a federal practice that matches all government practices including cost
accounting, security issues and program specific clauses. We currently engage in
long term proposal efforts and supplement the business with a team focused on
tactical selling activities. Our goal is to become a prime for AI/ML programs where
we could control the outcomes.*

*All personnel would need clearances. All work would have to take place in a
SCIF*

*Biggest issue is timing, lack of clearance. Small startups like ours don't have the
time to wait 6 months to see if they won a contract AFTER it took them 3 months
to write one. We are trying to do all business at <$250k so organizations can
sole-source our services. We are new to the DoD space.*



**Question 72.** The following factors influence my company's decision to engage with the Department of Defense. (107 answered)

| Answer Choice | Percentage |
|---|---|
| Ability to contribute to national security | 71.03 |
| Ability to work on challenging projects | 71.03 |
| Patriotism | 70.09 |
| Work on challenging issues | 66.36 |
| Funding | 58.88 |
| Profitability | 57.01 |
| Credibility of the customer | 56.07 |
| Budget certainty | 45.79 |
| Competitive advantage over other businesses | 44.86 |
| Gain experience | 44.86 |
| Access to data | 31.78 |
| Comfort with contracting process | 28.97 |
| Public perception | 26.17 |
| Favorable intellectual property terms | 25.23 |
| Ethical reasons | 24.30 |
| Access to secondary markets | 24.30 |
| Experienced government workforce | 20.56 |
| None of these factors | 0 |

The write-in responses are provided below (verbatim):

*The DoD, although sometimes cumbersome to work with, generally provides a more stable and long-term business relationship.*

*We want to provide the best-in-class technology to the warfighter.*

*Though much slower than a commercial customer, they tend to be stable and not influenced as strongly by external economics. Commercial opportunities follow economic trends*

*Who would not want to build amazing solutions to make our country better?*